\documentclass[twoside]{cernyrep}

\usepackage[utf8]{inputenc}
\DeclareUnicodeCharacter{2212}{-}

\usepackage{graphicx}
\usepackage{caption}
\usepackage{mdframed}
\usepackage{tikz}

\usepackage{epstopdf}

\usepackage{subfloat}
\usepackage{acronym}
\usepackage{subfig}
\usepackage{amsmath}
\usepackage{amsfonts}
\usepackage{color}
\usepackage{balance}
\usepackage{authblk}
\usepackage{fancyhdr}
\usepackage{multirow}
\usepackage{afterpage}
\usepackage{bbm}
\usepackage{tikz}

\usepackage{feynmf}

\usepackage[hang,flushmargin]{footmisc}
\fancypagestyle{plain}{}
\usepackage{emptypage}
\usepackage{lipsum}

\usepackage{texnames}
\usepackage{ctable}
\usepackage[T1]{fontenc}

\usepackage{blindtext}

\usepackage[bookmarks,colorlinks=true,linktoc=page,pdftex,linkcolor=blue,citecolor=blue,urlcolor=blue]{hyperref}
\usepackage{float}
\usepackage{xcolor}
\usepackage[toc,page]{appendix}
\usepackage[bookmarks,colorlinks=true,pdftex,linkcolor=blue,citecolor=blue,urlcolor=blue]{hyperref}

\usepackage[onehalfspacing]{setspace}
\newcommand{\chapabstract}[1]{
    \begin{quote}
        \singlespacing\small
        \rule{14cm}{1pt}
        #1
        \vskip-4mm
        \rule{14cm}{1pt}
\end{quote}}

\newcounter{savefootnote}
\newcounter{symfootnote}
\newcommand{\symfootnote}[1]{%
   \setcounter{savefootnote}{\value{footnote}}%
   \setcounter{footnote}{\value{symfootnote}}%
   \ifnum\value{footnote}>8\setcounter{footnote}{0}\fi%
   \let\oldthefootnote=\thefootnote%
   \renewcommand{\thefootnote}{\fnsymbol{footnote}}%
   \footnote{#1}%
   \let\thefootnote=\oldthefootnote%
   \setcounter{symfootnote}{\value{footnote}}%
   \setcounter{footnote}{\value{savefootnote}}%
}

\newcounter{box}
\newcommand*{\greybox}{\textbf{Box \arabic{box}. \addtocounter{box}{1}}}

\numberwithin{equation}{section}

\makeatletter
\newcommand{\thickhline}{%
    \noalign {\ifnum 0=`}\fi \hrule height 2pt
    \futurelet \reserved@a \@xhline
}
\newcolumntype{"}{@{\hskip\tabcolsep\vrule width 2pt\hskip\tabcolsep}}
\makeatother

\begin{document}


\chapter*{Lectures on Field Theory and the Standard Model: A Symmetry-Oriented 
Approach\symfootnote{To appear in 
Proceedings of the 2022 European School for High Energy Physics (Jerusalem, Israel, November 30th to
December 14th, 2022), 
eds. Markus Elsing and Alex Huss.}}

\vspace{-11mm}


\begin{flushleft}
\end{flushleft}


\label{sec:Alvarez}


\noindent
{\it Luis \'Alvarez-Gaum\'e$\,^{a}$ and Miguel~\'A. V\'azquez-Mozo$\,^{b}$\\[0.2cm]}
{$^{a}$ Simons Center for Geometry and Physics,
State University of New York Stony Brook,
NY-11794-3636, USA \& Theory Department CERN,
CH-1211 Geneva 23, Switzerland. \\ 
E-mail: \href{mailto:lalvarezgaume@scgp.stonybrook.edu}{lalvarezgaume@scgp.stonybrook.edu},
\href{mailto:Luis.Alvarez-Gaume@cern.ch}{Luis.Alvarez-Gaume@cern.ch}\\
$^{b}$ Departamento de F\'{\i}sica Fundamental, Universidad de Salamanca, Plaza de la Merced s/n, E-37008 Salamanca, Spain. \\
E-mail: \href{mailto:Miguel.Vazquez-Mozo@cern.ch}{Miguel.Vazquez-Mozo@cern.ch}}

\vspace{-7mm}

\chapabstract{
The standard model of particle physics represents the cornerstone of our understanding of the 
microscopic world. In these lectures we review its contents and structure, with a particular 
emphasis on the central role played by symmetries and their realization. This is 
not intended to be an exhaustive review but a 
discussion of selected topics that we find interesting, with the specific 
aim of clarifying some subtle points and potential misunderstandings. 
A number of more technical topics are discussed in separated boxes interspersed throughout the text.
}


\tableofcontents

\begin{fmffile}{diagrams}
\fmfcmd{%
style_def majorana expr p =
cdraw p;
cfill (harrow (reverse p, .5));
cfill (harrow (p, .5))
enddef;
style_def alt_majorana expr p =
cdraw p;
cfill (tarrow (reverse p, .55));
cfill (tarrow (p, .55))
enddef;}

\noindent

\section{Preliminaries}
\label{sec:preliminaries}

Quantum field theory (QFT) is the language in which we codify our knowledge about the fundamental laws of nature
in a manner compatible with quantum mechanics, relativity, and locality. Its 
most significant achievement has been formulating 
the standard model (SM) of strong, weak, and electromagnetic interactions. This theory summarizes
what we know about the physics of the fundamental constituents of matter. It also delineates
our ignorance, providing a glimpse 
of the known unknowns that will motivate future research.   
The story of QFT and the SM has been told many times
with various degrees of detail and depth (see~\cite{B_D,I_Z,Ramond,
Peskin,Weinberg,DeWitt,Maggiore,Nair,B_M,Paschos,C_G,Zee,Banks,AG_VM,Schwartz,
Kane,Goldberg,Raby} for a necessarily incomplete
sample of books on both topics). In the pages reserved for these lecture notes,
it is utterly impossible to 
provide a detailed account of the towering achievements accumulated since the discovery of
the electron by J.~J.~Thomson in 1897, whose most recent milestone was the announcement in 2012
of the discovery of the Higgs boson at CERN. 
Generations of physicists and engineers have made possible
the formulation of a theory describing the most fundamental laws of nature known so far. 

High energy physics is not the only arena in which QFT has shown its powers. In the
nonrelativistic regime, it leads to quantum many body theory, a mathematical framework
used in condensed matter physics to study phenomena such as superconductivity, 
superfluidity, and metals' thermal and electronic properties~\cite{AGD,Fetter_Walecka,Bruus_Frensberg}. 
Furthermore, in the last few decades QFT has also played a central role in understanding the formation
of the large scale structure in the universe~\cite{Burgess_LH,Baldauf_LH,snowmass_EFTC}.

Exciting as all these developments are, 
these lectures will focus on the applications of QFT to particle physics,
and particularly the construction of the SM. We will highlight symmetry arguments to show how virtually all 
known forms of symmetry realizations play a role in it. But even within this restricted scope,
space limitations require choosing not just the material to include 
but also the viewpoint to adopt. In explaining some of the ideas and 
techniques in 
our study of the SM, it is useful to focus on several 
key concepts, many of which are related to implementing symmetries
in a quantum system with infinite degrees of freedom. In doing so, we will encounter 
many surprises and some misconceptions to be clarified.  
Explaining physics can be compared to the performance of a well-known piece of music. 
Often the performer surprises the audience by accentuating some features of the work 
that only then are sufficiently 
appreciated. In such a vein, we will highlight some important fundamental aspects 
of the SM the reader may not have encountered previously, some of which also point to the limitations
of the theory. Although we will now shy away from diving into calculations when needed,
our aim here is less giving a detailed account of the technicalities involved than
providing the reader with both essential conceptual tools and inspiration to further deepen in the study 
of the topics to be presented. 

Having set our plan of action, we turn to physics and begin by reviewing the system of units to 
be used throughout the lectures. Since we are dealing with quantum relativistic systems, it
is natural to work with natural units where the speed of light and the Planck constant are
both set to one,~$c=\hbar=1$. Doing a bit of dimensional analysis, it is easy to see that  
setting these two 
fundamental constants to one means that of the three fundamental
dimensions~$L$ (length),~$T$ (time), and~$M$ (mass) only 
one is independent. 
Indeed, from~$[c]=LT^{-1}$ and~$[\hbar]=ML^{2}T^{-1}$
it follows that~$T=L$ and~$M=L^{-1}$, meaning that time have dimensions of length and masses
of~$(\mbox{length})^{-1}$. Alternatively, we may prefer to use energy~($E$)
as the fundamental dimension, as we will actually do in the following.
In this case, from~$[\mbox{energy}]=ML^{2}T^{-2}$ we see that both lengths and times have dimensions  
of~$(\mbox{energy})^{-1}$, while masses are measured in units of energy.
 
Using natural units simplifies expressions by eliminating factor 
of~$\hbar$ and~$c$ and brings other advantages. The most relevant for us is that it
provides a simple classification of the operators, or terms, appearing in the action or 
Hamiltonian defining a theory. As an example, let us consider the scalar field action
\begin{align}
S=\int d^{4}x\left({1\over 2}\partial_{\mu}\phi\,\partial^{\mu}\phi-{m^{2}\over 2}\phi^{2}
-{\lambda_{4}\over 4!}\phi^{4}-{\lambda_{6}\over 6!}\phi^{6}\right).
\label{eq:action_scalar_example}
\end{align}
Action is measured in the same units as~$\hbar$ (not by chance historically 
known as the quantum of action) and is therefore dimensionless in natural units. 
Taking into account that~$[d^{4}x]=E^{-4}$ and~$[\partial_{\mu}]=E$, we find from the 
kinetic term that~$[\phi]=E$, which in turn confirms that~$[m]=E$ as behooving a mass.
As for the coupling constants,~$\lambda_{4}$ is dimensionless 
while~$[\lambda_{6}]=E^{-2}$. 

Terms such as~$\phi^{6}$, whose coupling constants have negative energy dimension, 
are called higher-dimensional operators. 
In the modern (Wilsonian) view of QFT to be discussed in section~\ref{sec:renormalization}, they are seen as induced 
by physical processes 
above some energy scale~$\Lambda$, much higher than the energy at which we want 
to describe the physics using the corresponding action. 
The presence of higher-dimensional operators 
in the action signals
that we are dealing with a theory that is not fundamental, but some effective description 
valid at energies~$E\ll\Lambda$, that should be  
eventually replaced (completed) by some more fundamental theory at higher energies. 

\label{page:EFTs}
Although the action of an effective field theory (EFT) may contain an infinite number of 
higher-dimensional operators
of arbitrary high dimension, this does not make it any less 
predictive at low energies~\cite{Pich_EFT,Kaplan_EFT}. To understand this, let us look at 
a higher-dimensional operator~$\mathcal{O}_{n}$, with $[\mathcal{O}_{n}]=E^{n-4}$ for~$n>4$, 
entering in the action as 
\begin{align}
S\supset {g_{n}\over \Lambda^{n-4}}\int d^{4}x\,\mathcal{O}_{n},
\end{align}
where~$g_{n}$ is a dimensionless coupling. The corrections induced by this term to
processes occurring at energy~$E$ scales as~$(E/\Lambda)^{n-4}$, so for~$E\ll\Lambda$ there is
a clear hierarchy among the infinite set of higher-dimensional operators. The upshot is that using our EFT
to ask physical questions at sufficiently low energies, and taking into account 
the limited sensitivity of
our detectors, only a small number of 
higher-dimensional operators have to be considered in the 
computation of physical observables.

Applying the philosophy of EFT to the action~\eqref{eq:action_scalar_example} leads to identify 
the theory as an effective description valid at energies well below the scale set by~$\lambda_{6}$,
namely~$\Lambda\sim 1/\sqrt{\lambda_{6}}$. Nature
offers more interesting implementations of this scheme, some of which we will encounter later on
in the context of the SM. A 
particularly relevant case is that of general relativity (GR), that we discuss now
in some detail. We start with the Einstein-Hilbert 
action
\begin{align}
S={1\over 16\pi G_{N}}\int d^{4}x\,\sqrt{-g}R,
\label{eq:EH_action}
\end{align}
and consider fluctuations around the Minkowski metric (nonflat background metrics can also be 
used)
\begin{align}
g_{\mu\nu}=\eta_{\mu\nu}+2\kappa h_{\mu\nu},
\label{eq:metric_flucts}
\end{align}
where
\begin{align}
\kappa\equiv\sqrt{8\pi G_{N}}.
\label{eq:kappa_def}
\end{align}
Inserting~\eqref{eq:metric_flucts} into~\eqref{eq:EH_action} and expanding in powers of~$h_{\mu\nu}$
we get an action defining 
a theory of interacting gravitons propagating on flat spacetime~\cite{FLG,EA,Hamber}. 
Its interaction part contains an infinite number of terms with the structure
\begin{align}
S_{\rm int}&=\sum_{n=3}^{\infty}\kappa^{n-2}\int d^{4}x\,\mathcal{O}_{n+2}[h,\partial],
\label{eq:grav_intS}
\end{align}
where the operator $\mathcal{O}_{n+2}[h,\partial]$, which has energy dimension~$n+2$, contains~$n$ 
graviton fields and two derivatives, while
from eq.~\eqref{eq:kappa_def} we see that the coupling constant
has dimensions~$[\kappa]=E^{-1}$. In the spirit of EFT, this indicates 
that Einstein's gravity is not 
fundamental, but an effective description valid at energies below its natural energy 
scale set by the dimensionful gravitational constant, the so-called Planck scale
\begin{align}
\Lambda_{\rm Pl}\equiv \sqrt{\hbar c^{5}\over 8\pi G_{N}}=2.4\times 10^{18}\mbox{ GeV},
\label{eq:Planck_scale}
\end{align}
where we have restored powers of~$\hbar$ and~$c$. To get an idea of the size of this scale, 
let us just say it is about $10^{14}$ times the center-of-mass energy at which LHC currently operates.

The statement is occasionally encountered in the literature and the media 
that GR is impossible to quantize. This 
needs to be qualified. The effective action~\eqref{eq:grav_intS} can be consistently quantized provided
we restrict our physical questions to the range of energies where it can be 
used, namely~$E\ll\Lambda_{\rm Pl}$. In this regime,
the quantum fluctuations of the background metric shown 
in~\eqref{eq:metric_flucts} are of order~$E/\Lambda_{\rm Pl}$ 
and, therefore, small. Furthermore, powers of this same quantity suppress the induced corrections
and again, 
at the level of accuracy set by our experiments, only a small number of operators 
in~\eqref{eq:grav_intS} need to be retained to compute physical observables. In other words,
below the Planck energy scale quantum gravity is just a theory of weakly coupled gravitons
propagating on a regular background spacetime.

This state of affairs breaks down when the energy gets close to~$\Lambda_{\rm Pl}$. At this point
the quantum fluctuations of the geometry become large, and the hierarchy of terms in~\eqref{eq:grav_intS}
breaks down. Physically, what happens is that our gravitons become strongly coupled and therefore
cease to be the appropriate degrees of freedom to describe a quantum theory of gravity. 
Thus, the correct statement is not that there is no consistent theory of quantum 
gravity, but that we lack one {\em which remains valid at arbitrarily high energies}. 
The difference is crucial, since it is precisely the latter kind of theory needed
to analyze, for example, what happens close to spacetime singularities, where quantum effects 
are so large as to override the semiclassical description provided by GR. 
Viewed as an EFT, Einstein's (quantum) gravity 
is expected to be subsumed near~$\Lambda_{\rm Pl}$ into another theory, 
its ultraviolet (UV) completion,
which presumably remains valid to arbitrarily high energies. Among the particle physics community
string theory continues to be the favored candidate for such a 
framework (see for instance~\cite{I_U,Kiritsis} for a modern account).

The previous digression on EFTs leads us to the related issue of renormalizability, on which we will
further elaborate in section~\ref{sec:renormalization}.
All QFTs used in describing elementary particles, particularly the SM, 
lead to infinities when computing 
quantum corrections (terms of order~$\hbar$ or higher) to classical results. 
The origin of these divergences lies in the behavior of the theory at very high energies.
Quantum fluctuations of very short wavelength actually dominate the result, driving them 
to infinity. This problem was tackled 
already in the 1940s by the procedure of renormalization. To make a long story short, 
one begins by regularizing the
theory by setting a maximum energy~$\Lambda$, a cutoff, so fluctuations with wavelength smaller
than~$\Lambda^{-1}$ are ignored. This makes all results finite, albeit dependent on the otherwise
arbitrary cutoff. The key observation now is that 
the parameters in the action (field normalizations, masses, and coupling constants) 
can depend on~$\Lambda$, so physical observables are cutoff 
independent. For this to work, a further ingredient is needed: an operational definition
of masses and couplings, which serves to fix the dependence of the action parameters on the cutoff 
(for all the details see, for example, chapter~$8$ of ref.~\cite{AG_VM} or any other of the
QFT textbooks listed in the references).

In carrying out this program, two thing may happen. One is that divergences
can be removed with a finite number of operators in the action
(most frequently, just those already present in the classical
theory). This is the case of a renormalizable theory. The second situation arises when 
it is necessary to add an infinite number of 
new operators in order to absorb all the divergences in their corresponding couplings. 
The theory is then said to be nonrenormalizable. The SM belongs to the first type, while 
GR is an example of the second. As a rule of thumb, actions containing operators of dimension
equal or smaller than four define renormalizable theories, while the presence of higher-dimensional
operators renders the theory nonrenormalizable, at least when working in perturbation theory.

For decades, renormalizability was considered necessary for any decent theory of elementary particles.
The very formulation of the SM and, most particularly, its implementation of the 
Brout-Englert-Higgs (BEH) mechanism~\cite{Brout_Englert,Higgs1,Higgs2} through the Higgs boson was guided
by making the theory renormalizable. As a token of how important this requirement was
perceived to be at the time, let us mention that the electroweak sector of the SM developed 
by Sheldon L. Glashow, 
Steven Weinberg, 
and Abdus Salam~\cite{Glashow_SM,Weinberg_SM,Salam_SM} only started to be taken seriously by the particle physics community 
after Gerard 't Hooft and Martinus Veltman mathematically 
demonstrated its renormalizability~\cite{tHooft,tHooft_Veltman}. 

From a modern perspective, however, 
the condition that a theory must be renormalizable is regarded as too restrictive, equivalent to 
requiring that it remains valid at all energies. As a matter of fact, there is
no reason to exclude nonrenormalizable theories from our toolkit. They
can be interpreted as EFTs whose natural energy scale is set by the cutoff~$\Lambda$, giving
accurate results for processes involving 
energies~$E\ll\Lambda$. Furthermore, from this viewpoint, the cutoff 
ceases to be a mere mathematical artefact to be eventually hidden in the action parameters. 
Instead, it acquires a physical significance as the energy threshold of the unknown physics 
encoded in the higher dimensional operators of our EFTs. Otherwise expressed, nonrenormalizability
has lost its bad reputation and now is taken as a hint that some unknown physics is 
lurking at higher energies.

To make the previous discussion more transparent, let us look at the important case of
quantum chromodynamics (QCD), the theory describing the interaction of quarks and 
gluons. QCD is not just a renormalizable theory that it can be extrapolated at
arbitrary energies, but asymptotically free as well. \label{page:QCD_strong_weak_regimes}
This means that its coupling constant approaches zero as we go to higher energies, thus making perturbation
theory more and more reliable. The issue, however, is that when studying its low energy dynamics,
the QCD coupling grows as we decrease the energy
and the theory becomes strongly coupled. This has to be handled in a way somehow 
reminiscent of what we explained
when discussing quantum GR near the Planck scale:  
below certain energy scale~$\Lambda_{\rm QCD}$
we need to abandon the perturbative QCD (pQCD) description in terms of 
quarks and gluons, now strongly coupled, and find 
the ``right'', weakly coupled, degrees of freedom to build an operative~QFT. 
But, simultaneously, we have a huge advantage concerning the gravity case. 
There, the trouble arose in the unexplored region of 
extremely high
energies, where identifying the appropriate degrees of freedom, 
their interactions, or just the right framework remains anybody's guess (strings?
spin foam? causal sets?). By contrast, life
is much easier in QCD. The problematic regime happens at low energies, so 
to identify the weakly coupled degrees of freedom, we only need to ``look'', i.e., to
do experiments. From them, we 
learn that the physics has to be described in terms of mesons and baryons,
whose interactions are largely fixed by symmetries (an issue to which we  
will come back later). What is relevant for the present discussion is that the appropriate
framework, chiral perturbation theory~($\chi$PT), is a nonrenormalizable QFT whose
action contains a plethora of higher-dimensional operators. Its cutoff, however,
is not some arbitrary energy~$\Lambda$ whose role is just to make the theory finite, 
but the physical scale~$\Lambda_{\rm QCD}$ at which quarks and gluons get confined into hadrons. 
The theory of hadron interactions should then be understood as an EFT 
valid at energies~$E\ll\Lambda_{\rm QCD}$.

\begin{figure}[t]
\centerline{\includegraphics[scale=0.45]{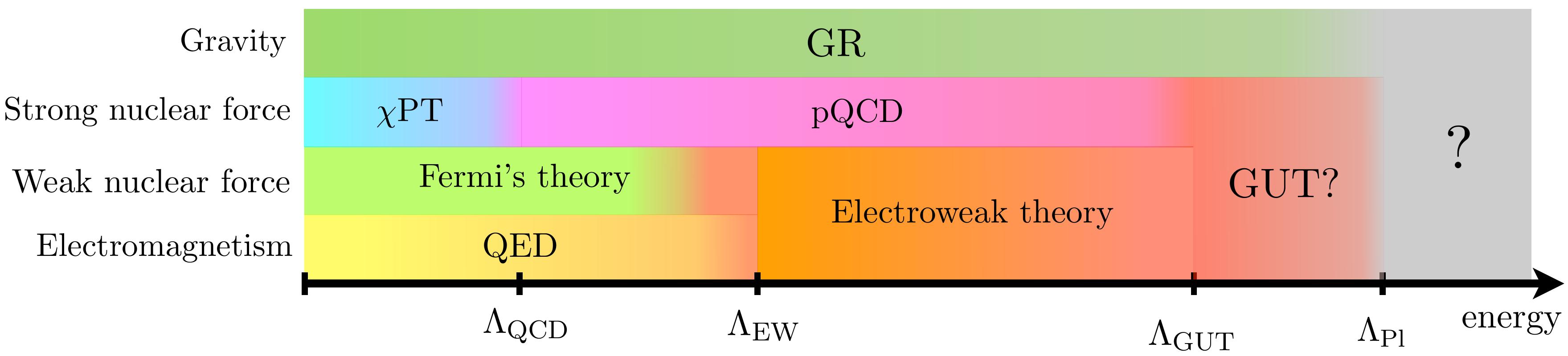}}
\caption[]{Simplified cartoon showing the network of EFTs behind our understanding of subatomic
physics.}
\label{fig:EFT_scheme}
\end{figure}

The existence of the Planck scale at which quantum gravity is expected to become the dominant
interaction has led to the realization that all quantum field theories have to be 
regarded as EFTs with a limited range of validity. This includes even renormalizable theories that, like
the SM, are well-defined in a wide range of energies. However,
explaining some experimental facts, such as nonzero neutrino 
masses, might require adding higher-dimensional operators to the theory, setting
the energy scale for new physics to be explored in future high-energy facilities. At this energy,
the SM will be superseded, maybe by some grand unified theory (GUT), which in turn is expected to 
break down at~$\Lambda_{\rm Pl}$. It is 
in this sense that EFTs provide the foundational framework to understand nature at the smallest 
length scales~(see fig.~\ref{fig:EFT_scheme}).
\label{page:EFTsf}

\section{From symmetry to physics}
\label{sec:intro}

Symmetry is a central theme of contemporary physics, although its tracks go back a long way in history. More or
less in disguise, symmetry-based arguments can be found in natural philosophy
since classical times. In his refutation of vacuum in the fourth 
book of {\em Physics} (215a), Aristotle used the homogeneity of empty space to conclude the 
principle of inertia, that he however regarded as an inconsistency since it contradicted 
his first principle of motion: 
whatever moves has to be moved by something else. Galileo Galilei's assumption that reversing
the velocity with which a free-rolling ball arrives at the basis of an inclined plane would make it climb 
exactly to the
height from which it was released can be also regarded as an early {\em de facto} 
application of time reversal 
symmetry.

Although the origins of the mathematical study of symmetry
are traced back to the first half of the 19th century with the
groundbreaking works on group theory of Evariste Galois and Niels Henrik Abel, 
its golden age was ushered in by
Felix Klein's 1872 Erlangen Program~\cite{klein,kline_book}. Its core idea is that 
different geometries can be fully derived from the knowledge of the group of transformations
preserving its objects (points, angles, figures, etc.). 
This establishes at the same time a hierarchy among geometries, determined by the relative generality 
of their underlying symmetry groups. In this way, Euclidean, affine, and hyperbolic geometries 
can be retrieved from projective geometry by restricting its group of transformations.

\begin{figure}[t]
\centerline{\includegraphics[scale=0.4]{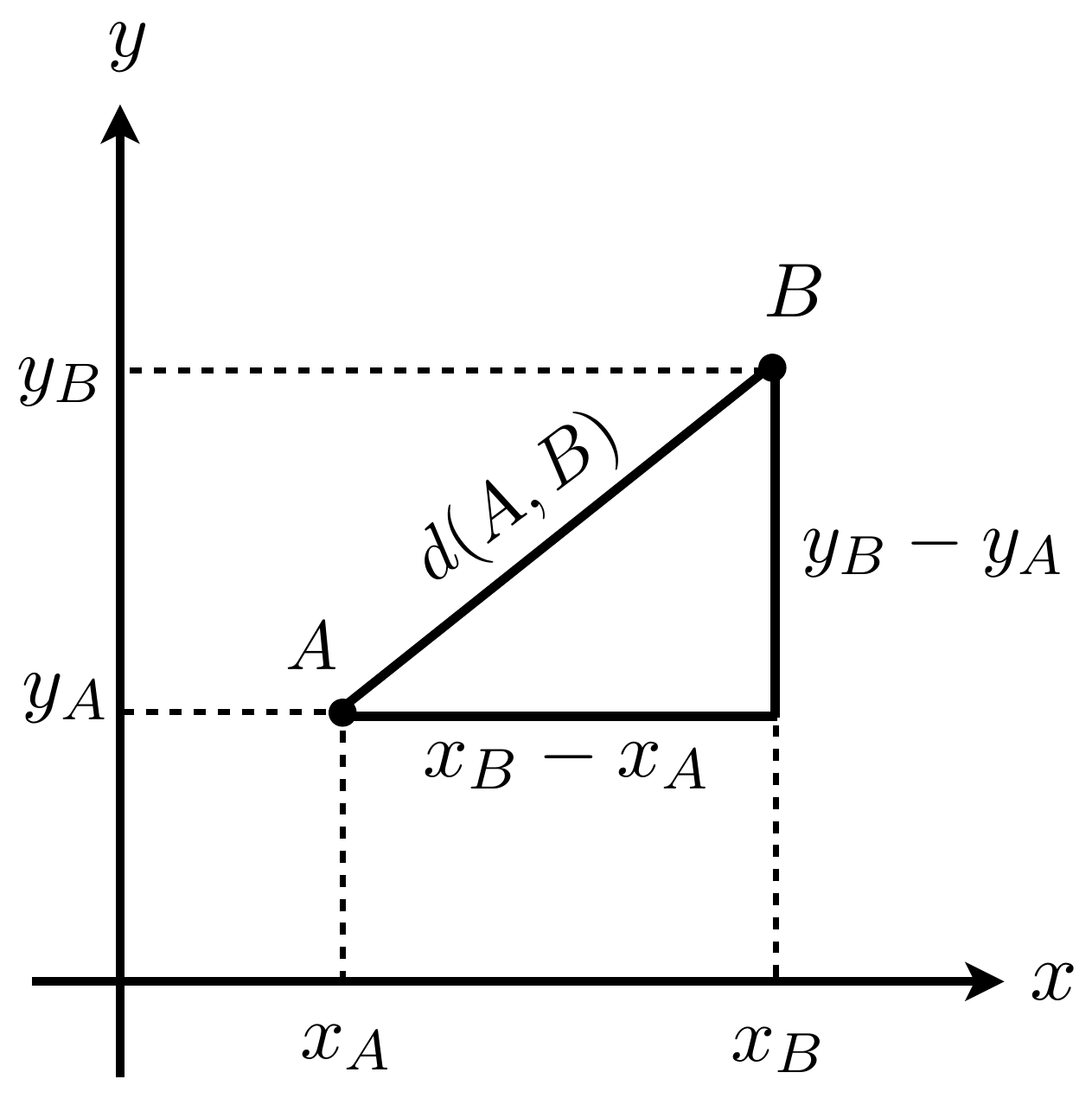}}
\caption[]{Euclidean distance between two points on the plane.}
\label{fig:euclidean_distance}
\end{figure}

As an example, the whole plane Euclidean geometry
emerges from the invariance under the combined action of rotations and rigid translations
\begin{align}
x'^{i}=R^{i}_{\,\,j}x^{j}+a^{i},
\end{align} 
where~$R^{i}_{\,\,j}\in \mbox{SO}(2)$ and~$a^{i}$ is and arbitrary two-dimensional vector. 
These two transformations
build together the Euclidean group~$E(2)\equiv \mbox{ISO}(2)$, leaving invariant the
Euclidean distance between two points~$A$ and~$B$ with Cartesian coordinates~$A=(x_{A},y_{A})$ 
and~$B=(x_{B},y_{B})$
\begin{align}
d(A,B)=\sqrt{(x_{B}-x_{A})^{2}+(y_{B}-y_{A})^{2}},
\label{eq:euclidean_distance}
\end{align}
which is just an application of 
the Pythagorean theorem (see fig.~\ref{fig:euclidean_distance}).
In a similar fashion, 
the geometry on the complex projective line~$\mathbb{CP}^{1}$ (a.k.a. the Riemann sphere)
follows from the invariance of geometrical objects 
under the projective linear group~$\mbox{PGL}(2,\mathbb{C})$, 
acting through M\"obius transformations on~$\mathbb{C}\cup\{\infty\}$
\begin{align}
z'={az+b\over cz+d},
\end{align}
where~$a,b,c,d\in\mathbb{C}$ and~$ad-bc\neq 0$. 
Among the invariants in this case are 
the four-point cross ratios associated with four points with complex
coordinates~$z_{1}$,~$z_{2}$,~$z_{3}$, and~$z_{4}$
\begin{align}
\mbox{CR}(z_{1},z_{2},z_{3},z_{4})\equiv {(z_{1}-z_{3})(z_{2}-z_{4})\over (z_{2}-z_{3})(z_{1}-z_{4})},
\end{align}
as well as the chordal distance between two points~$A$ and~$B$ on the Riemann sphere
\begin{align}
d(A,B)_{\rm chordal}&={2|z_{A}-z_{B}|\over \sqrt{(1+|z_{A}|^{2})(1+|z_{B}|^{2})}}.
\end{align}
M\"obius transformations preserve angles and maps circles to
circles, so  from a Kleinian 
point of view
they are {\em bona fide} geometrical objects on~$\mathbb{CP}^{1}$.

Klein's association of geometry and symmetry (i.e., group theory) revolutionized 
mathematics and became a game changer in physics. 
Beyond all early tacit uses, the systematic implementation of symmetry in physics
had to wait until the end of the 19th century. In 1894 Pierre Curie 
used group theoretical methods to study the role of spatial symmetries in physical 
phenomena~\cite{curie1894}, thus introducing mathematical tools so far only applied in crystallography. 
This inaugurated a trend taken up later by the emerging fields of relativity and atomic physics,
that led to key results like Emmy Noether's two celebrated theorems linking
symmetries with conserved charges~\cite{Noether} (see section~\ref{sec:Noether}).

\subsection{Relativity from geometry}

A beautiful example of geometry emerging from symmetry is provided by the geometrization of
special relativity carried out in 1908 by Hermann Minkowski\footnote{Einstein actually 
dubbed Minkowski's idea {\em \"uberflussige Gelehrsamkeit} (superfluous erudition)~\cite{pais}, 
although 
geometrization later turned out to be the basis of his general theory of relativity.}. 
Einstein's formulation of special relativity in terms of events occurring in some
instant~$t$ at some position~$\mathbf{r}$ 
(as measured by some inertial observer) leads naturally to introducing 
the four-dimensional space
of all potential events, each represented by a point with spacetime coordinates~$(t,\mathbf{r})$.
Although switching from one inertial observer to another changes the individual coordinates of the events, the invariance of
the speed of light implies the existence of an invariant. Given two arbitrary events 
taking place at
points~$\mathbf{r}$ and~$\mathbf{r}+\Delta\mathbf{r}$ and separated by a time lapse~$\Delta t$, its ``spacetime separation''
\begin{align}
\Delta s^{2}\equiv\Delta t^{2}-(\Delta\mathbf{r})^{2},
\label{eq:spacetime_separation}
\end{align}
remains the same for all inertial observers. The existence of this 
invariant with respect to the reference frame transformations introduced by 
Lorentz, Poincar\'e, and Einstein (and named after the first one) makes it natural to endow the space of events, or spacetime for short,
with the metric
\begin{align}
ds^{2}=dt^{2}-dx^{2}-dy^{2}-dz^{2}.
\label{eq:minkowski_metric}
\end{align}
This is how spacetime geometry originates from the postulate of invariance of the speed of light. 

We can take advantage of the language of tensors and write the line element~\eqref{eq:minkowski_metric}
in the form
\begin{align}
ds^{2}=\eta_{\mu\nu}dx^{\mu}dx^{\nu},
\label{eq:ds2_eta}
\end{align}
where~$(x^{0},x^{1},x^{2},x^{3})\equiv (t,x,y,z)$ and~$\eta_{\mu\nu}\equiv\mbox{diag}(1,-1,-1,-1)$ is 
the Minkowski metric.
The most general linear transformation leaving invariant~\eqref{eq:ds2_eta} 
[or~\eqref{eq:minkowski_metric}]
is written as
\begin{align}
x'^{\mu}=\Lambda^{\mu}_{\,\,\,\nu}x^{\nu}+a^{\mu},
\label{eq:Poincare_trans}
\end{align}
where~$\Lambda^{\mu}_{\,\,\,\nu}$ satisfies
\begin{align}
\eta_{\mu\nu}=\eta_{\alpha\beta}\Lambda^{\alpha}_{\,\,\,\mu}\Lambda^{\beta}_{\,\,\,\nu},
\label{eq:Lorentz_condition}
\end{align}
and~$a^{\mu}$ is an arbitrary constant vector. The linear coordinate change~\eqref{eq:Poincare_trans}
generates the Poincar\'e group,~$\mbox{ISO}(1,3)$, that includes all 
transformations~$\Lambda^{\mu}_{\,\,\nu}$ in the
Lorentz group~$\mbox{SO}(1,3)$ in addition to 
rigid translations. Notice that~$\Lambda^{\mu}_{\,\,\nu}$ is a $4\times 4$ matrix with 16 real components,
that the ten conditions~\eqref{eq:Lorentz_condition} reduce to six independent ones. They 
correspond to the three parameters of a three-dimensional rotation (e.g., the Euler 
angles) plus the three velocity components of a generic boost. Adding the four real numbers 
determining a spacetime translation, we conclude that the Poincar\'e transformation~\eqref{eq:Poincare_trans}
depends on ten independent real parameters. \label{page:lorentz_trans_generalities}

Besides the invariance of the speed of light,
Einstein's special relativity is also based on a second postulate, 
that all laws of physics take the same form for any inertial observer. This can
also be recast in geometric language by demanding
that all equations of physics be expressed as tensor identities 
with the structure
\begin{align}
T^{\mu_{1}\ldots\mu_{k}}_{\nu_{1},\ldots,\nu_{n}}(x)=0.
\end{align}
Under the generic Poincar\'e transformation~\eqref{eq:Poincare_trans}, the previous equation changes as
\begin{align}
T'^{\mu_{1}\ldots\mu_{k}}_{\nu_{1}\ldots\nu_{n}}(x')=\Lambda^{\mu_{1}}_{\,\,\,\,\alpha_{1}}
\ldots\Lambda^{\mu_{k}}_{\,\,\,\,\alpha_{k}}T^{\alpha_{1}\ldots\alpha_{k}}_{\beta_{1}\ldots\beta_{n}}(x)
\Lambda^{\beta_{1}}_{\,\,\,\nu_{1}}\ldots\Lambda^{\beta_{n}}_{\,\,\,\nu_{n}}=0,
\end{align} 
thus preserving the form~$T'^{\mu_{1}\ldots\mu_{k}}_{\nu_{1},\ldots,\nu_{n}}(x')=0$ 
it had for the original observer.

\begin{mdframed}[backgroundcolor=lightgray,hidealllines=true]
\vspace*{0.2cm}
\centerline{\greybox{\bf Retrieving Lorentz transformations}}
\vspace*{0.2cm}

\label{page:box_lorentz_trans}

It is a trivial exercise to recover the standard expression of a 
Lorentz transformations from the invariance of the 
line element~\eqref{eq:minkowski_metric}. For simplicity we consider a two-dimensional 
spacetime, equivalent to restricting to boosts along the~$x$-axis so the 
coordinates~$y'=y$ and~$z'=z$ remain unchanged. Implementing the coordinate change
\begin{align}
\left(
\begin{array}{c}
t' \\
x'
\end{array}
\right)=
\left(
\begin{array}{cc}
\Lambda^{0}_{\,\,0} & \Lambda^{0}_{\,\,1} \\
\Lambda^{1}_{\,\,0} & \Lambda^{1}_{\,\,1} 
\end{array}
\right)\left(
\begin{array}{c}
t \\
x
\end{array}
\right).
\end{align}
with the condition~$dt'^{2}-dx'^{2}=dt^{2}-dx^{2}$ implies
\begin{align}
(\Lambda^{1}_{\,\,0})^{1}-(\Lambda^{1}_{\,\,0})^{2}&=1, \nonumber \\[0.2cm]
(\Lambda^{2}_{\,\,1})^{1}-(\Lambda^{0}_{\,\,1})^{2}&=1, \\[0.2cm]
\Lambda^{0}_{\,\,0}\Lambda^{0}_{\,\,1}-\Lambda^{1}_{\,\,0}\Lambda^{1}_{\,\,1}&=0.
\nonumber
\end{align}
Using the properties of the hyperbolic functions we easily see that 
the first two identities are 
solved by~$\Lambda^{0}_{\,\,0}=\cosh\alpha$,~$\Lambda^{1}_{\,\,0}=\pm\sinh\alpha$ 
and~$\Lambda^{0}_{\,\,1}=\pm\sinh\beta$,~$\Lambda^{1}_{\,\,1}=\cosh\beta$, for
arbitrary~$\alpha$ and~$\beta$, with the third one requiring~$\beta=\alpha$. The sought transformation
is therefore parametrized as
\begin{align}
\left(
\begin{array}{c}
t' \\
x'
\end{array}
\right)=
\left(
\begin{array}{cc}
\cosh\alpha & -\sinh\alpha \\
-\sinh\alpha & \cosh\alpha 
\end{array}
\right)\left(
\begin{array}{c}
t \\
x
\end{array}
\right),
\label{eq:Lorentz_trans_alpha}
\end{align}
where the parameter~$\alpha$ is called the boost rapidity.
A comment on the signs is in order. First, we have taken~$\Lambda^{0}_{\,\,0}>0$ so the arrow of time 
points in 
the same direction for both observers (later in page~\pageref{pag:ortochr}
we will put a Greek name to this and call these transformations 
orthochronous). On the other hand, as we will see right away,
the parameter~$\alpha$ is related to the boost velocity. Choosing a
negative sign for the off-diagonal components of the matrix in~\eqref{eq:Lorentz_trans_alpha} 
means that~$\alpha>0$ corresponds to 
a boost in the direction of the positive~$x$-axis. 

To find the standard expression of the Lorentz transformation, we notice that the hyperbolic 
functions can be alternatively parametrized as
\begin{align}
\cosh\alpha&={1\over \sqrt{1-V^{2}}},\hspace*{1cm}
\sinh\alpha={V\over \sqrt{1-V^{2}}},
\end{align}
where the relation between the boost velocity and its rapidity is given by~$V=\tanh\alpha$. 
Plugging these expressions
into~\eqref{eq:Lorentz_trans_alpha}, we arrive at the well-known formulae
\begin{align}
t'={t-{Vx\over c^{2}}\over\sqrt{1-{V^{2}\over c^{2}}}}, \hspace*{1cm}
x'={x-Vt\over\sqrt{1-{V^{2}\over c^{2}}}},
\end{align}
where exceptionally we have restored powers of~$c$.

\label{page:box_lorentz_trans_end}
\end{mdframed}

\begin{figure}[t]
\centerline{\includegraphics[scale=0.4]{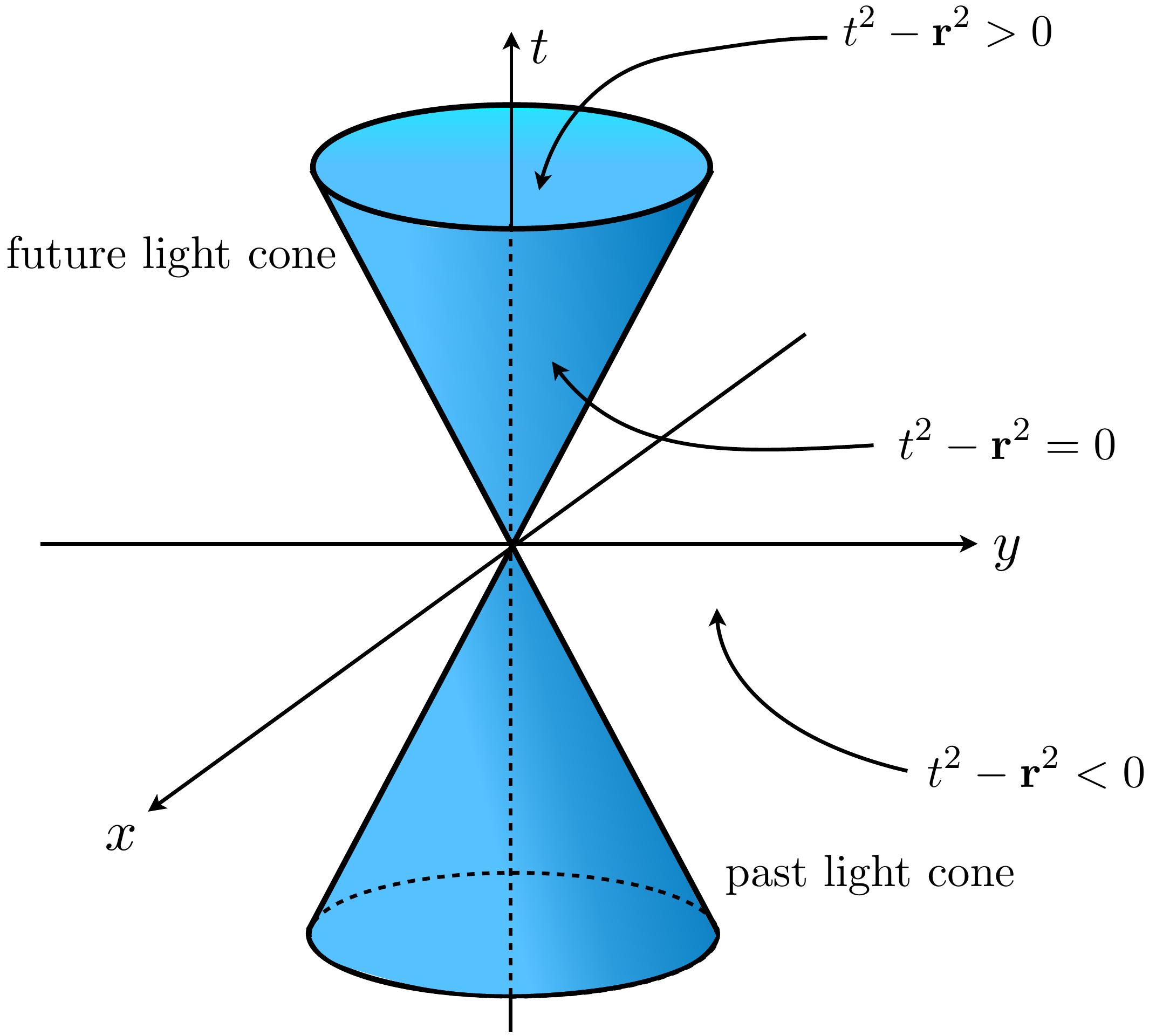}}
\caption[]{Representation of the light cone at the origin in a $(2+1)$-dimensional
spacetime.}
\label{fig:light_cone}
\end{figure}

Whereas the Euclidean distance~\eqref{eq:euclidean_distance} tells us about how far apart in space
two points lie, the spacetime geometry~\eqref{eq:minkowski_metric} contains information about the
causal relations between events. Let us consider an arbitrary event that, without lost of generality,
we place at the origin of our coordinate system~$x_{0}^{\mu}=(0,\boldsymbol{0})$. The question arises
as to whether some other event~$x^{\mu}=(t,\mathbf{r})$ may either influence what happens at~$x_{0}^{\mu}$
or be influenced by it. Since the speed of light is a universal velocity limit, the question is settled 
by checking
whether it is possible for a signal propagating with 
velocity~$v\leq 1$ to travel from~$(t,\mathbf{r})$ to~$(0,\mathbf{0})$, if~$t<0$, or vice-versa for
positive~$t$.
The condition for this to happen is
\begin{align}
{|\mathbf{r}|\over |t|}\leq 1 \hspace*{1cm} \Longrightarrow \hspace*{1cm} t^{2}-\mathbf{r}^{2}\geq 0.
\end{align}
The set of events satisfying this defines the interior and the surface of 
the light-cone associated with the event at~$(0,\mathbf{0})$, that we have depicted in 
fig.~\ref{fig:light_cone} for a~$(2+1)$-dimensional spacetime. 
Points in the causal past of the origin lie inside or on the past 
light-cone ($t<0$), whereas those on or inside the future light-cone~($t>0$) are causally reachable 
from~$(0,\mathbf{0})$. By contrast, events outside the light-cone cannot influence of be influenced
by the event at the origin, since this would require superluminal propagation. 
What we have said about the origin applies to any other event:
every point of the spacetime is endowed with its light-cone defining its area of casual influence.

Thus,
if two events lie outside each other's light-cones, they cannot influence one another.
Mathematically this is characterized by their spacetime separation satisfying~$\Delta s^{2}<0$, 
so they are said to be {\em spatially} separated. Interestingly, there always exists 
a reference frame in which both events happen at the same~$t$, i.e. they are 
simultaneous. This is not possible when one event is inside the other's light-cone, in which
case~$\Delta s^{2}>0$ and their separation is called {\em timelike}. 
Looking at~\eqref{eq:spacetime_separation} and remembering the invariant character 
of~$\Delta s^{2}$ we see that there can be no frame for which~$\Delta t=0$. Nonetheless, it is
always possible to find an inertial observer for which both events happen at the same point of 
space, i.e.~$\Delta\mathbf{r}=\mathbf{0}$. In this case~$\Delta s^{2}$ is just
the (squared) time elapsed between both events, as measured by the observer who is visiting
both. Notice for two events lying on each others light-cone there is no such possibility, since 
they can only be joined by signals propagating at the speed of light and no observer can 
travel at this velocity.

\begin{mdframed}[backgroundcolor=lightgray,hidealllines=true]
\vspace*{0.2cm}
\centerline{\greybox{\bf There is no twin paradox}}
\vspace*{0.2cm}

One of the most celebrated ``paradoxes'' associated with special relativity is that involving
two identical twins, one of which starts a round trip from Earth at very high speed 
while the second remains quietly behind. Relativistic time dilation implies that the clock carried by 
the traveling twin slows down with respect to the time set by a second clock on Earth, so at the
end of the trip the returning twin looks younger than the remaining sibling. So far, so good. 
However, applying the same argument to the frame of reference moving with the spaceship, the
conclusion seems to be the opposite: that the clock of the twin staying on Earth, that
is the one moving in the reference frame of the rocket, ticks slower and after the reunion 
it is the Earth twin the one looking younger.

To clarify this apparent ``paradox'' we have to keep in mind that special relativity is about
inertial observers. Thus, we are going to work with the reference frame of the twin standing on 
Earth, who follows the spacetime path (the
worldline) indicated in the following graph as~$1$ 
\begin{align*}
\includegraphics[width=7cm]{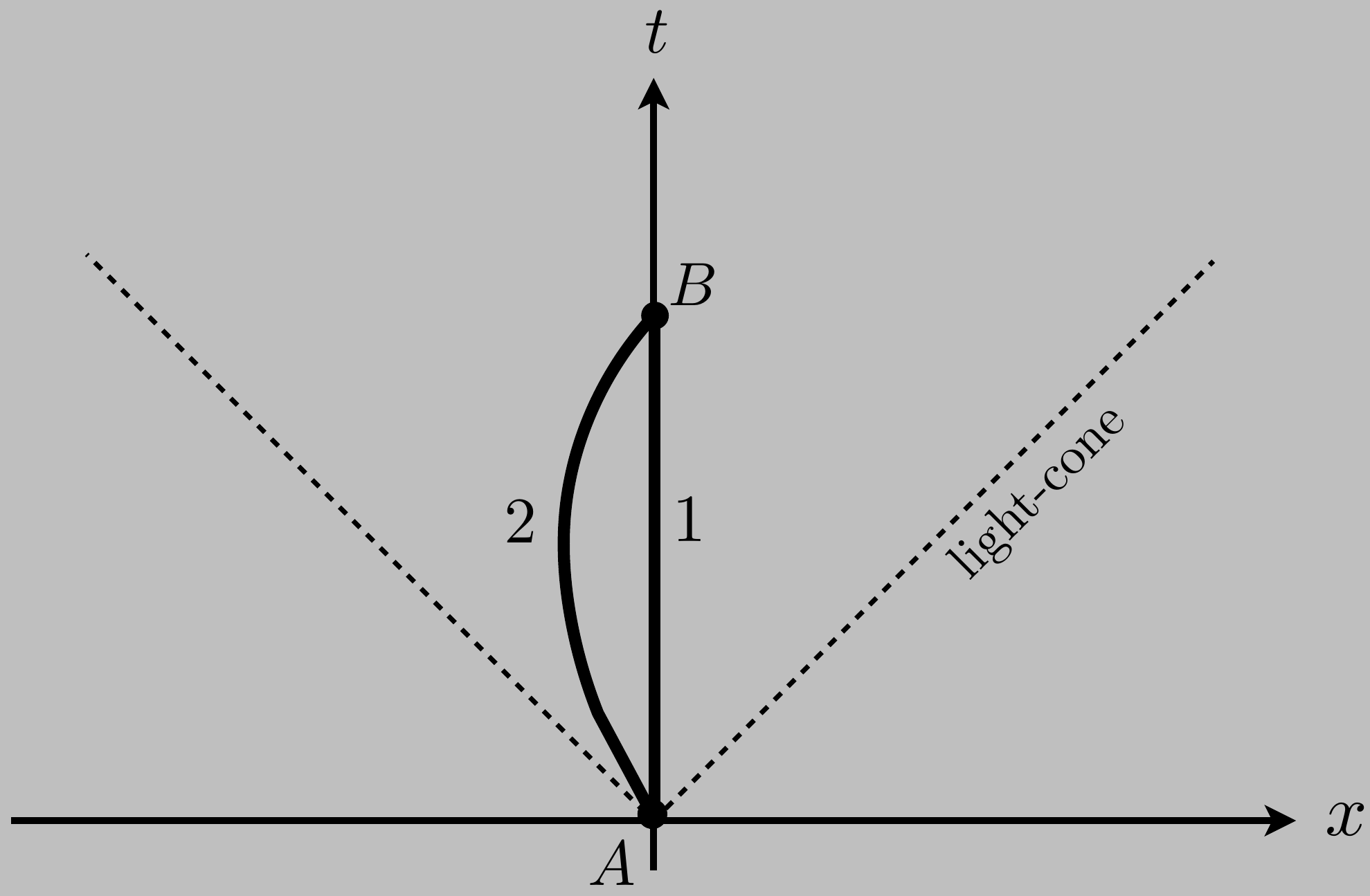}
\end{align*}
The travelling twin, on the other hand, follows the worldline labelled as~$2$, that starts and finishes on
Earth, moving back and forth along the $x$~direction. For simplicity, we restrict the movement of the
rocket to this coordinate, with the Earth located at~$x=0$.

Physical observers move along wordlines~$x^{\mu}(\lambda)$ 
whose tangent at any point defines a timeline 
vector~$\eta_{\mu\nu}\dot{x}^{\mu}(\lambda)\dot{x}^{\nu}(\lambda)>0$. 
The time elapsed between two events~$A$ and~$B$ as measured by the clock carried by the observer 
(called its proper time) equals
the spacetime length along the worldline~$\gamma_{AB}$
\begin{align}
\Delta s_{AB}=\int_{\gamma_{AB}}ds
=\int_{\lambda_{A}}^{\lambda_{B}}d\lambda\sqrt{\eta_{\mu\nu}\dot{x}^{\mu}(\lambda)\dot{x}^{\nu}(\lambda)}.
\label{eq:length_worldline_general}
\end{align}
A particularly convenient parametrization of the curve is provided by the coordinate time,~$x^{0}\equiv t$,
so writing~$x^{\mu}(t)=\big(t,\mathbf{R}(t)\big)$ the previous equation becomes
\begin{align}
\Delta s_{AB}=\int_{t_{A}}^{t_{B}}dt'\sqrt{1-\mathbf{v}(t')^{2}},
\end{align}
with~$\mathbf{v}(t)=\dot{\mathbf{R}}(t)$ the observer velocity satisfying~$|\mathbf{v}(t)|<1$.

Let us return to our twins. Both of them travel from~$A$ to~$B$, as shown in the 
graph above, but along different worldlines with different speeds. 
The one on Earth has~$\mathbf{v}=\mathbf{0}$, so the time
elapsed between the departure and arrival of the second twin is
\begin{align}
\Delta s_{AB}^{(1)}=t_{B}-t_{A}.
\end{align}
For the twin on the spaceship, by contrast, we do not even need to know anything 
about the details of the varying speed. It is enough to notice that
$0<\sqrt{1-\mathbf{v}(t)^{2}}<1$, implying
\begin{align}
\Delta s_{AB}^{(2)}<\Delta s_{AB}^{(1)}.
\label{eq:twin_paradox_solution}
\end{align} 
Consequently, after reunion, the traveling twin will be the younger.

A basic difference between the twins is that the one at rest is 
precisely the inertial observer for which the timelike separated events~$A$ and~$B$ happen at
the same point of space. In fact, the result~\eqref{eq:twin_paradox_solution} reflects
a property of this particular frame: its worldline represents 
the path of the longest proper time interpolating between
two given events.

As announced, 
the reason why there is no paradox is 
because only one of the twins is an inertial observer and their descriptions
cannot be simply interchanged without further ado. Seeing everything from the
point of view of the spaceship leads us to give up the Minkowski metric~\eqref{eq:minkowski_metric}.
Indeed, by changing the coordinates
\begin{align}
t'&=t, \nonumber \\[0.2cm]
\mathbf{r}'&=\mathbf{r}+\mathbf{R}(t),
\label{eq:change_twins}
\end{align}
the worldlines of both twins 
are respectively parametrized by~$x_{1}^{\mu}(t')=\big(t',-\mathbf{R}(t')\big)$ 
and~$x_{2}^{\mu}(t')=\big(t',\mathbf{0}\big)$, while
the spacetime metric now reads
\begin{align}
ds^{2}=\big[1-\mathbf{v}(t')^{2}\big]dt'^{2}+2\mathbf{v}(t')\cdot d\mathbf{r}'\,dt'-d\mathbf{r}'^{2},
\label{eq:metric_traveling_twin}
\end{align}
which is no longer the Minkowski metric.
To compute the proper time of both twins we use eq.~\eqref{eq:length_worldline_general},
replacing~$\eta_{\mu\nu}$ by the line element~\eqref{eq:metric_traveling_twin}. We then find
\begin{align}
\Delta s_{AB}^{(1)}&=\int_{t'_{A}}^{t'_{B}}dt'\sqrt{1-\mathbf{v}(t')^{2}+2\mathbf{v}(t')^{2}
-\mathbf{v}(t')^{2}}=t_{B}-t_{A}, 
\nonumber \\[0.2cm]
\Delta s_{AB}^{(2)}&=\int_{t'_{A}}^{t'_{B}}dt'\sqrt{1-\mathbf{v}(t')^{2}}<\Delta s_{AB}^{(1)},
\end{align}
which reproduce the results obtained above. The conclusion is that if properly analyzed, the descriptions 
from the points of view of both twins are absolutely consistent and no paradox arises.

\end{mdframed}

As time and space coordinates combine to label a point (event) in 
the four-dimensional Minkowski spacetime, 
so energy and momentum build up an energy-momentum four-vector~$p^{\mu}=(E,\mathbf{p})$. 
For a particle of mass~$m$
moving along an affinely paramerized worldline~$x^{\mu}(s)$, four-momentum is defined by
\begin{align}
p^{\mu}(s)\equiv m\dot{x}^{\mu}(s)=\left({m\over \sqrt{1-\mathbf{v}^{2}}},{m\mathbf{v}\over
\sqrt{1-\mathbf{v}^{2}}}\right),
\label{eq:relativistic_momentum}
\end{align}
with~$\mathbf{v}$ the particle's velocity. A first thing to be noticed here is that the particle's energy
is nonzero even when its velocity vanishes. Restoring powers of~$c$ 
\begin{align}
E\longrightarrow {E\over c}, \hspace*{1cm} m\longrightarrow mc 
\hspace*{1cm} \mathbf{v}\longrightarrow {\mathbf{v}\over c},
\end{align}
we get the famous equation~$E_{\rm rest}=mc^{2}$. On the other hand,
the particle's energy diverges as~$|\mathbf{v}|\rightarrow c$. This shows that the speed
of light a physical limiting velocity for any massive particle, since reaching~$|\mathbf{v}|=c$
would require pumping an infinite amount of energy into the system.
The transformation of energy and momentum among 
inertial observers is fixed by~$p^{\mu}$ being being a four-vector,
whose change under a Lorentz 
transformation~$\Lambda^{\mu}_{\,\,\,\nu}$ 
is given by~$p'^{\mu}=\Lambda^{\mu}_{\,\,\,\nu}p^{\nu}$. Considering a 
boost along the~$x$ direction with velocity~$V$ and using the expressions obtained in 
Box~1 in 
pages~\pageref{page:box_lorentz_trans}-\pageref{page:box_lorentz_trans_end}, we have
\begin{align}
E'={E-Vp_{x}\over \sqrt{1-V^{2}}}, \hspace*{1cm}
p_{x}'={p_{x}-V E\over \sqrt{1-V^{2}}},
\end{align} 
together with~$p_{y}'=p_{y}$ and~$p'_{z}=p_{z}$.

\begin{figure}[t]
\centerline{\includegraphics[scale=0.4]{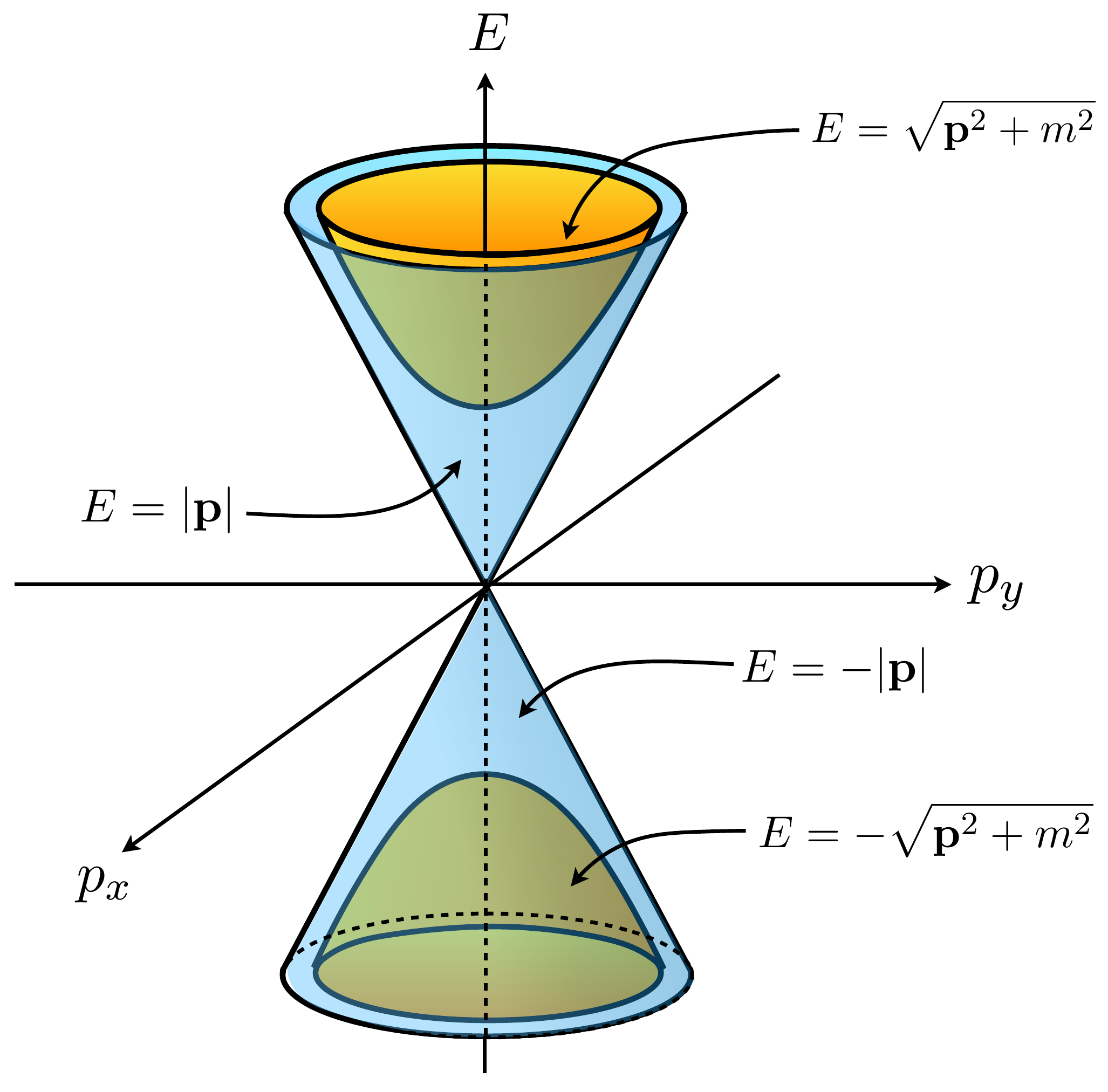}}
\caption[]{Energy-momentum hyperboloid for a particle of mass~$m\neq 0$ (orange). The energy-momentum
vector of a massless particle lies on the blue cone.}
\label{fig:hyperboloid}
\end{figure}

Equation~\eqref{eq:relativistic_momentum} also implies  
the mass-shell condition\footnote{In covariant terms, the 
mass-shell condition reads~$p_{\mu}p^{\mu}=m^{2}$ and follows from~\eqref{eq:relativistic_momentum}, 
remembering that the
particle's worldline is affinely parametrized,~$\eta_{\mu\nu}\dot{x}^{\mu}(s)\dot{x}^{\nu}(s)=1$.}
\begin{align}
E^{2}-\mathbf{p}^{2}=m^{2}.
\label{eq:disp_rel_generalSR}
\end{align}
In the four-dimensional energy-momentum space spanned by~$E$ and~$\mathbf{p}$, the particle's 
four-momentum~$p^{\mu}$ lies on the two-sheeted hyperboloid~$E=\pm\sqrt{\mathbf{p}^{2}+m^{2}}$, with the two 
signs corresponding to the upper and lower sheet. Interestingly, 
the mass-shell condition has a smooth limit as~$m\rightarrow 0$, where the hyperboloid degenerates into
the cone~$E^{2}=\mathbf{p}^{2}$, to which all massive hyperboloids asymptote for large spatial
momentum,~$|\mathbf{p}|\gg m$ (see fig.~\ref{fig:hyperboloid}). 
Unlike Newtonian mechanics, special relativity admits the existence
of zero-mass particles whose four-momenta have the form
\begin{align}
p^{\mu}=(|\mathbf{p}|,\mathbf{p}),
\end{align}
where we have chosen the positive energy solution. In terms of its energy and momentum,
the velocity of a massive particle is given by 
[cf.~\eqref{eq:relativistic_momentum} and~\eqref{eq:disp_rel_generalSR}]
\begin{align}
\mathbf{v}={\mathbf{p}\over \sqrt{\mathbf{p}^{2}+m^{2}}},
\end{align}
which as~$m\rightarrow 0$ gives~$|\mathbf{v}|=1$. Thus, massless particles necessarily propagate
at the speed of light.

\subsection{Relativity and quantum mechanics}

So far, our analysis has left out quantum effects. Special relativity can be combined with 
quantum mechanics to formulate relativistic wave equations plagued with 
trouble. An immediate problem arises from the energy hyperboloid depicted 
in fig.~\ref{fig:hyperboloid}. The existence of the lower sheet implies that the system of 
a relativistic quantum particle coupled to an electromagnetic field has no ground state, 
since the particle has infinitely many available 
states with arbitrary negative energy to which it could decay by radiating energy. 
This fundamental instability of the system is impossible to solve 
in the context of the
Klein-Gordon wave equation, while in the Dirac equation it can be avoided by ``filling'' all states in 
the lower sheet of the hyperboloid (the Dirac sea). The Pauli exclusion principle now prevents electrons
from occupying negative energy states, and the system is stable. 

The Dirac sea notwithstanding, the interpretation of the Dirac equation 
as a single-particle relativistic wave equation 
is problematic, leading to puzzling results such as the Klein paradox~\cite{Holstein,AG_VM}.
In fact, all the difficulties we run into when trying to marry quantum mechanics with special relativity
stem from insisting in a single-particle description, as can be seen from a simple heuristic arguments. 
As we know, 
Heisenberg's uncertainty principle correlates quantum fluctuations in the position and momentum of a particle
\begin{align}
\Delta x\Delta p_{x}\geq {\hbar\over 2}.
\end{align} 
Looking at physics at small distances requires 
taming spatial fluctuations below the scale of interest, which
in turn leads to large fluctuations in the particle's momentum. When the latter reaches the 
scale~$\Delta p_{x}\sim mc$, the corresponding energy fluctuations~$\Delta E\sim mc^{2}$ are large enough to
allow the creation of particles out of the vacuum and the single-particle description breaks down. 
Equivalently, localizing a particle 
below its Compton wavelength
\begin{align}
\Delta x\leq {\hbar \over 2mc},
\end{align}
leads to a quantum state 
characterized by an indefinite
number of them. Unlike what happens in nonrelativistic many body physics,
in the quantum-relativistic domain particle number is not conserved
and creation-annihilation of particles
is a central ingredient of the theory. Thus, the single-particle
description inherent to the relativistic wave equation is fundamentally wrong, as indicated
by the paradoxes and inconsistencies it leads to.

\begin{mdframed}[backgroundcolor=lightgray,hidealllines=true]
\vspace*{0.2cm}
\centerline{\greybox{\bf Antiparticles and causality}}
\vspace*{0.2cm}

One of the consequences of the Klein paradox alluded to above is the impossibility
of a consistent formulation of relativistic quantum mechanics without the inclusion of antiparticles.
We can reach the same conclusion by showing that antiparticles are the
unavoidable ingredient to preserve causality in a relativistic quantum theory.
To do so, let us consider a relativistic particle of mass~$m$ that 
at~$t=0$ is detected at the origin. Its quantum-mechanical propagator is given by 
\begin{align}
G(\tau,\mathbf{r})&\equiv \langle\mathbf{r}|e^{-i\tau\sqrt{\mathbf{p}^{2}+m^{2}}}|\mathbf{0}\rangle
=e^{-i\tau\sqrt{-\boldsymbol{\nabla}^{2}+m^{2}}}\delta^{(3)}(\mathbf{r}).
\end{align}
Physically, this quantity gives 
the probability amplitude of the particle being detected at a later time~$t=\tau$ at some
location~$\mathbf{r}$.
To explicitly evaluate the propagator, 
we Fourier transform the Dirac delta function and compute the resulting integral in terms of 
a modified Bessel function of the second kind
\begin{align}
G(\tau,\mathbf{r})&=\int{d^{3}k\over (2\pi)^{3}}e^{-i\tau\sqrt{\mathbf{k}^{2}+m^{2}}+i\mathbf{k}\cdot\mathbf{r}}
\nonumber \\[0.2cm]
&={1\over 2\pi^{2}|\mathbf{r}|}\int_{0}^{\infty}
kdk\,\sin(k|\mathbf{r}|)e^{-i\tau\sqrt{k^{2}+m^{2}}} \label{eq:propagator_Bessel}\\[0.2cm]
&=-{i\over 2\pi^{2}}{m^{2}t\over \tau^{2}-\mathbf{r}^{2}}K_{2}\left(im\sqrt{\tau^{2}-\mathbf{r}^{2}}
\right),
\nonumber 
\end{align}
where to write the last identity we regularized the momentum integral by 
analytical continuation~$\tau\rightarrow \tau-i\epsilon$. Naively, one would expect this 
propagator to vanish outside the light cone,~$\tau^{2}-\mathbf{r}^{2}<0$, since 
otherwise the particle would have a nonvanishing probability of being detected at points
spacelike separated from the origin, its location at~$t=0$. Were this to happen, it would imply 
a violation of causality.

Despite expectations, the modified Bessel function in~\eqref{eq:propagator_Bessel} 
is nonzero for both real and imaginary values of the argument and the propagator
spills out of the light-cone despite being derived from a relativistic Hamiltonian. 
The key point to understand what is going on 
is that when~$\mathbf{r}$ lies outside the light-cone at the origin there are
frames in which the detection of the particle at the position~$\mathbf{r}$ {\em precedes} its detection 
at the origin. 
In computing the propagator we should take this into account and consider the superposition of both
processes outside and inside the light-cone
\begin{align}
G(\tau,\mathbf{r})&=\left\{
\begin{array}{l}
\langle\mathbf{r}|e^{-i\tau\sqrt{\mathbf{p}+m^{2}}}|\mathbf{0}\rangle \hspace*{0.3cm}\mbox{when} \hspace*{0.3cm} \tau^{2}-\mathbf{r}^{2}>0 \\[0.2cm]
\langle\mathbf{r}|e^{-i\tau\sqrt{\mathbf{p}+m^{2}}}|\mathbf{0}\rangle
+\langle\mathbf{0}|e^{i\tau\sqrt{\mathbf{p}+m^{2}}}|\mathbf{r}\rangle 
\hspace*{0.3cm}\mbox{when} \hspace*{0.3cm} \tau^{2}-\mathbf{r}^{2}<0
\end{array}
\right.
\end{align}
Now, from the explicit 
expression~\eqref{eq:propagator_Bessel} we can check that~$\langle\mathbf{r}|e^{-i\tau\sqrt{\mathbf{p}+m^{2}}}|\mathbf{0}\rangle$ is 
purely imaginary when~$\tau^{2}-\mathbf{r}^{2}<0$. Since, on the other hand,
\begin{align}
\langle\mathbf{r}|e^{-i\tau\sqrt{\mathbf{p}+m^{2}}}|\mathbf{0}\rangle
+\langle\mathbf{0}|e^{i\tau\sqrt{\mathbf{p}+m^{2}}}|\mathbf{r}\rangle=
2{\rm Re\,}\langle\mathbf{r}|e^{-i\tau\sqrt{\mathbf{p}+m^{2}}}|\mathbf{0}\rangle,
\end{align}
we conclude that
\begin{align}
G(\mathbf{r},\tau)&=-{i\over 2\pi^{2}}{m^{2}t\over \tau^{2}-\mathbf{r}^{2}}K_{2}\left(im\sqrt{\tau^{2}-\mathbf{r}^{2}}
\right)\theta(\tau^{2}-\mathbf{r}^{2}),
\end{align}
and causality is consequently restored.

There exists an interesting interpretation of this cancellation mechanism
due to Ernst Stueckelberg~\cite{Stueckelberg_antiparticles} and 
Richard Feynman~\cite{Feynman_antiparticles,Feynman_antiparticles2}. 
Our propagator can be seen as the wave function of the particle of interest,~$\psi(\tau,\mathbf{r})\equiv G(\tau,\mathbf{r})$, 
satisfying the
boundary condition~$\psi(0,\mathbf{r})=\delta^{(3)}(\mathbf{r})$. We found that 
outside the light-cone there is a superposition of two processes: one in which
the particle is traveling from the origin to~$\mathbf{r}$ forward in time, and a second 
described by the wave function
\begin{align}
\psi(\tau,\mathbf{r})_{\Downarrow}\equiv \langle \mathbf{0}|e^{i\tau\sqrt{\mathbf{p}^{2}+m^{2}}}|\mathbf{r}\rangle
=\langle\mathbf{r}|e^{-i\tau\sqrt{\mathbf{p}^{2}+m^{2}}}|\mathbf{0}\rangle^{*}\equiv \psi(\tau,\mathbf{r})_{\Uparrow}^{*},
\label{eq:wavefunctions_up_down}
\end{align}
where the particle moves backwards in time from~$\mathbf{r}$ to the origin. Furthermore, writing
\begin{align}
\psi(\tau,\mathbf{r})_{\Downarrow}&=\int{d^{3}k\over (2\pi)^{3}}e^{i\tau\sqrt{\mathbf{k}^{2}+m^{2}}-i\mathbf{k}\cdot\mathbf{r}}
=\int{d^{3}k\over (2\pi)^{3}}e^{-i\tau(-\sqrt{\mathbf{k}^{2}+m^{2}})+i(-\mathbf{k})\cdot\mathbf{r}},
\end{align}
and comparing with the first line in eq.~\eqref{eq:propagator_Bessel}, we 
reinterpret~$\psi(\tau,\mathbf{r})_{\Downarrow}$
as describing
a state of mass~$m$ and momentum~$-\mathbf{k}$, lying in the lower sheet of the energy hyperboloid, 
and propagating forward in time. This represents a hole in the Dirac sea, i.e.  
an {\em antiparticle} of momentum~$\mathbf{k}$. 
Moreover, from~\eqref{eq:wavefunctions_up_down} we see that if our particle has
charge~$q$ with respect to some global~U(1), 
the antiparticle necessarily transforms with the opposite charge
\begin{align}
\psi(\tau,\mathbf{r})_{\Uparrow}\rightarrow e^{iq\theta}\psi(\tau,\mathbf{r})_{\Uparrow} \hspace*{1cm}
\Longrightarrow \hspace*{1cm} \psi(\tau,\mathbf{r})_{\Downarrow}\rightarrow e^{-iq\theta}\psi(\tau,\mathbf{r})_{\Downarrow}.
\end{align}

Antiparticles are therefore a necessary ingredient in a relativist theory of quantum processes if we want
to avoid superluminal effects. They automatically imply the possibility of creation/annihilation 
of particle-antiparticle pairs, turning what was intended as single-particle relativistic quantum
mechanics into a multiparticle theory where the number of particles is not even well defined.

\end{mdframed}

\begin{figure}[t]
\centerline{\includegraphics[scale=0.4]{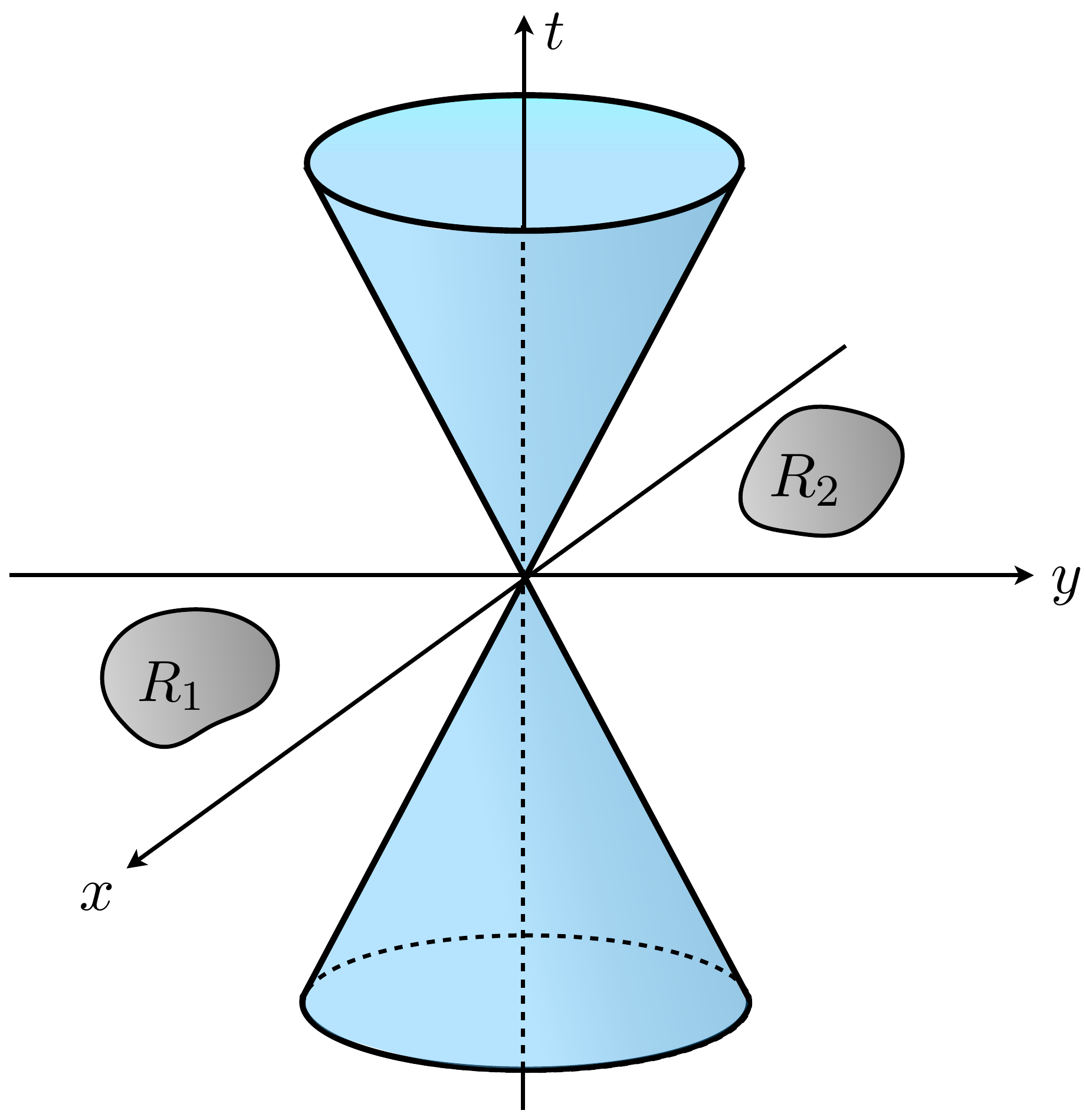}}
\caption[]{The two spacelike-separated regions~$R_{1}$ and~$R_{2}$ 
cannot causally influence one another.}
\label{fig:microcausality}
\end{figure}

A fundamental consequence of the causal structure of spacetime is that measurement of 
observables in regions that are spacelike separated cannot interfere with each other.
In the quantum theory these measurement are implemented by local operators~$\mathcal{O}(x)$
smeared over the spacetime region~$R$ where the measurement takes place
\begin{align}
\mathcal{O}(R)\equiv \int d^{4}x\,\mathcal{O}(x)f_{R}(x),
\end{align}
where
\begin{align}
f_{R}(x)&=\left\{
\begin{array}{ll}
1 & \mbox{if~$x\in R$} \\[0.2cm]
0 & \mbox{if~$x\notin R$}
\end{array}
\right.
\end{align} 
is the characteristic function associated with~$R$.
In mathematical terms, the noninterference of the measurements carried out
in spacelike separated regions~$R_{1}$ 
and~$R_{2}$ like those shown in fig.~\ref{fig:microcausality} is expressed by the vanishing
of the commutator of the associated operators
\begin{align}
[\mathcal{O}(R_{1}),\mathcal{O}(R_{2})]=0\hspace*{1cm} \mbox{if~$R_{1}$ and~$R_{2}$ are
spacelike separated},
\label{eq:microcausality_LQFT}
\end{align}
or equivalently
\begin{align}
[\mathcal{O}(x),\mathcal{O}(y)]=0 \hspace*{1cm} \mbox{if~$(x-y)^{2}<0$}.
\end{align}
This states the {\em principle of microcausality}, a profound form of locality
that has to be imposed on constructing 
any admissible QFT. To date no consistent theory has been formulated
violating this principle. This is why all theories to be encountered later in
these lecture will
be local quantum field theories (LQFTs) in the sense of eq.~\eqref{eq:microcausality_LQFT}.

\section{The importance of classical field theory}

Maxwell's electromagnetism is arguably the mother of all classical field theories. Despite its apparent
simplicity, the theory contains a number of symmetry and structures that underlie many other developments in QFT. 
This is the reason why it is worthwhile to spend some time extracting some lessons from classical electromagnetism 
that we will find useful later in our study of the SM and other theories. 

\subsection{The symmetries of Maxwell's theory}

Using Heaviside units, and keeping~$c=1$ all the way, the Maxwell's equations take the form
\begin{align}
\boldsymbol{\nabla}\cdot\mathbf{E}&={\color{blue} \rho_{e}}, \nonumber \\[0.2cm]
\boldsymbol{\nabla}\cdot\mathbf{B}&={\color{red} \rho_{m}}, \nonumber \\[0.2cm]
\boldsymbol{\nabla}\times\mathbf{E}&=-{\color{red} \mathbf{j}_{m}}-{\partial\mathbf{B}\over
\partial t} \label{eq:maxwell_eqs_general}\\[0.2cm]
\boldsymbol{\nabla}\times\mathbf{B}&={\color{blue} \mathbf{j}_{e}}+{\partial\mathbf{E}\over\partial t}.
\nonumber
\end{align}
Here we have introduced a color code signaling various layers of generality. Setting to zero all terms in blue and red we get the
vacuum Maxwell's equations governing the evolution of electromagnetic fields in the absence of any kind of matter. If we keep the terms
in blue but remove those in red, the resulting expressions describe the coupling of electric and magnetic fields to electrically charged matter,
where~${\color{blue} \rho_{e}}$ and~${\color{blue}\mathbf{j}_{e}}$ respectively represent the electric charge density and current.
These are the Maxwell's equations that can be found in most textbooks on classical electrodynamics 
(see, for example,~\cite{classicalED}). 

Let us postpone a little bit the discussion 
of the terms in red and concentrate on the second and third equations
\begin{align}
\boldsymbol{\nabla}\cdot\mathbf{B}&=0, \nonumber \\[0.2cm]
\boldsymbol{\nabla}\times\mathbf{E}&=-{\partial\mathbf{B}\over\partial t}.
\end{align}
They imply that the electric and magnetic fields can be written in terms of a scalar and a vector potential~$(\phi,\mathbf{A})$~as 
\begin{align}
\mathbf{B}&=\boldsymbol{\nabla}\times\mathbf{A},  \nonumber \\[0.2cm]
\mathbf{E}&=-\boldsymbol{\nabla}\phi-{\partial\mathbf{A}\over\partial t}.
\end{align}
These potentials, however, are not uniquely defined. The electric and magnetic fields remain unchanged if we replace
\begin{align}
\phi&\longrightarrow \phi+{\partial\epsilon\over\partial t}, \nonumber \\[0.2cm]
\mathbf{A}&\longrightarrow \mathbf{A}-\boldsymbol{\nabla}\epsilon,
\label{eq:gauge_transf_em_1}
\end{align}
with~$\epsilon(t,\mathbf{r})$ an arbitrary well-behaved function. This {\em gauge invariance} is
probably the most important of those structures of the electromagnetic theory that we said were of radical 
importance for QFT at large. Although at a classical level it might seem a mere technicality, it has profound implications
for the quantum theory and is the cornerstone of the whole SM. We explore its significance in some detail in the following. 
For computational purposes, it is convenient sometimes to (partially) fix the gauge freedom by imposing certain conditions on~$\phi$
and~$\mathbf{A}$. Two popular choices in classical electromagnetism are the Coulomb gauge~$\boldsymbol{\nabla}\cdot\mathbf{A}=0$ and 
the temporal (also called Weyl) 
gauge~$\phi=0$. These conditions still leave a residual invariance, generated in the first case
by harmonic functions~$\nabla^{2}\epsilon(t,\mathbf{r})=0$ and by time independent functions~$\epsilon(\mathbf{r})$ in the second. 
A covariant alternative is the Lorenz gauge
\begin{align}
\boldsymbol{\nabla}\cdot\mathbf{A}+{\partial\phi\over\partial t}=0,
\label{eq:Lorenz_gauge}
\end{align} 
preserved by gauge functions satisfying the wave equation,~$\Box\epsilon(t,\mathbf{r})=0$.

Gauge invariance introduces a {\em redundancy} in the description in terms of the 
electromagnetic potentials that however \label{pag:redundancy_gauge}
cannot reflect in physically measurable quantities such as the electric and magnetic fields. 
Although these are not the only 
gauge invariant quantities that can be constructed in terms of~$\varphi$ and~$\mathbf{A}$. 
There is also the Wilson loop, defined by
\begin{align}
U(\gamma)\equiv\exp\left(-i{\color{blue}e}\oint_{\gamma}d\mathbf{r}\cdot\mathbf{A}\right),
\end{align}
where~$\gamma$ is a closed path in space and~${\color{blue}e}$ the
electric charge. Implementing a gauge transformation on the vector potential
and using the Stokes theorem, we 
see that it is indeed gauge invariant
\begin{align}
\exp\left(-i{\color{blue}e}\oint_{\gamma}d\mathbf{r}\cdot\mathbf{A}\right)\longrightarrow 
\exp\left(-i{\color{blue}e}\oint_{\gamma}d\mathbf{r}\cdot\mathbf{A}+ie\oint_{\gamma}d\mathbf{r}\cdot\boldsymbol{\nabla}\epsilon\right)
=\exp\left(-i{\color{blue}e}\oint_{\gamma}d\mathbf{r}\cdot\mathbf{A}\right),
\label{eq:gauge_trans_em}
\end{align}
after taking into account that~$\gamma$ is closed. 
Whereas~$\mathbf{E}$ and~$\mathbf{B}$ 
are {\em local} observables depending on the spacetime point where they are measured, 
the Wilson loop is {\em nonlocal} since it ``explores'' the whole region enclosed by~$\gamma$. 

It is enlightening to study the consequences of gauge transformations for the dynamics of a quantum particle coupled to an electromagnetic field. 
In quantum mechanics the prescription of minimal coupling of a particle with 
electric charge~${\color{blue}e}$ to the electromagnetic field
\begin{align}
\mathbf{p}\longrightarrow \mathbf{p}-{\color{blue}e}\mathbf{A}, \hspace*{1cm} 
H\longrightarrow H+{\color{blue}e}\phi,
\end{align}
introduces an explicit dependence of the 
Schr\"odinger equation on the electromagnetic potentials
\begin{align}
i{\partial\psi\over\partial t}=\left[-{1\over 2m}\big(\boldsymbol{\nabla}-i{\color{blue}e}\mathbf{A}\big)^{2}+{\color{blue}e}\phi\right]\psi.
\end{align}
To preserve the gauge invariance of this equation, the transformations~\eqref{eq:gauge_trans_em} have to be supplemented by 
a phase shift of the wave function
\begin{align}
\psi(t,\mathbf{r})\longrightarrow e^{-i{\color{blue}e}\epsilon(t,\mathbf{r})}\psi(t,\mathbf{r}),
\label{eq:psi_gauge_trans}
\end{align}
which does not affect to the probability density~$|\psi(t,\mathbf{r})|^{2}$.
This shows that the gauge transformations in electromagnetism belong to the Abelian group~U(1)
of complex rotations, parametrized by elements
\begin{align}
U=e^{-i{\color{blue}e}\epsilon(t,\mathbf{r})},
\label{eq:U(1)_gen_trans}
\end{align}
in terms of which eq.~\eqref{eq:gauge_transf_em_1} reads
\begin{align}
\phi&\longrightarrow \phi+{i\over {\color{blue} e}}U^{-1}{\partial\over\partial t}U, 
\nonumber \\[0.2cm]
\mathbf{A}&\longrightarrow \mathbf{A}-{i\over {\color{blue} e}}U^{-1}\boldsymbol{\nabla}U. 
\end{align}

\begin{figure}[t]
\centerline{\includegraphics[scale=0.3]{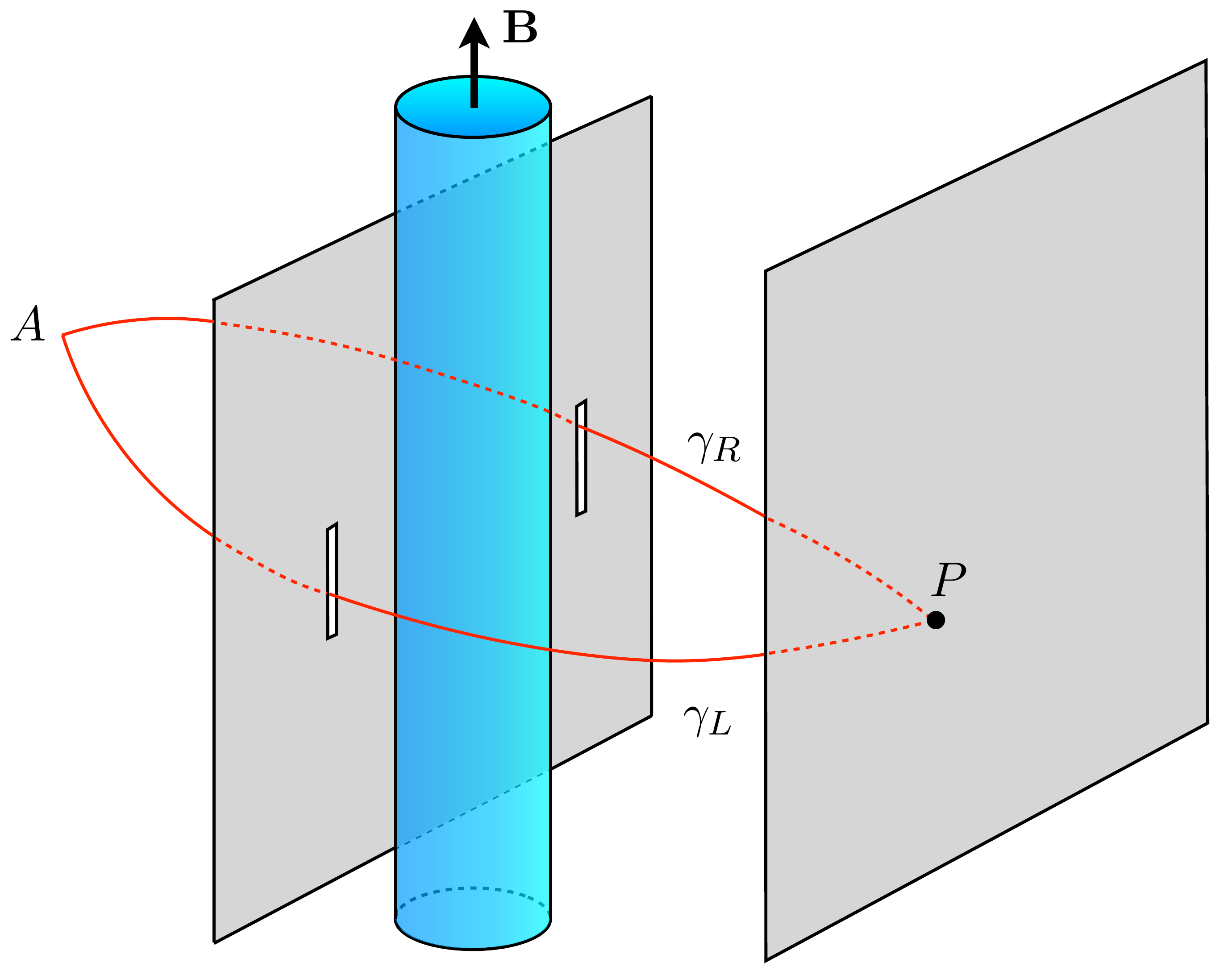}}
\caption[]{Experimental setup to exhibit the Aharonov-Bohm effect explained in Box~4.}
\label{fig:AB_exp}
\end{figure}

\begin{mdframed}[backgroundcolor=lightgray,hidealllines=true]
\vspace*{0.2cm}
\centerline{\greybox{\bf Wilson loops and quantum interference}}
\vspace*{0.2cm}

At the
classical level we can live with just local observables like the electric and magnetic fields, 
but not anymore when we introduce quantum effects. In this 
case the phase transformation of the wave function  
may give rise to observable interference phenomena. As we will see now, these are measured by a 
Wilson loop~$U(\gamma)$.

We work for simplicity in 
the temporal gauge~$\phi=0$. The action of a classical charged particle propagating
in the background of an electromagnetic potential~$\mathbf{A}(t,\mathbf{r})$ is given by
\begin{align}
S&={1\over 2}\int dt\,m\dot{\mathbf{r}}^{2}-{\color{blue}e}\int_{\gamma}d\mathbf{r}\cdot\mathbf{A},
\label{eq:action_charged_particle}
\end{align}
where~$\gamma$ is the particle trajectory and~${\color{blue}e}$ is the electron charge. 
An interesting property of the second term is that its value does not change
if we smoothly deform the path~$\gamma$ across any region where the magnetic field vanishes.
Let us consider two paths~$\gamma_{1}$ and~$\gamma_{2}$ joining two points~$A$ and~$B$ as shown here
\begin{align*}
\includegraphics[width=5cm]{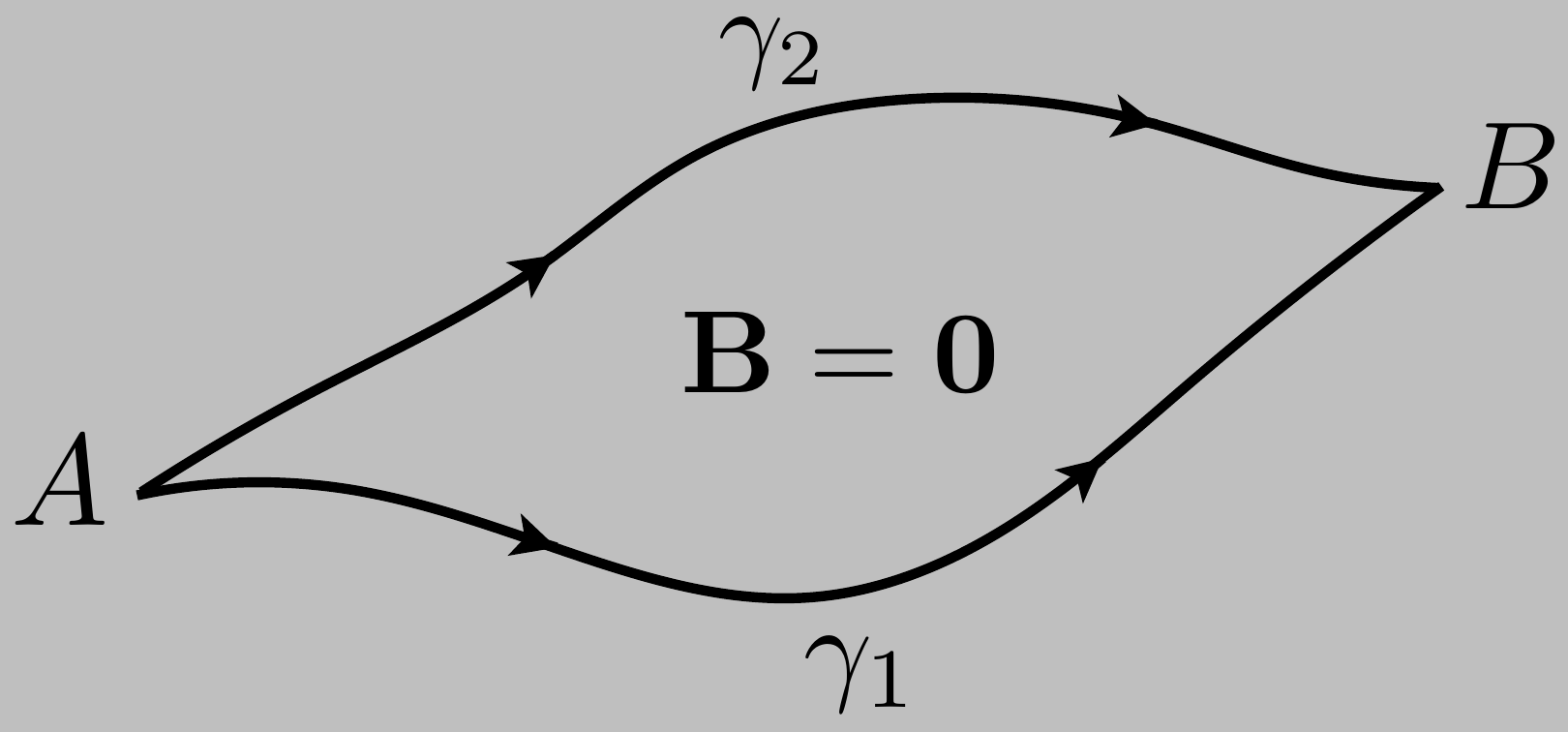}
\end{align*}
Computing the difference between the contributions of both paths, we find a Wilson loop
\begin{align}
\int_{\gamma_{1}}d\mathbf{r}\cdot\mathbf{A}-\int_{\gamma_{2}}d\mathbf{r}\cdot\mathbf{A}
&=\oint_{\gamma_{2}^{-1}\gamma_{1}}d\mathbf{r}\cdot\mathbf{A}=0,
\label{eq:topological_property_wilson_line}
\end{align}
where~$\gamma_{2}^{-1}\gamma_{1}$ represents the closed path from~$A$ to~$B$ following~$\gamma_{1}$ and
back to~$A$ along~$\gamma_{2}$. To see why this term is zero, let us denote by~$S$ any surface bounded 
by~$\gamma_{2}^{-1}\gamma_{1}$. Applying
the Stokes theorem, we have
\begin{align}
\oint_{\gamma_{2}^{-1}\gamma_{1}}d\mathbf{r}\cdot\mathbf{A}&=\int_{S}d\mathbf{S}\cdot(\boldsymbol{\nabla}\times\mathbf{A})=0,
\label{eq:AB_stokes_theorem}
\end{align}
since we assumed that~$\mathbf{B}=\boldsymbol{\nabla}\times\mathbf{A}=0$ in the
integration domain.

This topological property of the interaction term in~\eqref{eq:action_charged_particle} has important consequence in quantum mechanics, as pointed out by Yakir Aharonov and 
David Bohm\cite{AB}. 
Let us look at a double slit experiment performed with electrons 
in which behind the slitted screen we place a vertical solenoid confining a constant magnetic
field~$\mathbf{B}$ (see fig.~\ref{fig:AB_exp} in page~\pageref{page:AB_setup}). 
The amplitude for an electron emitted from~$A$ 
at~$t=0$ to be detected at a point~$P$ of the detection screen at~$t=\tau$ can be computed
as a coherent quantum superposition of all possible 
classical trajectories, expressed by the Feynman path integral
\begin{align}
G(\tau;\mathbf{r}_{A},\mathbf{r}_{P})&=\mathcal{N}\!\!\!\!\!\!\int\limits_{
\begin{array}{l} 
\!\!\scriptstyle \mathbf{r}(0)=\mathbf{r}_{A} \\[-0.2cm]
\!\!\scriptstyle \mathbf{r}(\tau)=\mathbf{r}_{P}
\end{array}
}\!\!\!\!\!\mathscr{D}\mathbf{r}\exp\left({i\over 2}\int_{0}^{\tau} dt\,m\dot{\mathbf{r}}^{2}
-i{\color{blue}e}\int_{\gamma}d\mathbf{r}\cdot\mathbf{A}\right),
\label{eq:path_integral_AB}
\end{align}
with~$\mathcal{N}$ a global normalization. The modulus squared of~$G(\tau;\mathbf{r}_{A},\mathbf{r}_{P})$
gives the probability of the electron being detected at the point~$P$ at time~$\tau$.

Recall that the magnetic field outside the solenoid is equal to zero and we can thus apply the 
topological property~\eqref{eq:topological_property_wilson_line} to conclude that 
the second term in the exponential of~\eqref{eq:path_integral_AB} takes the same value for {\em all}
trajectories~$\gamma_{L}$ passing through the left slit, and the same for {\em all} paths~$\gamma_{R}$ 
going through the right one. The total propagator can then be written as
\begin{align}
G(\tau;\mathbf{r}_{A},\mathbf{r}_{P})&=e^{-i{\color{blue}e}\int_{\gamma_{R}}d\mathbf{r}\cdot\mathbf{A}}
G_{R}(\tau;\mathbf{r}_{A},\mathbf{r}_{P})_{0}
+e^{-i{\color{blue}e}\int_{\gamma_{L}}d\mathbf{r}\cdot\mathbf{A}}
G_{L}(\tau;\mathbf{r}_{A},\mathbf{r}_{P})_{0}
\nonumber \\[0.2cm]
&=e^{-i{\color{blue}e}\int_{\gamma_{R}}d\mathbf{r}\cdot\mathbf{A}}\left[
G_{R}(\tau;\mathbf{r}_{A},\mathbf{r}_{P})_{0}
+e^{-i{\color{blue}e}\int_{\gamma_{L}}d\mathbf{r}\cdot\mathbf{A}
+i{\color{blue}e}\int_{\gamma_{R}}d\mathbf{r}\cdot\mathbf{A}}
G_{L}(\tau;\mathbf{r}_{A},\mathbf{r}_{P})_{0}\right],
\label{eq:propagator_AB_left_right}
\end{align}
where~$G_{R,L}(\tau;\mathbf{r}_{A},\mathbf{r}_{P})_{0}$ are the propagators for the electrons
going through the right (resp. left) slit in the absence of the solenoid. Now, although the global phase
disappears when computing the probability amplitude, 
the relative phase inside the brackets 
of the second line of~\eqref{eq:propagator_AB_left_right} contributes to the
interference pattern to be observed on the detection screen. Using the same arguments leading
to the result~\eqref{eq:topological_property_wilson_line}, we express this phase as the Wilson 
loop associated with the closed path~$\gamma_{R}^{-1}\gamma_{L}$
\begin{align}
\exp\left(-i{\color{blue}e}\int_{\gamma_{L}}d\mathbf{r}\cdot\mathbf{A}
+i{\color{blue}e}\int_{\gamma_{R}}d\mathbf{r}\cdot\mathbf{A}\right)
=\exp\left(-i{\color{blue}e}\oint_{\gamma_{R}^{-1}\gamma_{L}}d\mathbf{r}\cdot\mathbf{A}\right)
\equiv U(\gamma_{R}^{-1}\gamma_{L}).
\end{align}
It is important to keep in mind that~$\gamma_{R}^{-1}\gamma_{L}$ represents {\em any} closed path
going through both slits and enclosing the solenoid. 
To evaluate this Wilson loop let us take a bird's-eye view of the Aharonov-Bohm 
experimental setup in fig.~\ref{fig:AB_exp}, that we schematically represent as:
\begin{align*}
\includegraphics[width=7cm]{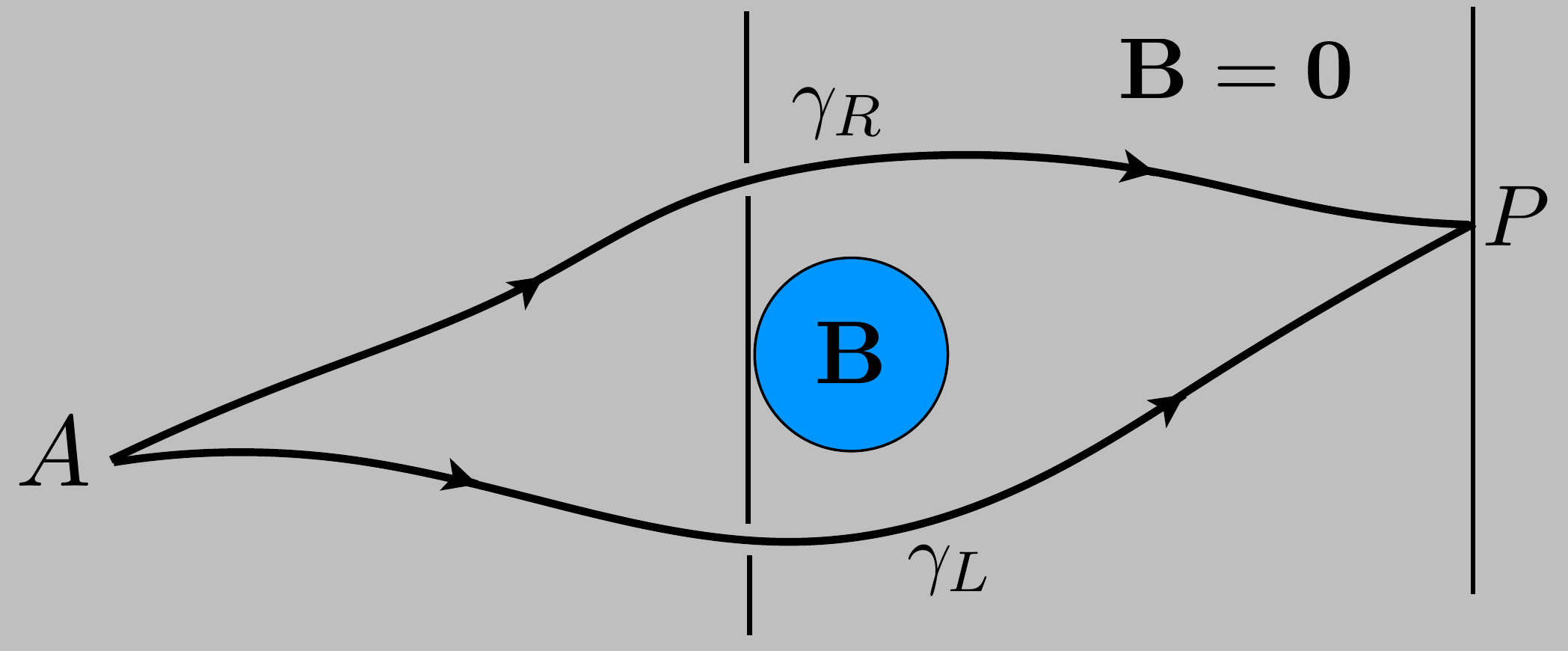}
\end{align*}
Should we apply the Stokes theorem to the calculation of~$U(\gamma_{R}^{-1}\gamma_{L})$ 
as we did in eq.~\eqref{eq:AB_stokes_theorem}, the resulting integral would not be zero anymore. 
As we see, the surface~$S$ enclosed by the loop is now pierced by the solenoid, and the magnetic 
field~$\mathbf{B}=\boldsymbol{\nabla}\times\mathbf{A}$ is not
zero everywhere. Instead
\begin{align}
\oint_{\gamma_{R}^{-1}\gamma_{L}}d\mathbf{r}\cdot\mathbf{A}=
\int_{S}d\mathbf{S}\cdot\mathbf{B}=\Phi,
\end{align}
where~$\Phi$ is the magnetic flux inside the solenoid and we have
\begin{align}
U(\gamma_{R}^{-1}\gamma_{L})=e^{-i{\color{blue}e}\Phi}\neq 1.
\end{align}
Hence, the presence of the solenoid modifies the interference pattern on the screen, even if the electrons
never enter the region where the magnetic field is nonzero. The reason is that
even if~$\mathbf{B}=\mathbf{0}$ outside,~$\mathbf{A}$ is not. Although no force is applied
to them, the electrons interact with
the vector potential whose global structure, codified in the nonlocal gauge-invariant
quantity~$U(\gamma_{R}^{-1}\gamma_{L})$, contains information about the confined magnetic field.

\end{mdframed}

\label{page:AB_setup}

Going back to the Maxwell's equations~\eqref{eq:maxwell_eqs_general}, we notice that the vacuum 
equations (with all blue and red terms removed) exhibit an interesting symmetry.
Combining the electric and magnetic fields into a single complex field~$\mathbf{E}+i\mathbf{B}$, 
the four equations can be summarized as
\begin{align}
\boldsymbol{\nabla}\cdot (\mathbf{E}+i\mathbf{B})&=0, \nonumber \\[0.2cm]
\boldsymbol{\nabla}\times (\mathbf{E}+i\mathbf{B})-i{\partial\over\partial t}(\mathbf{E}+i\mathbf{B})
&=0.
\label{eq:E+iB_vac_eqs}
\end{align}
Both identities remain invariant under the transformation
\begin{align}
\mathbf{E}+i\mathbf{B}\longrightarrow e^{i\theta}\big(\mathbf{E}+i\mathbf{B}\big),
\label{eq:em_duality_fields}
\end{align}
with~$\theta$ a real global angle. To be more specific, 
splitting the previous equation into its real and imaginary parts, we find
\begin{align}
\mathbf{E}&\longrightarrow \mathbf{E}\cos\theta-\mathbf{B}\sin\theta, \nonumber \\[0.2cm]
\mathbf{B}&\longrightarrow \mathbf{E}\sin\theta+\mathbf{B}\cos\theta,
\label{eq:em_duality_components}
\end{align}
which for~$\theta={\pi\over 2}$ interchanges electric and magnetic 
fields~$(\mathbf{E},\mathbf{B})\rightarrow (-\mathbf{B},\mathbf{E})$.

This electric-magnetic duality of the vacuum equations
is however broken by the source terms in the ``textbook'' Maxwell's equations 
[i.e., eq.~\eqref{eq:maxwell_eqs_general} without the terms in red]. The 
identities~\eqref{eq:E+iB_vac_eqs} are then recast as\label{page:monopoles}
\begin{align}
\boldsymbol{\nabla}\cdot (\mathbf{E}+i\mathbf{B})&={\color{blue}\rho_{e}}, \nonumber \\[0.2cm]
\boldsymbol{\nabla}\times (\mathbf{E}+i\mathbf{B})
-i{\partial\over\partial t}(\mathbf{E}+i\mathbf{B})
&=i{\color{blue} \mathbf{j}_{e}}.
\end{align}
Since~${\color{blue}\rho_{e}}$ and~${\color{blue} \mathbf{j}_{e}}$ are both real quantities, 
the only transformations preserving these equations are the trivial ones which either leave invariant
the electric and magnetic fields or reverse their signs (corresponding respectively 
to~$\theta=0,\pi$), the latter one also requiring the reversal of 
the sign of~${\color{blue}\rho_{e}}$ and~${\color{blue} \mathbf{j}_{e}}$.
Physically this makes sense, since as far as we know there is 
a fundamental asymmetry in nature between electric
and magnetic fields. While the first are sourced by point charges (electric monopoles)
at which field lines either begin or end, magnetic
fields are associated with the motion of electric charges and their field lines
always close on themselves. 
Restoring electric-magnetic duality in the Maxwell's equations 
requires treating the sources of both 
fields symmetrically, which means introducing magnetic charge density
and current. These are the terms in red in eq.~\eqref{eq:maxwell_eqs_general}, 
that we rewrite now as
\begin{align}
\boldsymbol{\nabla}\cdot (\mathbf{E}+i\mathbf{B})&={\color{blue}\rho_{e}}
+i{\color{red}\rho_{m}}, \nonumber \\[0.2cm]
\boldsymbol{\nabla}\times (\mathbf{E}+i\mathbf{B})
-i{\partial\over\partial t}(\mathbf{E}+i\mathbf{B})
&=i\big({\color{blue} \mathbf{j}_{e}}+i{\color{red}\mathbf{j}_{m}}\big).
\end{align}
These equations remain invariant under electric-magnetic duality~\eqref{eq:em_duality_fields}
when supplemented by a corresponding rotation of the sources
\begin{align}
{\color{blue}\rho_{e}}
+i{\color{red}\rho_{m}}&\longrightarrow e^{i\theta}\big({\color{blue}\rho_{e}}
+i{\color{red}\rho_{m}}\big), \nonumber \\[0.2cm]
{\color{blue} \mathbf{j}_{e}}+i{\color{red}\mathbf{j}_{m}}&\longrightarrow
e^{i\theta}\big({\color{blue} \mathbf{j}_{e}}+i{\color{red}\mathbf{j}_{m}}\big).
\label{eq:em_duality_charges}
\end{align}
For~$\theta={\pi\over 2}$ the interchange of electric and magnetic fields is accompanied by 
a swap of the electric and magnetic 
sources,~$({\color{blue}\rho_{e}},{\color{blue} \mathbf{j}_{e}})
\rightarrow (-{\color{red}\rho_{m}},-{\color{red}\mathbf{j}_{m}})$
and~$({\color{red}\rho_{m}},{\color{red}\mathbf{j}_{m}})\rightarrow 
({\color{blue}\rho_{e}},{\color{blue} \mathbf{j}_{e}})$. 

\begin{figure}[t]
\centerline{\includegraphics[scale=0.4]{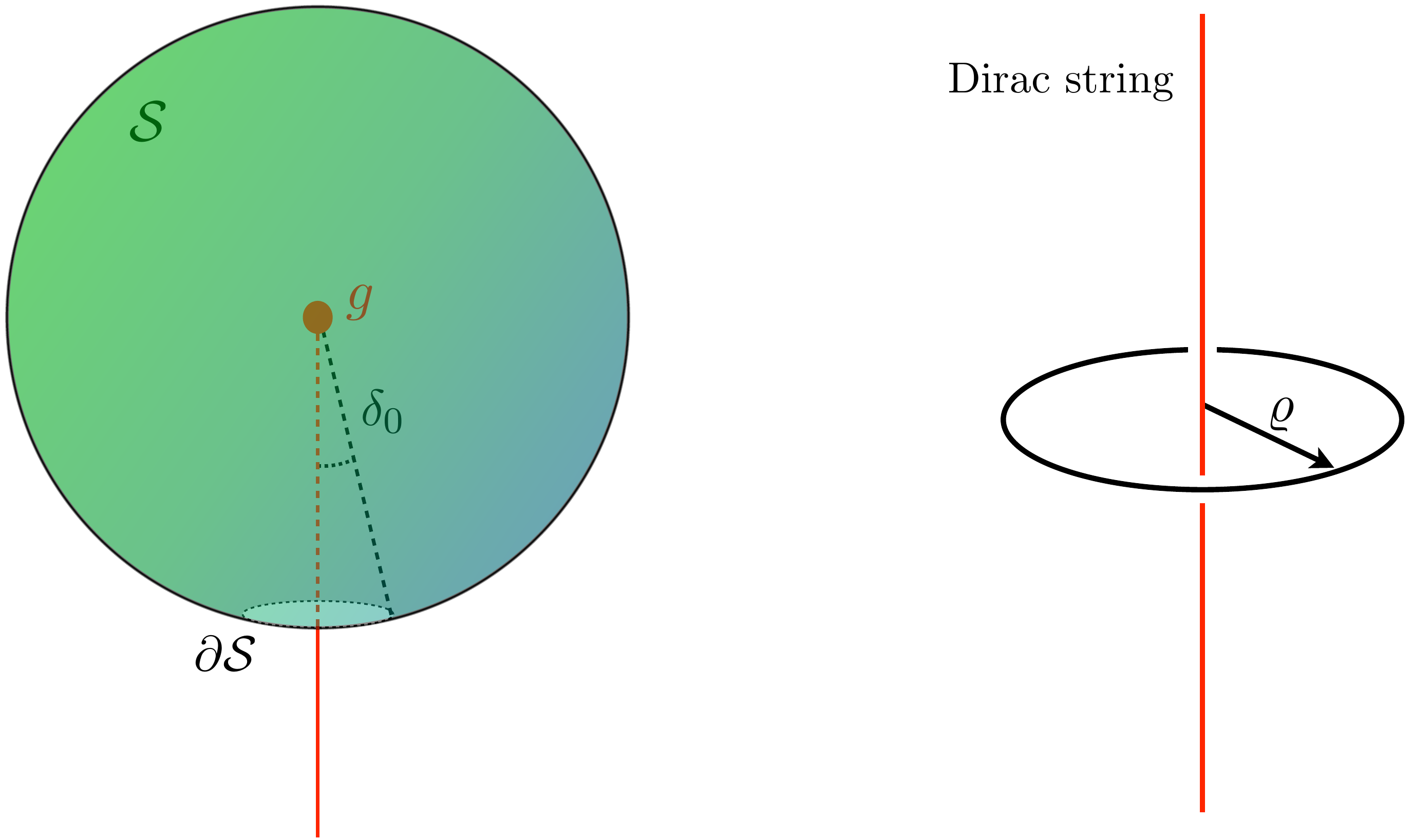}}
\caption[]{Left: section of a sphere around a Dirac magnetic monopole with charge~${\color{red} g}$,
resulting from cutting out a region around the south pole.
Its boundary~$\partial\mathcal{S}$ surrounds the singular Dirac string located along~$\theta=\pi$ (in red).
Right: closed path surrounding the Dirac string.}
\label{fig:monopole}
\end{figure}

The consequences of having particles with magnetic charge were first explored by
Dirac in~\cite{dirac_monopole}. Let us assume the existence of 
a point magnetic
source that for simplicity we locate at the 
origin,~${\color{red}\rho_{m}}={\color{red} g}\delta^{(3)}(\mathbf{r})$. The second equation
in~\eqref{eq:maxwell_eqs_general} leads to
\begin{align}
\boldsymbol{\nabla}\cdot\mathbf{B}={\color{red}g}\delta^{(3)}(\mathbf{r})
\hspace*{1cm} \Longrightarrow \hspace*{1cm} \mathbf{B}(\mathbf{r})={1\over 4\pi}{{\color{red}g}
\over r^{2}}\mathbf{u}_{r},
\label{eq:magnetic_field_monopole}
\end{align}
which would be a magnetic analog of the Coulomb field. An important point to consider is that, despite the source's presence, the magnetic field's divergence still vanishes everywhere except at the monopole's position. 
As a consequence, away from this point 
we can still write~$\mathbf{B}=\boldsymbol{\nabla}\times\mathbf{A}$, which is solved by
\begin{align}
\mathbf{A}(\mathbf{r})
={1\over 4\pi}{{\color{red} g}\over r}\tan\left({\theta\over 2}\right)\mathbf{u}_{\varphi},
\label{eq:vector_potential_monopole}
\end{align}  
where we are using spherical coordinates~$(r,\varphi,\theta)$. This vector potential is singular
not only at the monopole location at~$\mathbf{r}=0$, but all along the line~$\theta=\pi$ as well.
The existence of this singular Dirac string should not be a surprise. Were~$\mathbf{A}(\mathbf{r})$ 
be regular everywhere outside the origin, we could
apply the Stokes theorem to the integral giving the 
magnetic flux across a closed surface~$\mathcal{S}$ enclosing the monopole, 
to find
\begin{align}
\int_{\mathcal{S}}d\mathbf{S}\cdot\mathbf{B}&=\int_{\mathcal{S}}d\mathbf{S}\cdot(\boldsymbol{\nabla}\times
\mathbf{A})=\oint_{\partial \mathcal{S}}d\boldsymbol{\ell}\cdot\mathbf{A}=0,
\label{eq:intSdsB=0}
\end{align}
since~$\partial \mathcal{S}=\emptyset$. This would contradict the calculation of the same integral
applying Gauss' theorem
\begin{align}
\int_{\mathcal{S}}d\mathbf{S}\cdot\mathbf{B}=\int_{\mathcal{B}_{3}}\boldsymbol{\nabla}\cdot\mathbf{B}
={\color{red} g}\neq 0,
\label{eq:intSdsBGauss}
\end{align}
where~$\mathcal{B}_{3}$ denotes the three-dimensional region 
bounded by~$\mathcal{S}$ and containing the monopole. 
Notice that this second calculation is free of
trouble, since the magnetic field~\eqref{eq:magnetic_field_monopole} is regular everywhere 
on~$\mathcal{S}$.
The catch of course is that the vector potential is singular at~$\theta=\pi$ and
the surface~$\mathcal{S}$ in~\eqref{eq:intSdsB=0} cannot be closed.  
As shown on the left of fig.~\ref{fig:monopole}, its boundary is a circle surrounding the singularity
and the integral gives a nonzero result
\begin{align}
\oint_{\partial \mathcal{S}}d\boldsymbol{\ell}\cdot \mathbf{A}&=
{1\over 2}{{\color{red} g}}\sin{\delta_{0}}\tan\left({\delta_{0}\over 2}\right)
\stackrel{\delta_{0}\rightarrow 0}{\xrightarrow{\hspace{1cm}}} {\color{red} g},
\label{eq:ointdlAlimit}
\end{align}
where the last limit corresponds to shrinking the boundary to a point,
reproducing the result 
of eq.~\eqref{eq:intSdsBGauss}.

Even if mathematically unavoidable, the existence of a singularity is always a source of concern in 
physics. A way to restore our peace of mind in this case might be to make the Dirac string 
an artefact that somehow is rendered unobservable. One may think that a 
way to accomplish this is to apply a gauge transformation, 
since the vector potential is not uniquely defined. This, however, does not eliminate
the Dirac string, just changes its location. 

Let us look a bit closer at the vector potential~\eqref{eq:vector_potential_monopole} near
the Dirac string. Denoting by~$\varrho$ the linear distance to the string (see the right 
of fig.~\ref{fig:monopole}), 
in the limit~$\varrho\rightarrow 0$ we can write
\begin{align}
\mathbf{A}\approx {1\over 2\pi}{{\color{red} g}\over \varrho}\mathbf{u}_{\varphi}.
\label{eq:A_solenoid_Dirac}
\end{align}
This expression should be familiar from elementary electrodynamics, since it represents the 
vector potential outside an infinite solenoid. The Dirac string can be pictured then 
as an infinitely thin solenoid pumping magnetic flux into the monopole which, 
according to the limiting value
of the integral in eq.~\eqref{eq:ointdlAlimit}, is actually equal to 
the outgoing flux through a closed surface surrounding the monopole.

In Box~4 we learned a way to ``detect solenoids'' by their imprints on the wave function of 
charged quantum particles detectable by interference experiments. 
The Wilson loop of a particle with electric charge~${\color{blue} e}$ going around the
Dirac string is computed from the 
vector potential~\eqref{eq:A_solenoid_Dirac} and gives [see also eq.~\eqref{eq:ointdlAlimit}]
\begin{align}
U(\gamma)=\exp\left(-i{\color{blue} e}\oint_{\gamma}d\boldsymbol{\ell}\cdot\mathbf{A}\right)=e^{-i{\color{blue} e}{\color{red} g}}.
\end{align}
The absence of detectable interference requires this phase to be equal to one for any
electrically charged particle, which amounts to the condition
\begin{align}
{\color{blue} e}{\color{red} g}=2\pi n \hspace*{1cm} \Longrightarrow 
\hspace*{1cm} {\color{blue} e}={2\pi \over {\color{red} g}}n.
\label{eq:charge_quantization_cond}
\end{align}
with~$n$ an integer. This is a very interesting result, stating that the existence of 
a single magnetic monopole anywhere in the universe implies by consistency 
that electric charges have to be {\em quantized}. The quantization 
condition~\eqref{eq:charge_quantization_cond} remains invariant under electric-magnetic duality
with~$\theta={\pi\over 2}$.

Unconfirmed sightings in cosmic rays notwithstanding~\cite{cabrera_monopole,price_et_al_monopole}, 
no evidence exists of magnetically charged particles at the energies explored. They are, however,
an almost ubiquitous prediction of many theories beyond the SM, where they usually emerge as 
solitonic objects resulting from the spontaneous breaking in unified field theories 
leaving behind unbroken U(1)'s. Although they acquire masses of the order of the symmetry breaking scale, 
magnetic monopoles should have been 
created in huge amounts at the early stages of the universe's history. 
One of the original aims of cosmological inflation models was to dilute their presence in the
early universe, thus accounting for its apparent absence.

\begin{mdframed}[backgroundcolor=lightgray,hidealllines=true]
\vspace*{0.2cm}
\centerline{\greybox{\bf Magnetic monopoles from topology}}
\vspace*{0.2cm}
\label{pag:box_monopoles}

The origin of all our troubles with the Dirac monopole was after all {\em topological}: 
although the vector potential of the magnetic monopole is locally well defined anywhere away from
the origin, it cannot be extended globally to the 
sphere surrounding the monopole. There is however a way to avoid the singular Dirac string which was pointed out 
by Tai Tsun Wu and Chen Ning Yang~\cite{wu_yang}. When computing the flux integral~\eqref{eq:intSdsBGauss}, 
instead of covering the sphere with a single patch cutting out the region around the place where the Dirac string crosses the surface (in our 
case, the south pole), we can be more sophisticated and use two patches respectively centered at the north and south poles and overlapping at the equator. 
This is what we represent in the picture below, with~$D_{\pm}$ the upper and lower hemispheres glued together along their 
respective boundaries~$S_{\pm}^{1}$
\begin{align*}
\includegraphics[width=4.5cm]{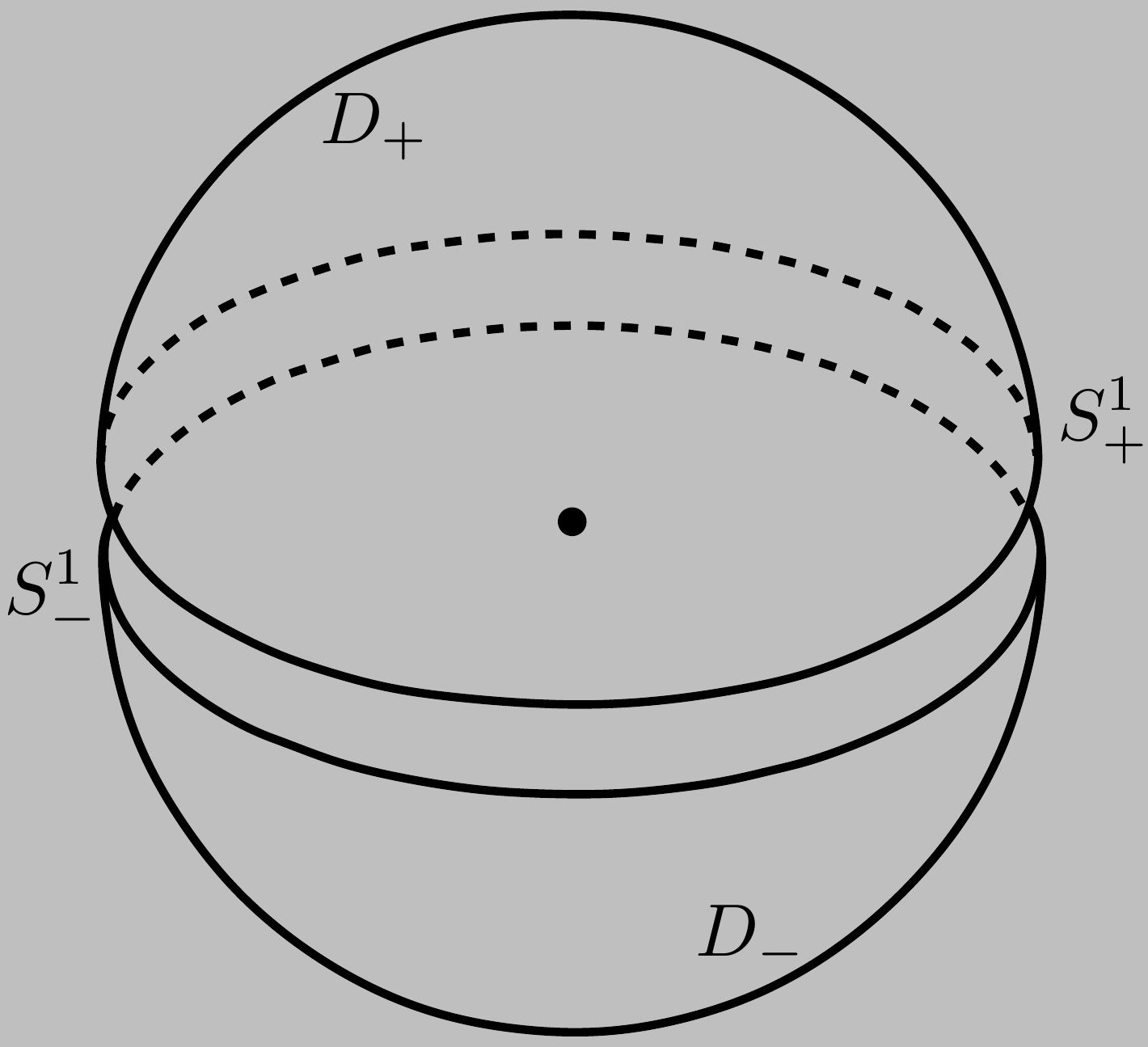}
\end{align*}
On both~$D_{+}$ and~$D_{-}$ we can write a vector potential whose curls reproduce the expression
of the monopole field~\eqref{eq:magnetic_field_monopole}
\begin{align}
\mathbf{A}(\mathbf{r})_{+}
&={1\over 4\pi}{{\color{red} g}\over r}\tan\left({\theta\over 2}\right)\mathbf{u}_{\varphi}
\hspace*{1.2cm} 0\leq \theta \leq {\pi\over 2},
\nonumber \\[0.2cm]
\mathbf{A}(\mathbf{r})_{-}
&=-{1\over 4\pi}{{\color{red} g}\over r}\cot\left({\theta\over 2}\right)\mathbf{u}_{\varphi}
\hspace*{1cm} {\pi\over 2}\leq \theta \leq \pi.
\end{align}
The important point here is that both expressions are perfectly regular in their respective domains, so our 
vector potential is regular everywhere on the sphere~$S^{2}=D_{+}\cup D_{-}$. An apparent 
obstacle arises 
in their overlap at the equator~$\theta={\pi\over 2}$, where the two expressions
do not agree 
\begin{align}
\mathbf{A}(\mathbf{r})_{+}\Big|_{S_{+}^{1}}
-\mathbf{A}(\mathbf{r})_{-}\Big|_{S_{-}^{1}}={1\over 2\pi}{{\color{red} g}\over r}\mathbf{u}_{\varphi}.
\end{align}
This is however not a problem, since as we know the vector potential is not uniquely defined. 
It is physically 
acceptable that the identification of the vector potentials at the equator is made modulo a gauge transformation,
which is indeed the case here
\begin{align}
\epsilon=-{{\color{red} g}\over 2\pi}\varphi \hspace*{1cm} \Longrightarrow 
\hspace*{1cm} \mathbf{A}(\mathbf{r})_{+}\Big|_{S_{+}^{1}}
=\mathbf{A}(\mathbf{r})_{-}\Big|_{S_{-}^{1}}-\boldsymbol{\nabla}\epsilon.
\label{eq:WY_deltaA_+-}
\end{align}
The magnetic flux
due to the magnetic monopole at its center can be evaluated using these expressions as
\begin{align}
\int_{\mathcal{S}^{2}}d\mathbf{S}\cdot\mathbf{B}&=\int_{D_{+}}d\mathbf{S}\cdot(\boldsymbol{\nabla}
\times\mathbf{A}_{+})+\int_{D_{-}}d\mathbf{S}\cdot(\boldsymbol{\nabla}
\times\mathbf{A}_{-})\nonumber \\[0.2cm]
&=\oint_{S_{+}^{1}}d\boldsymbol{\ell}\cdot\mathbf{A}_{+}+\oint_{S_{-}^{1}}d\boldsymbol{\ell}
\cdot\mathbf{A}_{-} \\[0.2cm]
&=\epsilon(2\pi)-\epsilon(0)={\color{red} g}. 
\nonumber
\end{align}
correctly reproducing~\eqref{eq:intSdsBGauss}. Notice that the two boundaries~$S^{1}_{\pm}
=\partial D_{\pm}$ have
opposite orientations, so using eq.~\eqref{eq:WY_deltaA_+-} the second line combines into a single 
integral of~$\epsilon'(\varphi)$ from~$0$ to~$2\pi$.

The gauge function~$\epsilon(\varphi)$ relating the vector potentials along the equator
is not single-valued on~$S^{1}$. This might pose a 
problem in the presence of quantum charged particles, since
their wave functions also change under gauge transformations 
[see eq.~\eqref{eq:psi_gauge_trans}]. In order to avoid multivaluedness of the wave function, we must
require
\begin{align}
e^{-i{\color{blue}e}\epsilon(0)}=e^{-i{\color{blue}e}\epsilon(2\pi)} 
\hspace*{1cm} \Longrightarrow \hspace*{1cm}
e^{i{\color{blue}e}{\color{red} g}}=1,
\end{align}
and the Dirac quantization condition~\eqref{eq:charge_quantization_cond} is retrieved. Alternatively,
we can also notice that under a gauge transformation the action of a particle
moving along the equator changes by~$\Delta S=-{\color{blue}e}{\color{red}g}$, 
as can be easily checked from eq.~\eqref{eq:action_charged_particle}.
This has no effect in the Feynman path integral 
provided~${\color{blue}e}{\color{red}g}=2\pi n$, with~$n\in\mathbb{Z}$, and the same
result is obtained.

The Wu-Yang construction highlights the topological structure underlying the magnetic monopole. 
Implementing the quantization 
condition~${\color{blue}e}{\color{red}g}=2\pi n$, the~U(1) 
transformation~\eqref{eq:WY_deltaA_+-}
relating the vector potential of both hemispheres takes the form [cf.~\eqref{eq:U(1)_gen_trans}]
\begin{align}
U=e^{in\varphi}.
\label{eq:U_monopole}
\end{align}
Since~U(1) is the multiplicative group of complex phases, it can be identified with the unit circle. 
As we move once along the equator and the azimuthal 
angle~$\varphi$ changes from~$0$ to~$2\pi$, the gauge transformation~\eqref{eq:U_monopole} 
wraps $n$~times around U(1), as we illustrate here for the particular case~$n=3$
\begin{align*}
\includegraphics[width=4.5cm]{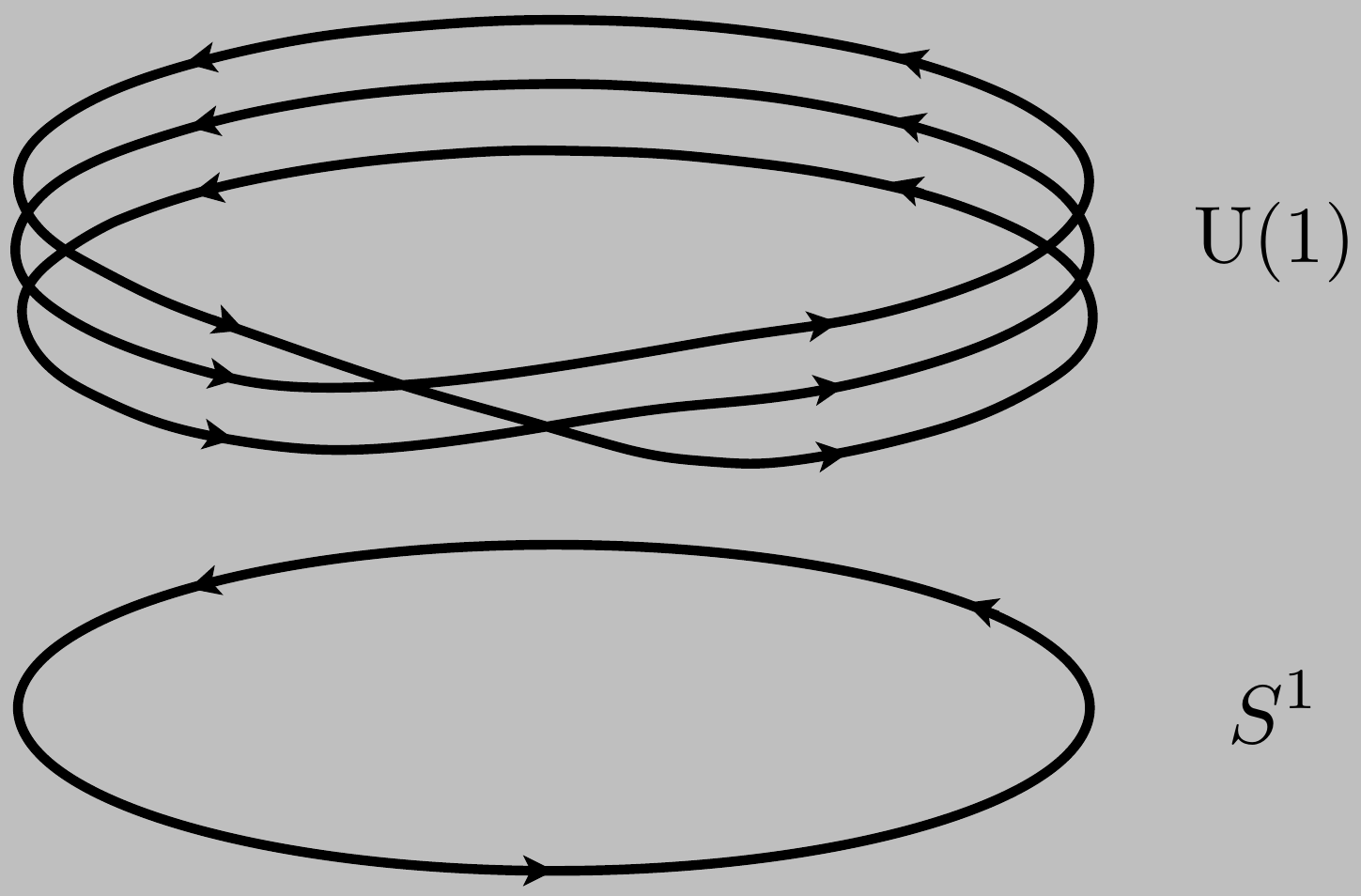}
\end{align*}
More technically speaking, when mapping the circle~$S^{1}$ onto~U(1) we encounter infinitely many
sectors that cannot be smoothly deformed into one another and are distinguished by
how many times the circle wraps around~U(1). The corresponding integer is an
element of the first homotopy group~$\pi_{1}[\mbox{U(1)}]=\mathbb{Z}$ classifying the 
continuous maps~$U:S^{1}\rightarrow\mbox{U(1)}$ (see, for 
example,~\cite{Azcarraga_Izquierdo,Nakahara,Nash,Frankel}
for physicist-oriented overviews of basic concepts in differential geometry).

This should not come as a surprise. After all, at face value, our insistence in expressing the magnetic field 
as the curl of the vector potential is incompatible with having a nonvanishing
value for $\boldsymbol{\nabla}\cdot\mathbf{B}$ as in eq.~\eqref{eq:magnetic_field_monopole}.
To reconcile these two facts we have to assume that although~$\mathbf{B}=\boldsymbol{\nabla}\times\mathbf{A}$ 
is valid on a contractible 
coordinate patch, there is no vector field~$\mathbf{A}$ globally defined on the sphere 
with this property. This is why in our case the topologically 
trivial configuration~$n=0$ corresponds to zero magnetic
charge and a vanishing magnetic field.

\end{mdframed}

Looking at the symmetries of classical electrodynamics, we notice one conspicuously absent
from the Maxwell's equations~\eqref{eq:maxwell_eqs_general}: Galilean invariance. It is amusing
that Maxwell composed a fully relativistic invariant field theory some forty years before Einstein's
formulation of special relativity. It took the latter's genius to realize that the tension between
classical mechanics and electrodynamics was to be solved giving full credit to the Maxwell's equations
and their spacetime symmetries. The price to pay was to modify Newtonian mechanics
to make it applicable
to systems involving velocities close to the speed of light. 

\subsection{Quantum electromagnetism}

The easiest way to show the relativistic invariance of the Maxwell's equations is to rewrite them
as tensor equations with respect to Poincar\'e transformations. To do so, we 
combine the scalar and vector electromagnetic
potentials into a single four-vector
\begin{align}
A^{\mu}\equiv (\phi,\mathbf{A}),
\end{align} 
while electric and magnetic fields are codified in the field strength two-tensor
\begin{align}
F_{\mu\nu}\equiv \partial_{\mu}A_{\nu}-\partial_{\nu}A_{\mu}.
\label{eq:field_strength_QED}
\end{align}
The latter can be explicitly computed to be
\begin{align}
F_{\mu\nu}=\left(
\begin{array}{cccc}
0 & E_{x} & E_{y} & E_{z} \\
-E_{x} & 0 & -B_{z} & B_{y} \\
-E_{y} & B_{z} & 0 & -B_{x} \\
-E_{z} & -B_{y} & B_{x} & 0
\end{array}
\right),
\label{eq:field_strength_matrix}
\end{align}
where~$\mathbf{E}=(E_{x},E_{y},E_{z})$ and~$\mathbf{B}=(B_{x},B_{y},B_{z})$.
The gauge transformations~\eqref{eq:gauge_transf_em_1} are now expressed in the more compact form
\begin{align}
A_{\mu}\longrightarrow A_{\mu}+\partial_{\mu}\epsilon,
\label{eq:gauge_freedom_cov}
\end{align}
which obviously leave~$F_{\mu\nu}$ invariant. It is also convenient to define the dual field
strength
\begin{align}
\widetilde{F}_{\mu\nu}={1\over 2}\epsilon_{\mu\nu\alpha\beta}F^{\alpha\beta},
\label{eq:dual_F_tensor_EM}
\end{align}
whose components are obtained from~\eqref{eq:field_strength_matrix} by 
replacing~$\mathbf{E}\rightarrow \mathbf{B}$ and~$\mathbf{B}\rightarrow -\mathbf{E}$.
Charge densities and currents are also merged into four-vectors
\begin{align}
{\color{blue} j_{e}^{\mu}}&\equiv({\color{blue} \rho_{e}},{\color{blue} \mathbf{j}_{e}}), \nonumber \\[0.2cm]
{\color{red} j_{m}^{\mu}}&\equiv({\color{red} \rho_{m}},{\color{red} \mathbf{j}_{m}}),
\end{align}
in terms of which the four Maxwell's 
equations~\eqref{eq:maxwell_eqs_general} are recast as
\begin{align}
\partial_{\mu}F^{\mu\nu}&={\color{blue} j_{e}^{\nu}}, \nonumber \\[0.2cm]
\partial_{\mu}\widetilde{F}^{\mu\nu}&={\color{red} j_{m}^{\nu}}.
\label{eq:Maxwell_covariant}
\end{align}
Some comments about the magnetic current are in order here. It should be noticed that the
definition~\eqref{eq:field_strength_QED} automatically implies the Bianchi identity
\begin{align}
\partial_{\mu}\widetilde{F}^{\mu\nu}
={1\over 2}\epsilon^{\nu\sigma\alpha\beta}\partial_{\sigma}F_{\alpha\beta}
=\epsilon^{\nu\sigma\alpha\beta}\partial_{\sigma}\partial_{\alpha}A_{\beta}=0,
\end{align}
contradicting the second equation in~\eqref{eq:Maxwell_covariant}.
In fact, we have already encountered this problem in its noncovariant version when discussing
magnetic monopoles: writing~$\mathbf{B}=\boldsymbol{\nabla}\times\mathbf{A}$ is incompatible
with having~$\boldsymbol{\nabla}\cdot\mathbf{B}\neq 0$. The solution given there is also applicable here.
What happens is that~\eqref{eq:field_strength_QED} is valid locally but {\em not globally}.
Magnetic monopoles can be described using the vector potential~$A_{\mu}$, 
but the gauge field configuration needs to be topologically nontrivial.

The tensors~$F_{\mu\nu}$ and~$\widetilde{F}_{\mu\nu}$ can be used to construct quantities that
are relativistic invariant. By contracting them, we find the two invariants
\begin{align}
F_{\mu\nu}F^{\mu\nu}&=\widetilde{F}_{\mu\nu}\widetilde{F}^{\mu\nu}
=-2\big(\mathbf{E}^{2}-\mathbf{B}^{2}\big), \nonumber \\[0.2cm]
F_{\mu\nu}\widetilde{F}^{\mu\nu}&=2\mathbf{E}\cdot\mathbf{B}.
\label{eq:two_invariants_EB}
\end{align}
This implies that the complex combinations
\begin{align}
\big(\mathbf{E}\pm i\mathbf{B}\big)^{2}=\mathbf{E}^{2}-\mathbf{B}^{2}\pm 2i\mathbf{E}\cdot\mathbf{B},
\end{align}
also remain invariant under the Lorentz group\footnote{They change however under electric-magnetic 
duality, which mixes the two quantities introduced in~\eqref{eq:two_invariants_EB}.}.
The present discussion is very relevant for building an action principle for classical
electrodynamics. In particular, noticing that~$F_{\mu\nu}\widetilde{F}^{\mu\nu}=2\partial_{\mu}
(A_{\nu}F^{\mu\nu})$ is a total derivative, the obvious choice is
\begin{align}
S&=\int d^{4}x\left(-{1\over 4}F_{\mu\nu}F^{\mu\nu}+j^{\mu}A_{\mu}\right) \nonumber \\[0.2cm]
&=\int dt d^{3}x\left[{1\over 2}\big(\mathbf{E}^{2}-\mathbf{B}^{2}\big)
+\rho\phi-\mathbf{j}\cdot\mathbf{A}\right],
\label{eq:maxwell_action}
\end{align}
which is also gauge invariant provided charge is conserved,~$\partial_{\mu}j^{\mu}=0$. 
Since from now on we will ignore the
presence of magnetic charges, we drop the color code used so far, as well as the subscript in 
the electric density and current. 

Although obtaining the Maxwell field equations from the action in~\eqref{eq:maxwell_action} 
is straightforward,
the canonical formalism is tricky. The reason is that~$\dot{\phi}$ does not appear in the 
action and as a consequence the momentum conjugate to~$A_{0}$ is identically zero. Thus, 
we have a constrained 
system that has to be dealt with using Dirac's formalism (see, for example,~\cite{AG_VM} for 
the details). At a practical level, we regard~$\mathbf{A}$ and~$\mathbf{E}$ as 
a pair of canonically conjugated variables
\begin{align}
\big\{A_{i}(t,\mathbf{r}),E_{j}(t,\mathbf{r}')\big\}_{\rm PB}=\delta_{ij}
\delta^{(3)}(\mathbf{r}-\mathbf{r}').
\label{eq:AEPB}
\end{align}
Using~$\dot{\mathbf{A}}=-\mathbf{E}
-\boldsymbol{\nabla}\phi$, we construct the Hamiltonian
\begin{align}
H&=\int dtd^{3}x\left[-\dot{\mathbf{A}}\cdot\mathbf{E}-{1\over 2}\big(\mathbf{E}^{2}-\mathbf{B}^{2}\big)
-\rho\phi+\mathbf{j}\cdot\mathbf{A}\right] \nonumber \\[0.2cm]
&=\int dt d^{3}x\left[{1\over 2}\big(\mathbf{E}^{2}+\mathbf{B}^{2}\big)+\phi\big(\boldsymbol{\nabla}
\cdot\mathbf{E}-\rho\big)+\mathbf{j}\cdot\mathbf{A}\right],
\label{eq:hamiltonian_electrodynamics}
\end{align}
where the term~$-\mathbf{E}\cdot\boldsymbol{\nabla}\phi$ has been integrated by parts and the
substitution~$\mathbf{B}=\boldsymbol{\nabla}\times\mathbf{A}$ is understood. 
Gauss' law~$\boldsymbol{\nabla}\cdot\mathbf{E}=\rho$ emerges as a constraint preserved by time evolution
\label{page:gauss_law_constraint}
\begin{align}
\big\{\boldsymbol{\nabla}\cdot\mathbf{E}-\rho,H\big\}_{\rm PB}
=-\boldsymbol{\nabla}\cdot\mathbf{j}-\dot{\rho}
\approx 0,
\label{eq:Gauss_law_timeindep_PB}
\end{align}
where we follow Dirac notation \label{page:Dirac_notation} and denote by~$\approx$ identities that are
satisfied after the equations of
motions are implemented.
It also generates the gauge transformations of the vector potential
\begin{align}
\delta \mathbf{A}(t,\mathbf{r})=\Big\{\mathbf{A}(t,\mathbf{r}),
\int d^{3}r'\epsilon(t,\mathbf{r}')\big[\boldsymbol{\nabla}\cdot\mathbf{E}(t,\mathbf{r}')
-\rho(t,\mathbf{r}')\big]\Big\}_{\rm PB}=-\boldsymbol{\nabla}\epsilon(t,\mathbf{r}).
\end{align}

Solving the vacuum field equations written in terms of the gauge potential
\begin{align}
\Box A_{\mu}-\partial_{\mu}\partial_{\nu}A^{\nu}=0,
\end{align}
requires fixing the gauge freedom~\eqref{eq:gauge_freedom_cov}. To preserve relativistic covariance
it is convenient to use the Lorenz gauge~$\partial_{\mu}A^{\mu}=0$ introduced 
in~\eqref{eq:Lorenz_gauge}, so the
gauge potential satisfies then wave equation~$\Box A_{\mu}=0$. Trying a plane wave ansatz
\begin{align}
A_{\mu}(x)\sim \varepsilon_{\mu}(k,\lambda)e^{-ik_{\mu}x^{\mu}},
\end{align}
the wave equation implies that the momentum vector~$k^{\mu}$ is null
\begin{align}
k_{\mu}k^{\mu}=0 \hspace*{1cm} \Longrightarrow \hspace*{1cm} k^{0}=\pm|\mathbf{k}|.
\end{align} 
The parameter~$\lambda$ in~$\varepsilon_{\mu}(k,\lambda)$
labels the number of independent polarization vectors, which  
the Lorenz gauge condition force to be transverse
\begin{align}
k^{\mu}\varepsilon_{\mu}(\mathbf{k},\lambda)=0.
\end{align}
Using this condition we elliminate the 
temporal polarization in terms of the other three 
\begin{align}
\varepsilon_{0}(\mathbf{k},\lambda)
={1\over |\mathbf{k}|}\mathbf{k}\cdot\boldsymbol{\varepsilon}(\mathbf{k},\lambda).
\label{eq:temporal_pol}
\end{align}
In addition,
there is a residual gauge freedom preserving the Lorenz condition
implemented on the plane wave solutions by shifts of the 
polarization vector proportional to the wave momentum
\begin{align}
\varepsilon_{\mu}(\mathbf{k},\lambda)\longrightarrow 
\varepsilon_{\mu}(\mathbf{k},\lambda)+\alpha(\mathbf{k}) k_{\mu}.
\label{eq:epsilon->epsilon+alphak}
\end{align}
Using this freedom to set~$\varepsilon_{0}(\mathbf{k},\lambda)$ to zero, we are left with just two
independent transverse
polarizations 
satisfying~$\mathbf{k}\cdot\boldsymbol{\varepsilon}(\mathbf{k},\lambda)=0$. The plane wave solution then reads
\begin{align}
\mathbf{A}(t,\mathbf{r})\sim\boldsymbol{\varepsilon}(\mathbf{k},\lambda)e^{-i|\mathbf{k}|t+i\mathbf{k}
\cdot\mathbf{r}},
\label{eq:A_plane_wave}
\end{align}
with~$A_{0}=0$ and~$\lambda=\pm 1$ labelling the two transverse 
polarizations, that in the following we will respectively identify with right-left
circular polarizations\footnote{For a massive vector field 
the Lorenz condition~$\partial_{\mu}A^{\mu}=0$ 
is still satisfied as an integrability condition of the equations of motion~$\partial_{\mu}F^{\mu\nu}
+m^{2}A^{\nu}=0$ and
eq.~\eqref{eq:temporal_pol} therefore holds. The key difference lies in that the residual
freedom~\eqref{eq:epsilon->epsilon+alphak} is absent and 
we have an additional longitudinal
polarization (i.e., aligned with~$\mathbf{k}$) in addition to the two transverse ones.\label{page:footnote_massivephoton}},~$\boldsymbol{\varepsilon}(\mathbf{k},\lambda)^{*}
=\boldsymbol{\varepsilon}(\mathbf{k},-\lambda)$. They moreover satisfy
\begin{align}
\boldsymbol{\varepsilon}(\mathbf{k},\lambda)
\cdot\big[\mathbf{k}\times \boldsymbol{\varepsilon}(\mathbf{k},\lambda')\big]
= i\lambda|\mathbf{k}|\delta_{\lambda,-\lambda'}.
\label{eq:cross_prod_polvectors}
\end{align}
This identity will be useful later on.

Since the field equations are linear a general solution can be written as a superposition
of the plane wave solutions~\eqref{eq:A_plane_wave} and their complex conjugates. 
Upon quantization the coefficients in this expansion become operators and we can write a general
expression for the gauge field operator
\begin{align}
\widehat{\mathbf{A}}(t,\mathbf{r})
&=\sum_{\lambda=\pm 1}\int{d^{3}k\over (2\pi)^{3}}{1\over 2|\mathbf{k}|}
\left[\boldsymbol{\varepsilon}(\mathbf{k},\lambda)\widehat{a}(\mathbf{k},\lambda)e^{-i|\mathbf{k}|t
+i\mathbf{k}\cdot\mathbf{r}}+\boldsymbol{\varepsilon}(\mathbf{k},\lambda)^{*}
\widehat{a}(\mathbf{k},\lambda)^{\dagger}
e^{i|\mathbf{k}|t
-i\mathbf{k}\cdot\mathbf{r}}\right],
\label{eq:A-hat_first}
\end{align}
where, with our gauge fixing,~$\widehat{A}_{0}(t,\mathbf{r})=0$.
The integration measure appearing in this expression results from integrating over all four-dimensional 
momenta lying on the upper light-cone in fig.~\ref{fig:hyperboloid}
\begin{align}
\int {d^{4}k\over (2\pi)^{4}}\delta(k_{\mu}k^{\mu})\theta(k^{0})[\ldots]=
\int {d^{3}k\over (2\pi)^{3}}{1\over 2|\mathbf{k}|}[\ldots],
\end{align}
and is by construction Lorentz invariant. The quantum states of the theory are vectors 
in the space of states 
the operator~\eqref{eq:A-hat_first} acts on. To determine it and therefore the excitations
of the quantum field, we establish first 
the algebra of operators and then find a representation. 
This is done by applying the canonical quantization prescription replacing classical
Poisson brackets with quantum commutators
\begin{align}
i\{\cdot,\cdot\}_{\rm PB}\longrightarrow [\cdot,\cdot].
\label{eq:heisenberg_prescription}
\end{align}
Using the definition~$\widehat{\mathbf{E}}=\partial_{0}
\widehat{\mathbf{A}}$, the electric field operator is computed to be
\begin{align}
\widehat{\mathbf{E}}(t,\mathbf{r})
&=-{i\over 2}\sum_{\lambda=\pm 1}\int{d^{3}k\over (2\pi)^{3}}
\left[\boldsymbol{\varepsilon}(\mathbf{k},\lambda)\widehat{a}(\mathbf{k},\lambda)e^{-i|\mathbf{k}|t
+i\mathbf{k}\cdot\mathbf{r}}
-\boldsymbol{\varepsilon}(\mathbf{k},\lambda)^{*}
\widehat{a}(\mathbf{k},\lambda)^{\dagger}
e^{i|\mathbf{k}|t
-i\mathbf{k}\cdot\mathbf{r}}\right],
\label{eq:electric_field_operator}
\end{align}
Classically, the electric field is canonically conjugate to 
the vector potential [see eq.~\eqref{eq:AEPB}], so the 
prescription~\eqref{eq:heisenberg_prescription} gives its equal-time commutator with the gauge field
\begin{align}
[A_{i}(t,\mathbf{r}),E_{i}(t,\mathbf{r}')]=i\delta_{ij}\delta^{(3)}(\mathbf{r}-\mathbf{r}').
\label{eq:EM_canonical_comm_rel}
\end{align}
that translates into the following commutation relations for the
operators~$\widehat{a}(\mathbf{k},\lambda)$ and their Hermitian conjugates
\begin{align}
[\widehat{a}(\mathbf{k},\lambda),\widehat{a}(\mathbf{k}',\lambda')^{\dagger}]
&=(2\pi)^{3}2|\mathbf{k}|\delta_{\lambda\lambda'}\delta^{(3)}(\mathbf{k}-\mathbf{k}'), \nonumber
\label{eq:photon_ca_opsalg}\\[0.2cm]
[\widehat{a}(\mathbf{k},\lambda),\widehat{a}(\mathbf{k}',\lambda')]
&=[\widehat{a}(\mathbf{k},\lambda)^{\dagger},\widehat{a}(\mathbf{k}',\lambda')^{\dagger}]=0.
\end{align}
This algebra is reminiscent of the one of creation-annihilation operators in the quantum harmonic
oscillator. Introducing a properly normalized vacuum state~$|0\rangle$ to be annihilated by 
all~$\widehat{a}(\mathbf{k};\lambda)$,
we define the vector
\begin{align}
|\mathbf{k},\lambda\rangle=\widehat{a}(\mathbf{k},\lambda)^{\dagger}|0\rangle,
\label{eq:single_photon}
\end{align}
representing a one-photon state with momentum~$\mathbf{k}$ and helicity~$\lambda$.
These states are covariantly normalized according to
\begin{align}
\langle \mathbf{k},\lambda|\mathbf{k}',\lambda'\rangle=
(2\pi)^{3}2|\mathbf{k}|\delta_{\lambda\lambda'}\delta^{(3)}(\mathbf{k}-\mathbf{k}'), 
\end{align}
as can be seen from eq.~\eqref{eq:photon_ca_opsalg}. Multiple photon states are obtained by
successive application of creation operators
\begin{align}
|\mathbf{k}_{1},\lambda_{1};\mathbf{k}_{2},\lambda_{2};\ldots;\mathbf{k}_{n},\lambda_{n}\rangle
=\widehat{a}(\mathbf{k}_{1},\lambda_{1})^{\dagger}\widehat{a}(\mathbf{k}_{2},\lambda_{2})^{\dagger}
\ldots \widehat{a}(\mathbf{k}_{n},\lambda_{n})^{\dagger}|0\rangle.
\label{eq:multiphoton_states}
\end{align}
From the commutation relation of creation operators
given in~\eqref{eq:photon_ca_opsalg} we see that the multi-photon state is even under the interchange of
whatever two photons, as it should be for bosons.

Although we have been talking about photons, we must check that the states~\eqref{eq:single_photon}
have the quantum numbers corresponding to these particles. So, first we compute their energy by 
writing the quantum Hamiltonian. Going back to eq.~\eqref{eq:hamiltonian_electrodynamics}, 
we set the sources to zero~($\rho=0$ and~$\mathbf{j}=\mathbf{0}$) and replace the electric and 
magnetic field for their corresponding operators. A first thing to notice is that the electric
field~\eqref{eq:electric_field_operator} satisfies the Gauss law~$\boldsymbol{\nabla}\cdot
\widehat{\mathbf{E}}=0$ as a consequence of the transversality condition of the polarizations vectors.
Computing in addition~$\mathbf{B}=\boldsymbol{\nabla}\times\mathbf{A}$ and after some algebra, we find
\begin{align}
\widehat{H}=\sum_{\lambda=\pm 1}\int {d^{3}k\over (2\pi)^{3}}{1\over 2|\mathbf{k}|}\,|\mathbf{k}|
\widehat{a}(\mathbf{k},\lambda)^{\dagger}\widehat{a}(\mathbf{k},\lambda)
+{1\over 2}\sum_{\lambda=\pm 1}\int d^{3}k\,|\mathbf{k}|\delta^{(3)}(\mathbf{0}).
\label{eq:hamiltonian_quantum_electromagnetism}
\end{align}
The second term on the right-hand side represents the energy of the 
vacuum state
\begin{align}
\widehat{H}|0\rangle=\left({1\over 2}\sum_{\lambda=\pm 1}d^{3}k\,|\mathbf{k}|\delta^{(3)}(\mathbf{0})\right)
|0\rangle,
\end{align}
and is doubly divergent. One infinity originates in the delta function and
comes about because we are working at infinite volume, a type of divergence that is
QFT are designated as {\em infrared} (IR). It can be 
regularized by setting our system in a box of volume~$V$, which  
replaces~$(2\pi)^{3}\delta^{(3)}(\mathbf{0})$. 
Proceeding in this way, we write the energy density of the vacuum
\begin{align}
\rho_{\rm vac}\equiv {E_{\rm vac}\over V}={1\over 2}\sum_{\lambda=\pm 1}\int 
{d^{3}k\over (2\pi)^{3}}\,|\mathbf{k}|.
\end{align} 
This expression has the obvious interpretation of being the result of adding 
the zero-point energies of infinitely many 
harmonic oscillators, each with frequency~$\omega=|\mathbf{k}|$. It
is still divergent, and since the infinity 
originates in the integration over arbitrary high momenta, it is 
called {\em ultraviolet} (UV). A way to get rid of it is
assuming that~$|\mathbf{k}|<\Lambda_{\rm UV}$, so after carrying out the integral 
the vacuum energy density is given by
\begin{align}
\rho_{\rm vac}={1\over 16\pi^{2}}\Lambda_{\rm UV}^{4}.
\label{eq:realscalarfield_rhovac}
\end{align}
In the spirit of effective field theory this UV cutoff is physically interpreted as the energy
scale at which our description of the electromagnetic field breaks down and has to be replaced by 
some more general theory. 

The vacuum energy 
density~\eqref{eq:realscalarfield_rhovac} is at the origin of the cosmological constant problem.
Due to its strong dependence with the UV cutoff, when we add the contributions of all known quantum field
to~$\rho_{\rm vac}$ 
the result is many orders of magnitude larger than the one measured through cosmological observations.
The way to handle this mismatch is by assuming the existence of a 
nonzero cosmological constant~$\Lambda_{c}$
contribution to the total vacuum energy of the universe as
\begin{align}
\rho_{\rm vac}={\Lambda_{c}\over 8\pi G_{N}}+\sum_{i}\rho_{{\rm vac},i},
\end{align}
where the sum is over all quantum fields in nature.
Identifying the UV cutoff with the Planck energy,~$\Lambda_{\rm UV}\simeq \Lambda_{\rm Pl}$, 
the cosmological constant 
has to be fine tuned over~120 orders of magnitude in order to cancel the excess contribution
of the quantum fields to the vacuum energy density of the universe (see, 
for example,~\cite{WeinbergCC,PadmanabhanCC,BoussoCC}
for comprehensive reviews).\label{pag:cosmological_constant}

Let us get rid of the vacuum energy for the time being by subtracting it from  
the Hamiltonian~\eqref{eq:hamiltonian_quantum_electromagnetism}. Acting with this
subtracted Hamiltonian on the multiparticle
states~\eqref{eq:multiphoton_states}, we find they are energy eigenstates
\begin{align}
\widehat{H}|\mathbf{k}_{1},\lambda_{1};\mathbf{k}_{2},\lambda_{2};\ldots;\mathbf{k}_{n},\lambda_{n}\rangle
=\big(|\mathbf{k}_{1}|+|\mathbf{k}_{2}|+\ldots+|\mathbf{k}_{n}|\big)
|\mathbf{k}_{1},\lambda_{1};\mathbf{k}_{2},\lambda_{2};\ldots;\mathbf{k}_{n},\lambda_{n}\rangle,
\end{align} 
with the eigenvalue giving the energy of~$n$ free photons with momenta~$\mathbf{k}_{1},\mathbf{k}_{2},
\ldots,\mathbf{k}_{n}$. The field momentum, on the other hand, is given by the Poynting operator
\begin{align}
\widehat{\mathbf{P}}
&=\int d^{3}r\,\mathbf{E}(t,\mathbf{r})\times \mathbf{B}(t,\mathbf{r}) \nonumber \\[0.2cm]
&=\sum_{\lambda=\pm 1}\int {d^{3}k\over (2\pi)^{3}}{1\over 2|\mathbf{k}|}\mathbf{k}\,
\widehat{a}(\mathbf{k},\lambda)^{\dagger}\widehat{a}(\mathbf{k},\lambda),
\end{align}
where, unlike the Hamiltonian, here there is no vacuum contribution due to the rotational
invariance of~$|0\rangle$. Its action on the states~\eqref{eq:multiphoton_states} gives
\begin{align}
\widehat{\mathbf{P}}
|\mathbf{k}_{1},\lambda_{1};\mathbf{k}_{2},\lambda_{2};\ldots;\mathbf{k}_{n},\lambda_{n}\rangle
=\big(\mathbf{k}_{1}+\mathbf{k}_{2}+\ldots+\mathbf{k}_{n}\big)
|\mathbf{k}_{1},\lambda_{1};\mathbf{k}_{2},\lambda_{2};\ldots;\mathbf{k}_{n},\lambda_{n}\rangle,
\end{align}
showing that the vector~$\mathbf{k}$ labelling the one-particle states~\eqref{eq:single_photon}
is rightly interpreted as the photon momentum. Finally, we compute the spin momentum operator
\begin{align}
\widehat{\mathbf{S}}
&=\int d^{3}x\,\widehat{\mathbf{A}}\times\widehat{\mathbf{E}} \nonumber \\[0.2cm]
&=i\sum_{\lambda,\lambda'=\pm 1}\int {d^{3}k\over (2\pi)^{2}}{1\over 2|\mathbf{k}|}
\boldsymbol{\varepsilon}(\mathbf{k},\lambda)\times \boldsymbol{\varepsilon}(\mathbf{k},\lambda')^{*}\,
\widehat{a}(\mathbf{k},\lambda)^{\dagger}\widehat{a}(\mathbf{k},\lambda).
\end{align}
Acting on a one-particle state~\eqref{eq:single_photon}, we find
\begin{align}
\widehat{\mathbf{S}}|\mathbf{k},\lambda\rangle
&=i\sum_{\lambda,\lambda'=\pm 1}\boldsymbol{\varepsilon}(\mathbf{k},\lambda)
\times \boldsymbol{\varepsilon}(\mathbf{k},\lambda')^{*}|\mathbf{k},\lambda\rangle.
\end{align}
We project now this expression on the direction of the photon's momentum, we find the 
helicity operator acting on the single photon state 
\begin{align}
\widehat{h}|\mathbf{k},\lambda\rangle
\equiv {\mathbf{k}\over |\mathbf{k}|}\cdot\widehat{\mathbf{S}}
|\mathbf{k},\lambda\rangle
={i\over |\mathbf{k}|}\sum_{\lambda,\lambda'=\pm 1}\mathbf{k}\cdot
\big[\boldsymbol{\varepsilon}(\mathbf{k},\lambda)
\times \boldsymbol{\varepsilon}(\mathbf{k},\lambda')^{*}\big]|\mathbf{k},\lambda\rangle.
\end{align} 
Using the relation~\eqref{eq:cross_prod_polvectors} to evaluate the mixed product inside the
sum, we arrive at 
\begin{align}
\widehat{h}|\mathbf{k},\lambda\rangle=\lambda |\mathbf{k},\lambda\rangle,
\end{align}
which shows that~$\lambda$ is indeed the helicity of the photon. We have convinced ourselves 
that our interpretation of the quantum numbers describing the Hamiltonian eigenstates was
correct, and they describe states with an arbitrary number of free photons of definite momenta and
helicities. Photons therefore emerge as the elementary excitations of the quantum electromagnetic 
field.

\subsection{Some comments on quantum fields}

The previous calculation also teaches an important lesson: the space of states of
a free quantum field (in this case the electromagnetic field) is in fact a Fock space, i.e., 
the direct sum of Hilbert spaces spanned by the~$n$-particle states~\eqref{eq:multiphoton_states},
\begin{align}
\mathscr{F}=\bigoplus_{n=0}^{\infty}\mathscr{H}_{n},
\label{eq:fock_space}
\end{align}
where we take~$\mathscr{H}_{0}=L\{|0\rangle\}$, the one-dimensional linear space generated by the vacuum
state~$|0\rangle$.
We have shown that the canonical commutation relations~\eqref{eq:EM_canonical_comm_rel}
admit a representation in the Fock space. Although we have
done this for the free sourceless Maxwell's theory, it is also the case for any other free field theory, as
we will see in other examples below.  
Including interactions does not change this, provided
they are sufficiently weak and to be treated in perturbation theory. Thus, the first step in 
describing a physical system is to identify the weakly coupled degrees of freedom, whose
multiparticle states span the Fock space representing the asymptotic states in scattering experiments
of the type carried out everyday 
in high energy facilities around the world. 
This is well illustrated by the case of QCD discussed in the Introduction
(see page~\pageref{page:QCD_strong_weak_regimes}), where while the asymptotic states are described by
hadrons, the fundamental interactions taking place are described in terms of weakly coupled
quarks and gluons\footnote{A technical caveat: Haag's theorem~\cite{Haag_onQFT}, however, 
states that for a general interacting QFT there exists no Fock space representation of the 
canonical commutation relation. This is usually interpreted as implying that full
interacting QFT is not a theory of particles~\cite{Streater_Wightman,Haag_book,Strocchi}.}.

\begin{mdframed}[backgroundcolor=lightgray,hidealllines=true]
\vspace*{0.2cm}
\centerline{\greybox{\bf Complex fields and antiparticles}}
\vspace*{0.2cm}
\label{page:box6_complex_fields}

The analysis presented for electrodynamics carries over to the quantization of other
free fields. A simple but particularly interesting example is provided by a complex scalar field, 
with action
\begin{align}
S=\int d^{4}x\,\Big(\partial_{\mu}\varphi^{*}\partial^{\mu}\varphi-m^{2}\varphi^{*}\varphi\Big).
\label{eq:action_free_scalar_field}
\end{align}
Life is now simpler since there is no gauge freedom and the Hamiltonian formalism is straightforward.
We compute the conjugate momentum and the canonical Poisson brackets
\begin{align}
\pi(t,\mathbf{r})={\delta S\over \delta \partial_{0}\varphi(t,\mathbf{r})}
=\partial_{0}\varphi(t,\mathbf{r})^{*} \hspace*{0.5cm} \Longrightarrow \hspace*{0.5cm}
\big\{\varphi(t,\mathbf{r}),\pi(t,\mathbf{r}')\big\}_{\rm PB}=\delta^{(3)}(\mathbf{r}-\mathbf{r}'),
\label{eq:poisson_bracket}
\end{align}
with the corresponding expression for the complex conjugate fields,~$\varphi(t,\mathbf{r})^{*}$
and~$\pi(t,\mathbf{r})^{*}$. 
The Hamiltonian is then given by
\begin{align}
H=\int d^{3}r\,\Big[\pi^{*}\pi+(\boldsymbol{\nabla}\varphi^{*})\cdot
(\boldsymbol{\nabla}\varphi)+m^{2}\varphi^{*}\varphi\Big].
\end{align}

The equation
of motion derived from the action~\eqref{eq:action_free_scalar_field} is the Klein-Gordon equation
\begin{align}
\big(\Box+m^{2}\big)\varphi=0,
\end{align}
which admits plane wave solutions of the form
\begin{align}
\varphi(x)\sim e^{ip_{\mu}x^{\mu}},
\end{align}
with~$p_{\mu}$ satisfying the mass-shell condition
\begin{align}
p_{\mu}p^{\mu}=m^{2} \hspace*{1cm} \Longrightarrow \hspace*{1cm} p^{0}\equiv \pm E_{\mathbf{p}}
=\pm \sqrt{\mathbf{p}^{2}+m^{2}}.
\end{align}
As with the electromagnetic field, the corresponding
quantum fields are operator-valued superposition of
plane waves
\begin{align}
\widehat{\varphi}(t,\mathbf{r})&=\int{d^{3}p\over (2\pi)^{3}}{1\over 2E_{\mathbf{p}}}
\left[\widehat{\alpha}(\mathbf{p})e^{-iE_{\mathbf{p}}t+i\mathbf{p}\cdot\mathbf{r}}
+\widehat{\beta}(\mathbf{p})^{\dagger}e^{iE_{\mathbf{p}}t-i\mathbf{p}\cdot\mathbf{r}}\right], 
\nonumber \\[0.2cm]
\widehat{\varphi}(t,\mathbf{r})^{\dagger}&=\int{d^{3}p\over (2\pi)^{3}}{1\over 2E_{\mathbf{p}}}
\left[\widehat{\beta}(\mathbf{p})e^{-iE_{\mathbf{p}}t+i\mathbf{p}\cdot\mathbf{r}}
+\widehat{\alpha}(\mathbf{p})^{\dagger}e^{iE_{\mathbf{p}}t-i\mathbf{p}\cdot\mathbf{r}}\right]
\end{align}
while the operator associated to the canonically conjugate momentum is given by
\begin{align}
\widehat{\pi}(t,\mathbf{r})&=-{i\over 2}\int{d^{3}p\over (2\pi)^{3}}
\left[\widehat{\beta}(\mathbf{p})e^{-iE_{\mathbf{p}}t+i\mathbf{p}\cdot\mathbf{r}}
-\widehat{\alpha}(\mathbf{p})^{\dagger}e^{iE_{\mathbf{p}}t-i\mathbf{p}\cdot\mathbf{r}}\right],
\nonumber \\[0.2cm]
\widehat{\pi}(t,\mathbf{r})^{\dagger}&={i\over 2}\int{d^{3}p\over (2\pi)^{3}}
\left[\widehat{\alpha}(\mathbf{p})e^{-iE_{\mathbf{p}}t+i\mathbf{p}\cdot\mathbf{r}}
-\widehat{\beta}(\mathbf{p})^{\dagger}e^{iE_{\mathbf{p}}t-i\mathbf{p}\cdot\mathbf{r}}\right].
\end{align}
The key observation here is that since~$\widehat{\varphi}$ is not Hermitian, the two
operators~$\widehat{\alpha}(\mathbf{p})$ and~$\widehat{\beta}(\mathbf{p})$ cannot be identified, 
as it was the case with the electromagnetic field.
Imposing the equal-time canonical commutation relations induced by the canonical Poisson brackets 
[see eq.~\eqref{eq:poisson_bracket}] leads to the following algebra of operators
\begin{align}
[\widehat{\alpha}(\mathbf{p}),\widehat{\alpha}(\mathbf{p}')^{\dagger}]
&=(2\pi)^{3}2E_{\mathbf{p}}\delta^{(3)}(\mathbf{p}-\mathbf{p}'), \nonumber \\[0.2cm]
[\widehat{\alpha}(\mathbf{p}),\widehat{\alpha}(\mathbf{p}')]
&=[\widehat{\alpha}(\mathbf{p})^{\dagger},\widehat{\alpha}(\mathbf{p}')^{\dagger}]=0,
\end{align}
and corresponding expressions for~$\widehat{\beta}(\mathbf{p})$ 
and~$\widehat{\beta}(\mathbf{p})^{\dagger}$, with both types of operators commuting with each other.
As with the photons, the Fock
space of states is built by acting with~$\widehat{\alpha}(\mathbf{p})^{\dagger}$'s
and~$\widehat{\beta}(\mathbf{p})^{\dagger}$'s on 
the vacuum state~$|0\rangle$, which is itself annihilated by~$\widehat{\alpha}(\mathbf{p})$'s
and~$\widehat{\beta}(\mathbf{p})$'s
\begin{align}
|\mathbf{p}_{1},\ldots,\mathbf{p}_{n};\mathbf{q}_{1},\ldots,\mathbf{q}_{m}\rangle=
\widehat{\alpha}(\mathbf{p}_{1})^{\dagger}
\ldots\widehat{\alpha}(\mathbf{p}_{n})^{\dagger}
\widehat{\beta}(\mathbf{q}_{1})^{\dagger}\ldots
\widehat{\beta}(\mathbf{q}_{1})^{\dagger}
|0\rangle,
\label{eq:complex_scalar_multipart}
\end{align}
where we have distinguished the momenta associated with the two kinds of creation operators.
Notice that since the operators on the right-hand side of this expression commute with 
each other, the order in which we list the momenta~$\mathbf{p}_{1},\ldots,\mathbf{p}_{n}$ 
and~$\mathbf{q}_{1},\ldots,\mathbf{q}_{m}$ is irrelevant, signalling that both types of
excitations are bosons.

The states constructed in~\eqref{eq:complex_scalar_multipart} in fact diagonalize the Hamiltonian
\begin{align}
\widehat{H}=\int{d^{3}p\over (2\pi)^{3}}{1\over 2E_{\mathbf{p}}}
E_{\mathbf{p}}\Big[\widehat{\alpha}(\mathbf{p})^{\dagger}
\widehat{\alpha}(\mathbf{p})+\widehat{\beta}(\mathbf{p})^{\dagger}
\widehat{\beta}(\mathbf{p})
\Big],
\end{align}
where here we have subtracted a UV and IR divergent vacuum contribution similar to the one 
encountered in eq.~\eqref{eq:hamiltonian_quantum_electromagnetism}. Indeed, itt is not difficult to show
that
\begin{align}
\widehat{H}|\mathbf{p}_{1},\ldots,\mathbf{p}_{n};&\mathbf{q}_{1},\ldots,\mathbf{q}_{m}\rangle
\nonumber \\[0.2cm]
&=
\big(E_{\mathbf{p}_{1}}+\ldots+E_{\mathbf{p}_{n}}+E_{\mathbf{q}_{1}}+\ldots
+E_{\mathbf{q}_{m}}\big)
|\mathbf{p}_{1},\ldots,\mathbf{p}_{n};\mathbf{q}_{1},\ldots,\mathbf{q}_{m}\rangle,
\end{align}
from where we conclude that the elementary excitations of the quantum real scalar field
are free scalars particles with well-defined energy and momentum. These particles, however, come in two
different types depending on whether they are created by~$\widehat{\alpha}(\mathbf{p})^{\dagger}$
or~$\widehat{\beta}(\mathbf{p})^{\dagger}$, although sharing the same dispersion 
relation have equal masses. 

The obvious question is what distinguish physically
one from the other. To answer
we have to study the symmetries of the classical theory. A look
at the action~\eqref{eq:action_free_scalar_field} shows 
that it is invariant under global phase rotations of the complex field
\begin{align}
\varphi(x)\longrightarrow e^{i\vartheta}\varphi(x), \hspace*{1cm} 
\varphi(x)^{*}\longrightarrow e^{-i\vartheta}\varphi(x),
\label{eq:phase_rotation_scalarfield}
\end{align}
with~$\vartheta$ a constant real parameter. Noether's theorem (see page~\pageref{page:noether_thm} below) 
states that associated to this symmetry
there must be a conserved current, whose expression turns out to be
\begin{align}
j^{\mu}=i\varphi^{*}\overleftrightarrow{\partial}^{\mu}\varphi
\equiv i\varphi^{*}\partial^{\mu}\varphi-i(\partial^{\mu}\varphi^{*})\varphi
\hspace*{0.5cm} \Longrightarrow \hspace*{0.5cm} \partial_{\mu}j^{\mu}=0.
\label{eq:conserved_current_complexfield}
\end{align}
In particular, the conserved charge is given by
\begin{align}
Q=\int d^{3}r\,\big(\varphi^{*}\pi^{*}-\pi\varphi\big),
\end{align}
and once classical fields are replaced by their operator counterparts (and complex
by Hermitian conjugation), we have the following form
for the charge operator
\begin{align}
\widehat{Q}=\int {d^{3}p\over (2\pi)^{3}}{1\over 2E_{\mathbf{p}}}\Big[
\widehat{\alpha}(\mathbf{p})^{\dagger}\widehat{\alpha}(\mathbf{p})
-\widehat{\beta}(\mathbf{p})^{\dagger}\widehat{\beta}(\mathbf{p})\Big].
\end{align}
By acting with it on one-particle states, we get
\begin{align}
\widehat{Q}|\mathbf{p};0\rangle&=|\mathbf{p};0\rangle, \nonumber \\[0.2cm]
\widehat{Q}|0;\mathbf{q}\rangle&=-|0;\mathbf{q}\rangle,
\end{align}
showing that the conserved charge distinguishes the excitations generated
by~$\widehat{\alpha}(\mathbf{p})^{\dagger}$ 
from those generated by~$\widehat{\beta}(\mathbf{p})^{\dagger}$.
Moreover, the complex scalar field can be coupled to the electromagnetic field by identifying the
current~\eqref{eq:conserved_current_complexfield} with the one appearing 
in the Maxwell action~\eqref{eq:maxwell_action}, its conservation guaranteeing 
gauge invariance of the combined action. Thus, the two kinds of particles with the same
mass and spin 
have oposite electric charges and are 
identified as particles and antiparticles. 
The complex (i.e., non-Hermitian) character of
the scalar field is crucial to have both particles and antiparticles. 
In the case of the gauge field~$\widehat{\mathbf{A}}$, 
Hermiticity identifies the operators associated with
positive and negative energy plane wave solutions as conjugate to each other,
making the photon its own antiparticle. 

\end{mdframed}

It is time we address another symmetry present in Maxwell's electrodynamics that is of 
pivotal importance for QFT as a whole: scale invariance. Looking at the free electromagnetic action
\begin{align}
S_{\rm EM}=-{1\over 4}\int d^{4}x\,F_{\mu\nu}F^{\mu\nu}.
\label{eq:maxwell_free_action}
\end{align}
we notice the absence of any dimensionful parameters, unlike 
in the case of 
the complex scalar field action~\eqref{eq:action_free_scalar_field} where we have a parameter~$m$ that turns
out to be the mass of its elementary quantum excitations. It seems that
the free Maxwell's theory should be invariant under changes of scale. 

To formulate the idea of scale invaraince
in more general and precise mathematical terms, let as assume a scale transformation 
of the coordinates
\begin{align}
x^{\mu}\longrightarrow \lambda x^{\mu},
\label{eq:conformalEM_1}
\end{align}
with~$\lambda$ a nonzero real parameter, combined with the following scaling of the fields in the theory
\begin{align}
\Phi(x)\longrightarrow \lambda^{-\Delta_{\Phi}}\Phi(\lambda^{-1}x),
\label{eq:conformalEM_2}
\end{align}
where~$\Delta_{\Phi}$ is called the field's scaling dimension.
Applying these transformations to particular case of 
the action~\eqref{eq:maxwell_free_action}, we find
\begin{align}
S_{\rm EM}\longrightarrow \lambda^{2-2\Delta_{A}}S_{\rm EM},
\end{align}
so by setting~$\Delta_{A}=1$ the action remains invariant under scale 
transformations. 

We will explore now whether the scale invariance of the free Maxwell's theory is preserved
by the coupling of the electromagnetic field to charged matter. 
As an example, let us consider the complex scalar field
we studied in Box~6 but now coupled to an electromagnetic field
\begin{align}
S&=\int d^{4}x\left\{\partial_{\mu}\varphi^{*}\partial^{\mu}\varphi-m\varphi^{*}\varphi
-{1\over 4}F_{\mu\nu}F^{\mu\nu}+ie\big[\varphi^{*}\partial_{\mu}\varphi
-(\partial_{\mu}\varphi^{*})\varphi\big]A^{\mu}+e^{2}\varphi^{*}\varphi A_{\mu}A^{\mu}
\right\} \nonumber \\[0.2cm]
&=\int d^{4}x\left[\big(\partial_{\mu}+ieA_{\mu}\big)\varphi^{*}\big(\partial^{\mu}-ieA^{\mu}\big)\varphi
-m^{2}\varphi^{*}\varphi-{1\over 4}F_{\mu\nu}F^{\mu\nu}\right].
\label{eq:complex_scalar_EM_action}
\end{align}
Here, besides the coupling~$j_{\mu}A^{\mu}$ suggested by the Maxwell's equations, 
we also have the term~$e^{2}\varphi^{*}\varphi A_{\mu}A^{\mu}$, that  
has to be added to preserve the invariance of the whole action under the 
gauge transformations\footnote{Notice that the combination~$(\partial_{\mu}-ieA_{\mu})\varphi$
appearing in the second line of eq.~\eqref{eq:complex_scalar_EM_action} transforms
as the complex scalar field itself. It defines the gauge covariant derivative of~$\varphi$,
its name reflecting its covariant transformation under gauge
transformations,~$D_{\mu}\varphi\rightarrow e^{ie\epsilon(x)}
D_{\mu}\varphi$.
\label{page:covariat_derivative_scalar}}
\begin{align}
\varphi\rightarrow e^{ie\epsilon(x)}\varphi^{*}, 
\hspace*{1cm} \varphi^{*}\rightarrow e^{-ie\epsilon(x)}\varphi^{*}, 
\hspace*{1cm} A_{\mu}\rightarrow A_{\mu}+\partial_{\mu}\epsilon(x).
\end{align}
Setting the scaling dimension of the scalar field to one,~$\Delta_{\varphi}=1$, we easily check that
the scale invariance of the action~\eqref{eq:complex_scalar_EM_action}
is only broken by the mass term of the scalar field
\begin{align}
m\int d^{4}x\,\varphi^{*}\varphi\longrightarrow \lambda^{2} m\int d^{4}x\,\varphi^{*}\varphi.
\end{align}
This confirms our intuition that 
classical scale invariance is incompatible with the presence of dimensionful parameters in the
action. It also shows 
that taking~$m=0$ the photon can be coupled to scalar charged
matter preserving the classical scale invariance of the free Maxwell theory. 
Several essential field theories sharing this property besides the example just
analyzed, most notably QCD once all quark masses are set to zero. 

The discussion above has emphasized the term {\em classical} whenever 
referring to scale invariance. 
The reason is that
this is a very fragile symmetry once quantum effects are
included. For example, 
let us go back to the action~\eqref{eq:complex_scalar_EM_action} but now take~$m=0$.
The classical scale
invariance is broken by quantum effects in the sense that, once the quantum corrections induced by
interactions are taken into account, physics depends on the energy scale
at which experiments are carried out. One way in which this happens is by the electric 
charge of the elementary excitations of 
the field depending on the energy at which it is measured\footnote{Incidentally,
most scale invariant QFTs are also invariant under the full conformal group, i.e., the group of coordinate transformations preserving the light cone.}. 
We will further elaborate on this phenomenon in section~\ref{sec:renormalization}.

\section{Some group theory and some more wave equations}

Scalars and vectors are relatively intuitive objects, which why we did not 
need to get into sophisticated mathematics to handle them. In nature, however, elementary scalar fields are
rare (as of today, we know just one, the Higgs field) and vector fields only describe interactions, 
not matter. To describe fundamental physics we need fields whose excitations are particles
with spin-${1\over 2}$, such as the electron, the muon, and the quarks. We have to 
plunge into group theory before we can formulate these objects rigurously.

\subsection{Special relativity and group theory}

\begin{figure}[t]
\centerline{\includegraphics[scale=0.4]{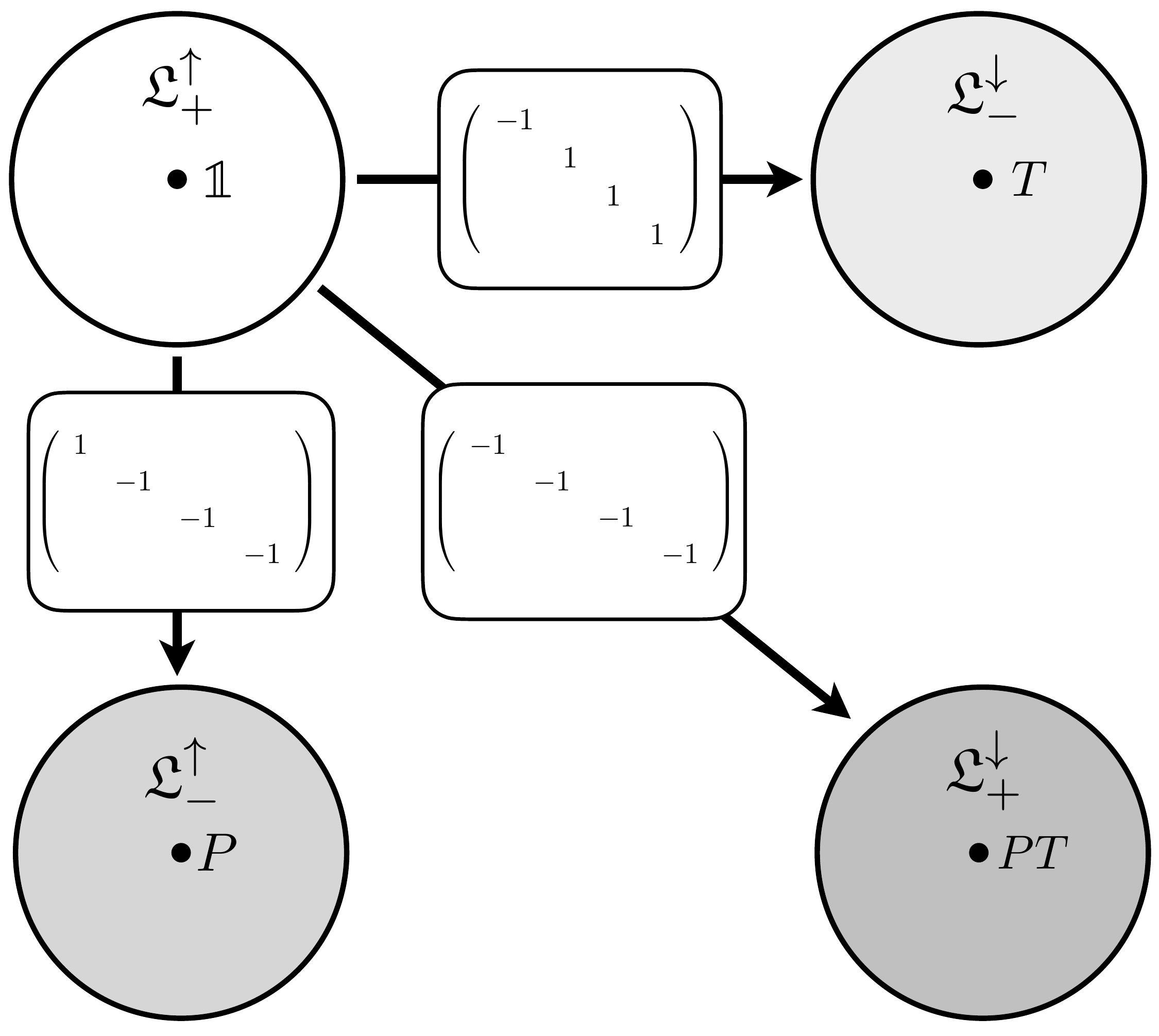}}
\caption[]{The four connected components of the Lorentz group. The matrices indicate the
transformations~$P$,~$T$, and~$PT$ 
mapping the connected component of the identity~$\mathfrak{L}^{\uparrow}_{+}$
to the other three.}
\label{fig:lorentz_group}
\end{figure}

Let us begin by giving a more technical picture of the Lorentz group. We have defined it as the 
set of linear transformations of the 
spacetime coordinates~$x'^{\mu}=\Lambda^{\mu}_{\,\,\,\nu}x^{\nu}$
satisfying~\eqref{eq:Lorentz_condition} and therefore preserving the Minkowski metric.
The first thing to be noticed is that this condition implies the inequality
\begin{align}
(\Lambda^{0}_{\,\,\,0})^{2}-\sum_{i=1}^{3}(\Lambda^{i}_{\,\,\,0})^{2}=1 \hspace*{1cm}
\Longrightarrow \hspace*{1cm} |\Lambda^{0}_{\,\,\,0}|\geq 1.
\end{align} 
The sign of~$\Lambda^{0}_{\,\,\,0}$ indicates whether or not the transformed time coordinate ``flows''
in the same direction as the original one, this being why transformations 
with~$\Lambda^{0}_{\,\,\,0}\geq 1$ are called {\em orthochronous}.\label{pag:ortochr} 
At the same time, 
eq.~\eqref{eq:Lorentz_condition} also implies\
\begin{align}
(\det\Lambda)^{2}=1 \hspace*{1cm} \Longrightarrow \hspace*{1cm} \det\Lambda=\pm 1.
\end{align}
Since it is not possible to change the signs of~$\Lambda^{0}_{\,\,\,0}$ or~$\det\Lambda$ by 
continuously deforming a Lorentz transformations, the full Lorentz group is seen to be composed
of four different connected components:\label{page:Lorentz_group_comps}
\begin{align}
\mathfrak{L}_{+}^{\uparrow}&:\mbox{ proper, orthochronous transformations with $\Lambda^{0}_{\,\,\,0}\geq 1$
and~$\det\Lambda=1$}, \nonumber \\[0.2cm]
\mathfrak{L}_{+}^{\downarrow}&:\mbox{ proper, non-orthochronous transformations with 
$\Lambda^{0}_{\,\,\,0}\leq -1$
and~$\det\Lambda=1$}, \nonumber \\[0.2cm]
\mathfrak{L}_{-}^{\uparrow}&:\mbox{ improper, orthochronous transformations with 
$\Lambda^{0}_{\,\,\,0}\geq 1$
and~$\det\Lambda=-1$}, \\[0.2cm]
\mathfrak{L}_{-}^{\downarrow}&:\mbox{ improper, non-orthochronous transformations with 
$\Lambda^{0}_{\,\,\,0}\leq -1$
and~$\det\Lambda=-1$},
\nonumber
\end{align}
The set of proper orthochronous transformations~$\mathfrak{L}_{+}^{\uparrow}$ 
contains the identity, while the remaining ones 
respectively include the time reversal operation~($\mbox{T}:x^{0}\rightarrow -x^{0}$), 
parity~($\mbox{P}:x^{i}\rightarrow -x^{i}$), and the composition of both. As indicated 
in fig.~\ref{fig:lorentz_group}, these discrete 
transformations also map the identity's connected component to the other three
\begin{align}
\mbox{T}:\mathfrak{L}_{+}^{\uparrow}\longrightarrow \mathfrak{L}_{-}^{\downarrow}, \hspace*{1cm}
\mbox{P}:\mathfrak{L}_{+}^{\uparrow}\longrightarrow \mathfrak{L}_{-}^{\uparrow}, \hspace*{1cm}
\mbox{PT}:\mathfrak{L}_{+}^{\uparrow}\longrightarrow \mathfrak{L}_{+}^{\downarrow}.
\end{align} 
Thus, to study the irreps of the Lorentz group it is enough to restrict
our attention to~$\mathfrak{L}_{+}^{\uparrow}\equiv \mbox{SO(1,3)}$.

As discussed in page~\pageref{page:lorentz_trans_generalities}, 
the proper group Lorentz~SO(1,3) is composed by two kinds of transformations: rotations with 
angle~$0\leq\phi<2\pi$ around
an axis defined by the unit vector~$\mathbf{u}$ and boosts with rapidity~$\lambda$
along the direction set by the unit vector~$\mathbf{e}$. Since we are on the connected component 
of the identity, the transformations can be written by exponentiation of the Lie algebra generators
\begin{align}
R(\phi,\mathbf{u})&=e^{-i\phi \mathbf{u}\cdot\mathbf{J}}, \nonumber \\[0.2cm]
B(\lambda,\mathbf{e})&=e^{-i\lambda\mathbf{e}\cdot\mathbf{M}},
\end{align}
where~$\mathbf{J}=(J_{1},J_{2},J_{3})$ and~$\mathbf{M}=(M_{1},M_{2},M_{3})$ are the generators
of rotations and boost respectively. They satisfy the 
algebra\footnote{The six generators~$(J_{i},M_{i})$ of the proper Lorentz group
can be fit into a rank-2 antisymmetric tensor with components~$\mathscr{J}_{0i}=M_{i}$
and~$\mathscr{J}_{ij}=\epsilon_{ijk}J_{k}$, satisfying the algebra~$[\mathscr{J}_{\mu\nu},
\mathscr{J}_{\alpha\beta}]=i\eta_{\mu\alpha}\mathscr{J}_{\nu\beta}-i\eta_{\mu\beta}\mathscr{J}_{\nu\alpha}
+i\eta_{\nu\beta}\mathscr{J}_{\mu\alpha}-i\eta_{\nu\alpha}\mathscr{J}_{\mu\beta}$.\label{pag:lorentz_algebra_cov}}
\begin{align}
[J_{i},J_{j}]&=i\epsilon_{ijk}J_{k}, \nonumber \\[0.2cm]
[J_{i},M_{j}]&=i\epsilon_{ijk}M_{k}, \label{eq:Lorentz_algebra1}\\[0.2cm]
[M_{i},M_{j}]&=-i\epsilon_{ijk}J_{k}.
\nonumber
\end{align}
Although the calculation leading to them is relatively easy, the previous commutation relations
can also be heuristically understood. The first commutator reproduces 
the usual algebra of infinitesimal rotations
familiar from elementary quantum mechanics. The second one is the simple statement that 
the generators of the boost along the three spatial directions transform as vectors under 
three-dimensional rotations. The third identity is the less obvious. It amounts to saying
that if we carry out two boosts along the directions set by unit vectors~$\mathbf{e}_{1}$ 
and~$\mathbf{e}_{2}$, the ambiguity in the order of the boost is equivalent to a three-dimensional
rotation with respect to the axis defined by~$\mathbf{e}_{1}\times\mathbf{e}_{2}$.

We could now try to find irreducible representations (irreps) 
of the algebra~\eqref{eq:Lorentz_algebra1}. 
Life get simpler if we relate this algebra to the one of a group we are more familiar with.
This can be done in this case by introducing the new set of generators
\begin{align}
J^{\pm}_{i}={1\over 2}\big(J_{i}\pm iM_{i}\big),
\end{align}
in terms of which, the algebra~\eqref{eq:Lorentz_algebra1} reads
\begin{align}
[J_{i}^{+},J_{j}^{+}]&=i\epsilon_{ijk}J^{+}_{k}, \nonumber \\[0.2cm]
[J_{i}^{-},J_{j}^{-}]&=i\epsilon_{ijk}J^{-}_{k}, \\[0.2cm]
[J_{i}^{+},J_{j}^{-}]&=0,
\nonumber
\end{align}
One thing we gain with this is that we have decoupled an algebra of six generators into two 
algebras of three generators each commuting with one another. 
But the real bonus here is that the individual algebras are 
those of SU(2), whose representation theory can be found in any quantum mechanics group.
Thus,~$\mbox{SO(1,3)}= \mbox{SU(2)}_{+}\times\mbox{SU(2)}_{-}$ and its
irreps are obtained by providing a pair of irreps
of~$\mbox{SU(2)}$, labelled by their total spins~$(\mathbf{s}_{+},\mathbf{s}_{-})$ 
with~$\mathbf{s}_{\pm}=\mathbf{0},\mathbf{1\over 2},\mathbf{1},\mathbf{3\over 2},\ldots$ 
Since~$J_{i}$ is a pseudovector, it does not change under parity transformations, whereas the
boost generators~$M_{i}$ do reverse sign
\begin{align}
\mbox{P}:J_{i}\longrightarrow J_{i}, \hspace*{1cm} \mbox{P}:M_{i}\longrightarrow -M_{i}.
\end{align}
As a consequence, parity interchanges the two SU(2) factors
\begin{align}
\mbox{P}:(\mathbf{s}_{+},\mathbf{s}_{-})\longrightarrow (\mathbf{s}_{-},\mathbf{s}_{+}).
\end{align}
Finally, the generators of the
group~\mbox{SO(3) $\approx$ SU(2)} of spatial rotations are given by
\begin{align}
J_{i}=J^{+}_{i}+J^{-}_{i},
\end{align}
so the irrep~$(\mathbf{s}_{+},\mathbf{s}_{-})$ descomposes into those of~SU(2) 
with~$j=\mathbf{s}_{+}+\mathbf{s}_{-},
\mathbf{s}_{+}+\mathbf{s}_{-}-1,\ldots,|\mathbf{s}_{+}-\mathbf{s}_{-}|$. 

Let us illustrate this general analysis with some relevant examples. We begin with the 
trivial irrep~$(\mathbf{s}_{+},\mathbf{s}_{-})=(\mathbf{0},\mathbf{0})$, 
whose generators are~$J_{i}^{\pm}=0$.
Fields transforming in this representation are scalar which under a Lorentz 
transformation~$x'^{\mu}=\Lambda^{\mu}_{\,\,\,\nu}x^{\nu}$ change according to
\begin{align}
\varphi'(x')=\varphi(x).
\end{align}
Another parity invariant representation is~$(\mathbf{s}_{+},\mathbf{s}_{-})
=(\mathbf{1\over 2},\mathbf{1\over 2})$, with 
generators~$J_{i}^{+}=J_{i}^{-}={1\over 2}\sigma^{i}$. Decomposing this irrep
with respect to those of spatial rotations, we see that they include a scalar~($j=0$) and
a three-vector~($j=1$). These correspond respectively to the zero and spatial components of a 
spin-one vector field~$V^{\mu}(x)$ transforming as
\begin{align}
V^{\mu}(x')=\Lambda^{\mu}_{\,\,\,\nu}V^{\nu}(x).
\end{align} 
Finally, we look at~$(\mathbf{s}_{+},\mathbf{s}_{-})=(\mathbf{1},\mathbf{1})$. 
This is decomposed in terms of three irreps
of~$\mbox{SU(2)}\approx\mbox{SO(3)}$ with~$j=2,1,0$. Together, they build a rank-two symmetric-traceless 
tensor field~$h^{\mu\nu}(x)=h^{\nu\mu}(x)$,~$\eta_{\mu\nu}h^{\mu\nu}(x)=0$ transforming as 
\begin{align}
h'^{\mu\nu}(x')=\Lambda^{\mu}_{\,\,\,\alpha}\Lambda^{\nu}_{\,\,\,\beta}h^{\alpha\beta}(x),
\end{align}
the three irreps of~SU(2) corresponding respectively 
to~$h^{ij}-{1\over 3}\delta^{ij}h^{00}$,~$h^{0i}=h^{i0}$, and~$h^{00}$.
This is a spin-two field like the one used to describe a graviton.

We look next with parity-violating representations, starting 
with~$(\mathbf{s}_{+},\mathbf{s}_{-})=(\mathbf{1\over 2},\mathbf{0})$.
Its generators are
\begin{align}
J^{+}_{k}={1\over 2}\sigma^{k}, \hspace*{1cm} J^{-}_{k}=0,
\end{align}
Hence, objects transforming in this representation have two complex components 
changing under rotations and boost according to
\begin{align}
\chi_{+}\longrightarrow e^{-{i\over 2}(\phi\mathbf{u}-i \boldsymbol{\lambda})\cdot\boldsymbol{\sigma}}
\chi_{+}.
\label{eq:positive_chirality_boosts+rot}
\end{align}
where~$\boldsymbol{\lambda}=(\lambda_{1},\lambda_{2},\lambda_{3})$ is the
boost's rapidity. In particular, we see that~$\chi_{+}$ transforms as a~SO(3) spinor. A field transforming 
in this representation is a {\em positive helicity} Weyl spinor. Very soon we will learn
the reason for its name.

\subsection{Chiral (and also nonchiral) fermions}

After all these group-theoretical considerations, it is time to start thinking about physics.
To construct an action principle for Weyl spinors, we need to build 
Lorentz invariant quantities from these fields. To begin with, we notice that 
the Hermitian conjugate spinor~$u_{+}^{\dagger}$ also transforms in 
the~$(\mathbf{1\over 2},\mathbf{0})$ representation of the Lorentz group, since
the representations of~$\mbox{SU(2)}$ are real. A general bilinear~$\chi_{+}^{\dagger}A\chi_{+}$, 
on the other hand, transforms
under the group~$\mbox{SO(3)}\approx\mbox{SU(2)}$ of three-dimensional rotations in the product 
representation~$\mathbf{1\over 2}\otimes\mathbf{1\over 2}=\mathbf{1}\otimes 
\mathbf{0}$. 
Computing the appropriate Clebsh-Gordan coefficients, we find
\begin{align}
\chi_{+}^{\dagger}\chi_{+}&\hspace*{0.5cm}\Longrightarrow\hspace*{0.5cm} j=0, \nonumber \\[0.2cm]
\chi_{+}^{\dagger}\sigma^{i}\chi_{x} &\hspace*{0.5cm}\Longrightarrow\hspace*{0.5cm} j=1.
\end{align}
They represent the time and spatial components of a four-vector 
\begin{align}
\chi_{+}^{\dagger}\sigma_{+}^{\mu}\chi_{+},
\label{eq:bilinear_vector_Weyl}
\end{align} 
where~$\sigma_{+}^{\mu}\equiv (\mathbbm{1},\sigma^{i})$. With this, we construct 
an action for the Weyl field as
\begin{align}
S_{+}=\int d^{4}x\,i\chi_{+}^{\dagger}\sigma_{+}^{\mu}\partial_{\mu}\chi_{+}.
\label{eq:positive_chir_action}
\end{align}
Notice that although~$\chi_{+}^{\dagger}\chi_{+}$ is invariant under rotations it does transform under
boosts. Therefore it is not a Lorentz scalar and cannot be added to the action as a mass term.

As for the
~$(\mathbf{s}_{+},\mathbf{s}_{-})
=(\mathbf{0},\mathbf{1\over 2})$ irrep of SO(1,3), a {\em negative helicity} Weyl spinor, 
the analysis is similar
to the one just presented and the corresponding expressions are obtained
from the ones derived above
by applying a parity transformation. In particular, we find its transformations under rotations
and boosts to be
\begin{align}
\chi_{-}\longrightarrow e^{-{i\over 2}(\phi\mathbf{u}+i \boldsymbol{\lambda})\cdot\boldsymbol{\sigma}}
\chi_{-},
\end{align}
showing that they also transform as~SO(3) spinors. Their free dynamics is derived from 
the action
\begin{align}
S_{-}=\int d^{4}x\,i\chi_{-}^{\dagger}\sigma_{-}^{\mu}\partial_{\mu}\chi_{-},
\label{eq:negative_chir_action}
\end{align}
where~$\sigma_{-}^{\mu}\equiv(\mathbbm{1},-\sigma^{i})$. 

Let us analyze in some more detail the physics of Weyl spinor fields.
The equations of motion derived from the actions~\eqref{eq:positive_chir_action}
and~\eqref{eq:negative_chir_action} are
\begin{align}
i\sigma_{\pm}^{\mu}\partial_{\mu}\chi_{\pm}=0 \hspace*{1cm} \Longrightarrow \hspace*{1cm}
\big(\partial_{0}\mp \boldsymbol{\sigma}\cdot\boldsymbol{\nabla}\big)\chi_{\pm}=0.
\label{eq:wave_eqs_weyl_spinors}
\end{align}
As in other cases, we search for positive energy~($k^{0}>0$) plane wave solutions of the form
\begin{align}
\chi_{\pm}(x)\sim u_{\pm}(\mathbf{k})e^{-ik\cdot x},
\label{eq:plane_wave_solutions_pmchirality}
\end{align}
where~$u_{\pm}(\mathbf{k})$ are~$(\mathbf{1\over 2},\mathbf{0})$ 
and~$(\mathbf{0},\mathbf{1\over 2})$ spinors normalized according
to
\begin{align}
u_{\pm}(\mathbf{k})^{\dagger}\sigma_{\pm}^{\mu}u_{\pm}(\mathbf{k})=2k^{\mu}\mathbbm{1}.
\label{eq:normalization_u_chiralpm}
\end{align}
Using this ansatz, the wave equations~\eqref{eq:wave_eqs_weyl_spinors} then takes the form
\begin{align}
\big(k_{0}\mp \mathbf{k}\cdot\boldsymbol{\sigma}\big)u_{\pm}(\mathbf{k})=0.
\label{eq:chiral_spinor_momentum_eq}
\end{align}
Multiplying by~$k_{0}\pm \mathbf{k}\cdot\boldsymbol{\sigma}$ on the left and
using~$k_{i}k_{j}\sigma^{i}\sigma^{j}=\mathbf{k}^{2}\mathbbm{1}$, we obtain the dispersion 
relation of a massless particle,~$k_{0}=|\mathbf{k}|$. Equation~\eqref{eq:chiral_spinor_momentum_eq}
implies the condition
\begin{align}
\left(\mathbbm{1}\mp {\mathbf{k}\over |\mathbf{k}|}\cdot\boldsymbol{\sigma}\right)u_{\pm}(\mathbf{k})=0 
\hspace*{1cm} \Longrightarrow \hspace*{1cm} \left({\mathbf{k}\over |\mathbf{k}|}\cdot\mathbf{s}\right)
u_{\pm}(\mathbf{k})=\pm{1\over 2}u_{\pm}(\mathbf{k}),
\end{align}
where~$\mathbf{s}\equiv{1\over 2}\boldsymbol{\sigma}$ is the spin operator. Helicity is defined as the
projection of the particle's spin on its direction of motion and the previous identity 
shows that~$u_{\pm}(k)$ are 
spinors with positive and negative helicity respectively. Since the generic Weyl spinors~$\chi_{\pm}$
can be written as a superposition of the plane wave solutions~\eqref{eq:plane_wave_solutions_pmchirality},
this explains the terminology introduced above.

To write a general positive (resp. negatively) helicity Weyl spinor, we also need to consider 
negative energy plane waves~$v_{\pm}(\mathbf{k})e^{-ik\cdot x}$, where~$k^{0}<0$. 
Imposing this to solve eq.~\eqref{eq:wave_eqs_weyl_spinors},
we find that~$v_{\pm}(\mathbf{k})$ satisfies
\begin{align}
\big(k^{0}\pm\mathbf{k}\cdot\boldsymbol{\sigma}\big)v_{\pm}(\mathbf{k})=0,
\end{align}
where we set the normalization
\begin{align}
v_{\pm}(\mathbf{k})^{\dagger}\sigma_{\pm}^{\mu}v_{\pm}(\mathbf{k})=2k^{\mu}\mathbbm{1}.
\end{align}
In addition, it can also be shown that the 
positive and negative energy solutions satisfy the orthogonality relations
\begin{align}
u(-\mathbf{k})^{\dagger}v(\mathbf{k})=v(-\mathbf{k})^{\dagger}u(\mathbf{k})=0.
\end{align}
These identities will be important later in determining the spectrum of excitation of the free quantum Weyl 
spinor field.

Classical Weyl spinors are complex fields and their 
actions~\eqref{eq:positive_chir_action} and~\eqref{eq:negative_chir_action} are invariant under
global phase rotations~$\chi_{\pm}\longrightarrow e^{i\vartheta}\chi_{\pm}$. 
The associated Noether currents~(see page~\pageref{page:noether_thm})
are the bilinear Lorentz vector constructed in eq.~\eqref{eq:bilinear_vector_Weyl}, and
the corresponding expression for negative helicity,
\begin{align}
j_{\pm}^{\mu}=\chi_{\pm}^{\dagger}\sigma_{\pm}^{\mu}\chi_{\pm}.
\label{eq:conserved_current_weyl}
\end{align}
Plugging this current into 
eq.~\eqref{eq:maxwell_action} we couple the Weyl spinors to the electromagnetic field
\begin{align}
S_{\pm}&=\int d^{4}x\left(i\chi_{\pm}^{\dagger}\sigma_{\pm}^{\mu}\partial_{\mu}\chi_{\pm}
+e\chi_{\pm}\sigma_{\pm}^{\mu}\chi_{\pm}A_{\mu}-{1\over 4}F_{\mu\nu}F^{\mu\nu}\right) \nonumber \\[0.2cm]
&=\int d^{4}x\,\left[i\chi_{\pm}^{\dagger}\sigma_{\pm}^{\mu}\big(\partial_{\mu}-ieA_{\mu}\big)\chi_{\pm}
-{1\over 4}F_{\mu\nu}F^{\mu\nu}\right],
\label{eq:QED_chiral_action}
\end{align}
where in the second line we find again the gauge covariant derivative first introduced
in eq.~\eqref{eq:complex_scalar_EM_action}. This action is invariant under gauge 
transformations, acting on the
Weyl spinor by {\em local} phase rotations~$\chi_{\pm}\longrightarrow e^{ie\epsilon(x)}\chi_{\pm}$. 
Moreover, given the absence of any dimensionful parameter in the action, we can expect the
classical theory to be scale invariant. This is indeed the case, with the Weyl spinors
having scaling dimension~$\Delta_{\chi}={3\over 2}$.

To quantize the Weyl field, we begin with the computation of the canonical Poisson algebra. 
The momentum canonically conjugate to the spinor is given by
\begin{align}
\pi_{\pm}\equiv{\delta S_{\pm}\over \delta \partial_{0}\chi_{\pm}}=i\chi_{\pm}^{\dagger} 
\end{align}
leading to
\begin{align}
\big\{\chi_{\pm,a}(t,\mathbf{r}),\chi_{\pm,b}(t,\mathbf{r}')^{\dagger}\big\}_{\rm PB}=-i\delta_{ab}
\delta^{(3)}(\mathbf{r}-\mathbf{r}'),
\end{align}
where~$a,b$ denote the spinor indices and all other Poisson brackets are equal to zero.
The Hamiltonian then reads
\begin{align}
H_{\pm}=\pm i\int d^{3}x\,\chi_{\pm}^{\dagger}(\boldsymbol{\sigma}\cdot\boldsymbol{\nabla})\chi_{\pm}.
\label{eq:Hpm_classical}
\end{align}

So much for the classical theory.
Quantum Weyl spinor fields are written as operator-valued superpositions of 
positive- and negative-energy plane wave 
solutions
\begin{align}
\widehat{\chi}_{\pm}(t,\mathbf{r})=\int {d^{3}k\over (2\pi)^{3}}{1\over 2|\mathbf{k}|}
\Big[\widehat{b}(\mathbf{k},\pm)u_{\pm}(\mathbf{k})e^{-i|\mathbf{k}|t+i\mathbf{k}\cdot\mathbf{r}}
+\widehat{d}(\mathbf{k},\pm)^{\dagger}v_{\pm}(\mathbf{k})^{*}
e^{i|\mathbf{k}|t-i\mathbf{k}\cdot\mathbf{r}}
\Big].
\label{eq:chiral_spinor_op_expansion}
\end{align}
It is important to remember that 
the previous
operator is not Hermitian. Similarly to what we learned from the analysis of the complex
scalar field, this implies that the operators~$\widehat{b}(\mathbf{k},\pm)$ 
and~$\widehat{d}(\mathbf{k},\pm)$ are independent and unrelated to each other 
by Hermitian conjugation. However, we need to be careful when constructing the algebra of 
field operators. For example, the spin-statistics theorem states that particles with 
half-integer spin are fermions, and their quantum states 
should be antisymmetric under the interchange
of two of them. To achieve this, the prescription~\eqref{eq:heisenberg_prescription} has to 
be modified and Poisson brackets are replaced by {\em anticommutators} instead of commutators
\begin{align}
i\{\cdot,\cdot\}_{\rm PB}\longrightarrow \{\cdot,\cdot\}.
\end{align}
Accordingly, we impose
\begin{align}
\big\{\widehat{\chi}_{\pm,a}(t,\mathbf{r}),
\widehat{\chi}_{\pm,b}(t,\mathbf{r}')^{\dagger}\big\}_{\rm PB}=\delta_{ab}
\delta^{(3)}(\mathbf{r}-\mathbf{r}'),
\end{align}
which, using the normalization~$u_{\pm}(\mathbf{k})^{\dagger}u_{\pm}(\mathbf{k})=2|\mathbf{k}|$
[cf.~\eqref{eq:normalization_u_chiralpm}],
leads to the operator algebra 
\begin{align}
\big\{\widehat{b}(\mathbf{k},\pm),\widehat{b}(\mathbf{k}',\pm)^{\dagger}\big\}
=(2\pi)^{3}2|\mathbf{k}|\delta_{ab}\delta^{(3)}(\mathbf{r}-\mathbf{r}'), \nonumber \\[0.2cm]
\big\{\widehat{d}(\mathbf{k},\pm),\widehat{d}(\mathbf{k}',\pm)^{\dagger}\big\}
=(2\pi)^{3}2|\mathbf{k}|\delta_{ab}\delta^{(3)}(\mathbf{r}-\mathbf{r}'),
\end{align}
with all remaining anticommutators equal to zero. As in the case of the complex scalar field
analyzed in Box~6, here we also get two types of particles generated by the 
two kinds of creation operators acting on the vacuum
\begin{align}
|\mathbf{k}_{1},\ldots,\mathbf{k}_{n};\mathbf{p}_{1},\dots,\mathbf{p}_{m}\rangle_{\pm}
=\widehat{b}(\mathbf{k}_{1},\pm)^{\dagger}\ldots \widehat{b}(\mathbf{k}_{n},\pm)^{\dagger}
\widehat{d}(\mathbf{p}_{1},\pm)^{\dagger}\ldots\widehat{d}(\mathbf{p}_{m},\pm)^{\dagger}|0\rangle.
\label{eq:multiWeyl_states}
\end{align}
As expected the state is antisymmetric under the interchange
of two particles of the same type, due to the anticommutation of the
creation operators. Similarly to the complex scalar field, the two types of particles are distinguished
by the charge operator defined by the conserved current~\eqref{eq:conserved_current_weyl}.  
\begin{align}
\widehat{Q}=\int d^{3}\mathbf{r}\,\widehat{\chi}_{\pm}(t,\mathbf{r})^{\dagger}
\widehat{\chi}_{\pm}(t,\mathbf{r})
\hspace*{1cm} \Longrightarrow \hspace{1cm}
\left\{
\begin{array}{l}
\widehat{Q}|\mathbf{k};0\rangle_{\pm}=|\mathbf{k};0\rangle_{\pm} \\[0.2cm]
\widehat{Q}|0;\mathbf{k}\rangle_{\pm}=-|0;\mathbf{k}\rangle_{\pm}
\end{array}
\right.,
\end{align}
so the states~$|0;\mathbf{k}\rangle_{\pm}$ are naturally identified as 
the antiparticles of~$|\mathbf{k};0\rangle_{\pm}$.

The calculation of the Hamiltonian operator follows the lines outlined in previous cases.
Replacing classical fields by operators in the Hamiltonian~\eqref{eq:Hpm_classical}, and
using the properties of the positive and negative energy solutions~$u(\mathbf{k})$ 
and~$v(\mathbf{k})$, we find after some algebra
\begin{align}
\widehat{H}_{\pm}&=\int {d^{3}k\over (2\pi)^{3}}{1\over 2|\mathbf{k}|}
\Big[|\mathbf{k}|\widehat{b}(\mathbf{k},\pm)^{\dagger}\widehat{b}(\mathbf{k},\pm)
+|\mathbf{k}|\widehat{d}(\mathbf{k},\pm)^{\dagger}\widehat{d}(\mathbf{k},\pm)\Big]
-\int d^{3}k\,|\mathbf{k}|\delta^{(3)}(\boldsymbol{0}).
\label{eq:hamiltonian_weyl}
\end{align}
We see from the first term on the right-hand side
that the multiparticle states~\eqref{eq:multiWeyl_states}
diagonalize the Hamiltonian, with particles and antiparticles having
zero mass,~$E_{\mathbf{k}}=|\mathbf{k}|$.
In this Hamiltonian we find once more the UV and IR divergent zero-point contribution, 
that once regularized gives a vacuum energy density
\begin{align}
\rho_{\rm vac}=-{1\over 8\pi^{2}}\Lambda_{\rm UV}^{4}.
\end{align} 
Although it will eventually be subtracted, it is
worthwhile to stop a moment and compare this with 
the expression~\eqref{eq:realscalarfield_rhovac}. A first thing 
meeting the eye is the relative factor of two in the Weyl spinor case. This reflects that while a
real scalar field has a single propagating degree of freedom, here we have 
two, associated with the complex field's real and imaginary parts. The second and physically
very relevant feature is the different sign, boiling down to having
anticommutators rather than commutators. It implies that bosons and fermions
contribute to the vacuum energy with oposite signs. This is the reason why supersymmetric theories,
which have as many bosonic as fermionic degrees of freedom and therefore zero vacuum energy, 
have been invoked to solve the problem
of the cosmological constant mentioned in page~\pageref{pag:cosmological_constant}, 
or at
least to ameliorate it\footnote{Since supersymmetry must be broken at low energies (after
all, we do not ``see'' the same number of bosons as fermions), 
there is still a nonvanishing contribution to the vacuum energy proportional to the fourth
power of the scale of supersymmetry breaking,~$\Lambda_{\rm SUSY}$, 
rather than the much higher~$\Lambda_{\rm Pl}$.}.

\begin{mdframed}[backgroundcolor=lightgray,hidealllines=true]
\vspace*{0.2cm}
\centerline{\greybox{\bf Dirac spinors}}
\vspace*{0.2cm}
\label{box:Dirac_spinors}

Although the theory of a single Weyl spinor violates parity, it is possible to construct
a parity-invariant theory by taking together two Weyl spinors with opposite chiralities. They can be
combined into a single object, a Dirac spinor 
\begin{align}
\psi\equiv \left(
\begin{array}{c}
\chi_{+} \\
\chi_{-}
\end{array}
\right),
\end{align}
which obviously transforms in the parity-invariant
reducible representation~$(\mathbf{1\over 2},\mathbf{0})\oplus (\mathbf{0},\mathbf{1\over 2})$.
The corresponding free action is obtained by adding the ones already written in eqs.~\eqref{eq:positive_chir_action}
and~\eqref{eq:positive_chir_action} for
Weyl spinors of different chiralities, namely
\begin{align}
S&=\int d^{4}x\,\Big(i\chi_{+}^{\dagger}\sigma_{+}^{\mu}\partial_{\mu}\chi_{+}
+i\chi_{-}^{\dagger}\sigma_{-}^{\mu}\chi_{-}\Big)
=i\int d^{4}x\,\psi^{\dagger}\left(
\begin{array}{cc}
\sigma_{+}^{\mu} & 0 \\
0 & \sigma_{-}^{\mu}
\end{array}
\right)\partial_{\mu}\psi.
\label{eq:action_pre_Dirac}
\end{align}
An important point to be taken into account now is that~$u_{\pm}$ and~$u_{\pm}^{*}$ do 
have opposite helicities. This is the reason 
why~$u_{\pm}^{\dagger}\sigma_{\pm}^{\mu}u_{\pm}\equiv u_{\pm, a}^{*}(\sigma_{\pm}^{\mu})_{ab}u_{\pm,b}$ 
defines a Lorentz vector, since~$(\mathbf{1\over 2},\mathbf{0})\otimes
(\mathbf{0},\mathbf{1\over 2})=(\mathbf{1\over 2},\mathbf{1\over 2})$
and~$(\sigma_{\pm}^{\mu})_{ab}$ are the Clebsh-Gordan coefficients decomposing the product representation into
its irreps. As a consequence,
whereas~$\psi^{*}$ does not transform in the same representation as~$\psi$, the spinor
\begin{align}
\overline{\psi}^{T}\equiv\left(
\begin{array}{c}
u_{-}^{*} \\
u_{+}^{*}
\end{array}
\right)=\left(
\begin{array}{cc}
0 & \mathbbm{1} \\
\mathbbm{1} & 0
\end{array}
\right)\psi^{*},
\end{align}
does. This suggest recasting the action~\eqref{eq:action_pre_Dirac} as
\begin{align}
S=i\int d^{4}x\,\overline{\psi}\left(
\begin{array}{cc}
0 & \mathbbm{1} \\
\mathbbm{1} & 0
\end{array}
\right)
\left(
\begin{array}{cc}
\sigma_{+}^{\mu} & 0 \\
0 & \sigma_{-}^{\mu}
\end{array}
\right)\partial_{\mu}\psi
=i\int d^{4}x\,\overline{\psi}
\left(
\begin{array}{cc}
0 & \sigma_{-}^{\mu} \\
\sigma_{+}^{\mu} & 0
\end{array}
\right)\partial_{\mu}\psi,
\label{eq:action_pre_Dirac2}
\end{align}
It seems natural to introduce a new set of~$4\times 4$ matrices, the {\em Dirac matrices},
defined by
\begin{align}
\gamma^{\mu}\equiv
\left(
\begin{array}{cc}
0 & \sigma_{-}^{\mu} \\
\sigma_{+}^{\mu} & 0
\end{array}
\right),
\label{eq:Dirac_matr_repschiral}
\end{align}
and satisfying the Clifford algebra
\begin{align}
\big\{\gamma^{\mu},\gamma^{\nu}\big\}=2\eta^{\mu\nu}\mathbbm{1},
\end{align}
as can be easily checked using the anticommutation relations of the Pauli matrices. The generators of the representation
of~$(\mathbf{1\over 2},\mathbf{0})\oplus (\mathbf{0},\mathbf{1\over 2})$ are then given in terms of the Dirac matrices 
by (see the footnote in 
page~\pageref{pag:lorentz_algebra_cov})
\begin{align}
\mathscr{J}^{\mu\nu}=-{i\over 4}[\gamma^{\mu},\gamma^{\nu}]\equiv \sigma^{\mu\nu}.
\label{eq:sigmamunu_def}
\end{align}
Denoting by~$\mathcal{U}(\Lambda)$ the matrix implementing the Lorentz transformation~$\Lambda^{\mu}_{\,\,\,\nu}$ on
Dirac spinors and using
the property~$\gamma^{\mu\dagger}=\gamma^{0}\gamma^{\mu}\gamma^{0}$, it is easy to show that~$\mathcal{U}(\Lambda)^{\dagger}
=\gamma^{0}\mathcal{U}(\Lambda)^{-1}\gamma^{0}$. This implies that 
while~$\psi\rightarrow\mathcal{U}(\Lambda)\psi$,
the conjugate spinor transforms contravariantly,~$\overline{\psi}\rightarrow \overline{\psi}\,
\mathcal{U}(\Lambda)^{-1}$, and 
the Dirac matrices themselves satisfy~$\mathcal{U}(\Lambda)^{-1}\gamma^{\mu}\mathcal{U}(\Lambda)
=\Lambda^{\mu}_{\,\,\,\nu}\gamma^{\nu}$. Let this serve as {\em a posteriori} 
justification of the introduction
of the conjugate field~$\overline{\psi}$.

The previous discussion shows that~$\overline{\psi}\psi$
is a Lorentz scalar that can be added to the Dirac 
action~\eqref{eq:action_pre_Dirac2}, that we now write in a much more compact form
\begin{align}
S=\int d^{4}x\big(i\overline{\psi}\gamma^{\mu}\partial_{\mu}\psi-m\overline{\psi}\psi\big).
\label{eq:Dirac_action1}
\end{align}
The associated field equations admit positive energy plane wave
solutions of the form~$\psi(x)\sim u(\mathbf{k},s)e^{-ik\cdot x}$, with~$s=\pm{1\over 2}$ labelling 
the two possible
values of the spin third component 
\begin{align}
\big(i\gamma^{\mu}\partial_{\mu}-m\big)\psi(x)=0 \hspace*{1cm} \Longrightarrow \hspace*{1cm}
\big({k\!\!\!/}-m\big)u(\mathbf{k},s)=0.
\label{eq:(kslash-m)u=0}
\end{align} 
Here we have introduced the Feynman slash notation~${a\!\!\!/}\equiv \gamma^{\mu}a_{\mu}$ that we will use throughout this lectures. Acting on
the equation to the right of~\eqref{eq:(kslash-m)u=0} with~${k\!\!\!/}+m$ and implementing the identity~${k\!\!\!/}{k\!\!\!/}=k^{2}\mathbbm{1}$, 
we find the 
massive dispersion relation~$k^{0}\equiv E_{\mathbf{k}}=\sqrt{\mathbf{k}^{2}+m^{2}}$.

To get a better idea about the role played by the mass term in the Dirac equation, it is instructive to write the 
equation~$\big({k\!\!\!/}-m\big)u(\mathbf{k},s)=0$ in terms of the two helicity components of the Dirac spinor
\begin{align}
\big(E_{\mathbf{k}}\mathbbm{1}-\mathbf{k}\cdot\boldsymbol{\sigma}\big)u_{+}(\mathbf{k},s)&=
m u_{-}(\mathbf{k},s), \nonumber \\[0.2cm]
\big(E_{\mathbf{k}}\mathbbm{1}+\mathbf{k}\cdot\boldsymbol{\sigma}\big)u_{-}(\mathbf{k},s)&=
m u_{+}(\mathbf{k},s).
\label{eq:pve_energy_sol_Dirac_comp1}
\end{align}
These expressions shows that the mass terms mixes the two helicities. Introducing the chirality matrix
\begin{align}
\gamma_{5}\equiv -i\gamma^{0}\gamma^{1}\gamma^{2}\gamma^{3}=
\left(
\begin{array}{cc}
\mathbbm{1} & 0 \\
0 & -\mathbbm{1}
\end{array}
\right),
\label{eq:chirality_matrix}
\end{align} 
the previous identity is rescast as
\begin{align}
\left(
\begin{array}{cc}
{\mathbf{k}\over |\mathbf{k}|}\cdot\mathbf{s} & 0 \\
0 & {\mathbf{k}\over |\mathbf{k}|}\cdot\mathbf{s}
\end{array}
\right)u(\mathbf{k},s)={1\over 2}\left({E_{\mathbf{k}}\over |\mathbf{k}|}\mathbbm{1}
-{m\over |\mathbf{k}|}\gamma^{0}\right)\gamma_{5}u(\mathbf{k},s),
\label{eq:chirality_vs_helicity_mneq0}
\end{align}
with~$\mathbf{s}={1\over 2}\boldsymbol{\sigma}$ the spin, so the matrix on the left-hand side of 
this expression
is the helicity operator~$h$ acting on a four-component Dirac spinor. 

The chirality matrix satisfies~$\gamma_{5}^{2}=\mathbbm{1}$ and 
anticommutes with all Dirac matrices,~$\{\gamma_{5},\gamma^{\mu}\}=0$. As a consequence,
its commutator with the Lorentz generators vanishes,~$[\gamma_{5},\sigma^{\mu\nu}]=0$, and by Schur's lemma this means that the
spinors~$P_{+}\psi$ and~$P_{-}\psi$ transform in different irreps of the Lorentz group, 
with~$P_{\pm}={1\over 2}(\mathbbm{1}\pm \gamma_{5})$ the projector onto
the two chiralities. The spinor's chirality is therefore a Lorentz invariant.

A look at eq.~\eqref{eq:chirality_vs_helicity_mneq0} shows that for a {\em massive} Dirac spinor
helicity (the projection of the spin onto the direction of motion) and chirality (the 
eigenvalue of the chirality matrix) are very different things. The former is not even a Lorentz invariant, 
since for a massive fermion with positive/negative helicity we can switch to a moving frame overcoming the particle and make 
the helicity negative/positive.   
Taking, however, the massless 
limit~$m\rightarrow 0$ we have~$E_{\mathbf{k}}\rightarrow |\mathbf{k}|$ and chirality and helicity
turn out to be equivalent
\begin{align}
h={1\over 2} \gamma_{5} \hspace*{1cm} (m=0).
\end{align}
This is why when dealing with massless spin-${1\over 2}$ fermions, 
both terms can be used indistinctly, although 
in the case of massive particles 
one should be very careful in using the one appropriate to the physical situation under analysis.

To quantize the theory, we write an expansion of the Dirac field operator into its positive and negative energy solutions
\begin{align}
\widehat{\psi}(t,\mathbf{r})=\sum_{s=\pm{1\over 2}}
\int {d^{3}k\over (2\pi)^{3}}{1\over 2E_{\mathbf{k}}}
\Big[\widehat{b}(\mathbf{k},s)u(\mathbf{k},s)e^{-i|\mathbf{k}|t+i\mathbf{k}\cdot\mathbf{r}}
+\widehat{d}(\mathbf{k},s)^{\dagger}v(\mathbf{k},s)^{*}
e^{i|\mathbf{k}|t-i\mathbf{k}\cdot\mathbf{r}}
\Big],
\label{eq:Dirac_op_bdscaop}
\end{align}
where the negative energy solutions~$v(\mathbf{k},s)$ 
are defined by the equation~$({k\!\!\!/}+m)v(\mathbf{k},s)=0$. The canonical anticommutation relations of the Dirac
field with its Hermitian conjugate
imply that~$\widehat{b}(\mathbf{k},s)$ and~$\widehat{b}(\mathbf{k},s)^{\dagger}$ are a system of fermionic creation-annihilation operators
for particles, while~$\widehat{d}(\mathbf{k},s)$ and~$\widehat{d}(\mathbf{k},s)^{\dagger}$ respectively annihilate and create antiparticles
out of the vacuum. The multiparticle states obtained by acting with creation operators on the Fock vacuum are eigenstates of the Dirac Hamiltonian, 
with the elementary 
excitations~$\widehat{b}(\mathbf{k},s)^{\dagger}|0\rangle$ and~$\widehat{d}(\mathbf{k},s)^{\dagger}|0\rangle$ representing spin~${1\over 2}$
particles (resp. antiparticles) of momentum~$\mathbf{k}$, energy~$E_{\mathbf{k}}=\sqrt{\mathbf{k}^{2}+m^{2}}$, 
and spin third component~$s$. The details of this analysis are similar to the ones presented above for 
Weyl fermions and can be found in any of the QFT textbooks listed in the references. 

Finally, let us mention that Dirac spinors can be coupled to the 
electromagnetic field as we did in eq.~\eqref{eq:QED_chiral_action} for the Weyl spinors.
The Dirac action~\eqref{eq:Dirac_action1} is invariant under global phase rotation of the 
spinor,~$\psi\rightarrow e^{i\alpha}\psi$, leading to the existence of a conserved current due
to the first Noether theorem (see page~\pageref{page:noether_thm})
\begin{align}
j^{\mu}=\overline{\psi}\gamma^{\mu}\psi.
\label{eq:vector_current_QED}
\end{align}
We can use this conserved current to couple fermions to the electromagnetic field and write the QED action
\begin{align}
S&=\int d^{4}x\,\left[-{1\over 4}F_{\mu\nu}F^{\mu\nu}+\overline{\psi}\big(i{\partial\!\!\!/}-m\big)\psi
+eA_{\mu}\overline{\psi}\gamma^{\mu}\psi\right] \nonumber \\[0.2cm]
&=\int d^{4}x\,\left[-{1\over 4}F_{\mu\nu}F^{\mu\nu}+\overline{\psi}\big(i{D\!\!\!\!/}-m\big)\psi\right],
\label{eq:QED_action}
\end{align}
where once again we encounter the covariant derivative~$D_{\mu}=\partial_{\mu}-ieA_{\mu}$ and 
the slash notation introduced in eq.~\eqref{eq:(kslash-m)u=0} is used. This  
action describes the interaction
of spinors with the electromagnetic field, that upon quantization is called
quantum electrodynamics (QED). It is an interacting theory of charged particles (e.g., electrons)
and photon that, unlike the free theories we have been dealing with so far, cannot be exactly solved. 
One particularly effective way to extract physical information is perturbation 
theory. This assumes that the coupling is sufficiently weak, so that physics can be reliably 
described in terms of
the interaction among the excitations of the free theory.

\end{mdframed}

\begin{table}[t]
\begin{center}
\begin{tabular}{c|l|c}
Representation & \hspace*{2.5cm} Field & Parity \\[0.2cm] \hline 
  & & \\[-0.3cm]
$(\mathbf{0},\mathbf{0})$ & Scalar & \checkmark\\[0.2cm]
$(\mathbf{1\over 2},\mathbf{0})$ & Positive helicity Weyl spinor & $\boldsymbol{\times}$ \\[0.2cm]
$(\mathbf{0},\mathbf{1\over 2})$ & Negative helicity Weyl spinor & $\boldsymbol{\times}$\\[0.2cm]
$(\mathbf{1\over 2},\mathbf{1\over 2})$ & Vector & \checkmark \\[0.2cm]
$(\mathbf{1\over 2},\mathbf{0})\oplus (\mathbf{0},\mathbf{1\over 2})$ & Dirac spinor & \checkmark\\[0.2cm]
$(\mathbf{1},\mathbf{0})$ & Self-dual rank-2 antisymmetric tensor & $\boldsymbol{\times}$\\[0.2cm]
$(\mathbf{0},\mathbf{1})$ & Anti-self-dual rank-2 antisymmetric tensor & $\boldsymbol{\times}$\\[0.2cm]
$(\mathbf{1},\mathbf{0})\oplus (\mathbf{0},\mathbf{1})$ & Antisymmetric rank-2 tensor 
& \checkmark\\[0.2cm]
$(\mathbf{1},\mathbf{1})$ & Symmetric-traceless rank-2 tensor & \checkmark \\[0.2cm]
\hline
\end{tabular}
\end{center}
\caption{Summary of some relevant representations of the Lorentz group and their parity properties.}
\label{table:Lorentz_reps}
\end{table}

Before closing our discussion of the irreps of the Lorentz group, let us mention some more
relevant examples. The representations~$(\mathbf{s}_{+},\mathbf{s}_{-})=(\mathbf{1},\mathbf{0})$ 
and~$(\mathbf{s}_{+},\mathbf{s}_{-})=(\mathbf{0},\mathbf{1})$ correspond to 
rank-2 antisymmetric tensor fields~$B_{\mu\nu}=B_{[\mu\nu]}$ 
respectively satisfying self-dual~$(+)$ and anti-self-dual~$(-)$ conditions
\begin{align}
B_{\mu\nu}=\pm {1\over 2}\epsilon_{\mu\nu\alpha\beta}B^{\alpha\beta}.
\label{eq:(anti)selfdual_cond_rank2}
\end{align}
An example of the~$(\mathbf{1},\mathbf{0})$ and~$(\mathbf{0},\mathbf{1})$ irreps 
are the complex combinations~$\mathbf{E}\pm i\mathbf{B}$ that we encountered
in our discussion of electric-magnetic duality in page~\pageref{page:monopoles}.
The two irreps can be added
to form the parity-invariant reducible representation~$(\mathbf{1},\mathbf{0})\oplus
(\mathbf{0},\mathbf{1})$, corresponding to a generic rank-2 antisymmetric tensor field such as the
electromagnetic field strength\footnote{Rank-2 antisymmetric tensor fields are ubiquitous in 
string theories, including those satisfying the (anti-)self-dual 
condition~\eqref{eq:(anti)selfdual_cond_rank2}.}. 

Finally, multiplying together two vector representations we have
\begin{align}
\left(\mathbf{1\over 2},\mathbf{1\over 2}\right)\otimes \left(\mathbf{1\over 2},\mathbf{1\over 2}\right)
=(\mathbf{1},\mathbf{1})\oplus\big[(\mathbf{1},\mathbf{0})\oplus(\mathbf{0},\mathbf{1})\big]\oplus 
(\mathbf{0},\mathbf{0}).
\end{align}
This is just group theory lingo to express 
the decomposition of the product~$V_{\mu}W_{\nu}$ of two four-vectors 
into its symmetric-traceless, antisymmetric,
and trace pieces
\begin{align}
V_{\mu}W_{\nu}=\left(V_{(\mu}W_{\nu)}-{1\over 4}\eta_{\mu\nu}V_{\alpha}W^{\alpha}\right)
+V_{[\mu}W_{\nu]}+{1\over 4}\eta_{\mu\nu}V_{\alpha}W^{\alpha}.
\end{align}
This leads to identify the~$(\mathbf{1},\mathbf{1})$ irrep as corresponding to a 
symmetric-traceless rank-2 tensor field. For the reader's benefit, we have 
summarized in table~\ref{table:Lorentz_reps}
the different representations of the Lorentz group discussed in this section, 
indicating as well whether or not they 
preserve parity.

\subsection{Some more group theory}

Having got some practice with the language of group theory, 
we close this section by enlarging our vocabulary with many 
important group-theoretic concepts that will become handy later on (see~\cite{Georgi,Ramond_groups}
for some physics oriented textbooks on group theory, or Appendix~B of~\cite{AG_VM} for a quick survey of basic
facts). Next, we focus 
on the relevant groups for the SM, namely~SU(3),~SU(2), and~U(1) associated with the strong
and electroweak interactions. The Abelian group~U(1) we have encountered when
discussing electromagnetism and learned there that it has a single generator, let us call it~$Q$,
so its elements are written as~$U(\vartheta)=e^{i\vartheta Q}$. This is the only 
irrep of this group, all others being reducible to a diagonal form.

Concerning~SU(2), its properties are well know from the theory of angular momentum in 
quantum mechanics and we have already used many of them in our analysis of the representations of the 
Lorentz group. Its 
three generators satisfy the algebra
\begin{align}
[T^{a}_{\mathbf{R}},T^{b}_{\mathbf{R}}]=i\epsilon^{abc}T^{c}_{\mathbf{R}},
\end{align}
where the subscript~$\mathbf{R}$ denotes the representation. Up to this point, 
we have labelled the irreps of SU(2) by their spin~$\mathbf{s}=\mathbf{0},
\mathbf{1\over 2},\mathbf{1},\ldots$ 
although they
are also frequently referred to by their dimension~$2\mathbf{s}+1$, as it is customary for all 
unitary groups~SU($N$). As an example,
the fundamental representations~$\mathbf{s}=\mathbf{1\over 2}$ is denoted by~$\mathbf{2}$ and
the adjoint~$\mathbf{s}=\mathbf{1}$ by~$\mathbf{3}$. In the former case 
the generators are written in terms of the three Pauli matrices
as~$T^{a}_{\mathbf{2}}={1\over 2}\sigma_{a}$, a fact we used when studying Weyl spinors.

As for the group~SU(3), less familiar from elementary physics, it 
has eight generators satisfying the Lie algebra
\begin{align}
[T_{\mathbf{R}}^{a},T_{\mathbf{R}}^{b}]=if^{abc}T_{\mathbf{R}}^{c} \hspace*{2cm} (a,b,c=1,\ldots,8),
\end{align}
where the structure constants are given by
\begin{align}
f^{123}=1,\hspace*{0.5cm} f^{147}=-f^{156}=f^{246}=f^{257}=f^{345}=-f^{367}={1\over 2},
\hspace*{0.5cm} f^{458}=f^{678}={\sqrt{3}\over 2},
\end{align}
the remaining ones being either zero or fixed from the ones just given by antisymmetry.
The group elements
are written as exponentials of linear combinations of the algebra generators
\begin{align}
U(\alpha)_{\mathbf{R}}=e^{i\alpha^{a}T^{a}_{\mathbf{R}}},
\end{align}
where the condition~$\det U(\alpha)_{\mathbf{R}}=1$ 
implies~${\rm tr\,}T^{a}_{\mathbf{R}}=0$ and the generators 
can be chosen to 
satisfy the orthogonality relations
\begin{align}
{\rm tr\,}\big(T^{a}_{\mathbf{R}}T^{b}_{\mathbf{R}}\big)=T_{2}(\mathbf{R})\delta^{ab}.
\end{align} 

Although similar in many aspects, there are however important differences between~SU(2) and~SU(3) 
concerning the character of their irreps. For any Lie algebra representation with
generators~$T^{a}_{\mathbf{R}}$ it is very easy to check that~$-T^{a*}_{\mathbf{R}}$ satisfy the same Lie algebra, defining the complex conjugate 
representation denoted by~$\overline{\mathbf{R}}$. A representation is said to be {\em real}
or {\em pseudoreal} whenever it is related to its complex conjugate irrep by a similarity~\label{page:complex_irreps}
transformation
\begin{align}
T^{a}_{\overline{\mathbf{R}}}\equiv-T^{a*}_{\mathbf{R}}=S^{-1}T^{a}_{\mathbf{R}}S,
\end{align}
with~$S$ either symmetric (real representation) or antisymmetric (pseudoreal representation).
For~SU(2) all irreps are real or pseudoreal. This is the reason why we only have one independent
irrep of a given dimension labelled by its spin. 
The group SU(3), on the other hand, has complex irreps. This 
is the case of the fundamental and an antifundamental 
representations,~$\mathbf{3}$ and~$\overline{\mathbf{3}}$, 
whose generators are given by
\begin{align}
T^{a}_{\mathbf{3}}={1\over 2}\lambda_{a} \hspace*{1cm} \mbox{and} \hspace{1cm} 
T^{a}_{\overline{\mathbf{3}}}=-{1\over 2}\lambda_{a}^{T},
\end{align}
where~$\lambda_{a}$ are the eight Gell-Mann matrices, given by
\begin{align}
\lambda_{1}&=\left(
\begin{array}{ccc}
0 & 1 & 0 \\
1 & 0 & 0 \\
0 & 0 & 0
\end{array}
\right), \hspace*{0.6cm}
\lambda_{2}=\left(
\begin{array}{ccc}
0 & -i & 0 \\
i & 0 & 0 \\
0 & 0 & 0
\end{array}
\right), \hspace*{0.6cm}
\lambda_{3}=\left(
\begin{array}{ccc}
1 & 0 & 0 \\
0 & -1 & 0 \\
0 & 0 & 0
\end{array}
\right), \nonumber \\[0.3cm]
\lambda_{4}&=\left(
\begin{array}{ccc}
0 & 0 & 1 \\
0 & 0 & 0 \\
1 & 0 & 0
\end{array}
\right), \hspace*{0.6cm}
\lambda_{5}=\left(
\begin{array}{ccc}
0 & 0 & -i \\
0 & 0 & 0 \\
i & 0 & 0
\end{array}
\right), \hspace*{0.6cm}
\lambda_{6}=\left(
\begin{array}{ccc}
0 & 0 & 0 \\
0 & 0 & 1 \\
0 & 1 & 0
\end{array}
\right), 
\label{eq:appendix:gmmatrices}\\[0.3cm]
\lambda_{7}&=\left(
\begin{array}{ccc}
0 & 0 & 0 \\
0 & 0 & -i \\
0 & i & 0
\end{array}
\right), \hspace*{0.4cm}
\lambda_{8}=\left(
\begin{array}{ccc}
{1\over \sqrt{3}} & 0 & 0 \\
0 & {1\over \sqrt{3}} & 0 \\
0 & 0 & -{2\over\sqrt{3}}
\end{array}
\right). \nonumber 
\end{align}

Two instances of the group~SU(3) exists in the SM. One is the color gauge symmetry
of QCD, which we will study in some detail
in later sections. The second is the
global~SU(3)$_{f}$ flavor symmetry of the eightfold way, originally formulated by 
Murray Gell-Mann~\cite{Gell-Mann_SU3} and
Yuval Ne'eman~\cite{Neeman_SU3}. With the hindsight provided by the quark model, this classification
scheme is based on the assumption that strong nuclear force does not distinguish 
among different quark flavors\footnote{Quarks were proposed as hadron constituents
in~\cite{Gell-Mann_quarks,Zweig_quarks}, some
three years after the formulation of the eightfold way. The name, as with quarks, 
was invented by Gell-Mann
drawing this time not from James Joyce but
from the Noble Eightfold Path of Buddhism: Right View, Right Intention, Right Speech, 
Right Conduct, Right Livelihood, Right Effort, Right Mindfulness, and Right Meditation.}.
Let us consider the action for three quark flavors~$q_{i}$ ($i=1,2,3$)\label{page:eightfold_way}
\begin{align}
S&=\sum_{i=u,d,s}\int d^{4}x\,\overline{q}_{i}\big(i{\partial\!\!\!/}-m_{i}\big)q_{i}
+S_{\rm int} \nonumber \\[0.2cm]
&=\int d^{4}x\,\overline{\boldsymbol{q}}\big(i{\partial\!\!\!/}\mathbbm{1}
-\boldsymbol{m}\big)\boldsymbol{q}+S_{\rm int},
\label{eq:quark_action1}
\end{align} 
where~$S_{\rm int}$ represent interaction terms that we will not care about for the time being and
in the second line we have grouped the quarks into a triplet~$\boldsymbol{q}$ and rewrote the
action in matrix notation, with~$\boldsymbol{m}={\rm diag}(m_{u},m_{d},m_{s})$.
Under~$\mbox{SU(3)}_{f}$ the quark triplet transforms in the fundamental 
irrep~$\boldsymbol{3}$ as~$\boldsymbol{q}\rightarrow U\boldsymbol{q}$. 
This result in the following transformation of the free action 
\begin{align}
\int d^{4}x\,\overline{\boldsymbol{q}}\big(i{\partial\!\!\!/}\mathbbm{1}
-\boldsymbol{m}\big)\boldsymbol{q}
\longrightarrow 
\int d^{4}x\,\overline{\boldsymbol{q}}\big(i{\partial\!\!\!/}\mathbbm{1}-
U^{\dagger}\boldsymbol{m}U\big)\boldsymbol{q},
\end{align}
where~$\boldsymbol{m}={\rm diag}(m_{u},m_{d},m_{s})$. 
Since all three quark masses are different,~$\boldsymbol{m}$ is not proportional
to the identity and~$U^{\dagger}\boldsymbol{m}U\neq \boldsymbol{m}$, and 
the mass term breaks the global~SU(3)$_{f}$
invariance. Moreover, the strong interaction does not distinguishes quark flavors and~$S_{\rm int}$ 
remains invariant. Thus, we conclude that~SU(3)$_{f}$
is an approximate symmetry of QCD that becomes exact
in the limit of equal, in particular 
zero quark masses (also called, for obvious reasons, the chiral limit). 

Mesons are bound states of a quark and an antiquark, the later transforming in 
the antifundamental~$\overline{\mathbf{3}}$ irrep. Their classification into~SU(3)$_{f}$ multiplets 
follows from decomposing into irreps the product of the fundamental and the antifundamental
\begin{align}
\mathbf{3}\otimes\overline{\mathbf{3}}=\mathbf{8}\oplus\mathbf{1}.
\label{eq:octet_SU(3)_mesons}
\end{align} 
The octet contains the~$\pi^{0}$, $\pi^{\pm}$, $K^{0}$, $\overline{K}^{0}$,
$K^{\pm}$, and~$\eta_{8}$ mesons, while the singlet is the $\eta_{1}$ meson. 
In fact, the~$\eta_{1}$ and~$\eta_{8}$ mesons
mix together into the~$\eta$ and the~$\eta'$ mesons, which are the interaction eigenstates 
in the
electroweak sector of the SM. A similar classification scheme works for the baryons. 
Being composed of three quarks, the baryon multiplets emerge from decomposing the product of three fundamental
representations
\begin{align}
\mathbf{3}\otimes\mathbf{3}\otimes\mathbf{3}=\mathbf{10}\oplus\mathbf{8}\oplus\mathbf{8}\oplus\mathbf{1}.
\end{align}
The proton and the neutron are in one of the octets, together with the~$\Sigma^{0}$, $\Sigma^{\pm}$,
$\Xi^{0}$, and $\Xi^{-}$ 
particles of nonzero strangeness. 
Were SU(3)$_{f}$ an exact symmetry, the masses of all
hadrons within a single multiplet would be equal. However, the differences
in the quark masses induce a mass split, which in the case of the octet containing 
the proton and the neutron, is about~$30\%$ of the average mass. By contrast, the mass split
between the proton and the neutron is only $0.1\%$ of their average mass. The wider mass gap with
the other octet members results from the larger mass of the strange 
quark,~$m_{s}>m_{u}\sim m_{d}$.

\section{A tale of many symmetries}
\label{sec:tale_of_symmetries}

Symmetry is probably the most important heuristic principle at our disposal in fundamental physics. The formulation of
particle physics models starts with selecting those symmetries/invariances to be implemented
in the theory,
what usually restrict drastically the types of interactions allowed. 
In the SM gauge, for example, invariance plus the condition that the action
only contains operators of dimension four or less fixes the action, up to
a relatively small number of numerical parameters to be experimentally measured in high energy facilities.

\subsection{The symmetries of physics}

Our approach to symmetry up to here 
has been rather casual. 
It is time to be more precise, 
beginning with a discussion of the types of symmetries we encounter in QFT and how they are implemented.
\begin{enumerate}

\item[i)]
{\bf Kinematic (or spacetime) symmetries}. They act on the spacetime coordinates 
and field indices. This class of symmetries includes Lorentz,
Poincar\'e, scale, and conformal transformations that we already encountered in previous sections.

\item[ii)]
{\bf Discrete symmetries.} They
include parity~P, charge conjugation~C, time reversal~T, 
and the compositions~CP and~CPT. If gravity and electromagnetism were the only interactions in 
nature,
the universe would be invariant under C, P, and T separately. However, 
nuclear (both weak and strong) interactions break P, C, T and CP in different degrees. 

CPT, however, turns out to be a symmetry of QFT forced upon us by
the basic requirements of Poincar\'e invariance and locality. Moreover, it is a completely general result
that can be demonstrated without relying on the specific form of any Hamiltonian (for a detailed
proof of this result, called the CPT theorem, see chapter~11 of~\cite{AG_VM}).

\item[iii)]
{\bf Global continuous symmetries}. These are transformations depending on a 
continuous constant parameter. One example is the invariance of the complex scalar field 
action~\eqref{eq:action_free_scalar_field}
under spacetime constant phase rotation~\eqref{eq:phase_rotation_scalarfield}. 
The current view in QFT is that
global symmetries are accidental properties of the low energy theories, whereas, in the UV, all fundamental symmetries should be 
local (see next).

\item[iv)]
{\bf Local (gauge) invariance}. Unlike the previous case, the theory is invariant under a set
of continuous transformations that vary from point to point in 
spacetime. The archetypical example is the gauge invariance of the
Maxwell's equations found in~\eqref{eq:gauge_transf_em_1}.
Unlike standard quantum mechanical symmetries, gauge invariance does not map one physical state into another, but 
represents a redundancy in the labelling the physical states.\label{pag:redundancy_gauge_page_2}
This is 
the price we pay to describe fields  
with spin one and two in a way that manifestly preserves locality and Lorentz invariance. To highlight 
this fundamental feature, we will refrain from talking about gauge symmetry and stick to gauge 
invariance (we will qualify this statement below). 

\item[v)]
{\bf Spontaneously/softly broken symmetries.} In all instances discussed above, we have assumed that 
the symmetries/invariances are
realized at the action level and in the spectrum of the quantum theory.
Classically, it is possible that the symmetries of the action are not reflected in their solutions which
implies that in the quantum theory, the spectrum does not remain invariant under the symmetry. When this
happens, we say that the symmetry (or invariance) is {\em spontaneously broken}. 
Since the breaking takes place
by the choice of vacuum, it does not affect the UV behavior of the theory. 
Another situation when this also happens is when adding terms to the action that, 
explicitly break the symmetry but does not modify the UV behavior of the theory (e.g., mass terms).
In this case, the symmetry is {\em softly broken}.

\item[vi)]
{\bf Anomalous symmetries.}
Usually, symmetries are identified in the classical action and then implemented in the quantum theory.
This tacitly assumes that all classical symmetries remain after quantization, and this 
is not always the case. Sometimes, the classical symmetry is impossible to implement quantum mechanically, 
and it is said to be {\em anomalous}. Anomalies are originated in very profound
mathematical properties of QFT and they have important physical consequences.

\end{enumerate}

Let us see now how symmetries are implemented in QFT. We know from quantum mechanics
that symmetries are maps among rays in the theory's Hilbert space that preserve probability amplitudes.
More precisely, for two arbitrary states~$|\alpha\rangle$ and~$|\beta\rangle$, a symmetry is
implemented by some operator~$U$ acting as
\begin{align}
|\alpha\rangle\longrightarrow |U\alpha\rangle, \hspace*{1cm} 
|\beta\rangle\longrightarrow |U\beta\rangle.
\end{align}
and satisfying the condition that probability amplitudes are preserved
\begin{align}
|\langle\alpha|\beta\rangle|=|\langle U\alpha|U\beta\rangle|.
\label{eq:Wigner_eq1}
\end{align}
There are two ways in which this last condition can be achieved. One is that
\begin{align}
\langle\alpha|\beta\rangle=\langle U\alpha|U\beta\rangle,
\end{align}
implying that the operator~$U$ is {\em unitary}. 
But there also exists a second
alternative to fullfil eq.~\eqref{eq:Wigner_eq1}
\begin{align}
\langle U\alpha|U\beta\rangle=\langle\alpha|\beta\rangle^{*}.
\end{align}
In this case the operator~$U$ is said to be {\em antiunitary}. Notice that consistency requires 
that in this case the operator~$U$ implementing the symmetry 
should be antilinear:
\begin{align}
U\big(a|\alpha\rangle+b|\beta\rangle\big)=a^{*}|U\alpha\rangle+b^{*}|U\beta\rangle,
\end{align}
for any two states~$|\alpha\rangle$ and~$|\beta\rangle$ and~$a,b\in\mathbb{C}$.

Our discussion has led us to 
Wigner's theorem~\cite{Wigner_th}: symmetries are implemented
quantum-mechanically either by unitary or antiunitary operators.
In fact, continuous symmetries are always implemented by the first kind. 
This can be understood by thinking that a family of operators~$U(\lambda)$, depending
on a continuous parameter, can always
be smoothly deformed to the identity, a linear and not an antilinear operator. 
On the other hand, there are two critical discrete symmetries implemented by antiunitary operators: 
time reversal~T and~CPT.

\subsection{Noether's two theorems}
\label{sec:Noether}

In the case of continuous symmetries, we have the celebrated theorem due to Noether linking
them to the existence of conserved quantities~\cite{Noether}. \label{page:noether_thm}
What is often called ``the'' 
Noether theorem is actually the first of two theorems, dealing with 
the consequences of 
{\em global} and {\em local} symmetries respectively. 
Let us begin with the first one considering a classical field theory of~$n$ fields
whose field equations remain invariant 
under infinitesimal variations~$\phi_{i}\rightarrow\phi_{i}+\delta_{\epsilon}\phi_{i}$ linearly
depending on $N$~continuous parameters $\epsilon_{A}$. There are two essential things about the
transformations we are talking about. First, they form a group, as can be seen by noticing
that the composition of two symmetries is itself a symmetry and, that for each transformation, there exists
its inverse obtained by reversing the signs of~$\epsilon_{A}$. The second fact is that the infinitesimal
transformations can be exponentiated to cover all transformations that can be continuously connected
to the identity. The latter statement is rather subtle in the case of diffeomorphisms (i.e., coordinate transformations), but we will not worry about them here.

Since the transformations leave invariant the field equations, the theory's 
Lagrangian density 
must change at most by a total derivative, namely
\begin{align}
S=\int d^{4}x\,\mathcal{L}(\phi_{i},\partial_{\mu}\phi_{i}) \hspace*{1cm} \Longrightarrow \hspace*{1cm}
\delta_{\epsilon}S=\int d^{4}x\,\partial_{\mu}K^{\mu},
\label{eq:general_variationN1}
\end{align}
where~$K^{\mu}$ is linear in the~$\epsilon_{A}$'s.
At the same time, a general variation of the action can be written as
\begin{align}
\delta_{\epsilon}S
=\int d^{4}x\,\left\{\left[{\partial\mathcal{L}\over\partial \phi_{i}}
-\partial_{\mu}\left({\partial\mathcal{L}\over \partial\, \partial_{\mu}\phi_{i}}\right)
\right]\delta_{\epsilon}\phi_{i}+
\partial_{\mu}\left({\partial\mathcal{L}\over \partial\,\partial_{\mu}\phi_{i}}
\delta_{\epsilon}\phi_{i}\right)\right\},
\label{eq:general_variation2_N1}
\end{align}
so equating expressions~\eqref{eq:general_variationN1} 
and~\eqref{eq:general_variation2_N1}, we find
\begin{align}
\int d^{4}x\,\left\{\left[{\partial\mathcal{L}\over\partial \phi_{i}}
-\partial_{\mu}\left({\partial\mathcal{L}\over \partial\, \partial_{\mu}\phi_{i}}\right)
\right]\delta_{\epsilon}\phi_{i}+
\partial_{\mu}\left({\partial\mathcal{L}\over \partial\,\partial_{\mu}\phi_{i}}
\delta_{\epsilon}\phi_{i}-K^{\mu}\right)\right\}=0,
\label{eq:general_variation3_N1}
\end{align}
which is valid for arbitrary~$\epsilon$. From this equation we identify the conserved current
\begin{align}
j^{\mu}(\epsilon)={\partial\mathcal{L}\over \partial\,\partial_{\mu}\phi_{i}}
\delta_{\epsilon}\phi_{i}-K^{\mu}
\hspace*{1cm} \Longrightarrow \hspace*{1cm}
\partial_{\mu}j^{\mu}(\epsilon)=
\left[\partial_{\mu}\left({\partial\mathcal{L}\over \partial\, \partial_{\mu}\phi_{i}}\right)
-{\partial\mathcal{L}\over\partial \phi_{i}}\right]\delta_{\epsilon}\phi_{i}\approx 0,
\label{eq:current_N1}
\end{align}
where again we used the Dirac notation first
introduced in page~\pageref{page:Dirac_notation}. Notice that since the expression of the current
is linear in the parameters~$\epsilon_{A}$ the current can be written as~$j^{\mu}(\epsilon)
=\epsilon_{A}j^{\mu}_{A}$, and~\eqref{eq:current_N1} is satisfied for 
arbitrary values of~$\epsilon_{A}$,
we conclude that there are a total of~$N$ conserved currents~$\partial_{\mu}j^{\mu}_{A}$.
An important point glaring in the previous analysis is that current conservation
happens {\em on-shell}, i.e., once the equations of motion are implemented\footnote{A
note of warning: the
term on-shell is employed in physics with at least two different meanings. In the one used here we say that
an identity is valid on-shell whenever it holds after the equations of motion are implemented.
The second use applies to the four-momentum of a particle with mass~$m$. 
The momentum~$p^{\mu}$ (or the particle carrying it)
is said to be on-shell if it satisfies~$p^{2}=m^{2}$. As an example,
particles running in loops in Feynman diagrams are off-shell in this sense.}.

The second Noether theorem \label{page:Noether_second} deals with local symmetries depending on a number of
point-dependent parameters~$\epsilon_{A}(x)$. It is important to keep in mind that the first theorem remains
valid in this case, in the sense that there exists a current $j_{\mu}$ whose divergence is
proportional to the equations of motion. To simplify expressions, let us denote the latter as
\begin{align}
E_{i}(\phi)\equiv \partial_{\mu}\left({\partial\mathcal{L}\over\partial\,\partial_{\mu}\phi_{i}}\right)
-{\partial\mathcal{L}\over \partial \phi_{i}},
\end{align}
and consider that our theory is invariant under field transformations involving 
only $\epsilon_{A}(x)$ and their 
first derivatives
\begin{align}
\delta_{\epsilon}\phi_{i}=R_{i,a}(\phi_{k})\epsilon_{A}
+R^{\mu}_{i,A}(\phi_{k})\partial_{\mu}\epsilon_{A}.
\label{eq:local_trans_N2}
\end{align}
This includes, for example, the gauge transformations of 
electromagnetism,~$\delta_{\epsilon}A_{\mu}=\partial_{\mu}\epsilon$
(the argument here can be 
easily generalized to include transformations 
depending up to the $k$-th derivative of the gauge functions).
The general variation of the action~$\delta_{\epsilon}S$ has the same structure shown in 
eq.~\eqref{eq:general_variation3_N1}
\begin{align}
\int d^{4}x\,\Big[-E_{i}(\phi)\delta_{\epsilon}\phi_{i}+\partial_{\mu}j^{\mu}(\epsilon)\Big]=0.
\label{eq:general_variation_new_notationN2}
\end{align}
with~$\delta_{\epsilon}\phi_{i}$ given in~\eqref{eq:local_trans_N2} and~$j^{\mu}$ the 
Noether current implied by the first theorem and defined in eq.~\eqref{eq:current_N1}.
A crucial difference now is that since~$j^{\mu}(\epsilon)$ is linear in~$\epsilon_{A}$, 
when these parameters vanish at infinity 
the boundary term on the right-hand side
appearing when integrating by parts is zero
\begin{align}
\delta_{\epsilon}S&=
-\int d^{4}x\,\epsilon_{A}(x)\Big\{R_{i,A}(\phi_{k})E_{i}(\phi_{k})
-\partial_{\mu}\Big[R_{i,A}^{\mu}(\phi_{k})
E_{i}(\phi_{k})\Big]\Big\}.
\end{align} 
Thus, if this is a symmetry.~$\delta_{\epsilon}S=0$ for any~$\epsilon_{A}(x)$, we obtain the identities
\begin{align}
R_{i,A}(\phi_{k})E_{i}(\phi_{k})-\partial_{\mu}\Big[R_{i,A}^{\mu}(\phi_{k})E_{i}(\phi_{k})\Big]=0,
\label{eq:Delta(E)}
\end{align}
where we should remember 
that~$A=1,\ldots,N$, with~$N$ the number of gauge functions (i.e., the dimension of the
symmetry's Lie algebra). This results is Noether's second theorem:
invariance of a field theory under local transformations implies the existence of several 
differential identities among the field equations, meaning that some
are redundant. 

As to the existence of conserved currents associated with local
invariance, using eq.~\eqref{eq:Delta(E)} it can be shown that
\begin{align}
\partial_{\mu}\Big[\epsilon_{A}(x)R^{\mu}_{i,A}(\phi_{k})E_{i}(\phi_{k})\Big]
=E_{i}(\phi_{k})\delta_{\epsilon}\phi_{i},
\end{align}
from where we read the conserved current
\begin{align}
S^{\mu}(\epsilon)\equiv\epsilon_{A}(x)R^{\mu}_{i,A}(\phi_{k})E_{i}(\phi_{k}) 
\hspace*{1cm} \Longrightarrow
\hspace*{1cm} \partial_{\mu}S^{\mu}(\epsilon)=E_{i}(\phi_{k})\delta_{\epsilon}\phi_{i}\approx 0.
\label{eq:SmuN2}
\end{align}
This quantity is however trivial, in the sense that it vanishes on-shell,~$S^{\mu}(\epsilon)\approx 0$. 
Notice, however, that
the conserved current obtained as the result of the first Noether theorem also applies to the gauge case.
Indeed, considering
transformations such that~$\epsilon_{A}(x)$ does not vanish at infinity, we find 
from~\eqref{eq:general_variation_new_notationN2}
\begin{align}
\partial_{\mu}j^{\mu}(\epsilon)=E_{i}(\phi_{k})\delta_{\epsilon}\phi_{i}\approx 0,
\label{eq:jmu-SmuN2}
\end{align}
where~$j^{\mu}$ is explicitly given by the expression on the left of eq.~\eqref{eq:current_N1}. 
This shows that for theories with local invariances the only nontrivial conserved currents are the 
ones provided by Noether's first theorem, associated with transformations that do not vanish 
at infinity (see also the discussion in Box~9 below).

Together with the conserved current from the first Noether theorem, 
there exists a conserved charge defined by its time component
\begin{align}
Q(\epsilon)=\int_{\Sigma}d^{3}r\,j^{0}(\epsilon),
\label{eq:chargeQgeneraldef}
\end{align}
where~$\Sigma$ is a three-dimensional spatial section of spacetime. Using current conservation 
it is easy to see that the time derivative of the charge vanishes on-shell
\begin{align}
\dot{Q}(\epsilon)\approx-\int_{\Sigma}d^{3}r\,\boldsymbol{\nabla}\cdot\mathbf{j}(\epsilon)
=\int_{\partial\Sigma}
d\mathbf{S}\cdot\mathbf{j}(\epsilon)=0,
\label{eq:dotQ=0}
\end{align}
provided the spatial components of the current~$\mathbf{j}(\epsilon)$ is zero at~$\partial\Sigma$ 
or, equivalently, there is no flux of charge entering or leaving the spatial sections at infinity.

Applying the first Noether theorem to different symmetries, we get a number of conserved quantities:

\begin{itemize}
\item[-] 
The energy-momentum tensor~$T^{\mu}_{\,\,\,\,\nu}$ is the conserved current associated with 
the invariance of field theories under
spacetime translations,~$x^{\mu}\rightarrow x^{\mu}+a^{\mu}$. Its general expression 
is
\begin{align}
T^{\mu}_{\,\,\,\,\nu}={\partial\mathcal{L}\over \partial\,\partial_{\mu}\phi_{i}}\partial_{\nu}\phi_{i}
-\delta^{\mu}_{\nu}\mathcal{L},
\end{align}
with~$\partial_{\mu}T^{\mu}_{\,\,\,\,\nu}=0$.
Notice that this canonical is not necessarily symmetric
as, for example, in Maxwell's electrodynamics
\begin{align}
T^{\mu}_{\,\,\,\,\nu}=-F^{\mu\alpha}\partial_{\nu}A_{\alpha}+{1\over 4}\delta^{\mu}_{\nu}
F_{\alpha\beta}F^{\alpha\beta}.
\label{eq:em_tensor_canonical}
\end{align} 
It can nevertheless be symmetrized by adding a term of the
form~$\partial_{\sigma}K^{\sigma\mu}_{\,\,\,\,\,\,\,\,\nu}$, with~$K^{\sigma\mu}_{\,\,\,\,\,\,\,\,\nu}
=-K^{\mu\sigma}_{\,\,\,\,\,\,\,\,\nu}$, that does not spoil its 
conservation~\cite{Belinfante,Rosenfeld}. In the case of the electromagnetism, the resulting
Belinfante-Rosenfeld energy-momentum tensor reads
\begin{align}
K^{\mu\nu}_{\,\,\,\,\,\,\,\,\sigma}=F^{\mu\nu}A_{\sigma} \hspace*{1cm}
\Longrightarrow \hspace*{1cm} \widetilde{T}^{\mu}_{\,\,\,\,\nu}=-F^{\mu\alpha}F_{\nu\alpha}
+{1\over 4}\delta^{\mu}_{\nu}F_{\alpha\beta}F^{\alpha\beta}.
\end{align}
This modified energy-momentum tensor not only is symmetric 
but, unlike~\eqref{eq:em_tensor_canonical}, also gauge invariant. 
Notice that since conserved currents are 
quantities evaluated on-shell, we can apply the vacuum field equations~$\partial_{\mu}F^{\mu\nu}=0$.

\item[-]
Invariance under infinitesimal Lorentz transformations~$\delta x^{\mu}=\omega^{\mu}_{\,\,\,\nu}x^{\nu}$,
with~$\omega_{\mu\nu}=-\omega_{\nu\mu}$, implies the conservation of the total angular momentum
\begin{align}
J^{\mu}_{\,\,\,\,\nu\sigma}=T^{\mu}_{\,\,\,\,\nu}x_{\sigma}-T^{\mu}_{\,\,\,\,\sigma}x_{\nu}
+S^{\mu}_{\,\,\,\,\nu\sigma},
\end{align}
where~$J^{\mu}_{\,\,\,\,\nu\sigma}=-J^{\mu}_{\,\,\,\,\sigma\nu}$ 
and~$\partial_{\mu}J^{\mu}_{\,\,\,\,\nu\sigma}=0$.
The first two terms on the right-hand side
represent the ``orbital'' contribution induced by the Lorentz variation 
of the spacetime coordinates, while~$S^{\mu}_{\,\,\,\,\nu\sigma}$ is the ``intrinsic'' angular momentum
(or spin)
coming from the spacetime transformation properties of the field 
itself. For a scalar field this last part vanishes\footnote{To connect with the notation employed in our discussion of the first Noether theorem, let us indicate
that the conserved current~\eqref{eq:current_N1} 
associated to the invariance under spacetime translations 
is written by~$j^{\mu}(a^{\sigma})=T^{\mu}_{\,\,\,\,\nu}a^{\nu}$, 
whereas~$j^{\mu}(\omega^{\alpha\beta})=J^{\mu}_{\,\,\,\,\nu\sigma}\omega^{\nu\sigma}$
is the current whose conservation follows from Lorentz invariance.}.

\item[-]
As a further application, let us mention the invariance of complex fields under phase rotation,
already anticipated in various examples in previous pages. For instance, in the case
of the complex scalar field studied in Box~6, applying~\eqref{eq:current_N1}
to infinitesimal
variations~$\delta_{\vartheta}\phi=i\vartheta\phi$,~$\delta_{\vartheta}\phi^{*}=-i\vartheta\phi^{*}$
leads to the conserved 
current~\eqref{eq:conserved_current_complexfield}. The corresponding analysis for 
Weyl spinors gives~\eqref{eq:conserved_current_weyl}.  

\end{itemize}

\subsection{Quantum symmetries: to break or not to break (spontaneously)}

In the quantum theory symmetries are realized on the Hilbert space of
physical states. In particular, the charge~\eqref{eq:chargeQgeneraldef} is 
promoted to a Hermitian operator~$\widehat{Q}(\epsilon)$ implementing infinitesimal transformations on
the fields
\begin{align}
\delta_{\epsilon}\widehat{\phi}_{k}=-i[\widehat{Q}(\epsilon),\widehat{\phi}_{k}],
\end{align}
whereas, due to the conservation equation~\eqref{eq:dotQ=0}, it
commutes with the Hamiltonian,~$[\widehat{Q}(\epsilon),\widehat{H}]=0$. 
In the case of rigid transformations,
the parameters~$\epsilon_{A}$ can be taken outside the integral in~\eqref{eq:chargeQgeneraldef} to
write~$\widehat{Q}(\epsilon)=\epsilon_{A}\widehat{Q}^{A}$. Finite transformations in the 
connected component of the identity are obtained then
by exponentiating the charge operator
\begin{align}
\widehat{\mathscr{U}}(\epsilon)=e^{i\epsilon_{A}\widehat{Q}^{A}} \hspace*{1cm} \Longrightarrow
\hspace*{1cm} \widehat{\mathscr{U}}(\epsilon)^{\dagger}
\widehat{\phi}_{k}(x)\widehat{\mathscr{U}}(\epsilon)=\mathscr{U}_{k\ell}(\epsilon)\widehat{\phi}_{\ell}(x),
\label{eq:transformation_phi_group}
\end{align}
where~$\mathscr{U}_{k\ell}(\epsilon)$ is the representation of the symmetry group acting
on the field indices and
the Hermiticity of~$\widehat{Q}$ guarantees the unitarity 
of~$\widehat{\mathscr{U}}(\epsilon)$. The implication for the free theory is that the creation-annihilation
operators transform covariantly under the symmetry. Consequently, to determine the action 
of~$\widehat{\mathscr{U}}(\epsilon)$ on the Fock space of the theory, we need to know 
how the charge acts on the vacuum. Here, we 
may have two possibilities corresponding to different realization of the symmetry.

\noindent
{\bf Wigner-Weyl realization:} the vacuum state is left invariant by the symmetry
\begin{align}
\widehat{\mathscr{U}}(\epsilon)|0\rangle=|0\rangle \hspace*{1cm} \Longrightarrow \hspace*{1cm} 
\widehat{Q}_{a}|0\rangle=0.
\end{align}
If this is the case, the symmetry is manifest in the spectrum, falling into representations
of the symmetry group. Since the whole Fock space is generated by succesive application of 
the fields~$\widehat{\phi}_{k}(x)$ on the vacuum, it is enough to know how the symmetry acts on the 
states~$|\phi_{k}\rangle\equiv \widehat{\phi}_{k}(x)|0\rangle$
\begin{align}
\widehat{\mathscr{U}}(\epsilon)|\phi_{k}\rangle
=\mathscr{U}_{k\ell}(\epsilon)|\phi_{\ell}\rangle.
\end{align}
where~$\mathscr{U}_{k\ell}(\epsilon)$ is the representation of the symmetry group 
introduced in~\eqref{eq:transformation_phi_group}.

This is what happens, for example, in the hydrogen atom. Its ground state has~$j=0$ and therefore
remains invariant a generic rotation labelled by the Euler angles~$\phi$, $\theta$, and $\psi$
\begin{align}
\widehat{\mathscr{R}}(\phi,\theta,\psi)|0,0,0\rangle=|0,0,0\rangle,
\end{align}
while the other states transform in irreps of the rotation group~$\mbox{SO(3)}\simeq\mbox{SU(2)}$
\begin{align}
\widehat{\mathscr{R}}(\phi,\theta,\psi)|n,j,m\rangle=\sum_{m'=-j}^{j}\mathscr{D}^{(j)}_{mm'}(\phi,\theta,\psi)
|n,j,m'\rangle,
\end{align}  
where~$\mathscr{D}^{(j)}_{mm'}(\phi,\theta,\psi)$ is the spin~$j$ rotation matrix~\cite{Merzbacher}.
From this point of view, the angular momentum and magnetic quantum numbers introduced
to account for certain properties of atomic spectra 
are just group theory labels indicating how the atomic state transforms under spatial rotations.
Symmetries in quantum mechanical systems with finite degrees of freedom 
are usually realized \`a la Wigner-Weyl, since tunneling among different vacua results in 
an invariant ground state.
We will return to this issue on page~\pageref{page:SSB_infinite_volume}.

\noindent
{\bf Nambu-Goldstone realization:} the vacuum state is not invariant under the symmetry. This means
that the conserved charge do not annihilate the vacuum
\begin{align}
\widehat{Q}(\epsilon)|0\rangle \neq 0.
\label{eq:NG_charge_general}
\end{align}
Whenever this happens, the symmetry is said to be {\em spontaneously broken}.
Notice that the previous equation does not imply 
that~$\widehat{Q}_{a}|0\rangle\neq 0$ for all~$a$. 
There might be a subset of charges satisfying~$\widehat{Q}_{A}|0\rangle=0$, 
with~$\{A\}\subset\{a\}$
that we refer
to as {\em unbroken} generators. It is easy to see that 
since~$[\widehat{Q}_{A},\widehat{Q}_{B}]|0\rangle=0$,
they must form a closed subalgebra under commutation. 

Let us illustrate this mode of realization of the symmetry with the example of~$N$
real scalar fields~$\varphi^{i}$ with action
\begin{align}
S=\int d^{4}x\,\left[{1\over 2}\partial_{\mu}\varphi^{i}\partial^{\mu}\varphi^{i}-V(\varphi^{i}
\varphi^{i})\right].
\label{es:action_scalarth_SSB}
\end{align}
This theory is invariant under global infinitesimal transformations
\begin{align}
\delta_{\epsilon}\varphi^{i}=\epsilon_{a}(T_{\mathbf{f}}^{a})^{i}_{\,\,\,j}\varphi^{j},
\end{align}
with~$T_{\mathbf{f}}^{a}$ the generators in the fundamental representation of~SO($N$). 
Using the standard procedure, we compute the associated Hamiltonian
\begin{align}
H=\int d^{3}x\left[{1\over 2}\pi^{i}\pi^{i}+{1\over 2}
(\boldsymbol{\nabla}\varphi^{i})\cdot (\boldsymbol{\nabla}\varphi^{i})+V(\varphi^{i}\varphi^{i})\right],
\end{align}
with~$\pi^{i}=\partial_{0}\varphi^{i}$ the conjugate momenta. From this expression we read
the~SO($N$)-invariant potential energy
\begin{align}
\mathscr{V}(\varphi^{i})=\int d^{3}x\,\left[{1\over 2}
(\boldsymbol{\nabla}\varphi^{i})\cdot (\boldsymbol{\nabla}\varphi^{i})
+V(\varphi^{i}\varphi^{i})\right].
\end{align}
Its minimum is attained for spatially constant configurations~$\boldsymbol{\nabla}\varphi^{i}=0$
lying at the bottom of the potential~$V(\varphi^{i}\varphi^{i})$. This is known as the
{\em vacuum expectation value} (vev) of the field and it is represented as~$\langle\varphi^{i}\rangle$.
Its value is determined by 
\begin{align}
\left.{\partial V\over \partial \varphi^{i}}\right|_{\varphi^{k}=\langle\varphi^{k}\rangle}=0.
\label{eq:minimum_cond_mp}
\end{align}
Once the vev~$\langle\varphi^{i}\rangle$ is known, we can expand the fields around it 
by writing~$\varphi^{i}=\langle\varphi^{i}\rangle+\xi^{i}$. Substituting 
in~\eqref{es:action_scalarth_SSB} we obtain the 
the action for the
fluctuations~$\xi^{i}$ whose quantization gives the elementary excitations (particle) of
the field in this vacuum.

Here we may encounter two possible situations. One is that the vev of the field is~SO($N$)
invariant,~$(T_{\mathbf{f}})^{i}_{\,\,\,j}\langle\varphi^{j}\rangle=0$. In this case the action
of the fluctuations~$\xi^{i}$ inherits the global symmetry of the parent theory 
that is then realized \`a la Wigner-Weyl. Here we want to explore the second alternative, 
the vev breaks at least part of the symmetry. Let us split the~SO($N$) generators 
into~$T_{\mathbf{f}}^{a}=\{K_{\mathbf{f}}^{\alpha},H_{\mathbf{f}}^{A}\}$, such that
\begin{align}
(K^{\alpha}_{\mathbf{f}})^{i}_{\,\,\,j}\langle\varphi^{j}\rangle\neq 0, \hspace*{1cm}
(H^{A}_{\mathbf{f}})^{i}_{\,\,\,j}\langle\varphi^{j}\rangle=0,
\end{align}
and the global symmetry~SO($N$) is spontaneously broken.
As argued after eq.~\eqref{eq:NG_charge_general}, the generators preserving the symmetry must
form a Lie subalgebra generating the unbroken subgroup~$H\subset\mbox{SO($N$)}$ and
we have the spontaneous symmetry breaking (SSB) pattern~$\mbox{SO($N$)}\rightarrow H$.

Generically, the action for the field fluctuations around the vev can be written as
\begin{align}
S=\int d^{4}x\,\left({1\over 2}\partial_{\mu}\xi^{i}\partial^{\mu}\xi^{i}
-{1\over 2}M^{2}_{ij}\xi^{i}\xi^{j}+\ldots\right),
\end{align}
where the ellipsis stands for interactions terms and the mass-squared matrix~$M^{2}_{ij}$ is given by
\begin{align}
M^{2}_{ij}\equiv \left.{\partial^{2}V\over\partial\varphi^{i}\partial\varphi^{j}}\right|_{\varphi^{k}
=\langle\varphi^{k}\rangle}.
\label{eq:mass_matrix_mp}
\end{align}
The~$\mbox{SO($N$)}$ invariance of the potential~$\delta_{\epsilon}V=0$ implies 
\begin{align}
\epsilon_{a}{\partial V\over\partial \varphi^{i}}(T^{a}_{\mathbf{f}})^{i}_{\,\,\,j}\varphi^{j}=0
\hspace*{1cm} \Longrightarrow \hspace*{1cm}
\epsilon_{a}{\partial^{2}V\over\partial \varphi^{k}\partial\varphi^{i}}(T^{a}_{\mathbf{f}})^{i}_{\,\,\,j}
\varphi^{j}+\epsilon_{a}{\partial V\over\partial \varphi^{i}}(T^{a}_{\mathbf{f}})^{i}_{\,\,\,k}=0,
\end{align}
where in the equation on the right we have taken a further derivative with respect to~$\varphi^{k}$.
Evaluating this expression at the vev, and taking into account eqs.~\eqref{eq:minimum_cond_mp}
and~\eqref{eq:mass_matrix_mp}, we find
\begin{align}
M_{ik}(T^{a}_{\mathbf{f}})^{k}_{\,\,\,j}\langle\varphi^{j}\rangle=0.
\end{align}
This equation is trivially satisfied for the unbroken generators~$H^{A}_{\mathbf{f}}$, 
but has
very nontrivial physical implications
for~$K^{\alpha}_{\mathbf{R}}$. It states that there are as many zero eigenvalues of the mass matrix 
as broken generators, i.e., the theory contains one massless particle for each generator 
not preserving the vacuum. This result is the Goldstone theorem~\cite{Goldstone,GSW}, and
the corresponding massless particles emerging as the result of
spontaneous symmetry breaking are known as Nambu-Goldstone (NG) modes~\cite{Nambu_NG,NJL}.
Although obtained here using a particular example and in a classical setup, 
the result is also valid quantum mechanically and applicable
to any field theory with a global symmetry group~$G$ spontaneously broken down to a
subgroup~$H\subset G$, where the broken part of the symmetry is the coset space~$G/H$. 
One way to prove the Goldstone theorem in the quantum theory is by considering
instead of the classical action the quantum effective action
and replacing~$V(\varphi^{i}\varphi^{i})$
with the effective potential, including all interactions among the scalar fields
resulting from resumming quantum effects. It can also
be shown that the NG modes always have zero spin, also known as NG bosons. 

Although we are mostly concerned with applications to particle physics, the idea of
SSB, in general, and the Goldstone
theorem, in particular, 
have critical applications to nonrelativistic systems, particularly in condensed matter
physics\footnote{It should be stressed that historically the very notion of SSB
and of NG bosons was inspired by solid state physics, as it is clear in the seminal works
by Yoichiro Nambu~\cite{Nambu_NG} and Jeffrey Goldstone~\cite{Goldstone}. 
Another example of this cross-fertilization between the fields of 
condensed matter and high energy physics can be found in the formulation of the 
Brout-Englert-Higgs mechanism to be discussed in section~\ref{sec:BEH_mechanism}.}. 
In particular, 
the notion of SSB is intimately related to the theory of phase
transitions~\cite{Annett,Goldenfeld,Schakel}. It is frequently the 
case that the phase change is associated with the system changing its ground state.
For example, the translational symmetry present in a 
liquid is spontaneously broken at its freezing point when
the full group of three-dimensional translation is broken down to the crystalographic group 
preserving the lattice in the solid phase. 
The corresponding NG bosons are the three species of acoustic phonons. 
These are massless quasiparticles in the sense that their dispersion relation at low
momentum takes the form~$E_{\mathbf{k}}\simeq c_{s}|\mathbf{k}|$, 
with~$c_{s}$ the speed of sound, so it has no mass gap.
Another well-known
example is a ferromagnet below the Curie point. The rotationally symmetric 
ground state at high temperature is replaced by a lowest energy configuration where 
atomic magnetic moments align, generating a macroscopic magnetization that spontaneously
breaks rotational symmetry. Magnetic waves, called magnons, are the associated NG gapless modes. 

Besides their intrinsic physical interest, these 
condensed matter examples are useful in bringing home a very important aspect of NG bosons:
they do not need to be elementary states. Indeed, phonons and magnons
are quasiparticles and, therefore, collective excitations of the system. But also in high energy
physics we encounter situations where the NG bosons are bound states of elementary 
constituents. The most relevant example is the pions, appearing as NG bosons 
associated with the spontaneous breaking of chiral symmetry in QCD (see Box~8 below).

It is frequently stated that systems with SSB present vacuum degeneracy.
Although technically the theory might possess various vacua, there are important subtleties 
involved in the infinite volume limit preventing quantum transitions among them, 
that would restore the broken symmetry through tunneling. Let us consider a theory
at finite volume~$V$ and with a family of degenerate
vacua labelled by a properly normalized real parameter~$\xi$. It can be shown
that the overlap between any two of 
these vacua 
is exponentially suppressed but nonzero (see chapter 7 of~\cite{AG_VM} for a more detailed analysis)
\label{page:SSB_infinite_volume}
\begin{align}
|\langle \xi'|\xi\rangle|=e^{-{1\over 4}(\xi'-\xi)^{2}V^{2\over 3}}|\langle\xi|\xi\rangle|.
\end{align}
This means that transitions among Fock states built on different 
vacua are allowed, resulting in a unique ground state invariant under the original symmetry.
As a consequence, no SSB can happen at finite volume and symmetries are usually realized
\`a la Wigner-Weyl. 

The situation is radically different in the~$V\rightarrow\infty$  limit
when the overlap between any two vacua
vanishes~$\langle \xi'|\xi\rangle\rightarrow 0$. This means that the 
Fock space of states builds on different vacua that
are mutually orthogonal,
and no transition among them can occur. At a more heuristic level, what happens is that 
at infinite volume switching from one vacuum to another requires a nonlocal operation
acting at each spacetime point. 
Notice, however, that at a practical level if the volume is ``large enough'' compared with the
system's microscopic characteristic 
scale we can consider the vacua as orthogonal for all purposes. This is why
we see SSB in finite samples, as illustrated by the examples of ferromagnets and superconductors.

\begin{mdframed}[backgroundcolor=lightgray,hidealllines=true]
\vspace*{0.2cm}
\centerline{\greybox{\bf Of quarks, chiral symmetry breaking, and pions}}
\vspace*{0.2cm}
\label{eq:box_chiral_SB}

The SM offers a very important implementation of SSB as a consequence 
of quark low-energy dynamics. Let us consider a generalization
of the action in eq.~\eqref{eq:quark_action1}, now 
with~$N_{f}$ different quark flavors. 
Writing~$\boldsymbol{q}^{T}=(q_{1},\ldots,q_{N_{f}})$, the
action reads
\begin{align}
S&=\int d^{4}x\,\overline{\boldsymbol{q}}\big(i{\partial\!\!\!/}\mathbbm{1}-\boldsymbol{m}\big)
\boldsymbol{q}+S_{\rm int} \nonumber \\[0.2cm]
&=\int d^{4}x\,\Big(i\overline{\boldsymbol{q}}_{R}{\partial\!\!\!/}\boldsymbol{q}_{R}
+i\overline{\boldsymbol{q}}_{L}{\partial\!\!\!/}\boldsymbol{q}_{L}
-\overline{\boldsymbol{q}}_{R}\boldsymbol{m}\boldsymbol{q}_{L}
-\overline{\boldsymbol{q}}_{L}\boldsymbol{m}\boldsymbol{q}_{R}
\Big)+S_{\rm int},
\end{align}
where in the second line we split the quark fields into its right- and left-handed 
chiralities and 
in~$S_{\rm int}$ we include all interaction terms. This theory is 
invariant under global $\mbox{U($N_{f}$)}$ transformations acting on the fermion fields as
\begin{align}
\boldsymbol{q}_{R,L}\rightarrow \mathscr{U}(\alpha)\boldsymbol{q}_{R,L}
\hspace*{1cm} \mbox{where} \hspace*{1cm} 
\mathscr{U}(\alpha)=e^{i\alpha^{A}T_{\mathbf{R}}^{A}},
\end{align}
and~$(T_{\mathbf{R}}^{A})^{i}_{\,\,\,j}$, with $A=1,\ldots,N_{f}^{2}$, are the U($N_{f}$) generators
in the representation~$\mathbf{R}$ with dimension~$N$.
We observe that it is the presence of the mass term, mixing right- and left-handed quarks, what forces the 
two chiralities to transform under the same transformation of~U($N_{f}$). This is why in the chiral limit
(i.e., zero quark masses~$\boldsymbol{m}\rightarrow 0$) 
the global symmetry is enhanced from~U($N_{f}$) to~$\mbox{U($N_{f}$)}_{R}\times
\mbox{U($N_{f}$)}_{L}$, acting independently on the two chiralities
\begin{align}
\boldsymbol{q}_{R}\rightarrow \mathscr{U}(\alpha_{R})\boldsymbol{q}_{R}, 
\hspace*{1cm}
\boldsymbol{q}_{L}\rightarrow \mathscr{U}(\alpha_{L})\boldsymbol{q}_{L},
\end{align}
where~$\alpha^{a}_{R}$ and~$\alpha^{a}_{L}$ are independent.
Thus, there are two independent Noether currents
\begin{align}
j^{\mu}_{R}(\alpha)=\alpha_{R}^{A}\overline{\boldsymbol{q}}_{R}
\gamma^{\mu}T_{\mathbf{R}}^{A}\boldsymbol{q}_{R}, \hspace*{1cm}
j^{\mu}_{L}(\alpha)=\alpha_{L}^{A}\overline{\boldsymbol{q}}_{L}
\gamma^{\mu}T_{\mathbf{R}}^{A}\boldsymbol{q}_{L}
\end{align}
as well as~$2\times N_{f}^{2}$ conserved charges
\begin{align}
Q^{A}_{R}=\int d^{3}x\,\boldsymbol{q}_{R}^{\dagger}T_{\mathbf{R}}^{A}\boldsymbol{q}_{R},
\hspace*{1cm} Q^{A}_{L}=\int d^{3}x\,\boldsymbol{q}_{L}^{\dagger}T_{\mathbf{R}}^{A}\boldsymbol{q}_{L}.
\end{align}
Upon quantization, these charges are replaced by the corresponding operators~$\widehat{Q}_{R,L}^{A}$
whose commutator realizes the algebra of generators of~$\mbox{U($N_{f}$)}_{R}\times\mbox{U($N_{f}$)}_{L}$.

Taking into account that~$\mbox{U($N_{f}$)}=\mbox{U(1)}\times
\mbox{SU($N_{f}$)}$, the theory's global symmetry group can be written as
\begin{align}
\mbox{U($N_{f}$)}_{R}\times\mbox{U($N_{f}$)}_{L}=\mbox{U(1)}_{B}\times\mbox{U(1)}_{A}\times
\mbox{SU($N_{f}$)}_{R}\times\mbox{SU($N_{f}$)}_{L}.
\end{align}
The first two factors on the right-hand side act on the quark fields respectively as
\begin{align}
\boldsymbol{q}\rightarrow e^{i\alpha}\boldsymbol{q}, \hspace*{1cm}
\boldsymbol{q}\rightarrow e^{i\beta\gamma_{5}}\boldsymbol{q},
\label{eq:V_AV_trans_quarks}
\end{align}
the former symmetry leading to baryon number conservation (hence the subscript). 
The~$\mbox{U(1)}_{A}$ factor 
is an axial vector transformation acting on the two chiralities with opposite phases
and is broken by anomalies (more on this in section~\ref{sec:anomalies}). The action of the two~$\mbox{SU($N_{f}$)}_{R,L}$ 
factors, on the other hand, is defined by
\begin{align}
\mbox{SU($N_{f}$)}_{R}:\left\{
\begin{array}{rrl}
\boldsymbol{q}_{R}&\rightarrow& U_{R}\boldsymbol{q}_{R} \\[0.2cm]
\boldsymbol{q}_{L}&\rightarrow&  \boldsymbol{q}_{L}
\end{array}
\right. \hspace*{1cm}
\mbox{SU($N_{f}$)}_{L}:\left\{
\begin{array}{rrl}
\boldsymbol{q}_{R}&\rightarrow&  \boldsymbol{q}_{R} \\[0.2cm]
\boldsymbol{q}_{L}&\rightarrow& U_{L} \boldsymbol{q}_{L}
\end{array}
\right.
\label{eq:SUN_LR}
\end{align}
with
\begin{align}
U_{R,L}\equiv e^{i\alpha_{L,R}^{I}t^{I}_{\mathbf{f}}}
\end{align}
and~$t^{I}_{\mathbf{f}}$ ($I=1,\ldots,N_{f}^{2}-1)$
the generators of the fundamental irrep of~SU($N_{f}$).

At low energies the strong quark dynamics triggers quark condensation, giving a 
non-zero vev to the scalar quark bilinear~$\overline{q}_{i}q_{j}$
\begin{align}
\langle 0| \overline{q}_{i}q_{j}|0\rangle
\equiv \langle 0|\big(\overline{q}_{i,R}q_{j,L}+\overline{q}_{i,L}q_{i,R}\big)|0\rangle
=\Lambda_{\chi\rm SB}^{3}\delta_{ij},
\label{eq:chira_SSB_vev}
\end{align}
where~$\Lambda_{\chi\rm SB}$ is the energy scale associated with the condensation.
This vev, however, is only invariant under the ``diagonal'' subgroup of
the~$\mbox{SU($N_{f}$)}_{R}\times\mbox{SU($N_{f}$)}_{L}$ transformations~\eqref{eq:SUN_LR}
consisting of transformations with~$U_{R}=U_{L}$. 
What happens is that the global~$\mbox{SU($N_{f}$)}_{R}\times\mbox{SU($N_{f}$)}_{L}$
chiral symmetry is spontaneously broken down to its vector subgroup
\begin{align}
\mbox{U(1)}_{B}\times
\mbox{SU($N_{f}$)}_{R}\times\mbox{SU($N_{f}$)}_{L} \longrightarrow
\mbox{U(1)}_{B}\times\mbox{SU($N_{f}$)}_{V},
\end{align}
Goldstone's theorem implies that associated with each spontaneously broken generator
there should be a massless NG boson. In our case there
are~$N_{f}^{2}-1$ broken generators corresponding to the~$\mbox{SU($N_{f}$)}_{A}$ factor. 
Excitations around the vev~\eqref{eq:chira_SSB_vev}
are parametrized by the field~$\Sigma_{ij}(x)$ defined by
\begin{align}
\overline{q}_{i}(x)q_{j}(x)=\Lambda_{\chi\rm SB}^{3}\Sigma_{ij}(x).
\end{align} 
This in turn can be written in terms of the NG matrix
field~$\boldsymbol{\pi}(x)\equiv \pi^{A}(x)t^{A}_{\bf f}$ as
\begin{align}
\boldsymbol{\Sigma}(x)\equiv e^{{i\sqrt{2}\over f_{\pi}}\boldsymbol{\pi}(x)},
\label{eq:NG_bosons_field_general}
\end{align}
with~$f_{\pi}$ 
a constant with dimensions of energy called the pion decay constant for
reasons that will eventually be clear. Mathematically speaking, the
field~$\boldsymbol{\Sigma}$ parametrizes the coset
\begin{align}
{\mbox{SU($N_{f}$)}_{R}\times
\mbox{SU($N_{f}$)}_{L}
\over \mbox{SU($N_{f}$)}_{V}},
\end{align}
leading to the following transformation under~$\mbox{SU($N_{f}$)}_{R}\times
\mbox{SU($N_{f}$)}_{L}$
\begin{align}
\boldsymbol{\Sigma}\longrightarrow U_{R}\boldsymbol{\Sigma}U_{L}^{\dagger}.
\label{eq:NG_matrix_trans}
\end{align}

We specialize the analysis now to the case~$N_{f}=2$, with only have the~$u$ and~$d$ quarks.
The unbroken~$\mbox{SU(2)}_{V}$ symmetry is just the good old isospin interchanging both quarks, while the 
NG bosons are the three pions~$\pi^{\pm}$ and~$\pi^{0}$ 
\begin{align}
\boldsymbol{\pi}={1\over \sqrt{2}}\left(
\begin{array}{cc}
\pi^{0} & \sqrt{2}\pi^{+} \\
\sqrt{2}\pi^{-} & -\pi^{0}
\end{array}
\right).
\end{align}
The objection might be raised
that pions are not massless particles as the Goldstone theorem requires. 
Our analysis has ignored the
nonvanishing quark masses, explicitly breaking the~$\mbox{SU(2)}_{R}\times\mbox{SU(2)}_{L}$ global
chiral symmetry. Since the~$u$ and~$d$ quarks are relatively light, 
we have instead three {\em pseudo}-NG bosons whose masses are not zero but 
still lighter than
other states in the theory. It is precisely the strong mass hierarchy between the pions
and the remaining hadrons what identifies them as the 
pseudo-NG bosons associated with chiral symmetry breaking. In the~$N_{f}=3$ case, where we add
the strange quark to the two lightest ones, $\mbox{SU(3)}_{V}$ is Gell-Mann's eightfold way 
discussed on page~\pageref{page:eightfold_way} and
the set of pseudo-NG bosons is enriched by the four kaons and the $\eta$-meson 
in the octet appearing on the right-hand 
side of eq.~\eqref{eq:octet_SU(3)_mesons}. 

As mentioned in the introduction, quarks and gluons do not exist as asymptotic states and QCD at low energies is a theory 
of hadrons. The lowest lying particles are the pion triplet, whose interactions can be obtained from symmetry 
considerations alone
playing the EFT game. 
The question is how to write the simplest action for NG bosons containing operators with the
lowest energy dimension and
compatible at the same time with all the symmetries of the theory. For terms with just two derivatives, 
the solution is
\begin{align}
S_{\rm NG}&={f_{\pi}^{2}\over 4}
\int d^{4}x\,{\rm tr\,}\Big(\partial_{\mu}\boldsymbol{\Sigma}^{\dagger}\partial^{\mu}\boldsymbol{\Sigma}\Big)
\nonumber \\[0.2cm]
&=\int d^{4}x\,\left[{1\over 2}
{\rm tr\,}\Big(\partial_{\mu}\boldsymbol{\pi}\partial^{\mu}\boldsymbol{\pi}\Big)
-{1\over 3f_{\pi}^{2}}{\rm tr\,}\Big(\partial_{\mu}\boldsymbol{\pi}[\boldsymbol{\pi},[\boldsymbol{\pi},
\partial^{\mu}\boldsymbol{\pi}]\Big)
+\ldots\right].
\label{eq:pion_action}
\end{align} \label{page:chiral_lagrangian}
This {\em chiral effective action}  contains an infinite sequence of higher-dimensional operators 
suppressed by increasing powers of the dimensionful 
constant~$f_{\pi}$. It determines how pions couple among themselves at low
energies. Its coupling to the electromagnetic field is obtained by 
replacing~$\partial_{\mu}\boldsymbol{\Sigma}$ by
the adjoint covariant derivative~$D_{\mu}\boldsymbol{\Sigma}=\partial_{\mu}\boldsymbol{\Sigma}
-iA_{\mu}[Q,\boldsymbol{\Sigma}]$ where the charge matrix is 
given by~$Q=e\sigma^{3}$. This, however, does not exhausts all their electromagnetic interaction.
Neutral pions couple to photons as a consequence of 
the anomalous realization of the~$\mbox{U(1)}_{A}$ symmetry, resulting in 
the~$\pi^{0}\rightarrow 2\gamma$ decay (see section~\ref{sec:anomalies}).

In our analysis of chiral symmetry breaking we encountered two energy 
scales:~$\Lambda_{\chi\rm SSB}$ appearing in~\eqref{eq:chira_SSB_vev}
as a consequence of the quark condensate having dimensions 
of~$\mbox{(energy)}^{3}$, and~$f_{\pi}$ needed to give the pion fields their proper dimensions
in eq.~\eqref{eq:NG_bosons_field_general}. 
Both of them have to be experimentally measured. In the pion EFT it 
is~$f_{\pi}$ what determines the 
relative size of the infinite terms in the effective action~\eqref{eq:pion_action}.
Operators weighted by~$f_{\pi}^{-n}$ 
typically give contributions or order~$(E/f_{\pi})^{n}$ with~$E$ the characteristic energy
of the process under study. In the spirit of EFT, working at a given experimental precision 
only a finite number of terms in the chiral Lagrangian have to be retained, making the 
theory fully predictive (see~\cite{Pich,Scherer_Schindler} for comprehensive reviews of
chiral perturbation theory).

\end{mdframed}

\subsection{The Brout-Englert-Higgs mechanism}
\label{sec:BEH_mechanism}

Besides the ones already discussed, 
a further instance of SSB in condensed matter connecting 
with one of the key concepts in the formulation of the SM,
the Brout-Englert-Higgs (BEH) mechanism. In the 
Bardeen-Cooper-Schrieffer~(BCS) theory of superconductivity the transition from the normal to
the superconductor phase is triggered by the condensation of Cooper pairs, collective excitations
of two electrons bound together by phonon exchange. Having net electric charge, 
the Cooper pair wave function transform under electromagnetic U(1) phase rotations and their condensation
spontaneously breaks this invariance. The physical consequence of this is a
screening of magnetic fields inside the superconductor, the Meissner effect, 
physically equivalent to the electromagnetic vector
potential~$\mathbf{A}(t,\mathbf{r})$ acquiring an effective nonzero mass~\cite{Anderson}.

The main difference between the BCS example and the ones discussed above is that 
this is not about spontaneously breaking some global symmetry, but gauge invariance itself. 
This might look like risky 
business, since we know that preserving gauge invariance is crucial to get rid
of unwanted physical states that otherwise would pop up in the theory's physical spectrum
destroying its consistency. As we will see, due to the magic of SSB gauge invariance is in fact not
lost, only hidden. That is why, even if not manifest, it still protects the theory.

\begin{figure}[t]
\centerline{\includegraphics[scale=0.2]{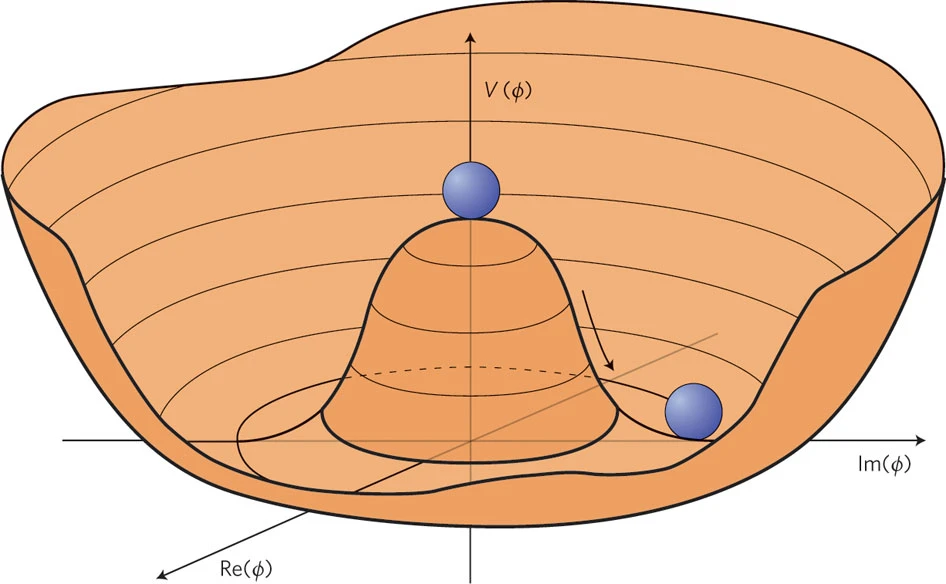}}
\caption[]{Illustration from ref.~\cite{AG_E} depicting the celebrated Mexican hat potential
shown in eq.~\eqref{eq:mexican_hat_U(1)}.}
\label{fig:higgs_potential}
\end{figure}

Let us analyze spontaneous symmetry breaking triggered by 
a complex scalar coupled to the electromagnetic field. We start with the action
\begin{align}
S=\int d^{4}r\,\left[-{1\over 4}F_{\mu\nu}F^{\mu\nu}+(D_{\mu}\phi)^{*}(D^{\mu}\phi)
-{\lambda\over 4}\left(\phi^{*}\phi-{v^{2}\over 2}\right)^{2}\right],
\label{eq:action_abelian_higgs_1}
\end{align}
where~$D_{\mu}=\partial_{\mu}-ieA_{\mu}$ is the covariant derivative already introduced 
in the footnote of page~\pageref{page:covariat_derivative_scalar}. This action 
is invariant under U(1) gauge transformations acting as
\begin{align}
\phi(x)\longrightarrow e^{ie\epsilon(x)}\phi(x), \hspace*{1cm}
\phi(x)^{*}\longrightarrow e^{-ie\epsilon(x)}\phi(x)^{*}, \hspace*{1cm}
A_{\mu}(x)\longrightarrow A_{\mu}(x)+\partial_{\mu}\epsilon(x).
\label{eq:gauge_trans_abelian_higgs}
\end{align}
As shown in fig.~\ref{fig:higgs_potential},
the scalar field potential
\begin{align}
V(\phi^{*}\phi)={\lambda\over 4}\left(\phi^{*}\phi-{v^{2}\over 2}\right)^{2},
\label{eq:mexican_hat_U(1)}
\end{align}
has the celebrated Mexican hat shape with a valley of 
minima located at~$\phi^{*}\phi={v^{2}\over 2}$. When the scalar field takes a nonzero
vev
\begin{align}
\langle\phi\rangle={v\over\sqrt{2}} e^{i\vartheta_{0}},
\label{eq:vev_U(1)_higgs}
\end{align}
U(1)~invariance is spontaneously broken,
since~$\langle\phi\rangle$ does not remain 
invariant,~$\langle\phi\rangle\rightarrow e^{ie\epsilon}\langle\phi\rangle$.
The dynamics of the fluctuations around the vev~\eqref{eq:vev_U(1)_higgs} is obtained by 
plugging
\begin{align}
\phi(x)={1\over \sqrt{2}}\big[v+h(x)\big]e^{i\vartheta(x)},
\label{eq:AHMphi_fluctuations}
\end{align}
into the~\eqref{eq:action_abelian_higgs_1}. The resulting action is
\begin{align}
S&=\int d^{4}x\left[-{1\over 4}F_{\mu\nu}F^{\mu\nu}+{e^{2}v^{2}\over 2}\left(A_{\mu}+{1\over e}
\partial_{\mu}\vartheta\right)
\left(A^{\mu}+{1\over e}\partial^{\mu}\vartheta\right)
+{1\over 2}\partial_{\mu}h\partial^{\mu}h-{\lambda v^{2}\over 4}h^{2} \right.\nonumber \\[0.2cm]
&-\left.{\lambda v\over 4} h^{3}-{\lambda\over 16}h^{4}
+{e^{2}\over 2}\left(A_{\mu}+{1\over e}
\partial_{\mu}\vartheta\right)
\left(A^{\mu}+{1\over e}\partial^{\mu}\vartheta\right)(2vh+h^{2})
\right],
\label{eq:non_fixed_higgsaction}
\end{align}
which remains invariant under~U(1) gauge transformations, now acting as
\begin{align}
A_{\mu}\longrightarrow A_{\mu}+\partial_{\mu}\epsilon, \hspace*{1cm}
\vartheta\longrightarrow \vartheta-e\epsilon, \hspace*{1cm} h\longrightarrow h.
\label{eq:gauge_trans_abHiggs_pre}
\end{align}
In fact, the phase field~$\vartheta(x)$ is the NG boson resulting from 
the spontaneous breaking of the U(1) symmetry by the vev in eq.~\eqref{eq:vev_U(1)_higgs}. 

At this stage, we still keep a photon with two polarizations while the two real degrees of freedom
of the complex field~$\phi$ have been recast in terms of the field~$h$ and the NG boson~$\vartheta$. We can 
fix the gauge freedom~\eqref{eq:gauge_trans_abHiggs_pre} 
by setting~$\vartheta=0$. In doing so, the disappearing NG boson 
transmutes into the longitudinal component of~$A_{\mu}$, as befits a massive gauge field (see 
the footnote on page~\pageref{page:footnote_massivephoton}). We then arrive at the gauged-fixed action
\begin{align}
S&=\int d^{4}x\left(-{1\over 4}F_{\mu\nu}F^{\mu\nu}+{e^{2}v^{2}\over 2}A_{\mu}A^{\mu}
+{1\over 2}\partial_{\mu}h\partial^{\mu}h-{\lambda v^{2}\over 4}h^{2} \right.\nonumber \\[0.2cm]
&-\left.{\lambda v\over 4} h^{3}-{\lambda\over 16}h^{4}
+e^{2}vA_{\mu}A^{\mu}h+{e^{2}\over 2}A_{\mu}A^{\mu}h^{2}\right),
\end{align}
where the photon has acquired a nonzero mass\footnote{The same result can be obtained noticing that
the action~\eqref{eq:non_fixed_higgsaction} contains a term~$ev^{2}A^{\mu}\partial_{\mu}\vartheta$
mixing the NG boson and the gauge field. Physically, this means that as the photon propagates 
it transmutes into the NG boson and vice versa. Resumming these transmutations results in
the mass term for~$A^{\mu}$.}
\begin{align}
m_{\gamma}=ev.
\end{align} 
The real scalar field~$h$ gets massive as well
\begin{align}
m_{h}=v\sqrt{\lambda\over 2},
\end{align}
and has cubic and quartic self-interactions terms, besides coupling to the photon through terms
involving two gauge field and one scalar and two gauge fields and two scalars.
As we see, no degree of freedom has gone amiss. We
ended up with a massive photon with three physical polarizations
and a real scalar, making up for the four real degrees of freedom we started with. 
SSB has just
rearranged the theory's degrees of the theory.

Here we have been only concerned with giving mass to the photon. Imagine now that we would
have two chiral fermions~$\psi_{R}$,~$\psi_{L}$ such that they transform
differently under~$\mbox{U(1)}$
\begin{align}
\psi_{L}(x)\longrightarrow e^{ie\epsilon(x)}\psi_{L}(x), \hspace*{1cm}
\psi_{R}(x)\longrightarrow \psi_{R}(x).
\end{align}
Due to the theory's chiral nature, 
a mass term of the form~$\overline{\psi}_{L}\psi_{R}+\overline{\psi}_{R}\psi_{L}$ would
not be gauge invariant, so it seems that we need to keep our fermions massless for the sake of 
consistency. Using the Higgs field, however, there is a way to construct an action
where the fermions couple to the complex scalar field in a gauge invariant way \label{page:Yukawa_coupling}
\begin{align}
S_{\rm fermion}=\int d^{4}x\,\Big(i\overline{\psi}_{R}{D\!\!\!\!/\,}\psi_{R}
+i\overline{\psi}_{L}{D\!\!\!\!/\,}\psi_{+}-c\phi \overline{\psi}_{L}\psi_{R}
-c\phi^{*}\overline{\psi}_{R}\psi_{L}\Big),
\end{align}
where~$c$ is some dimensionless constant.
This particular form of the coupling between~$\phi$ and the fermions is called a
Yukawa coupling, since it is similar to the one introduced by Hideki Yukawa in his 1935 theory
of nuclear interactions between nucleons and mesons~\cite{Yukawa}. The interest of this construction
is that once the field~$\phi$ acquires
the vev~\eqref{eq:vev_U(1)_higgs}, and after gauging away the field~$\vartheta$, the 
fermion action takes the form
\begin{align}
S_{\rm fermion}&=\int d^{4}x\,\left[i\overline{\psi}_{R}{D\!\!\!\!/\,}\psi_{R}
+i\overline{\psi}_{L}{D\!\!\!\!/\,}\psi_{L}-{cv\over \sqrt{2}}\big(\overline{\psi}_{L}\psi_{R}
-\overline{\psi}_{R}\psi_{L}\big)\right.\nonumber \\[0.2cm]
&\left.-{c\over \sqrt{2}}h\overline{\psi}_{L}\psi_{R}
-{c\over \sqrt{2}}h\overline{\psi}_{R}\psi_{L}
\right].
\end{align}
Thus, the same mechanism giving mass to the photon also results in a mass for the
fermion field
\begin{align}
m_{f}={cv\over \sqrt{2}},
\end{align}
also generated without a explicit breaking of gauge invariance, hidden
due to the choice of vacuum of the complex scalar field. Notice that, owing to symmetry breaking,
the now massive Dirac fermion couples to the remaining scalar degree of freedom~$h$ withan 
strength controlled by the dimensionless constant~${c\over\sqrt{2}}={m_{f}\over v}$. This indicates
that the higher the mass of the fermion, the stronger it couples to the Higgs field. This feature, as we will see,
has important experimental consequences for the~SM.
 
This Abelian Higgs model illustrates the basic features of the BEH mechanism
responsible for giving masses to the SM particles, with the scalar field~$h$
corresponding to the Higgs boson discovered at CERN in 2012~\cite{Higgs_ATLAS,Higgs_CMS}. 
In its nonrelativistic version 
it also provides the basis for the
Ginzburg-Landau analysis of the BCS theory of superconductivity, where the 
free energy in the broken phase 
has the same structure as the potential terms in the action~\eqref{eq:action_abelian_higgs_1}
\begin{align}
\mathscr{F}_{\rm BCS}&=\int d^{3}r\,\left\{{1\over 2\mu}(\boldsymbol{\nabla}\times\mathbf{A})^{2}
+{1\over 2m_{*}}\big|\boldsymbol{\nabla}\phi-ie_{*}\mathbf{A}\phi\big|^{2}
+{\lambda(T)\over 4}\left[\phi^{*}\phi-{v(T)\over 2}^{2}\right]^{2}\right\}.
\end{align}
Here~$\phi(\mathbf{r})$ is the Cooper pair condensate,~$\mu$ the magnetic permeability
of the medium, and~$m_{*}$ and~$e_{*}$ the effective
mass and charge of the quasiparticles. For~$T>T_{c}$ we have~$v(T)=0$, so at temperatures about
the critical one the only minimum of the free energy is at~$\langle\phi\rangle=0$. When~$T<T_{c}$, on the
other hand,~$v(T)\neq 0$ and the~U(1) invariance of the theory is spontaneously broken 
at~$|\langle\phi\rangle|=v(T)$ minima, while the former one at~$\langle\phi\rangle=0$ becomes
a local maximum. As in the case studied earlier, this results in a nonzero
mass for the vector potential~$\mathbf{A}(\mathbf{r})$ given by~$m(T)=e_{*}v(T)$. This provides 
the order parameter of the transition and physically accounts for
the Meissner effect inside the superconductor~\cite{Annett}. The system also contains a 
scalar massive excitation, the condensed matter equivalent of the Higgs 
boson~\cite{Littlewood_Varma,Shimano_Tsuji}.

\begin{mdframed}[backgroundcolor=lightgray,hidealllines=true]
\vspace*{0.2cm}
\centerline{\greybox{\bf ``Large'' vs. ``small'' gauge transformations}}
\vspace*{0.2cm}
\label{box:large_vs_small}

We return briefly to the discussion of Noether's second theorem on page~\pageref{page:Noether_second}.
There we paid attention to gauge transformations in the connected component of the identity and
made an important distinction among those approaching the identity at the spacetime boundary  
($\epsilon_{A}\rightarrow 0$)
and those that do not. Let us call them ``small''
and ``large'' gauge transformations respectively. To understand the physical difference between them, 
we compare~\eqref{eq:jmu-SmuN2} 
with~\eqref{eq:SmuN2} to see that~$j^{\mu}-S^{\mu}$ is conserved even off-shell, 
namely that~$\partial_{\mu}(j^{\mu}-S^{\mu})$ is {\em identically zero}. 
This means that we can write
\begin{align}
j^{\mu}=S^{\mu}+\partial_{\nu}k^{\mu\nu}\approx \partial_{\nu}k^{\mu\nu},
\end{align}
where~$k^{\mu\nu}$ is an antisymmetric tensor and we have
applied that~$S^{\mu}$ vanishes on-shell. 
This peculiar structure of the gauge theory current implies that
the gauge charge is determined by an integral over the {\em boundary} of the spatial sections
\begin{align}
Q\approx \int_{\Sigma}dV\,\partial_{i}k^{0i}=\int_{\partial\Sigma}dS_{i}\,k^{0i}.
\end{align}
Since the current, and therefore also~$k^{\mu\nu}$, is linear in the gauge 
functions~$\epsilon_{A}(x)$, we conclude that the charge vanishes for ``small'' gauge transformations
\begin{align}
Q_{\rm small}\approx 0.
\end{align}
This is not the case of ``large'' transformations, 
the ones determining the value of~$Q$.

A very important fact to remember about ``small'' gauge transformations is that they are the ones
leading to the Noether identities~\eqref{eq:Delta(E)} that, as we indicated, express the redundancy intrinsic to 
gauge theories. Quantum mechanically, invariance under these transformations is 
mandatory in order to get rid of the spurious states that we introduced as the price of 
maintaining locality and Lorentz covariance. They cannot be spontaneously broken 
or affected by anomalies without rendering the theory inconsistent. However, no such
restriction exists for ``large'' transformations, that can be broken without
disastrous consequences. 

To connect with the discussion of the Abelian Higgs model, let us look at the
case of Maxwell's electrodynamics in the temporal gauge~$A_{0}=0$.
In the quantum theory, the vacuum Gauss law constraint~$\boldsymbol{\nabla}\cdot\mathbf{E}=0$ 
is implemented by the corresponding operator annihilating physical states, namely (to keep
notation simple, we drop hats to denote operators)
\begin{eqnarray}
\boldsymbol{\nabla}\cdot\mathbf{E}|{\rm phys}\rangle=0.
\label{eq:gausslaw_q}
\end{eqnarray}
Finite gauge transformations preserving the temporal gauge condition~$A_{0}=0$ 
are generated by time-independent gauge functions and implemented in the space of states
by the operator
\begin{eqnarray}
\mathscr{U}_{\epsilon}=\exp\left[i\int d^{3}r\,\mathbf{E}(t,\mathbf{r})\cdot\boldsymbol{\nabla}\epsilon(\mathbf{r})\right].
\end{eqnarray}
Using the canonical commutation relations~\eqref{eq:EM_canonical_comm_rel}, we readily compute
\begin{eqnarray}
\mathscr{U}_{\epsilon}A_{0}(t,\mathbf{r})\mathscr{U}_{\epsilon}^{-1}&=& 0, \nonumber \\[0.2cm]
\mathscr{U}_{\epsilon}\mathbf{A}(t,\mathbf{r})\mathscr{U}_{\epsilon}^{-1}&=& \mathbf{A}(t,\mathbf{r})+\boldsymbol{\nabla}\epsilon(\mathbf{r}).
\end{eqnarray}
At the same time, the operator~$\mathscr{U}_{\epsilon}$ leaves the physical states invariant
\begin{eqnarray}
\mathscr{U}_{\epsilon}|{\rm phys}\rangle&=& \exp\left[i\int d^{3}x\,\mathbf{E}(t,\mathbf{r})\cdot\boldsymbol{\nabla}\epsilon(\mathbf{r})\right]
|{\rm phys}\rangle \nonumber \\[0.2cm]
&=&\exp\left[-i\int d^{3}x\,\epsilon(\mathbf{r})\boldsymbol{\nabla}\cdot\mathbf{E}(t,\mathbf{r})\right]
|{\rm phys}\rangle=|{\rm phys}\rangle,
\end{eqnarray}
where in the second line it is crucial that the gauge function $\epsilon(\mathbf{r})$ 
{\em vanishes at infinity} so that after integrating by parts we do not pick up a boundary
term. This means that~$\mathscr{U}_{\epsilon}\rightarrow \mathbbm{1}$ as~$|\mathbf{r}|\rightarrow \infty$.

We have shown that invariance of the physical states under ``small'' gauge transformations 
follows from Gauss' law~\eqref{eq:gausslaw_q} annihilating them,
precisely the condition that factors out the spurious degrees of freedom.
The conclusion is that ``large'' gauge transformations are not necessary to eliminate the gauge redundancy
and can be broken without jeopardizing the consistency of the theory. This is precisely
how the BEH mechanism works. The nonvanishing vacuum expectation value of the 
complex scalar field breaks ``large'' gauge transformations
without spoiling Gauss' law.
This is the reason why we need to qualify our
statement in pages~\pageref{pag:redundancy_gauge} and~\pageref{pag:redundancy_gauge_page_2} that
gauge invariance is just a redundancy in state labelling:
``small'' gauge transformations are indeed redundancies, but ``large'' gauge transformations are
{\em bona fide} symmetries.

\end{mdframed}

\section{Some more gauge invariances}
\label{sec:more_gauge_invariances}

So far the only gauge theory we dealt with was Maxwell's electrodynamics, although here and there
we hinted at its non-Abelian generalizations. It is about time to introduce these in a more
systematic fashion. We start with a set of fermions~$\boldsymbol{\psi}^{T}=(\psi_{1},\ldots,\psi_{N})$
transforming in some representation~$\mathbf{R}$ of the gauge group~$G$
\begin{align}
\boldsymbol{\psi}\longrightarrow e^{i\alpha^{a}T^{a}_{\mathbf{R}}}\boldsymbol{\psi}
\equiv g(\alpha)\boldsymbol{\psi}.
\label{eq:finite_gauge_trans_NA}
\end{align}
By now, we know very well how to construct an action that has this symmetry
\begin{align}
S=\int d^{4}x\,\overline{\boldsymbol{\psi}}\big(i{\partial\!\!\!/}-m\big)\boldsymbol{\psi}.
\label{eq:NAYM_firstfermionaction}
\end{align} 
The problem arises when we want to make~$G$ a local invariance. In this case, the action we just
wrote fails to be invariant due to the nonvanishing derivatives of~$\alpha^{a}(x)$
\begin{align}
\partial_{\mu}\boldsymbol{\psi}\longrightarrow 
g\partial_{\mu}\boldsymbol{\psi}+i\partial_{\mu}g
\boldsymbol{\psi}=g\big(\partial_{\mu}\boldsymbol{\psi}
+ig^{-1}\partial_{\mu}g\big)
\boldsymbol{\psi},
\end{align}
where, to avoid cluttering expressions, we have omitted the dependence of the group element~$g$
on the parameters~$\alpha^{a}$.

To overcome
this problem we have to find a covariant derivative~$D_{\mu}$, 
similarly to the one we introduced for Maxwell's theory,
with the transformation
\begin{align}
D_{\mu}\boldsymbol{\psi}\longrightarrow gD_{\mu}\boldsymbol{\psi}.
\label{eq:cov_der_nonabelian}
\end{align}
A reasonable ansatz turns out to be
\begin{align}
D_{\mu}\boldsymbol{\psi}=\big(\partial_{\mu}-iA_{\mu}\big)\boldsymbol{\psi},
\label{eq:covDer_NAYM}
\end{align}  
where we omitted the identity multiplying~$\partial_{\mu}$ and~$A_{\mu}\equiv A_{\mu}^{a}T^{a}_{\mathbf{R}}$ 
is a field taking values
in the algebra of generators of~$G$. 
In order to get the transformations~\eqref{eq:cov_der_nonabelian},~$A_{\mu}$ has to transform according to
\begin{align}
A_{\mu}\longrightarrow A'_{\mu}=ig^{-1}\partial_{\mu}g+g^{-1}A_{\mu}g.
\label{eq:trans_ANA_noindices}
\end{align}
With this we 
can turn~\eqref{eq:NAYM_firstfermionaction} into a locally invariant action by replacing
$\partial_{\mu}$ with~$D_{\mu}$ defined in eq.~\eqref{eq:covDer_NAYM}. In addition, we must include the
dynamics of the new field~$A_{\mu}$ adding a suitable kinetic term 
that preserves the gauge invariance of the fermionic action.
The Abelian-informed choice~$\partial_{\mu}A_{\nu}-\partial_{\nu}A_{\mu}$ for the gauge field 
strength will not do, since it does not transform covariantly
\begin{align}
\partial_{\mu}A_{\nu}-\partial_{\nu}A_{\mu}
\longrightarrow &\,\, g^{-1}\big(\partial_{\mu}A_{\nu}-\partial_{\nu}A_{\mu}\big)g
+i\big[g^{-1}\partial_{\mu}g,g^{-1}\partial_{\nu}g\big] \nonumber \\[0.2cm]
&+\big[g^{-1}A_{\mu}g,g^{-1}\partial_{\nu}g\big]
+\big[g^{-1}\partial_{\mu}g,g^{-1}A_{\nu}g\big].
\end{align}
This however suggests a wiser choice
\begin{align}
F_{\mu\nu}=\partial_{\mu}A_{\nu}-\partial_{\nu}A_{\mu}+i\big[A_{\mu},A_{\nu}\big],
\label{eq:field_strength_NAYM}
\end{align}
with the much nicer (i.e., covariant) transformation
\begin{align}
F_{\mu\nu}\longrightarrow F'_{\mu\nu}=g^{-1}F_{\mu\nu}g.
\label{eq:FNA_gauge_trans}
\end{align}
Notice that, similar to~$A_{\mu}$, the field strenght~$F_{\mu\nu}$ takes values in 
the algebra of generators, so we can write~$F_{\mu\nu}=F^{a}_{\mu\nu}T^{a}_{\mathbf{R}}$, with
the components given by
\begin{align}
F_{\mu\nu}^{a}=\partial_{\mu}A^{a}_{\nu}-\partial_{\nu}A^{a}_{\mu}+f^{abc}A^{b}_{\mu}A^{c}_{\nu}.
\label{eq:field_strengthNAYM_indices}
\end{align}
where~$f^{abd}$ are the structure constants of the Lie algebra of generators,~$[T^{a}_{\mathbf{R}},
T^{b}_{\mathbf{R}}]=if^{abc}T^{c}_{\mathbf{R}}$.

We denote by~$\mathscr{G}$ the set of gauge transformations acting on the fields.
Although to fix ideas here, we have considered transformations~\eqref{eq:finite_gauge_trans_NA} 
in the connected component of the identity~$\mathscr{G}_{0}$,
the derived expressions remain valid for {\rm all}
transformations in~$\mathscr{G}$, even if they lie in disconnected 
components (we saw an example of this in the case of the Lorentz group studied in
page~\pageref{page:Lorentz_group_comps}). For transformations in~$\mathscr{G}_{0}$, 
we can write their infinitesimal form
\begin{align}
g(\alpha)\simeq \mathbbm{1}+i\alpha^{a}T^{a}_{\mathbf{R}},
\end{align}
to write the first order transformation of both the gauge field and its field strength
\begin{align}
\delta_{\alpha}A^{a}_{\mu}&=\partial_{\mu}\alpha^{a}+if^{abc}\alpha^{b}A_{\mu}^{c}
\equiv (D_{\mu}\alpha)^{a}, \nonumber \\[0.2cm]
\delta_{\alpha}F^{a}_{\mu\nu}&=if^{abc}\alpha^{b}F^{c}_{\mu\nu},
\label{eq:AFNAYM_indices}
\end{align}
where in the first line we expressed
the variation of the gauge field in terms of the (adjoint) covariant derivative
of the gauge function. The field strength, in turn, can be also recast as the commutator of two
covariant derivatives,~$F_{\mu\nu}=[D_{\mu},D_{\nu}]$.

After all these preliminaries, we can write a gauge invariant action for fermions coupled to non-Abelian
gauge fields
\begin{align}
S_{\mbox{\tiny YM}}
&=\int d^{4}x\left[-{1\over 2g_{\mbox{\tiny YM}}^{2}}{\rm tr\,}\big(F_{\mu\nu}F^{\mu\nu}\big)
+\overline{\boldsymbol{\psi}}\big(i{D\!\!\!\!/}-m\big)\boldsymbol{\psi}\right] \nonumber \\[0.2cm]
&=\int d^{4}x\left[-{1\over 4g_{\mbox{\tiny YM}}^{2}}F^{a}_{\mu\nu}F^{a\mu\nu}
+\overline{\boldsymbol{\psi}}\big(i{\partial\!\!\!/}-m\big)\boldsymbol{\psi}
+A^{a}_{\mu}\overline{\boldsymbol{\psi}}\gamma^{\mu}T^{a}_{\mathbf{R}}\boldsymbol{\psi}\right],
\label{eq:YM_action}
\end{align}
where~$g_{\mbox{\tiny YM}}$ is the only coupling constant of the theory\footnote{The factors 
of~$g_{\mbox{\tiny YM}}$ in front of the first term in the action can be removed by 
a rescale~$A_{\mu}\rightarrow g_{\mbox{\tiny YM}}A_{\mu}$. In doing so, an inverse power of the coupling
constant appears in the derivative terms in~\eqref{eq:trans_ANA_noindices} and the first 
identity in~\eqref{eq:AFNAYM_indices}, while the commutator in~\eqref{eq:field_strength_NAYM}
acquires a power of $g_{\mbox{\tiny YM}}$, as well as the structure constant term
in~\eqref{eq:field_strengthNAYM_indices}.}. 
This non-Abelian generizations
of QED was first formulated by C.~N.~Yang and Robert~L. Mills~\cite{YM}.
Yang-Mills (YM) theories are the backbone of our understanding of elementary 
particle physics.
Although the action~$S_{\mbox{\tiny YM}}$ reduces to that of 
QED in eq.~\eqref{eq:QED_action} 
for~$G=U(1)$, it displays a much richer structure for non-Abelian gauge groups. For starters,
the commutator in the field strength~\eqref{eq:field_strength_NAYM} is nonzero and 
the~$F_{\mu\nu}^{a}F^{a\mu\nu}$  term
in~\eqref{eq:YM_action} contains cubic and quartic gauge field self-interaction terms. This indicates that, 
unlike the photon, non-Abelian gauge bosons are never free particles even if uncoupled to 
matter.

The general 
analysis of gauge invariance follows in many aspects the Abelian case. The corresponding electric and
magnetic fields are defined in terms of the gauge potential~$A_{\mu}^{a}\equiv (A^{a}_{0},
-\mathbf{A}^{a})$ by
\begin{align}
\mathbf{E}^{a}&=-\boldsymbol{\nabla}A_{0}^{a}-{\partial\mathbf{A}^{a}\over\partial t}
+f^{abc}A_{0}^{a}\mathbf{A}^{b},  
\nonumber \\[0.2cm]
\mathbf{B}^{a}&=\boldsymbol{\nabla}\times\mathbf{A}^{a}+f^{abc}\mathbf{A}^{b}\times\mathbf{A}^{c},
\end{align}
and, unlike their Abelian counterparts, they are not gauge invariant. 
The electric field~$\mathbf{E}^{a}$ is in fact the momentum canonically conjugate 
to~$\mathbf{A}^{a}$
\begin{align}
\big\{A^{a}_{i}(t,\mathbf{r}),E^{b}_{j}(t,\mathbf{r}')\big\}_{\rm PB}
=\delta_{ij}\delta^{ab}\delta^{(3)}(\mathbf{r}-\mathbf{r}'),
\label{eq:ET_PB_AE_NA}
\end{align}
and the Hamiltonian reads
\begin{align}
H=\int d^{3}x\,\left[{1\over 2}\mathbf{E}^{a}\cdot\mathbf{E}^{a}
+{1\over 2}\mathbf{B}^{a}\cdot\mathbf{B}^{a}
+A_{0}^{a}\big(\mathbf{D}\cdot\mathbf{E}\big)^{a}
\right].
\end{align}
Similarly to Maxwell's electrodynamics,~$A_{0}^{a}$ plays 
the role of a Lagrange multiplier enforcing the Gauss law constraint,
now reading
\begin{align}
(\mathbf{D}\cdot\mathbf{E})^{a}\equiv\boldsymbol{\nabla}\cdot\mathbf{E}^{a}
+f^{abc}\mathbf{A}^{b}\times\mathbf{E}^{c}=0.
\label{eq:Gauss_law_NA_version}
\end{align}

In the quantum theory, classical fields are replaced by operators. Using the
non-Abelian version of the temporal 
gauge,~$A_{0}^{a}=0$, residual gauge transformations correspond to 
time-independent gauge functions~$\alpha^{a}(\mathbf{r})$ and are generated 
by~$\mathbf{D}\cdot\mathbf{E}$
\begin{align}
\delta_{\alpha}\mathbf{A}(t,\mathbf{r})
&=i\left[\int d^{3}r\,\alpha^{a}(\mathbf{r})(\mathbf{D}\cdot\mathbf{E})^{a},
\mathbf{A}(t,\mathbf{r})\right] \nonumber \\[0.2cm]
&=\boldsymbol{\nabla}\alpha^{a}+if^{abc}\alpha^{b}\mathbf{A}^{c} \equiv (\mathbf{D}\alpha)^{a},
\end{align}
where we have used the canonical commutation relations derived from~\eqref{eq:ET_PB_AE_NA} and
to avoid boundary terms after integration by parts we need to restrict to 
``small'' gauge transformations where~$\alpha^{a}(\mathbf{r})$ vanishes when~$|\mathbf{r}|\rightarrow
\infty$. Those in 
the connected component of the identity~$\mathscr{G}_{0}$
are therefore implemented on the space of physical states by the operator
\begin{align}
\mathscr{U}(\alpha)=\exp\left[i\int d^{3}r\,\alpha^{a}(\mathbf{r})(\mathbf{D}\cdot\mathbf{E})^{a}\right].
\end{align}
As in the Abelian case discussed in Box~9 (see page~\pageref{box:large_vs_small}),
the invariance under these ``small'' gauge transformations has to be preserved at all expenses to
avoid unphysical states entering the theory's spectrum.
To achieve this, we require that the Gauss law annihilates physical states
\begin{align}
(\mathbf{D}\cdot\mathbf{E})^{a}|\mbox{phys}\rangle=0.
\label{eq:Gauss_law_NA_op}
\end{align}
In the presence of non-Abelian sources,~$(\mathbf{D}\cdot\mathbf{E})^{a}$ gets replaced
by~$(\mathbf{D}\cdot\mathbf{E})^{a}-\rho^{a}$, with~$\rho^{a}$ the matter charge density operator.

We should not forget about ``large''
gauge transformations whose gauge parameter~$\alpha^{a}(\mathbf{r})$ does not vanish 
when~$|\mathbf{r}|\rightarrow\infty$. Notice that 
any transformation of this kind can be written
as
\begin{align}
g(\mathbf{r})_{\rm large}=hg(\mathbf{r})_{\rm small},
\end{align}
where~$h\neq\mathbbm{1}$ is a rigid transformations such that~$g(\mathbf{r})_{\rm large}
\rightarrow h$ as~$|\mathbf{r}|\rightarrow \infty$. They build up what can be called a 
copy of the group at infinity,~$G_{\infty}$, the global invariance
leading to charge conservation by the first Noether theorem. This is a real symmetry
that quantum mechanically can be realized either \`a la Wigner-Weyl or \`a la Nambu-Goldstone. 
For the SM gauge group~$\mbox{SU(3)}\times\mbox{SU(2)}\times\mbox{U(1)}$, the
color~$\mbox{SU(3)}_{\infty}$ symmetry remains unbroken by the vacuum, whereas due to the BEH mechanism
the electroweak 
factor~$[\mbox{SU(2)}\times\mbox{U(1)}]_{\infty}$ is partially realized \`a la Nambu-Goldstone, 
with a preserved~$\mbox{U(1)}_{\infty}$ corresponding to the global invariance of
electromagnetism\footnote{As we will see shortly, the unbroken~U(1) generator is a mixture
of the two generators of the Cartan subalgebra of the electroweak~$\mbox{SU(2)}\times\mbox{U(1)}$
gauge group factor.}.  

\section{Anomalous symmetries}
\label{sec:anomalies}

In section~\ref{sec:tale_of_symmetries}, we mentioned the possibility that classical symmetries or invariances
could somehow turn out to be incompatible with the process of quantization but so far did not
elaborate any further. Since anomalous symmetries are crucial in our understanding of a number of
physical phenomena, it is about time to look into anomalies in some detail
(see~\cite{AG_anomalies,Bertlmann,Fujikawa_Suzuki,AG_VM_GS} for some reviews on the topic).

\subsection{Symmetry vs. the quantum}

Let us go back to the QED action~\eqref{eq:QED_action}. We have already discussed the
global phase invariance leading by the first Noether theorem to the
conserved current~\eqref{eq:vector_current_QED}. In addition, we can also consider the transformations
\begin{align}
\psi\longrightarrow e^{i\alpha \gamma_{5}}\psi, \hspace*{1cm}
\overline{\psi}\longrightarrow \overline{\psi}e^{i\alpha\gamma_{5}},
\label{eq:axial_vector_trans}
\end{align}
where~$\gamma_{5}$ is the chirality matrix defined in eq.~\eqref{eq:chirality_matrix}. Unlike the
transformation~$\psi\rightarrow e^{i\vartheta}\psi$ rotating the positive and
negative chirality components of the Dirac spinor by the same phase, 
in~\eqref{eq:axial_vector_trans} they change by opposite phases. In what follows, 
we refer to the first
type as {\em vector} transformations, while the second we dub as {\em axial-vector}. 
The latter, however, are not a symmetry of the QED action for~$m\neq 0$, 
since~$\overline{\psi}\psi\rightarrow \overline{\psi}e^{2i\alpha\gamma_{5}}\psi\neq \overline{\psi}\psi$,
whereas~$\overline{\psi}\gamma^{\mu}\partial_{\mu}\psi$ invariant. 
In fact, using the Dirac field equations it can be shown that the axial-vector current
\begin{align}
j^{\mu}_{5}=\overline{\psi}\gamma_{5}\gamma^{\mu}\psi,
\label{eq:AV_currentQED}
\end{align}
satisfies the relation
\begin{align}
\partial_{\mu}j_{5}^{\mu}=2im\overline{\psi}\gamma_{5}\psi,
\label{eq:AV_WardIdQED}
\end{align}
and for~$m=0$ gives the conservation equation associated with the invariance of massless QED
under axial-vector transformations. Similar to what we found on Box~8 for the
flavor symmetry of QCD, in this limit the global~$\mbox{U(1)}_{V}$ symmetry of QED 
gets enhanced to~$\mbox{U(1)}_{V}\times\mbox{U(1)}_{A}$.

In the quantum theory Noether currents are constructed as products
of field operators evaluated at the same spacetime point. 
These quantities are typically divergent and it is necessary to introduce some regularization in order
to make sense of them. In the case of QED
one way to handle the vector current~$j^{\mu}(x)=
\overline{\psi}(x)\gamma^{\mu}\psi(x)$ is by using point splitting
\begin{align}
j^{\mu}(x,\epsilon)_{\rm reg}\equiv
\overline{\psi}\left(x-{1\over 2}\epsilon\right)\gamma^{\mu}\psi\left(x+{1\over 2}\epsilon\right)
\exp\left(ie\int_{x-{1\over 2}\epsilon}^{x+{1\over 2}\epsilon}dx^{\mu}A_{\mu}\right),
\label{eq:axial_current_point_splitting}
\end{align}
where the divergences appear as poles in~$\epsilon=0$.
Notice that since the phases introduced by the gauge transformations of the two fields are evaluated
at different points, an extra Wilson line term is needed to restore gauge invariance of the regularized
current.
Alternatively, we can use 
Pauli-Villars (PV) regularization, where a number of spurious fermion fields of masses~$M_{i}$
are added to the action
\begin{align}
S_{\rm reg}=\int d^{4}x\left[-{1\over 4}F_{\mu\nu}F^{\mu\nu}
+\overline{\psi}(i{D\!\!\!\!/}-m)\psi+\sum_{i=1}^{n}
c_{k}\overline{\Psi}_{k}(i{D\!\!\!\!/}-M_{k})\Psi_{k}\right],
\end{align}
with~$n$ and~$c_{k}$ chosen so the limit
\begin{align}
j^{\mu}(x)_{\rm reg}\equiv\lim_{x'\rightarrow x}
\left[\overline{\psi}(x')\gamma^{\mu}\psi(x)
+\sum_{k=1}^{n}c_{k}\overline{\Psi}_{k}(x')\gamma^{\mu}
\Psi_{k}(x)
\right],
\end{align}
remains finite (i.e., all poles at~$x-x'=0$ cancel). 
An important feature of the PV regularization is that it explicitly preserves gauge invariance. 
The masses~$M_{k}$ act as
regulators, since in the limit~$M_{k}\rightarrow\infty$ the PV fermions decouple and 
the original divergences reapear. 
 
The need to make sense of composite operators is at the bottom of the potential problems with
current conservation in the quantum domain. The regularization procedure might collide with some
of the classical symmetries of the theory, resulting in its breaking after divergences are properly 
handled. This is why 
our discussion of the regularization of the current operator in QED has been conspicuously concerned with
the issue of gauge invariance of the vector current. The existence of gauge invariant regularization
schemes guarantees that the current coupling to the gauge field 
can be defined in the quantum theory without spoiling its 
conservation~$\partial_{\mu}j^{\mu}=0$ at operator level. 
Otherwise, we would be in serious trouble, as we can see by 
applying the quantization prescription~\eqref{eq:heisenberg_prescription} to the 
stability condition of the Gauss law~\eqref{eq:Gauss_law_timeindep_PB}
\begin{align}
[G,H]=-i\partial_{\mu}j^{\mu},
\end{align}
where we have defined~$G\equiv \boldsymbol{\nabla}\cdot\mathbf{E}-j^{0}$.
If~$\partial_{\mu}j^{\mu}\neq 0$,
the Gauss law condition ensuring the factorization of redundant states would not be preserved by 
time evolution. Indeed, imposing the constraint at~$t=0$ on some 
state,~$G|\Psi(0)\rangle=0$,
we would have at first order in~$\delta t$
\begin{align}
G|\Psi(\delta t)\rangle
&=-i\delta tGH|\Psi(0)\rangle 
=-\delta t\partial_{\mu}j^{\mu}|\Psi(0)\rangle \neq 0,
\end{align}
so the constraint is no longer satisfied and unphysical states enter the spectrum.
Another sign that something goes wrong when implementing the Gauss law constraint 
in theories with gauge anomalies appears when computing the commutator of two~$G$'s evaluated at
different points. 
In the presence of a gauge anomaly, it is no longer zero~\cite{Faddeev,Nelson_AG,KSS}
\begin{align}
[G(\mathbf{r}),G(\mathbf{r}')]=c\mathbf{B}(\mathbf{r})\cdot \boldsymbol{\nabla}\delta^{(3)}(\mathbf{r}
-\mathbf{r}'),
\label{eq:commutator_Gauss_law}
\end{align}
where~$c\neq 0$ is a constant determined by the value of~$\partial_{\mu}j^{\mu}$.
This result implies that~$G(\mathbf{r})|\mbox{phys}\rangle=0$ 
cannot be consistently imposed, since this condition
would imply~$[G(\mathbf{r}),G(\mathbf{r}')]|\mbox{phys}\rangle=0$ 
whereas the right-hand side of~\eqref{eq:commutator_Gauss_law} gives a 
nonzero result when acting on the 
state\footnote{Something similar happens in the case of non-Abelian gauge theories
that we will discuss in the next section. There, the commutator
of two Gauss law operators acquires a central extension,~$[G^{a}(\mathbf{r}),G^{b}(\mathbf{r}')]
=if^{abc}G^{c}(\mathbf{r})\delta^{(3)}(\mathbf{r}-\mathbf{r}')+\mathcal{A}^{ab}(\mathbf{r},\mathbf{r}')$,
with~$G^{a}\equiv (\mathbf{D}\cdot\mathbf{E})^{a}-j^{a0}$ in this case.}. 
This being the case, spurious states cannot be factored out from the spectrum, with the upshot
that the theory becomes inconsistent.

This shows that in constructing QFTs, gauge anomalies cannot emerge. This condition
is a very powerful constraint in model building, since it limits both the type of fields that 
can be allowed in the actions and also their couplings. As we will see
in Box~13 in page~\pageref{pag:box_anomaliesSM}, in the SM this requirement
completely fixes the hypercharges of quarks and leptons, up to a global normalization 
(see~\cite{AG_VM_GS} for examples of anomaly cancellation in the SM and beyond).

After this digression, we go back to the quantum mechanical definition of the
axial-vector current~\eqref{eq:AV_currentQED} and the fate of its 
(pseudo)conservation~\eqref{eq:AV_WardIdQED}. To simplify things, we consider the
massless case where axial-vector transformations~\eqref{eq:axial_vector_trans} 
are a symmetry of the classical action.
A very convenient way to study this problem
is to treat the gauge field as a classical external source coupling to the 
quantum Dirac field. This is made clear by denoting gauge fields and field strengths 
using calligraphic fonts as~$\mathscr{A}_{\mu}$ and~$\mathscr{F}_{\mu\nu}$ respectively.
Instead of working with operators, we deal with their vacuum expectation values in the
presence of the background field and 
compute~$\langle J^{\mu}_{5}\rangle_{\mathscr{A}}\equiv\langle 0|J^{\mu}_{5}|0\rangle$ 
together with its
divergence. This can be done using either the regularized operators introduced above
(see, for example,~\cite{Jackiw_ABJ} for a calculation using point-splitting regularization) or
diagrammatic techniques. In the latter case, we need to compute the celebrated triangle diagrams
\begin{align}
\nonumber \\[-0.4cm]
\parbox{40mm}{
\begin{fmfgraph*}(100,65)
\fmfleft{i1,i2,i3}
\fmfright{o1,o2}
\fmf{phantom}{i1,v1,v2,o1}
\fmf{phantom}{i3,v4,v3,o2}
\fmf{phantom,tension=0}{v1,v5,v2}
\fmf{phantom,tension=0}{v2,v3}
\fmf{photon,tension=0.6}{v3,o2}
\fmf{photon,tension=0.6}{v2,o1}
\fmf{phantom}{i2,v5}
\fmfdot{v5}
\fmffreeze
\fmf{fermion,tension=0.1}{v5,v3,v2,v5}
\fmflabel{$J_{5}^{\mu}$}{v5}
\fmflabel{$\mathscr{A}_{\alpha}$}{o2}
\fmflabel{$\mathscr{A}_{\beta}$}{o1}
\end{fmfgraph*}
}+\hspace*{1cm}
\parbox{40mm}{
\begin{fmfgraph*}(100,65)
\fmfleft{i1,i2,i3}
\fmfright{o1,o2}
\fmf{phantom}{i1,v1,v2,o1}
\fmf{phantom}{i3,v4,v3,o2}
\fmf{phantom,tension=0}{v1,v5,v2}
\fmf{phantom,tension=0}{v2,v3}
\fmf{phantom}{i2,v5}
\fmfdot{v5}
\fmffreeze
\fmf{photon,tension=0.6}{v3,o1}
\fmf{photon,tension=0.6,rubout}{v2,o2}
\fmf{fermion,tension=0.1}{v5,v3,v2,v5}
\fmflabel{$J_{5}^{\mu}$}{v5}
\fmflabel{$\mathscr{A}_{\alpha}$}{o2}
\fmflabel{$\mathscr{A}_{\beta}$}{o1}
\end{fmfgraph*}
}\label{eq:triangle_diagrams}
\\[-0.5cm]
\nonumber
\end{align}
where in the left vertex of both diagrams (indicated by a dot) 
an axial-vector current is inserted, whereas 
the other two are coupled to the externa gauge field through the 
vector gauge currents. Since in this lectures we are not 
entering into the computation of Feynman graphs, we will not elaborate on how 
to calculate these ones. Details can be found in chapter~9 of ref.~\cite{AG_VM} or 
in~\cite{Bertlmann}. Here we just give the final result for the anomaly of the axial-vector current
\begin{align}
\partial_{\mu}\langle J^{\mu}_{5}\rangle_{\mathscr{A}}
=-{e^{2}\hbar\over 16\pi^{2}}\epsilon^{\mu\nu\alpha\beta}
\mathscr{F}_{\mu\nu}\mathscr{F}_{\alpha\beta}.
\label{eq:axial_anomaly}
\end{align}
Despite having used all the time natural units with~$\hbar=1$, in this 
expression we have restored the powers of
the Planck constant to make explicit the fact that the anomaly is a pure quantum effect.

This crucial result has a long history. The diagrams in eq.~\eqref{eq:triangle_diagrams} were
computed in 1949 
by Jack Steinberger~\cite{Steinberger} and later by Julian Schwinger in 1951~\cite{Schwinger}, in 
both cases in the
context of the electromagnetic decay of neutral mesons\footnote{Other early calculations
of the triangle diagrams were carried out in 1949 
by Hiroshi Fukuda and Yoneji Miyamoto~\cite{Fukuda_Miyamoto}
and by S. Ozaki, S. Oneda, and S. Sasaki~\cite{OOS}.}. Almost two decades later,
the consequences of the triangle diagram for the quantum realization of the
axial-vector symmetry of QED were pointed out by Stephen Adler~\cite{Adler} and John S. Bell and Roman
Jackiw~\cite{Bell_Jackiw} in what are
considered today the foundational papers of the subject of quantum anomalies.

There are some very important issues that should be mentioned concerning the calculation of
the axial anomaly~\eqref{eq:axial_anomaly}. We have stressed how the anomaly could be seen as originated
by the need to regularize UV (i.e., short distance) divergences in the definition of the current or,
alternatively,
in the computation of the triangle diagrams. Nevertheless, using either method, we find a 
regular result in the limit in which the regulator is removed. In the language of
QFT, we do not need to subtract and renormalize divergences to find the anomaly of the axial 
current. At the level of diagrams, what happens is that 
although the integrals are linearly divergent this
only results in an ambiguity in their value that is fixed by requiring the gauge (vector) current
to be conserved. In the case of the point splitting calculation, introducing a
Wilson line similar to the one inserted in eq.~\eqref{eq:axial_current_point_splitting} in the regularized 
definition of the axial-vector current to preserve gauge invariance
we are led to the axial anomaly after taking the~$\epsilon\rightarrow 0$ limit.

Another important point to be stressed is a {\em tension} between the 
conservation of the gauge and the axial-vector currents: we can impose the conservation of either
of the two, {\em but not of both simultaneously}. After the above discussion of the dire consequences
of violating gauge current conservation, the choice is clear enough. 

\subsection{The physical power of the anomaly}

When studying the global symmetries of QCD, we have also encountered axial transformations
[see Box~8, and in particular eq.~\eqref{eq:V_AV_trans_quarks}] and mentioned that it is anomalous.
Now we can be more explicit.
The axial-vector current of interest in this case is given by
\begin{align}
J_{5}^{\mu}&=\overline{\boldsymbol{q}}\gamma_{5}\gamma^{\mu}\boldsymbol{q},
\label{eq:singlet_current_QCD}
\end{align}
where a sum over color indices should be understood.
Its anomaly comes from triangle diagrams 
similar to the ones shown in fig.~\eqref{eq:triangle_diagrams}, 
this time with quarks running in the loop. But, together with the triangles coupling to the  
electromagnetic external potential~$\mathscr{A}_{\mu}$, 
we also have a pair of triangles where the vertices on the right
couple to external gluon field~$\mathcal{A}_{\mu}^{a}$ (for this, we also use calligraphic
fonts to indicate that we are dealing with classical sources). This results in the anomaly
\begin{align}
\partial_{\mu}\langle J^{\mu}_{5}\rangle_{\mathscr{A},\mathcal{A}}
=-{N_{c}\over 16\pi^{2}}\bigg(\sum_{f=1}^{N_{f}}q_{f}^{2}\bigg)\epsilon^{\mu\nu\alpha\beta}
\mathscr{F}_{\mu\nu}\mathscr{F}_{\alpha\beta}
-{N_{f}\over 16\pi^{2}}\epsilon^{\mu\nu\alpha\beta}
\mathcal{F}_{\mu\nu}^{a}\mathcal{F}_{\alpha\beta}^{a},
\end{align} 
where~$\mathcal{F}_{\mu\nu}^{a}$ is the non-Abelian field strength associated with the
external gluon field and~$N_{c}$ is the number of colors. 
The coefficient of the first term is obtained by summing the 
expression of the axial anomaly given in~\eqref{eq:axial_anomaly} to all quarks running in the loop.
As for the second, the quarks couple to the gluon fields through the 
gauge current
\begin{align}
J^{\mu a}=\overline{\boldsymbol{q}}\gamma^{\mu}\tau^{a}\boldsymbol{q},
\end{align}
where~$\tau^{a}$ are the generators of the fundamental representation of SU(3) acting on the color indices 
of each component of~$\boldsymbol{q}$. Since the axial current does not act on color indices, the prefactor
is proportional to~$({\rm tr\,}\mathbbm{1})({\rm tr\,}\{\tau^{a},\tau^{b}\})=N_{f}\delta^{ab}$, with~$\mathbbm{1}$
the identity in flavor space.

Anomalies can also affect the global non-Abelian~$\mbox{SU($N_{f}$)}_{L}\times\mbox{SU($N_{f}$)}_{R}$ symmetry 
defined in~\eqref{eq:SUN_LR}. This global symmetry group 
can be rearranged in terms of vector and axial transformations
~$\mbox{SU($N_{f}$)}_{L}\times\mbox{SU($N_{f}$)}_{R}=\mbox{SU($N_{f}$)}_{V}\times\mbox{SU($N_{f}$)}_{A}$
acting on the quark fields as
\begin{align}
\mbox{SU($N_{f}$)}_{V}:
\boldsymbol{q}&\rightarrow e^{i\alpha_{V}^{I}t^{I}_{\mathbf{f}}} \boldsymbol{q}, \hspace*{2cm}
\mbox{SU($N_{f}$)}_{A}:
\boldsymbol{q}\rightarrow e^{i\alpha_{A}^{I}t^{I}_{\mathbf{f}}\gamma_{5}} \boldsymbol{q},
\end{align}
with~$\boldsymbol{q}_{R}$ and~$\boldsymbol{q}_{L}$ transforming respectively with
the same or opposite~$\mbox{SU($N_{f}$)}$ parameters\footnote{A warning note here. Unlike the
Abelian U(1)$_{A}$, transformations in~$\mbox{SU($N_{f}$)}_{A}$ do not close and therefore do not form
a group. This can be checked by composing two of them and applying the Baker-Campbell-Hausdorff
formula. Our notation has to be understood in a formal sense.}. 
Vector currents, however, are always anomaly-free.
A simple way to come to this conclusion is to notice that the PV regularization method introduced above
preserved all vector symmetries, since these remain unbroken by fermion mass terms\footnote{This 
argument also applies to the SU(3) gauge invariance of QCD, which cannot be anomalous since it
acts in the same way on quarks of both chiralities. As a consequence, the theory can be regularized in 
a gauge invariant way.}. We thus focus on the chiral~$\mbox{SU($N_{f}$)}_{A}$ factor,
whose associated axial-vector current is
\begin{align}
J_{5}^{I\mu}=\overline{\boldsymbol{q}}\gamma_{5}\gamma^{\mu}t_{\mathbf{f}}^{I}\boldsymbol{q},
\end{align}
where, again, there is a tacit sum over the quark color index.
As in the case of the singlet current~\eqref{eq:singlet_current_QCD}, there are contribution coming 
from 
the photon and gluon couplings of the quarks. Taking into account that, unlike photons, 
gluons are flavor-blind, we find
\begin{align}
\partial_{\mu}\langle J^{I\mu}_{5}\rangle_{\mathscr{A},\mathcal{A}}
=-{N_{c}\over 16\pi^{2}}\bigg[\sum_{f=1}^{N_{f}}q_{f}^{2}(t_{\mathbf{f}}^{I})_{ff}\bigg]\epsilon^{\mu\nu\alpha\beta}
\mathscr{F}_{\mu\nu}\mathscr{F}_{\alpha\beta}
-{N_{f}\over 16\pi^{2}}\big({\rm tr\,}t_{\mathbf{f}}^{I}\big)\epsilon^{\mu\nu\alpha\beta}
\mathcal{F}_{\mu\nu}^{a}\mathcal{F}_{\alpha\beta}^{a}.
\end{align}
Since all generators of~$\mbox{SU($N_{f}$)}$ are traceless the second term is zero but the first one
does not necessarily vanishes.

Let be focus on the dynamics of the two lightest quarks~$u$ and~$d$, 
where~$q_{u}={2\over 3}e$ and~$q_{d}=-{1\over 3}e$. 
In this case~$N_{f}=2$ and the flavor group is generated 
by~$t^{I}_{\mathbf{f}}={1\over 2}\sigma^{I}$, with~$\sigma^{I}$ the Pauli matrices. We have then
\begin{align}
\sum_{f=1}^{2}q_{f}^{2}(t_{\mathbf{f}}^{1})_{ff}=\sum_{f=1}^{2}q_{f}^{2}(t_{\mathbf{f}}^{2})_{ff}=0, \hspace*{1cm}
\sum_{f=1}^{2}q_{f}^{2}(t_{\mathbf{f}}^{3})_{ff}={e^{2}\over 6},
\end{align}
where~$N_{c}$ is the number of quark colors. 
This means that~$J^{3\mu}_{5}$ is anomalous
\begin{align}
\partial_{\mu}\langle J^{3\mu}_{5}\rangle_{\mathscr{A},\mathcal{A}}
=-{e^{2}N_{c}\over 48\pi^{2}}\epsilon^{\mu\nu\alpha\beta}\mathscr{F}_{\mu\nu}\mathscr{F}_{\alpha\beta}.
\label{eq:axial_anomaly_3rdflavorcomp}
\end{align}
The physical importance of this result lies in that after chiral symmetry breaking (see Box~8 in 
page~\pageref{eq:box_chiral_SB}), the operator~$\partial_{\mu}J^{a\mu}_{5}$ becomes the
interpolating field for pions, creating them out of the vacuum\footnote{The first identity 
follows from~$\langle \pi^{a}(p)|J^{b\mu}_{5}(x)|0\rangle\sim p^{\mu}\delta^{ab}e^{-ip\cdot x}$, a 
direct consequence
of the Goldstone theorem~\cite{GSW}.}
\begin{align}
\langle \pi^{a}(p)|\partial_{\mu}J^{a\mu}_{5}(x)|0\rangle=f_{\pi}m_{\pi}\delta^{ab}e^{-ip\cdot x}
\hspace*{1cm} \Longrightarrow \hspace*{1cm} 
\pi^{a}(x)={1\over f_{\pi}m_{\pi}}\partial_{\mu}J^{a\mu}_{5}(x),
\label{eq:overlap_interp_pions}
\end{align}
where~$m_{\pi}$ is the pion mass and~$f_{\pi}$ the pion decay constant introduced in 
eq.~\eqref{eq:NG_bosons_field_general} to parametrize the matrix of NG bosons resulting from 
chiral symmetry breaking.
Although to compute the anomaly~\eqref{eq:axial_anomaly_3rdflavorcomp} we took the electromagnetic
field to be a classical source, the corresponding operator identity implies the existence of a nontrivial 
overlap between the neutral pion state and the state with two photons
\begin{align}
\langle \mathbf{k}_{1},\lambda_{1};\mathbf{k}_{2},\lambda_{2}|\pi^{0}(p)\rangle
={e^{2}N_{c}\over 12\pi^{2}f_{\pi}}(2\pi)^{4}\delta^{(4)}(p-k_{1}-k_{2})
\epsilon_{\mu\nu\alpha\beta}k_{1}^{\mu}k_{2}^{\nu}\epsilon^{\alpha}(\mathbf{k}_{1})
\epsilon^{\beta}(\mathbf{k}_{2}).
\end{align}
The width of the process can be computed from this result to be \label{page:pion_decay}
\begin{align}
\Gamma(\pi^{0}\rightarrow 2\gamma)={\alpha^{2}N_{c}^{2}m_{\pi}^{3}\over 576\pi^{3}f_{\pi}^{2}}
=7.73\,\mbox{eV},
\label{eq:neutral_pion_width}
\end{align}
which is perfectly consistent with experimental measurements~\cite{pion_width}
\begin{align}
\Gamma(\pi^{0}\rightarrow 2\gamma)_{\rm exp}=7.798\pm 0.056\mbox{ (stat.)}\,
\pm 0.109\mbox{ (syst.)}
\mbox{ eV}.
\end{align}
Incidentally, the presence of~$f_{\pi}=93\,\mbox{MeV}$ in eq.~\eqref{eq:neutral_pion_width} 
give a rationale for it
being called the pion decay constant.

\begin{figure}[t]
\centerline{\includegraphics[scale=0.45]{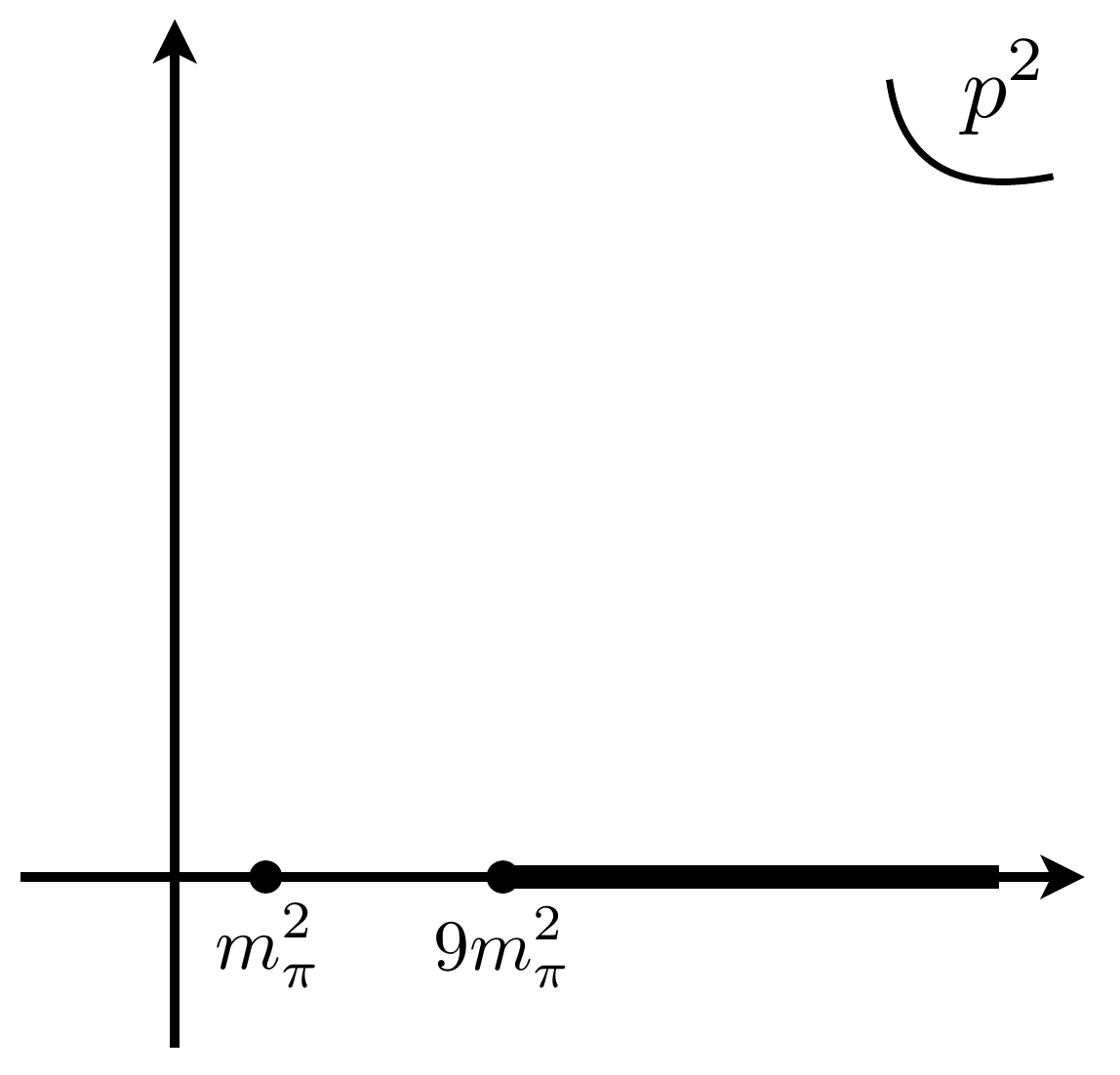}}
\caption[]{Complex~$p^{2}$-plane showing the structure of singularities of the function~$f(p^{2})$
in eq.~\eqref{eq:PCAC_amplitude}: a pole at~$p^{2}=m_{\pi}^{2}$ and a branch cut
beginning at~$p^{2}=9m_{\pi}^{2}$.}
\label{fig:f(p2)-plane}
\end{figure}

The electromagnetic decay of the neutral pion is a direct consequence of the existence of 
the axial anomaly. On general grounds, it can be argued that the amplitude for the decay process
of the~$\pi^{0}$ into two photons has the structure
\begin{align}
\langle \mathbf{k}_{1},\lambda_{1};\mathbf{k}_{2},\lambda_{2}|\pi^{0}(p)\rangle
=i{p^{2}-m_{\pi}^{2}\over f_{\pi}m_{\pi}^{2}}p^{2}f(p^{2})(2\pi)^{4}\delta^{(4)}(p-k_{1}-k_{2})
\epsilon_{\mu\nu\alpha\beta}k_{1}^{\mu}k_{2}^{\nu}\epsilon^{\alpha}(\mathbf{k}_{1})
\epsilon^{\beta}(\mathbf{k}_{2}),
\label{eq:PCAC_amplitude}
\end{align}
with~$f(p^{2})$ a function of the pion squared momentum. We could naively assume~$f(p^{2})$ to be 
well-behaved, with a pole singularity at~$p^{2}=m_{\pi}^{2}$ and
a branch cut starting at~$9m_{\pi}^{2}$ signalling multi-pion production (see fig.~\ref{fig:f(p2)-plane}).
Were this the case, the amplitude would be suppressed in the~$p^{2}\rightarrow 0$ limit. 
Historically, this result was known as the
Sutherland-Veltman theorem~\cite{Sutherland,Veltman} and essentially ruled out the existence
of the process~$\pi^{0}\rightarrow 2\gamma$, that was nevertheless observed. The catch lies in 
that the regularity hypothesis concerning~$f(p^{2})$, called partial conservation 
of the axial current (PCAC), is wrong due to the axial anomaly.
The calculation of the triangle diagrams~\eqref{eq:triangle_diagrams} shows that this function 
is not regular at zero momentum, but actually 
has a pole
\begin{align}
f(p^{2})\sim {i e^{2}N_{c}\over 12\pi}{1\over p^{2}} \hspace*{1cm} \mbox{as \hspace*{0.5cm} $p\rightarrow 0$}.
\end{align}
This singularity 
is precisely responsible for compensating the low-momentum suppression of 
the amplitude~\eqref{eq:PCAC_amplitude},
giving the nonzero result accounting for the~$\pi^{0}\rightarrow 2\gamma$ decay.
It is somewhat fascinating that the anomaly, that we identified from the start as 
resulting from UV ambiguities in the definition of the current, is also associated with an IR pole
and determined by its residue. This reflects the profound topological connexions
of QFT anomalies~\cite{AG_anomalies,Bertlmann,Fujikawa_Suzuki,AG_VM_GS}.
 
\begin{mdframed}[backgroundcolor=lightgray,hidealllines=true]
\vspace*{0.2cm}
\centerline{\greybox{\bf The path integral way to the anomaly}}
\vspace*{0.2cm}
\label{box:anomalies}

There are many different roads leading to the chiral anomaly. 
For our presentation above we have chosen the perturbative approach, involving the
computation of the two one-loop triangle diagrams shown in eq.~\eqref{eq:triangle_diagrams}. 
But the anomaly can also be computed using path integrals, where it appears as a result  
of the noninvariance of the functional measure under chiral rotations of
the Dirac fermions. 
  
To see how this comes about, let us 
consider again a Dirac fermion coupled to an external electromagnetic field~$\mathscr{A}_{\mu}$ 
that we treat as a classical source. Its action is given by
\begin{align}
S[\psi,\overline{\psi},\mathscr{A}_{\mu}]
&=\int d^{4}x\overline{\psi}\gamma^{\mu}\big(i\partial_{\mu}+e\mathscr{A}_{\mu}\big)\psi 
\nonumber \\[0.2cm]
&=\int d^{4}x\Big[\overline{\psi}_{R}\big(i\partial_{\mu}+e\mathscr{A}_{\mu}\big)\psi_{R}
+\overline{\psi}_{L}\big(i\partial_{\mu}+e\mathscr{A}_{\mu}\big)\psi_{L}\Big],
\end{align}
where in the second line we split the Dirac fermion into its two chiralities.
A quantum effective action~$\Gamma[\mathscr{A}_{\mu}]$ for the external field
can be defined by integrating out the fermions
\begin{align}
e^{i\Gamma[\mathscr{A}_{\mu}]}=\int \mathscr{D}\overline{\psi}\mathscr{D}\psi
\,e^{iS[\psi,\overline{\psi},\mathscr{A}_{\mu}]}.
\label{eq:fermion_part_funct}
\end{align} 
The important point in this expressions is that the Dirac fields are dummy 
variables that can be modified without
changing the value of the functional integral. In particular, we can implement the following
``change of variables''
\begin{align}
\psi=e^{i\alpha\gamma_{5}}\psi' \hspace*{1cm} 
\Longrightarrow \hspace*{1cm} \psi_{R,L}=e^{\pm i\alpha}\psi'_{R,L},
\label{eq:chov_chiral}
\end{align}
writing the original
Dirac field in terms of its chiral-transformed [see eq.~\eqref{eq:axial_vector_trans}].
As we know, in the absence of a Dirac mass term the fermion action does not change
\begin{align}
S[\psi,\overline{\psi},\mathscr{A}_{\mu}]=S[\psi',\overline{\psi}',\mathscr{A}_{\mu}],
\end{align}
reflecting the classical chiral invariance of the massless theory.

However, we have to be careful 
when implementing this change in the integral~\eqref{eq:fermion_part_funct}. 
The reason is that we have to properly transform the fermion integration measure, which
in principle might pick up a nontrivial Jacobian. Since the transformation is linear in the 
fermions, this Jacobian can only depend
on the external sources, as well as on the transformations parameter~$\alpha$
\begin{align}
\mathscr{D}\overline{\psi}\mathscr{D}\psi=J[\mathscr{A}_{\mu}]
\mathscr{D}\overline{\psi}'\mathscr{D}\psi'.
\end{align}
Taking this into account, we go back to~\eqref{eq:fermion_part_funct} that
now reads
\begin{align}
e^{i\Gamma[\mathscr{A}_{\mu}]}&=\int \mathscr{D}\overline{\psi}'\mathscr{D}\psi'
\,e^{iS[\psi',\overline{\psi}',\mathscr{A}_{\mu}]+\log J[\mathscr{A}_{\mu}]}
\equiv\int \mathscr{D}\overline{\psi}'\mathscr{D}\psi'
\,e^{iS'[\psi',\overline{\psi}',\mathscr{A}_{\mu}]}.
\end{align}
Thus, the effective action can be computed in the new variables provided
we use the new fermion action~$S'[\psi',\overline{\psi}',\mathscr{A}_{\mu}]$ including an additional term
\begin{align}
S'[\psi',\overline{\psi}',\mathscr{A}_{\mu}]=
\int d^{4}x\,\overline{\psi}'\gamma^{\mu}\big(i\partial_{\mu}+e\mathscr{A}_{\mu}\big)\psi'
-i\log J[\mathscr{A}_{\mu}],
\end{align}
that, coming from the functional measure, is obviously a pure quantum effect.
A convenient way to compute the Jacobian is by expanding the Dirac fermions 
in a basis of Dirac operator~${D\!\!\!\!/}\,(\mathscr{A})\equiv \gamma^{\mu}(\partial_{\mu}
-ie\mathscr{A}_{\mu})$ eigenstates.  
Using a regularization method preserving gauge invariance, a finite result is 
obtained~\cite{Fujikawa,Fujikawa2,Fujikawa_Suzuki}. 
\begin{align}
-i\log J[\mathscr{A}_{\mu}]={e^{2}\alpha\over 16\pi^{2}}\int d^{4}x\,\epsilon^{\mu\nu\alpha\beta}
\mathscr{F}_{\mu\nu}\mathscr{F}_{\alpha\beta}.
\end{align}
Notice that in the case of massive fermions 
the change~\eqref{eq:chov_chiral} also introduces, besides the quantum anomalous term, a complex phase
in the mass which has a classical origin
\begin{align}
S'[\psi',\overline{\psi}',\mathscr{A}_{\mu}]&=
\int d^{4}x\,\Big[\overline{\psi}_{R}'\gamma^{\mu}\big(i\partial_{\mu}+e\mathscr{A}_{\mu}\big)\psi_{R}'
+\overline{\psi}_{L}'\gamma^{\mu}\big(i\partial_{\mu}+e\mathscr{A}_{\mu}\big)\psi_{L}' \nonumber \\[0.2cm]
&+m e^{2i\alpha}\big(\overline{\psi}_{R}'\psi_{L}'
+\overline{\psi}_{L}'\psi_{R}'\big)\Big]
+{e^{2}\alpha\over 16\pi^{2}}\int d^{4}x\,\epsilon^{\mu\nu\alpha\beta}
\mathscr{F}_{\mu\nu}\mathscr{F}_{\alpha\beta}.
\end{align}
The last term associated to the nonzero Jacobian is just the integrated form of the
chiral anomaly found in~\eqref{eq:axial_anomaly}. The analysis just presented will be useful
in analyzing the strong CP problem in the next section.

\end{mdframed}

\section{The strong CP problem and axions}

When studying magnetic monopoles in Box~5 (see page~\pageref{pag:box_monopoles}),
we discussed the possibility of having nontrivial gauge field topologies.
In this section, we are going to look deeper into the role played by topology in non-Abelian gauge field 
theories and study how nonequivalent topological gauge field configurations define 
different vacua of the theory. 

\subsection{The (infinitely) many vacua of QCD}

To fix ideas, let us consider pure YM theory in the temporal
gauge~$A^{a}_{0}=0$, preserved by the set~$\mathscr{G}$ of
time-independent gauge transformations~$g(\mathbf{r})$. Adding
to the Euclidean space~$\mathbbm{R}^{3}$ the point at infinity, it gets 
compatified to a three-sphere,~$\mathbbm{R}^{3}\cup\{\infty\}\simeq S^{3}$.
Thus, the residual gauge transformations in~$\mathscr{G}$
define maps from~$S^{3}$ onto the gauge 
group\footnote{At a more physical level, the compactification of~$\mathbbm{R}^{3}$
to~$S^{3}$ amounts to requiring
that all fields, as well as gauge transformations, have well-defined limits 
as~$|\mathbf{r}|\rightarrow\infty$, independent of the direction along which the limit
is taken.}
\begin{align}
\mathscr{G}:S^{3}\longrightarrow G.
\end{align}
The space~$\mathscr{G}$ consists of infinitely topological nonequivalent 
sectors classified by the third-homotopy group~$\pi_{3}(G)$~\cite{Azcarraga_Izquierdo,Nakahara,Nash,Frankel}.
As an example, let us consider a gauge theory with group~$G=\mbox{SU(2)}$. 
This Lie group is topologically equivalent to 
a three-dimensional sphere~$S^{3}$, as can be seen by writing
\begin{align}
g=n^{0}\mathbbm{1}+i\mathbf{n}\cdot\boldsymbol{\sigma},
\end{align}
with~$n^{0}$ and~$\mathbf{n}=(n^{1},n^{2},n^{3})$ real. 
Both unitarity
\begin{align}
g^{\dagger}g=gg^{\dagger}=\big[(n^{0})+\mathbf{n}^{2}\big]\mathbbm{1}
=\mathbbm{1},
\end{align}
and the requirement of unit determinant
\begin{align}
\det{g}=(n^{0})^{2}+\mathbf{n}^{2}=1,
\end{align}
lead to the condition
\begin{align}
(n^{0})^{2}+\mathbf{n}^{2}=1,
\end{align}
so~$(n^{0},\mathbf{n})$ 
parametrizes the unit three-sphere~$S^{3}$. Since~$\pi_{3}(S^{3})=\mathbbm{Z}$, 
the set of time-independent SU(2)~gauge 
transformations decomposes into topological nonequivalent sectors
\begin{align}
\mathscr{G}=\bigcup_{n\in\mathbbm{Z}}\mathscr{G}_{n},
\end{align}
where~$n$ is
the winding number of the map~$S^{3}\rightarrow S^{3}$.
For a gauge transformation~$g(\mathbf{r})$, its winding number can be shown to be
\begin{align}
n&={1\over 24\pi^{2}}\int_{S^{3}}d^{3}r\,\epsilon_{ijk}{\rm tr\,}\Big[
(g^{-1}\partial_{i}g)(g^{-1}\partial_{j}g)(g^{-1}\partial_{k}g)\Big].
\label{eq:winding_number_gauge_conf}
\end{align}
Moreover, two gauge 
transformations can be continuously deformed into one another only when they share the same
winding number, with~$\mathscr{G}_{0}$ the identity's connected
component. Additivity is an important property of the winding number. Given~$g\in\mathscr{G}_{n}$
and~$g'\in\mathscr{G}_{n'}$, their product~$gg'$ has
winding number
\begin{align}
n_{gg'}=n_{g}+n_{g'},
\end{align}
and in particular~$n_{g^{-1}}=-n_{g}$. This, together with
the fact that~$\mathbbm{1}\in\mathscr{G}_{0}$, 
shows that~$\mathscr{G}_{0}$ 
is the only sector forming a subgroup.

From the discussion in section~\ref{sec:more_gauge_invariances},
we learn that physical states are preserved by ``small'' gauge transformations
in~$\mathscr{G}_{0}$ provided they satisfy the Gauss law~\eqref{eq:Gauss_law_NA_op}.
As for transformations in~$\mathscr{G}_{n}$ with~$n\neq 0$, keeping in mind that quantum 
states are rays in a Hilbert space defined up to a global complex phase
we conclude that physical invariance under a transformation~$g_{1}\in\mathscr{G}_{1}$ requires
\begin{align}
g_{1}|{\rm phys}\rangle=e^{i\theta}|{\rm phys}\rangle,
\end{align}
for some~$\theta\in\mathbbm{R}$. This number should be independent of the state, 
since otherwise gauge transformations
would give rise to observable interference. 
Another relevant fact to notice is that the value of~$\theta$ is also
independent of the transformation in~$\mathscr{G}_{1}$. To see this, let us 
consider~$g_{1},g_{1}'\in \mathscr{G}_{1}$ and assume that
\begin{align}
g_{1}|\mbox{phys}\rangle=e^{i\theta}|\mbox{phys}\rangle, 
\hspace*{1cm}
g_{1}'|\mbox{phys}\rangle=e^{i\theta'}|\mbox{phys}\rangle.
\end{align}
Since by additivity of the winding 
number~$g_{1}'g_{1}^{-1}\in\mathscr{G}_{0}$, and transformations in the connected
component of the identity leave the physical states invariant without any complex phase, 
we immediately conclude that~$\theta'=\theta$. Using a similar argument it is straightforward to show that 
for~$g_{n}\in\mathscr{G}_{n}$
\begin{align}
g_{n}|\mbox{phys}\rangle=e^{in\theta}|\mbox{phys}\rangle.
\label{eq:phase_theta_states}
\end{align}
The conclusion is that 
a single actual number~$\theta$ determines the action of all gauge transformations on physical 
states.

\begin{figure}[t]
\centerline{\includegraphics[scale=0.45]{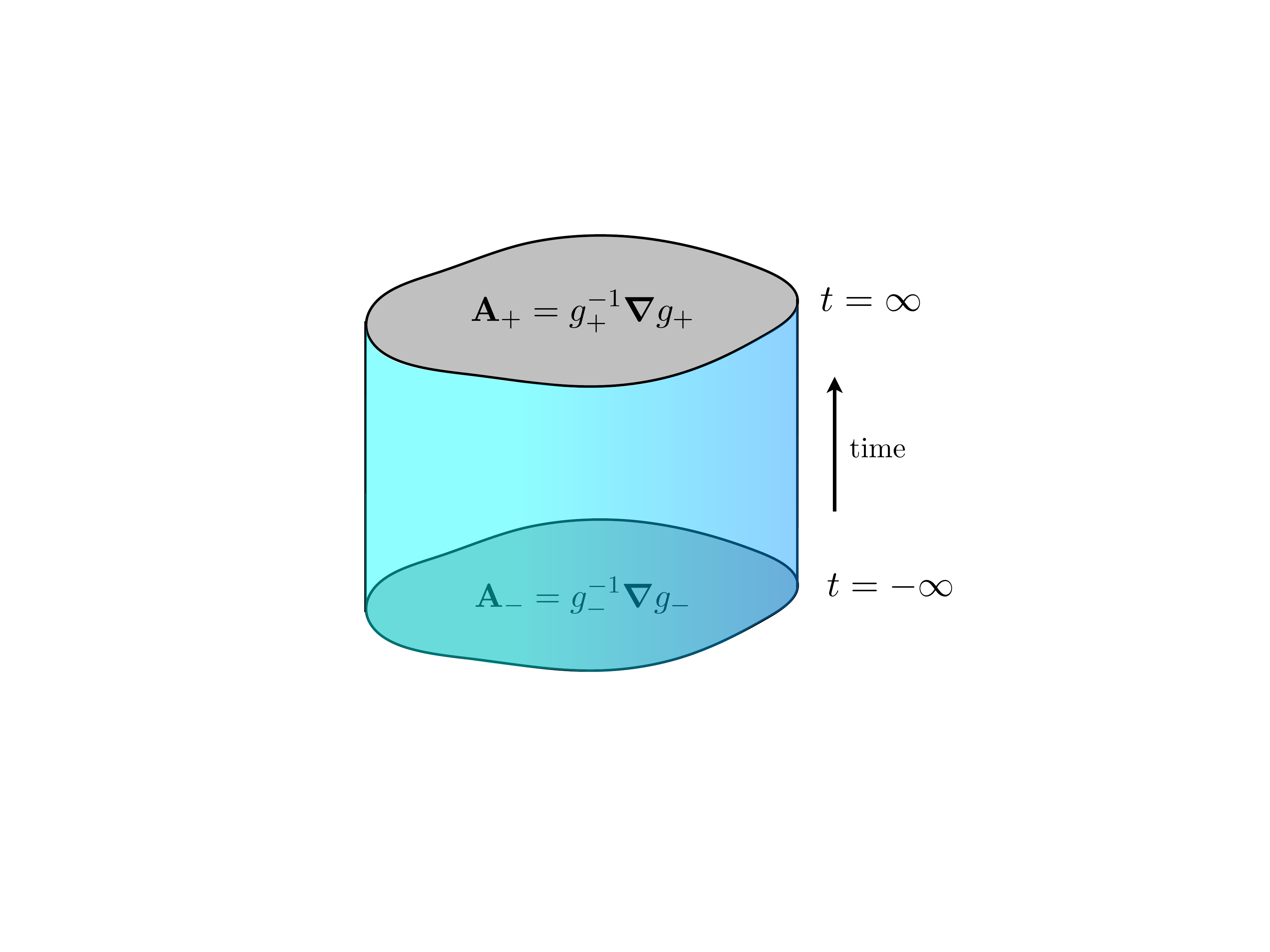}}
\caption[]{Representation of the spacetime interpolating between two pure gauge
configurations~$\mathbf{A}_{\pm}=g_{\pm}\boldsymbol{\nabla}g_{\pm}$ at~$t=\pm\infty$,
in the~$A_{0}=0$ gauge.}
\label{fig:theta_term_stdiag}
\end{figure}

We can reach the same conclusion about the vacuum structure of YM theories
in a different way. Besides the gauge kinetic term in 
the action~\eqref{eq:YM_action}, there is also a second admissible gauge invariant term
\begin{align}
S_{\theta}
&=-{\theta\over 32\pi^{2}}\int d^{4}x\,F^{a}_{\mu\nu}\widetilde{F}^{a\mu\nu} \nonumber \\[0.2cm]
&=-{\theta\over 8\pi^{2}}\int d^{4}x\,\mathbf{E}^{a}\cdot\mathbf{B}^{a},
\label{eq:theta_term_YM}
\end{align}
where~$\widetilde{F}^{a}_{\mu\nu}$ is the non-Abelian analog of the dual tensor field
introduced in eq.~\eqref{eq:dual_F_tensor_EM}, defined as
\begin{align}
\widetilde{F}^{a}_{\mu\nu}={1\over 2}\epsilon_{\mu\nu\alpha\beta}F^{a\alpha\beta}.
\end{align}
What makes the $\theta$-term~\eqref{eq:theta_term_YM} interesting is that it is the integral of
a total derivative 
\begin{align}
\epsilon^{\mu\nu\alpha\beta}F_{\mu\nu}^{a}F_{\alpha\beta}^{a}&=\partial_{\mu}\mathscr{J}^{\mu},
\end{align}
and therefore does not contribute to the field equations. The current on the right-hand side
of the previous equation takes the form (see Box~11 below for a rather simple 
derivation of this result)
\begin{align}
\mathscr{J}^{\mu}&=4\epsilon^{\mu\nu\alpha\beta}\left(A_{\nu}^{a}\partial_{\alpha}A_{\beta}^{a}
+{1\over 3}f^{abc}A_{\nu}^{a}A_{\alpha}^{b}A_{\beta}^{c}\right).
\end{align}
In the~$A_{0}^{a}=0$ gauge, we have
\begin{align}
\epsilon^{\mu\nu\alpha\beta}F_{\mu\nu}^{a}F_{\alpha\beta}^{a}
&=4{\partial\over\partial t}\left[\mathbf{A}^{a}\cdot\big(\boldsymbol{\nabla}\times\mathbf{A}^{a}\big)
+{1\over 3}f^{abc}\mathbf{A}^{a}\cdot\big(\mathbf{A}^{b}\times\mathbf{A}^{c}\big)\right],
\label{eq:theta_term_total_der_comps}
\end{align}
which once integrated and with the proper normalization gives the following
expression of the $\theta$-term
\begin{align}
S_{\theta}
&=-{\theta\over 8\pi^{2}}\left\{\left.\int d^{3}r\,\left[
\mathbf{A}^{a}\cdot\big(\boldsymbol{\nabla}\times\mathbf{A}^{a}\big)
+{1\over 3}f^{abc}\mathbf{A}^{a}\cdot\big(\mathbf{A}^{b}\times\mathbf{A}^{c}\big)
\right]\right|_{t=\infty}\right.
\nonumber \\[0.2cm]
&\left.\left.-\int d^{3}r\,
\left[\mathbf{A}^{a}\cdot\big(\boldsymbol{\nabla}\times\mathbf{A}^{a}\big)
+{1\over 3}f^{abc}\mathbf{A}^{a}\cdot\big(\mathbf{A}^{b}\times\mathbf{A}^{c}\big)\right]
\right|_{t=-\infty}\right\}.
\label{eq:Stheta_tempgauge1}
\end{align}
To ensure finiteness, we take the gauge field~$\mathbf{A}=\mathbf{A}^{a}T^{a}_{\mathbf{R}}$ 
to approach 
pure-gauge configurations~$\mathbf{A}_{\pm}=g_{\pm}^{-1}\boldsymbol{\nabla}g_{\pm}$ at~$t=\pm\infty$
(see fig.~\ref{fig:theta_term_stdiag}). 
It is easy to see that the integrands in eq.~\eqref{eq:Stheta_tempgauge1}
are not gauge invariant and therefore the $\theta$-term is nonzero (again, a derivation is outlined in Box~11)
\begin{align}
S_{\theta}
&={\theta\over 24\pi^{2}}\int d^{3}r\,{\rm tr\,}\Big\{(g_{+}^{-1}\boldsymbol{\nabla}g_{+})\cdot\Big[
(g^{-1}_{+}\boldsymbol{\nabla}g_{+})\times(g_{+}^{-1}\boldsymbol{\nabla}g_{+})\Big]\Big\} \nonumber \\[0.2cm]
&-{\theta\over 24\pi^{2}}\int d^{3}r\,{\rm tr\,}\Big\{(g_{-}^{-1}\boldsymbol{\nabla}g_{-})\cdot\Big[
(g_{-}^{-1}\boldsymbol{\nabla}g_{-})\times(g_{-}^{-1}\boldsymbol{\nabla}g_{-})\Big]\Big\}.
\end{align}
Comparing with eq.~\eqref{eq:winding_number_gauge_conf}, we identify the winding numbers~$n_{\pm}$
of the asymptotic gauge transformations~$g_{\pm}$, to write
\begin{align}
S_{\theta}
&=(n_{+}-n_{-})\theta.
\label{eq:Stheta_integer}
\end{align}
Thus, non-Abelian gauge field configurations are classified into topological sectors
interpolating between early and late time configurations of definite winding number~$n_{\pm}$.
These sectors are labelled by the integer~$n=n_{+}-n_{-}$, and when summing
in the Feynman path integral over all gauge configurations we also have to include all possible
sectors. Each one is weighted by the same phase
\begin{align}
e^{iS_{\theta}}=e^{in\theta},
\end{align}
that we encountered in eq.~\eqref{eq:phase_theta_states}.

\begin{mdframed}[backgroundcolor=lightgray,hidealllines=true]
\vspace*{0.2cm}
\centerline{\greybox{\bf Gauge fields and differential forms}}
\vspace*{0.2cm}

The analysis of YM theories gets very much simplified in the language of differential
forms~\cite{Azcarraga_Izquierdo,Nakahara,Nash,Frankel}. 
The gauge field~$A_{\mu}=A_{\mu}^{a}T_{\mathbf{R}}^{a}$ can
be recast
as the Lie algebra valued one-form
\begin{align}
A=-iA_{\mu}dx^{\mu},
\label{eq:A-one-form}
\end{align} 
while the two-form field strength is given by
\begin{align}
F\equiv -{i\over 2}F_{\mu\nu}dx^{\mu}\wedge dx^{\nu}=dA+A\wedge A,
\label{eq:F-two-form}
\end{align}
where in the second term on the right-hand side a matrix multiplication of the one-forms is also 
understood (in the Abelian case the matrices commute and the term vanishes due to the anticommutativity
of the wedge product).
The factor of~$-i$ in both eqs.~\eqref{eq:A-one-form} and~\eqref{eq:F-two-form} is introduced
to avoid cluttering expressions with powers of~$i$.

Gauge transformations are determined 
by a zero-form~$g\in\mathscr{G}$ acting on the gauge field one-form as
[cf.~\eqref{eq:trans_ANA_noindices}]
\begin{align}
A\longrightarrow A'=g^{-1}dg+g^{-1}Ag.
\label{eq:A_form_gauge_trans}
\end{align}
This leads to the corresponding transformation of the field strength
\begin{align}
F\longrightarrow F'&=dA'+A'\wedge A' \nonumber \\[0.2cm]
&=g^{-1}Fg,
\end{align}
that once written in components agrees with the one given in eq.~\eqref{eq:FNA_gauge_trans}.
In fact, given an adjoint~$p$-form field
\begin{align}
\Phi_{p}=-{i\over p!}\Phi_{\mu_{1}\ldots\mu_{p}}
dx^{\mu_{1}}\wedge\ldots\wedge dx^{\mu_{p}} \hspace*{0.5cm} \Longrightarrow \hspace*{0.5cm}
\Phi_{p}\rightarrow \Phi_{p}'=g^{-1}\Phi_{p}g,
\end{align} 
a covariant exterior derivative is defined acting as
\begin{align}
D\Phi_{p}\equiv d\Phi_{p}+A\wedge\Phi_{p}-(-1)^{p}\Phi_{p}\wedge A
\hspace*{0.5cm} \Longrightarrow \hspace*{0.5cm}
(D\Phi_{p})'=g^{-1}(D\Phi_{p})g,
\end{align}
satisfying the Leibniz rule
\begin{align}
D(\Phi_{p}\wedge \Psi_{q})=(D\Phi_{p})\wedge \Psi_{q}+(-1)^{p}\Phi_{p}\wedge(D\Psi_{q}).
\end{align}
Using these definitions and properties, 
it is easy to check that the field strength two-form~\eqref{eq:F-two-form} verifies
the Bianchi identity~$DF=0$.

In four dimensions there are two gauge invariant four-forms that can be constructed from the
field-strength two-form. The first one is
\begin{align}
{\rm tr\,}(F\wedge \star F),
\label{eq:trFstarF}
\end{align} 
where~$\star$ denotes the 
Hodge dual, acting on a~$p$-form field as~\cite{Nakahara}
\begin{align}
\star \Phi_{p}=-{i\over p!(4-p)!}\epsilon^{\mu_{1}\ldots\mu_{p}}_{\hspace*{0.9cm}\nu_{1}\ldots\nu_{4-p}}
\Phi_{\mu_{1}\ldots\mu_{p}}dx^{\nu_{1}}\wedge\ldots\wedge dx^{\nu_{4-p}}.
\end{align}
Since this operation commutes with the multiplication by a zero-form, the gauge invariance 
of~\eqref{eq:trFstarF} follows directly from applying the cyclic property of the trace. 
In addition, we can also construct a second gauge invariant four-form 
\begin{align}
{\rm tr\,}(F\wedge F),
\label{eq:FwedgeF_term}
\end{align}
so the action of pure YM theory without matter couplings can be written
as
\begin{align}
S_{\mbox{\tiny YM}}={1\over 2g_{\mbox{\tiny YM}}^{2}}\int_{\mathcal{M}_{4}}{\rm tr\,}(F\wedge\star F)
+{\theta\over 8\pi^{2}}\int_{\mathcal{M}_{4}}{\rm tr\,}(F\wedge F),
\label{eq:form_versionSY_theta}
\end{align}
where~$\mathcal{M}_{4}$ represents the four-dimensional spacetime. The two terms 
correspond respectively to the
kinetic and~$\theta$ terms given in components in
eqs.~\eqref{eq:YM_action} and~\eqref{eq:theta_term_YM}.
Incidentally, 
notice that while the term inside the first integral 
is always a maximal form in any dimension, the one in the second 
term is only maximal in~$D=4$. 
In fact, no analog of the~$\theta$-term exits in odd-dimensional spacetimes.

Although in these lectures we are restricting our attention 
to (flat) Minkowski spacetime, QFTs can also be defined in curved spacetimes. In this respect, 
the action~\eqref{eq:form_versionSY_theta} written in terms of differential forms is
also valid for non-flat metrics. An interesting difference between the two terms is that, while
the first one depends on the spacetime metric the $\theta$-term does not and is therefore topological. 
Metric dependence 
is actually signaled by the presence of the Hodge dual in the action. 

Another relevant fact that can be easily shown using differential forms is that the
$\theta$-term is a total derivative, as we saw in
eq.~\eqref{eq:theta_term_total_der_comps}.
Indeed, eq.~\eqref{eq:FwedgeF_term} can be explicitly written in terms of the gauge field one-form as
\begin{align}
{\rm tr\,}(F\wedge F)&={\rm tr\,}(dA\wedge dA+2dA\wedge A\wedge A
+A\wedge A\wedge A \wedge A) \nonumber \\[0.2cm]
&=d\,{\rm tr\,}\left(A\wedge dA+{2\over 3}A\wedge A\wedge A\right),
\end{align}
where we have used that~${\rm tr\,}(A\wedge A\wedge A \wedge A)=0$, as a result of the 
anticommutativity of one-forms and the trace's cyclic property.
Using the properties of the Hodge dual operator, we finally write
\begin{align}
\star{\rm tr\,}(F\wedge F)=d^{\dagger}J,
\end{align}
where~$d^{\dagger}\equiv \star d\star$ is the adjoint exterior derivative~\cite{Nakahara} 
and~$J$ is the current one form
\begin{align}
J=\star {\rm tr\,}\left(A\wedge dA+{2\over 3}A\wedge A\wedge A\right).
\label{eq:current_CS}
\end{align}
Once expressed in components we retrieve eq.~\eqref{eq:theta_term_total_der_comps}.

The trace on the right-hand side of~\eqref{eq:current_CS} defines the {\em Chern-Simons form}.
Applying~\eqref{eq:A_form_gauge_trans} and after some algebra we obtain its gauge transformation
\begin{align}
\omega_{3}(A)\equiv {\rm tr\,}&\left(A\wedge dA+{2\over 3}A\wedge A\wedge A\right) \nonumber \\[0.2cm]
&\longrightarrow \hspace*{0.5cm}\omega_{3}(A)-{1\over 3}{\rm tr\,}\Big[(g^{-1}dg)\wedge(g^{-1}dg)\wedge(g^{-1}dg)\Big].
\label{eq:CS_form_gauge_trans}
\end{align}
The Chern-Simons form is a very interesting object for many reasons. One is that
with it gives rise to the action
\begin{align}
S_{\rm CS}=-{k\over 4\pi}\int_{\mathcal{M}_{3}}
{\rm tr\,}\left(A\wedge dA+{2\over 3}A\wedge A\wedge A\right),
\label{eq:CS_action}
\end{align} 
where~$\mathcal{M}_{3}$ is a three-dimensional spacetime and 
$k$ is a constant known as the Chern-Simons level. Although~\eqref{eq:CS_form_gauge_trans} implies that
the action is not gauge invariant
\begin{align}
S_{\rm CS}\longrightarrow S_{\rm CS}+{k\over 12\pi}\int_{\mathcal{M}_{3}}
{\rm tr\,}\Big[(g^{-1}dg)\wedge (g^{-1}dg)\wedge (g^{-1}dg)\Big].
\label{eq:CS_action_transformation}
\end{align} 
the extra term equals~$2\pi nk$, with~$n$
the winding number of the gauge transformation
defined in eq.~\eqref{eq:winding_number_gauge_conf}. Since the quantum theory can be formulated using
functional integrals involving~$\exp(iS_{\rm CS})$, 
this gauge variance is not a problem 
provided the Chern-Simons level~$k$ is an integer. The action~\eqref{eq:CS_action} 
defines a topological field theory 
appearing in many contexts in physics,
ranging from quantum gravity~\cite{Achucarro_Townsend,Witten_CS} 
to condensed matter, where it has found important aplications in the theory of the quantum Hall 
effect~\cite{Schakel,Ezawa}.

To conclude this discussion, 
let us also mention that the four-form~\eqref{eq:FwedgeF_term} is also related to the axial 
anomaly studied in section~\ref{sec:anomalies}. 
Defined on a Euclidean spacetime, the integrated anomaly of the
axial-vector current can be shown to be~\cite{Fujikawa,Fujikawa2,Fujikawa_Suzuki}
\begin{align}
\int_{\mathcal{M}_{4}} d^{4}x\,\partial_{\mu}\langle J_{A}^{\mu}(x)\rangle=
-2i\big(N_{+}-N_{-}),
\label{eq:Fujikawa_id}
\end{align} 
and~$N_{\pm}$ are the number of positive/negative chirality solutions to the 
equation~${D\!\!\!\!/}\,(A)\psi=0$, 
with~${D\!\!\!\!/}\,(A)\equiv\gamma^{\mu}(\partial_{\mu}-iA_{\mu})$ the Dirac operator on the 
Euclidean manifold~$\mathcal{M}_{4}$. The difference~$N_{+}-N_{-}$ appearing on the right-hand side 
of eq.~\eqref{eq:Fujikawa_id} is in fact a topological invariant called the {\em index} of the Dirac operator.
This quantity can be computed using the Atiyah-Singer index 
theorem~\cite{Azcarraga_Izquierdo,Nakahara,Nash,Frankel} 
and in four dimensions it is given by
the integral of the four-form~\eqref{eq:FwedgeF_term}
\begin{align}
\mbox{ind\,}{D\!\!\!\!/}=-{1\over 8\pi^{2}}\int_{\mathcal{M}_{4}}F\wedge F.
\end{align}
which, as explained above,  
is itself a topological quantity. By substituting this result into~\eqref{eq:Fujikawa_id},
we retrieve the known form of the anomaly, apart from a global factor 
of~$i$ that is the consequence of working in Euclidean signature.

\end{mdframed}

\subsection{Breaking CP strongly}
\label{sec:CP_violation}

An significant feature of the $\theta$-term~\eqref{eq:theta_term_YM} is that it violates both
parity and CP, the
combination of parity and charge conjugation,
\begin{align}
\mbox{CP}:\left\{
\begin{array}{rll}
\mathbf{E}^{a}(t,\mathbf{r})&\longrightarrow &\mathbf{E}^{a}(t,-\mathbf{r})  \\[0.2cm]
\mathbf{B}^{a}(t,\mathbf{r})&\longrightarrow &-\mathbf{B}^{a}(t,-\mathbf{r})
\end{array}
\right. \hspace*{1cm} \Longrightarrow \hspace*{1cm} 
\mbox{CP}: S_{\theta}\longrightarrow 
-S_{\theta}.
\end{align}
To understand these transformations heuristically, we can use the analogy with Maxwell's electric and
magnetic fields to conclude that~$\mathbf{E}^{a}$ 
is reversed by both parity and charge conjugation, whereas the 
pseudovector~$\mathbf{B}^{a}$ is preserved by the former and reversed by the latter.
Notice that since CPT is a symmetry of QFT, a breaking of CP is equivalent to a violation
of time reversal T.

\begin{figure}[t]
\centerline{\includegraphics[scale=0.35]{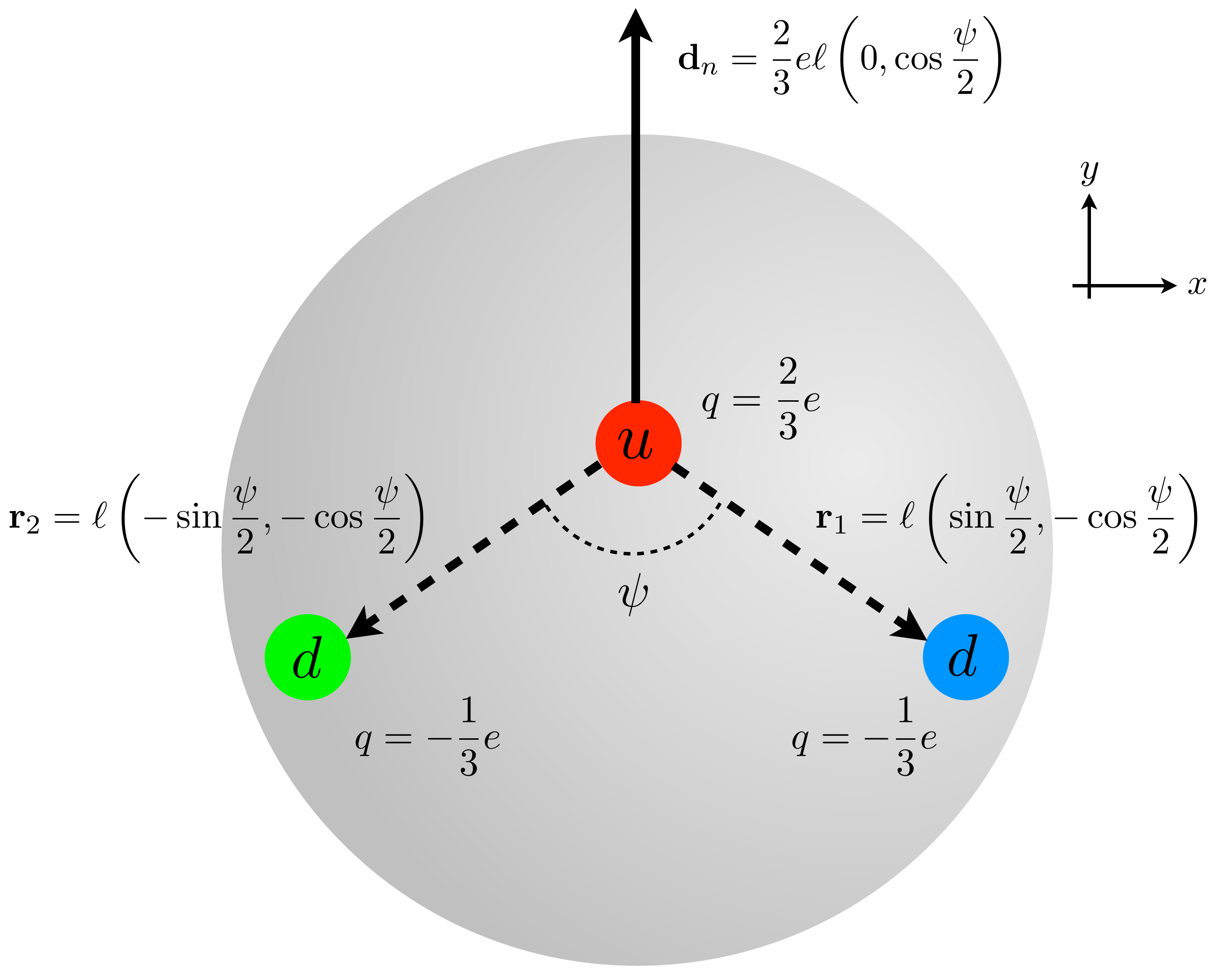}}
\caption[]{Classical depiction of the neutron and its
electric dipolar moment~$\mathbf{d}_{n}$. The components of the $d$~quarks position
vectors~$\mathbf{r}_{1}$ and~$\mathbf{r}_{2}$ are
written using the coordinate axes shown in the picture, with origin on the position of the $u$~quark.}
\label{fig:neutron}
\end{figure}

Among the phenomena where CP (or T) violation can manifest in QCD
is the existence of a nonvanishing
electric dipole moment of the neutron (see, for example,~\cite{Villadoro,Hook} for reviews). 
To be clear, 
were neutrons elementary, we would not expect them to have an 
electric dipolar moment. But being composed of three valence quarks with different charges,
a nonvanishing value may appear depending on the quark distribution. To estimate its size, 
let us consider a classical picture of the neutron assuming
a structure similar to the water molecule (see fig.~\ref{fig:neutron}): 
the two $d$~quarks are located at a distance~$\ell$ of the $u$~quark and their position 
vectors~$\mathbf{r}_{1}$ and~$\mathbf{r}_{2}$ span 
an angle~$\psi$ with each other. Taking coordinates on the plane
defined by the three quarks, 
the modulus of the electric dipole moment~$\mathbf{d}_{n}$ is readily computed to be
\begin{align}
|\mathbf{d}_{n}|={2\over 3}e\ell\cos{\psi\over 2}\equiv {2\over 3}e\ell \sin{\theta\over 2},
\label{eq:d_n_general}
\end{align}
where we have introduced the 
angle~$\theta\equiv \pi-\psi$, controlling the amount of CP violation.
To estimate the prefactor in eq.~\eqref{eq:d_n_general}, 
we recall that the distance~$\ell$ between the quarks is of the order of the pion's Compton wavelength
\begin{align}
\ell\simeq{\hbar\over m_{\pi}c},
\end{align}
where for computational purposes
we have restored powers of~$\hbar$ and~$c$. Noticing 
that~$\hbar c\simeq 200\,\mbox{MeV}\cdot\mbox{fm}$ and~$m_{\pi}c^{2}\simeq 135\,\mbox{MeV}$, 
we find
\begin{align}
|\mathbf{d}_{n}|\simeq 10^{-13}\sin{\theta\over 2}\,e\cdot\mbox{cm}.
\end{align}
A comparison with experimental measurements of the neutron electric 
dipole~\cite{PDG,Snowmass_EDM}
\begin{align}
|\mathbf{d}_{n}|_{\rm exp}\lesssim 10^{-26}\,e\cdot\mbox{cm},
\label{eq:EDM_exp_bound}
\end{align}
leads then to the bound
\begin{align}
\theta\lesssim 10^{-13}.
\label{eq:theta_angle_molecule}
\end{align}
This means that the angle~$\psi=\pi-\theta$ in fig.~\ref{fig:neutron} is extremely close
to~$\pi$, making the quark configuration inside the neutron 
look like a CO$_{2}$ rather than a water molecule.

This cartoon calculation exhibits the basic feature of the so-called {\em strong CP problem}: 
the stringent
experimental bound for the neutron electric dipole moment implies the existence of a dimensionless parameter
that is extremely small without any dynamical reason. Once we rephrase the problem in the 
correct language of QCD, we will see that this parameter is precisely the $\theta$~coupling 
introduced in eq.~\eqref{eq:theta_term_YM}.

From a QFT point of view the neutron electric dipole emerges from the dimension-five 
nonminimal coupling of the neutron to the electromagnetic field
\begin{align}
S\supset -{i\over 2}|\mathbf{d}_{n}|\int d^{4}x\,\overline{n}\sigma^{\mu\nu}\gamma_{5}n F_{\mu\nu},
\label{eq:DEM_neutron_op}
\end{align}
where~$n$ is the neutron field
and~$\sigma^{\mu\nu}$ has been defined in eq.~\eqref{eq:sigmamunu_def}. This term is explicitly
gauge invariant but breaks parity, as follows from the presence of~$\gamma_{5}$. It is, however, invariant
under charge conjugation, which preserves the neutron and gauge fields, and therefore
it breaks~CP. The operator~\eqref{eq:DEM_neutron_op} is in fact an effective interaction emerging from 
loops diagrams in the EFT of pions and nucleons described by an extension of the 
action~\eqref{eq:pion_action}.
To construct this theory, 
let us consider QCD with the two light flavors~$u$ and~$d$. Written in terms of the chiral isospin doublets
\begin{align}
\boldsymbol{q}_{R,L}=\left(
\begin{array}{c}
u_{R,L} \\
d_{R,L}
\end{array}
\right),
\end{align}
the microscopic action takes the form
\begin{align}
S&=\int d^{4}x\left(i\overline{\boldsymbol{q}}_{R}{D\!\!\!\!/\,}\boldsymbol{q}_{R}
+i\overline{\boldsymbol{q}}_{L}{D\!\!\!\!/\,}\boldsymbol{q}_{L}
+\overline{\boldsymbol{q}}_{L}M\boldsymbol{q}_{R}
+\overline{\boldsymbol{q}}_{R}M^{T}\boldsymbol{q}_{L}
-{\theta\over 32\pi^{2}}\epsilon^{\mu\nu\alpha\beta}F^{a}_{\mu\nu}
F^{a}_{\alpha\beta}
+\ldots\right),
\label{eq:S_ud_quarks}
\end{align}
where~$D_{\mu}=\partial_{\mu}-iA_{\mu}^{a}T^{a}$ denotes the gauge covariant derivative and the mass matrix is given by
\begin{align}
M=\left(
\begin{array}{cc}
m_{u} & 0 \\
0 & m_{d}
\end{array}
\right).
\label{eq:quark_mass_matrix}
\end{align}
We have included the $\theta$-term, while the ellipsis indicates other terms not important for
the argument. In writing the action~\eqref{eq:pion_action} we assumed that 
quarks are massless, and also the NG bosons associated with chiral SSB, but we now relax this condition. 
Although the chiral~$\mbox{SU(2)}_{R}\times\mbox{SU(2)}_{L}$ transformations
\begin{align} 
\boldsymbol{q}_{R,L}\longrightarrow U_{R,L}\boldsymbol{q}_{R,L},
\end{align}
do not leave the quark action~\eqref{eq:S_ud_quarks} invariant, we can restore the symmetry
promoting the mass matrix~$M$ to a spurion field transforming as
\begin{align}
M\longrightarrow U_{L}MU_{R}^{\dagger}.
\end{align}
Thus, the original action can be seen as one where chiral symmetry is spontaneously broken 
by~$M$ taking the value in 
eq.~\eqref{eq:quark_mass_matrix}.
The transformation of~$M$, 
together with eq.~\eqref{eq:NG_matrix_trans}, provides the basic clue to incorporate masses into the 
NG action~\eqref{eq:pion_action}. An invariant 
mass term can be built by taking the trace of the product of
the mass and the NG boson matrices
\begin{align}
S_{\rm NG}&=\int d^{4}x\,\left[{f_{\pi}^{2}\over 4}
{\rm tr\,}\big(D_{\mu}\boldsymbol{\Sigma}^{\dagger}D^{\mu}\boldsymbol{\Sigma}\big)
+f_{\pi}^{3}B_{0}{\rm tr\,}\big(M^{\dagger}\boldsymbol{\Sigma}
+\boldsymbol{\Sigma}^{\dagger}M\big)\right].
\label{eq:pion_eff_massterm}
\end{align}
Here~$D_{\mu}\boldsymbol{\Sigma}=\partial_{\mu}\boldsymbol{\Sigma}-iA_{\mu}[Q,\boldsymbol{\Sigma}]$,
with~$Q=e\sigma^{3}$ the pion charge matrix, is
the electromagnetic covariant derivative 
and~$B_{0}$ is a numerical constant that cannot be determined within the EFT 
framework\footnote{The pion effective action~$S_{\rm NG}$ also contains terms induced by 
the anomalous global symmetries of QCD, which are fully determined by the mathematical structure
of the anomaly~(see, for example, \cite{AG_anomalies}). An example is the term 
proportional 
to~$\big({\rm tr\,}\log\boldsymbol{\Sigma}-{\rm tr\,}\log\boldsymbol{\Sigma}^{\dagger}\big)
F_{\mu\nu}\widetilde{F}^{\mu\nu}$, accounting for the 
electromagnetic decay of the neutral pion discussed in page~\pageref{page:pion_decay}.
\label{page:anomaly_pion_footnote}
}. 
Substituting the explicit expressions of~$M$ and~$\boldsymbol{\Sigma}$ and expanding in powers of the
pion fields, we find the mass term
\begin{align}
\Delta S_{\rm NG}=-f_{\pi}B_{0}(m_{u}+m_{d})\int d^{4}x\Big[(\pi^{0})^{2}+2\pi^{+}\pi^{-}\Big],
\end{align}
from where we read the pion mass
\begin{align}
m_{\pi}^{2}=2f_{\pi}B_{0}(m_{u}+m_{d}) \hspace*{1cm} \Longrightarrow \hspace*{1cm}
B_{0}={m_{\pi}^{2}\over 2f_{\pi}(m_{u}+m_{d})}.
\end{align}
Within this approximation, neutral and charged pions have the same mass. 

Nucleons can also be added to the chiral Lagrangian (see~\cite{BKM,Georgi_WI} for reviews). 
They are introduced through the isospin doublet
\begin{align}
N=\left(
\begin{array}{c}
p \\
n
\end{array}
\right),
\end{align}
transforming under~$\mbox{SU(2)}_{R}\times\mbox{SU(2)}_{L}$ as~\cite{CWZ,CCWZ,GSS}
\begin{align}
N \longrightarrow K(U_{R},U_{L},\boldsymbol{\Sigma})N.
\label{eq:nucleon_chiral_trans}
\end{align}
The so-called compensating field~$K(U_{R},U_{L},\boldsymbol{\Sigma})$ is a SU(2)-valued matrix
depending on the 
NG boson matrix~$\boldsymbol{\Sigma}(x)$, and through it on the spacetime point. 
It is defined
by~$K(U_{R},U_{L},\boldsymbol{\Sigma})=\boldsymbol{u}'(x)^{-1}U_{R}\boldsymbol{u}(x)$, 
where~$\boldsymbol{u}(x)^{2}\equiv \boldsymbol{\Sigma}(x)$ 
and~$\boldsymbol{u}'(x)^{2}\equiv \boldsymbol{\Sigma}'(x)=
U_{R}\boldsymbol{\Sigma}(x)U_{L}^{\dagger}$, thus providing 
a nonlinear realization of the~$\mbox{SU(2)}_{R}\times\mbox{SU(2)}_{L}$ global chiral symmetry
acting on the nucleon isospin doublet.

Having established the transformation of nucleons, we add to the effective action the term
\begin{align}
\Delta S_{\pi N}=\int d^{4}x\,\overline{N}\Big[i{D\!\!\!\!/\,}
-f(\boldsymbol{\Sigma})\Big]N.
\label{eq:piNaction}
\end{align}
with~$f(\boldsymbol{\Sigma})$ a matrix-valued function depending on the NG boson matrix and
such that~${\mathcal{D}\!\!\!\!/\,}\equiv {D\!\!\!\!/\,}
+if(\boldsymbol{\Sigma})$ defines a covariant derivative with respect to the local
transformation~\eqref{eq:nucleon_chiral_trans}, ${\mathcal{D}\!\!\!\!/\,}\rightarrow 
K{\mathcal{D}\!\!\!\!/\,}K^{\dagger}$.
At linear order in the pion fields, it includes the pion-nucleon vertices
\begin{align}
f(\boldsymbol{\Sigma})&=m_{N}\mathbbm{1}
+{g_{A}\over 2f_{\pi}}\gamma^{\mu}\gamma_{5}\partial_{\mu}\boldsymbol{\pi}
+\mathcal{O}(\boldsymbol{\pi}^{2}) \nonumber \\[0.2cm]
&=m_{N}\mathbbm{1}
+{g_{A}\over 2\sqrt{2}f_{\pi}}\big(\overline{n}\gamma^{\mu}\gamma_{5}n-\overline{p}\gamma^{\mu}
\gamma_{5}p\big)\partial_{\mu}\pi^{0}+{g_{A}\over 2f_{\pi}}\big(\overline{n}\gamma^{\mu}\gamma_{5}p\,
\partial_{\mu}\pi^{-}+\overline{p}\gamma^{\mu}\gamma_{5}n\,\partial_{\mu}\pi^{+}\big)
,
\label{eq:f(sigma)notheta}
\end{align}
where~$m_{N}$ is the nucleon mass. Incidentally, substituting 
this expression of~$f(\boldsymbol{\Sigma})$ into the 
action~\eqref{eq:piNaction} we can integrate by parts and move the derivative 
from~$\boldsymbol{\pi}$ to~$N$ and~$\overline{N}$. For scattering processes with
on-shell nucleons the Dirac equation~${i\partial\!\!\!/}N=m_{N}N$ can be implemented to write 
the nucleon-pion interaction term as~$ig_{\pi\! N\!N}\overline{N}t_{\mathbf{f}}^{I}N \pi^{I}$, 
with~$t_{\mathbf{f}}^{I}$ the generators in the fundamental representation of~$\mbox{SU(2)}$. 
Furthermore, the coupling constant~$g_{\pi\! N\!N}$ satisfies by
the Goldberger-Treiman relation~\cite{Goldberger_Treiman}
\begin{align}
f_{\pi} g_{\pi\!N\!N}=g_{A}m_{N}.
\end{align} 
Notice that since~$g_{A}$ is real, the couplings in eq.~\eqref{eq:f(sigma)notheta} preserve CP.

We would like to study the effects in the chiral Lagrangian of adding the $\theta$-term 
to the quark action.
At this point we should invoke the analysis presented in Box~10 
(see page~\pageref{box:anomalies}) where we saw how, due to the chiral anomaly, 
implementing a chiral rotation of the fermions 
induces a~$\theta$-term in the action. More precisely, performing a chiral rotation of
the $u$-quark
\begin{align}
u_{R,L}\longrightarrow e^{\pm i\alpha}u_{R,L},
\end{align}
results in shifting the value of the theta angle 
\begin{align}
S&=\int d^{4}x\left(i\overline{\boldsymbol{q}}_{R}{D\!\!\!\!/\,}\boldsymbol{q}_{R}
+i\overline{\boldsymbol{q}}_{L}{D\!\!\!\!/\,}\boldsymbol{q}_{L}
+\overline{\boldsymbol{q}}_{L}M\boldsymbol{q}_{R}
+\overline{\boldsymbol{q}}_{R}M^{\dagger}\boldsymbol{q}_{L}
-{\theta-2\alpha\over 32\pi^{2}}\epsilon^{\mu\nu\alpha\beta}F^{a}_{\mu\nu}
F^{a}_{\alpha\beta}
+\ldots\right),
\end{align}
and a complex mass matrix
\begin{align}
M=
\left(
\begin{array}{cc}
e^{2i\alpha}m_{u} & 0 \\
0 & m_{d}
\end{array}
\right).
\end{align}
In particular, 
setting~$\alpha={1\over 2}\theta$ the $\theta$-term cancels and all dependence 
on~$\theta$ is shifted to a phase in the mass matrix~$M$.
In more physical terms, we have transferred the source of CP violation in the quark action from 
the $\theta$-term to a complex coupling\footnote{In fact, 
it is easy to prove that the quantity~$\overline{\theta}\equiv
\theta+\arg\det{M}$ remains invariant under chiral transformations of the quarks.}.

It might seem that, at 
the level of the chiral effective field theory, the phase in the mass matrix~$M=\mbox{diag}(e^{i\theta}
m_{u},m_{d})$ could be removed by an appropriate chiral transformation of the NG 
field~$\boldsymbol{\Sigma}(x)$. In doing so, however, we introduce a $\theta$-dependence 
in~$f(\boldsymbol{\Sigma},\theta)$ defined in~\eqref{eq:piNaction}, inducing additional 
nucleon-pion couplings. In particular, besides the neutron-proton-pion
vertex in eq.~\eqref{eq:f(sigma)notheta}, there is a new CP violating vertex contributing
to the dimension-five non-minimal electromagnetic coupling in eq.~\eqref{eq:DEM_neutron_op}
\begin{align}
\nonumber\\[-0.7cm]
\parbox{35mm}{
\begin{fmfgraph*}(90,90)
\fmfleft{i1,i2}
\fmfright{o1}
\fmf{heavy,tension=1}{i1,v1,i2}
\fmfv{decoration.shape=square,decoration.size=.15w,decoration.filled=full}{v1}
\fmf{photon,tension=1.2}{v1,o1}
\fmflabel{$\gamma$}{o1}
\fmflabel{$n$}{i1}
\fmflabel{$n$}{i2}
\end{fmfgraph*}
}\hspace*{0.5cm}=\hspace*{0.3cm}
\parbox{35mm}{
\begin{fmfgraph*}(95,90)
\fmfleft{i1,i2}
\fmfright{o1,o2,o3}
\fmf{heavy,tension=1}{i1,v1}
\fmf{phantom,tension=1}{v2,v4,o3}
\fmf{phantom,tension=1}{v1,v3,o1}
\fmf{phantom,tension=1}{v3,v5,v4}
\fmf{dashes,right=0.4,label=$\pi^{-}$,tension=0.1}{v1,v5,v2}
\fmf{heavy,tension=1}{v2,i2}
\fmf{heavy,label=$p$,tension=0}{v1,v2}
\fmfblob{.15w}{v1}
\fmfv{decoration.shape=circle,decoration.size=.08w,decoration.filled=full}{v2}
\fmfv{decoration.shape=circle,decoration.size=.10w,decoration.filled=30}{v5}
\fmf{photon,tension=1.2}{v5,o2}
\fmflabel{$\gamma$}{o2}
\fmflabel{$n$}{i1}
\fmflabel{$n$}{i2}
\end{fmfgraph*}
}
\hspace*{0.5cm}+\hspace*{0.2cm}
\parbox{35mm}{
\begin{fmfgraph*}(95,90)
\fmfleft{i1,i2}
\fmfright{o1,o2,o3}
\fmf{heavy,tension=1}{i1,v1}
\fmf{phantom,tension=1}{v2,v4,o3}
\fmf{phantom,tension=1}{v1,v3,o1}
\fmf{phantom,tension=1}{v3,v5,v4}
\fmf{dashes,right=0.4,tension=0.1,label=$\pi^{-}$}{v1,v5,v2}
\fmf{heavy,tension=1}{v2,i2}
\fmf{heavy,label=$p$,tension=0}{v1,v2}
\fmfv{decoration.shape=circle,decoration.size=.08w,decoration.filled=full}{v1}
\fmfblob{.15w}{v2}
\fmfv{decoration.shape=circle,decoration.size=.10w,decoration.filled=30}{v5}
\fmf{photon,tension=1.2}{v5,o2}
\fmflabel{$\gamma$}{o2}
\fmflabel{$n$}{i1}
\fmflabel{$n$}{i2}
\end{fmfgraph*}
} \\[-0.5cm]
\nonumber
\end{align}
The black dots in the diagrams on 
the right-hand side represent the CP-violating vertex, whereas the lined blobs indicate the 
neutron-pion coupling in~\eqref{eq:f(sigma)notheta}. The chiral loop integrals
are logarithmically divergent and once evaluated give the following contribution to the neutron 
electric dipole moment~\cite{Crewther_et_al}
\begin{align}
|\mathbf{d}_{n}|={1\over 4\pi^{2}}{|g_{\pi\!N\!N}\overline{g}_{\pi\!N\!N}|\over m_{N}}
\log\left({m_{N}\over m_{\pi}}\right),
\end{align}
where
\begin{align}
|\overline{g}_{\pi\!N\!N}|\approx 0.027|\theta|, 
\end{align}
is the coupling of the CP-violating vertex and, 
in the spirit of EFT, integrals have been cut off at~$\Lambda=m_{\pi}$.
Substituting the value for the CP-preserving pion-nucleon coupling and implementing
the experimental bound~\eqref{eq:EDM_exp_bound}, we find
\begin{align}
|\theta|\lesssim 10^{-11}.
\end{align}
We see that the amount of fine tuning in the $\theta$~parameter 
needed to explain experiments is not very far off
the one obtained for the angle~$\theta$ in~\eqref{eq:theta_angle_molecule} 
in the classical toy model of the neutron (not by accident 
both quantities were denoted by the same Greek letter).

\begin{figure}[t]
\centerline{\includegraphics[scale=0.3]{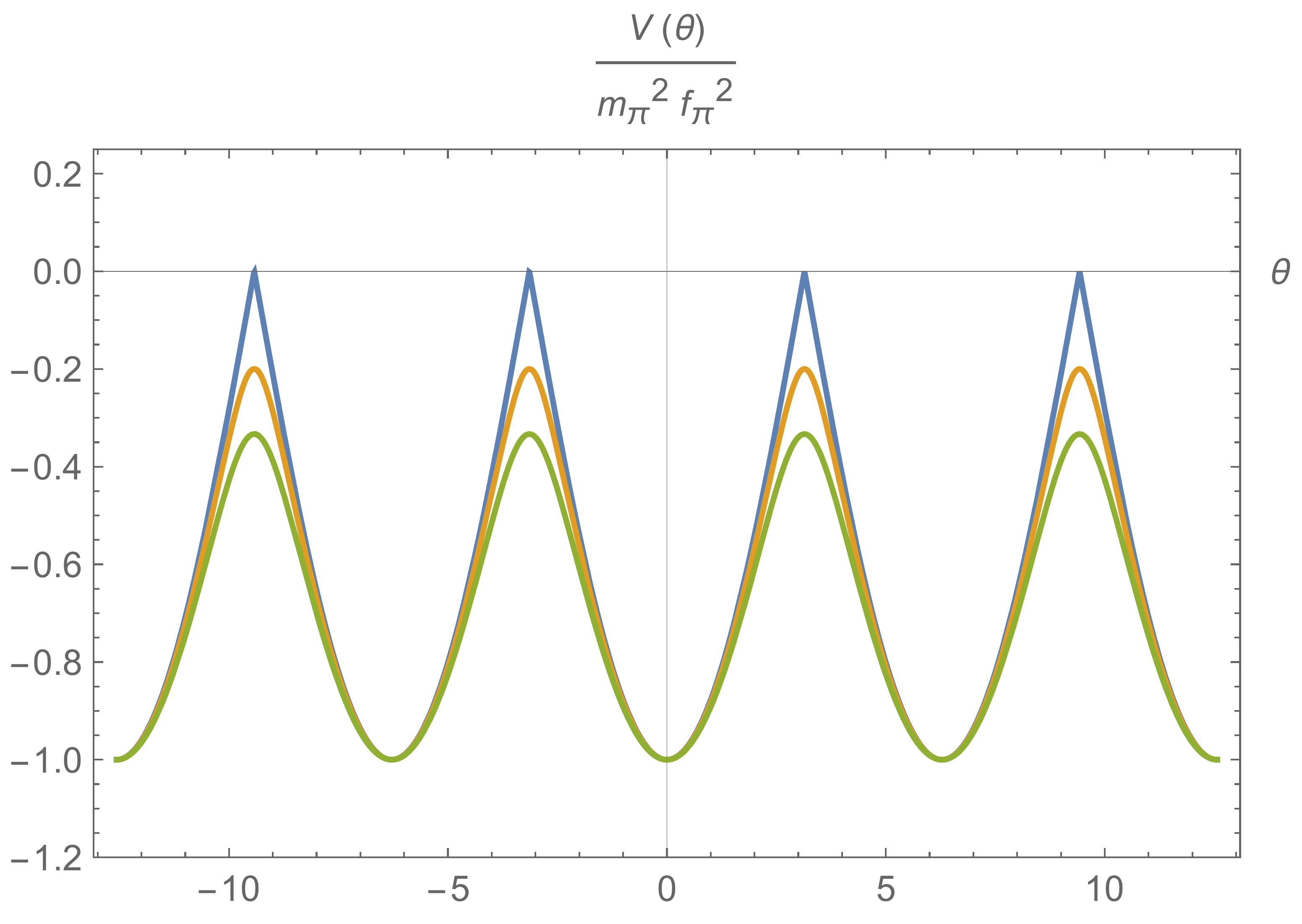}}
\caption[]{Plot of~$V(\theta)$ in eq.~\eqref{eq:V(theta)_gen} 
for three different values of the~${m_{u}\over m_{d}}$ ratio: 1~(blue), 0.3~(orange), and 0.5~(green).}
\label{fig:theta_plot}
\end{figure}

\begin{mdframed}[backgroundcolor=lightgray,hidealllines=true]
\vspace*{0.2cm}
\centerline{\greybox{\bf A ``potential'' for~$\theta$}}
\vspace*{0.2cm}

We would like to understand how the energy of the ground state of QCD depends on the
parameter~$\theta$. There are a number of things that can be said about this quantity, that 
we denote by~$V(\theta)$. As we learned above [see eq.~\eqref{eq:Stheta_integer}], 
the $\theta$-term is a topological object and any 
physical quantity depending on it like~$V(\theta)$ should be periodic in~$\theta$ with
period equal to~$2\pi$
\begin{align}
V(\theta+2\pi)=V(\theta).
\end{align}
Moreover, there exists a very elegant 
argument showing that energy is minimized for~$\theta=0$~\cite{Vafa_Witten}
\begin{align}
V(0)\leq V(\theta).
\end{align}
To go beyond these general considerations and find 
an explicit expression of~$V(\theta)$ in QCD, we consider the potential energy in the 
pion effective action~\eqref{eq:pion_eff_massterm}
\begin{align}
\mathcal{V}(\boldsymbol{\Sigma})
=-{m_{\pi}^{2}f_{\pi}^{2}\over 2(m_{u}+m_{d})}{\rm tr\,}\big(M^{\dagger}\boldsymbol{\Sigma}+M
\boldsymbol{\Sigma}^{\dagger}\big),
\label{eq:potential_theta_pi}
\end{align}
where~$M$ is given by
\begin{align}
M=\left(
\begin{array}{cc}
e^{i\theta}m_{u} & 0 \\
0 & m_{d}
\end{array}
\right).
\end{align}
To find the vacuum energy, 
we look for a NG boson matrix configuration minimizing~$\mathcal{V}(\boldsymbol{\Sigma})$.

In fact, since the mass matrix is diagonal it can be seen that the trace 
in~\eqref{eq:potential_theta_pi} only depends on the diagonal components of~$\boldsymbol{\Sigma}$.
This means that in order to minimize the potential it is enough consider NG matrices of the 
form~$\boldsymbol{\Sigma}=\mbox{diag}(e^{i\varphi_{1}},e^{i\varphi_{2}})$. Furthermore, the
dependence on~$\theta$ in the mass matrix can be shifted to the NG boson matrix by 
the field redefinition
\begin{align}
\boldsymbol{\Sigma}\longrightarrow 
\widetilde{\boldsymbol{\Sigma}}\equiv
\left(
\begin{array}{cc}
e^{-{i\theta\over 2}} & 0 \\
0 & 1
\end{array}
\right)
\boldsymbol{\Sigma}
\left(
\begin{array}{cc}
e^{-{i\theta\over 2}} & 0 \\
0 & 1
\end{array}
\right)=
\left(
\begin{array}{cc}
e^{i(\varphi_{1}-\theta)} & 0 \\
0 & e^{i\varphi_{2}}
\end{array}
\right).
\end{align}
Imposing the condition~$\det\widetilde{\boldsymbol{\Sigma}}=1$, 
we have~$\varphi_{1}+\varphi_{2}=\theta\mbox{ mod }2\pi$. 

Substituting the redefined NG matrix field~$\widetilde{\Sigma}$ 
into~\eqref{eq:potential_theta_pi} with~$M=\mbox{diag}(m_{u},m_{d})$, we arrive at the potential
\begin{align}
\mathcal{V}(\varphi_{1},\varphi_{2})
=-{m_{\pi}^{2}f_{\pi}^{2}\over m_{u}+m_{d}}\big(m_{u}\cos\varphi_{1}
+m_{d}\cos\varphi_{2}\big),
\label{eq:potential_theta_pre}
\end{align} 
that has to be minimized subject to the constraint~$\varphi_{1}+\varphi_{2}=\theta$. 
The equation to be solved is
\begin{align}
m_{u}\sin\varphi_{1}=m_{d}\sin(\theta-\varphi_{1}),
\end{align}
that, after a bit of algebra, gives
\begin{align}
\cos^{2}{\varphi_{1}}={(m_{u}+m_{d}\cos\theta)^{2}\over m_{u}^{2}+m_{d}^{2}+2 m_{u}m_{d}\cos\theta},
\nonumber \\[0.2cm]
\cos^{2}{\varphi_{2}}={(m_{d}+m_{u}\cos\theta)^{2}\over m_{u}^{2}+m_{d}^{2}+2 m_{u}m_{d}\cos\theta}.
\end{align}
Substituting these results into~\eqref{eq:potential_theta_pre}, we arrive at the expression of the 
QCD vacuum energy as a function of~$\theta$
\begin{align}
V(\theta)=-{m_{\pi}^{2}f_{\pi}^{2}\over m_{u}+m_{d}}
\sqrt{m_{u}^{2}+m_{d}^{2}+2 m_{u}m_{d}\cos\theta}.
\label{eq:V(theta)_gen}
\end{align}
In fig.~\ref{fig:theta_plot} we have represented this function for various values
of the ratio~$m_{d}/m_{u}$, from where we see that, as announced, the minimum occurs at~$\theta=0$.
We also see that when~$m_{u}=m_{d}$ there are cusps at the maxima located at~$\theta=(2n+1)\pi$, that
are smoothed out when the quarks have different masses. Being an experimental fact
that~$\theta$ is very small, we can expand~$V(\theta)$ around~$\theta=0$ to find
\begin{align}
V(\theta)=-m_{\pi}^{2}f_{\pi}^{2}+{1\over 2}m_{\pi}^{2}f_{\pi}^{2}{m_{u}m_{d}\over (m_{u}+m_{d})^{2}}
\theta^{2}.
\end{align} 
This expression will become handy later on when it will be reinterpreted as the potential 
for the axion field.

Since~$m_{s}\gg m_{u},m_{d}$ we have restricted our attention to QCD with the two lightest flavors, 
although 
the analysis can be easily extended to any~$N_{f}\geq 2$. The resulting 
expression of the ground state energy~$V(\theta;m_{1},\ldots,m_{f})$ for small~$\theta$
is symmetric under permutations of the quark masses
and satisfies a recursion relation
\begin{align}
V(\theta;m_{1},\ldots,m_{f-1})=\lim_{m_{f}\rightarrow \infty} V(\theta;m_{1},\ldots,m_{f}),
\end{align}
implementing the decoupling of the~$f$-th flavor.

\end{mdframed}

\noindent
\subsection{Enters the axion}

We would like to understand the smallness of~$\theta$ in a natural way, i.e., either as following from
some symmetry principle or by finding out some dynamical reason for its 
value\footnote{The fact that in the CO$_{2}$ molecule the angle~$\theta$ is zero is a consequence of
the dynamics of the atomic orbitals and is therefore ``natural''.}. One possible explanation 
would be that~$m_{u}=0$, so a chiral rotation of the $u$-quark field
would get rid of the $\theta$-term without introducing CP-violating phases in the chiral 
Lagrangian. This is however no good, since all experimental evidences indicate that the
$u$-quark is not massless.

A very popular solution to the CP problem is the one proposed by Roberto Peccei and Helen 
Quinn~\cite{Peccei_Quinn1,Peccei_Quinn2} consisting in making the $\theta$-parameter the vev
of a pseudoscalar field~$a(x)$, the {\em axion}~\cite{Weinberg_axion,Wilczek_axion},
whose potential would drive it 
to~$\langle 0|a(x)|0\rangle=0$. To be more precise, let us consider the action
\begin{align}
S&=\int d^{4}x\left(i\overline{\boldsymbol{q}}_{R}{D\!\!\!\!/\,}\boldsymbol{q}_{R}
+i\overline{\boldsymbol{q}}_{L}{D\!\!\!\!/\,}\boldsymbol{q}_{L}
+\overline{\boldsymbol{q}}_{L}M\boldsymbol{q}_{R}
+\overline{\boldsymbol{q}}_{R}M^{\dagger}\boldsymbol{q}_{L}
-{1\over 32\pi^{2}f_{a}}a F^{a}_{\mu\nu}\widetilde{F}^{a\mu\nu}\right),
\label{eq:pre-axion_action}
\end{align}
where~$f_{a}$ is an energy scale introduced so the axion field has canonical dimenensions
of energy. We can now play the old game of shifting the last term in the action~\eqref{eq:pre-axion_action}
to a complex phase in the mass matrix. In the low-energy effective field theory, this phase 
can be absorbed into the NG bosons matrix by the field redefinition (cf. the analysis presented
in Box~12)
\begin{align}
\boldsymbol{\Sigma}\longrightarrow 
\left(
\begin{array}{cc}
e^{-{ia\over 2f_{a}}} & 0 \\
0 & 1
\end{array}
\right)
\boldsymbol{\Sigma}
\left(
\begin{array}{cc}
e^{-{ia\over 2f_{a}}} & 0 \\
0 & 1
\end{array}
\right).
\label{eq:axion_field_red}
\end{align}
In the absence of a mass term for the NG bosons,~$\boldsymbol{\Sigma}$ only has derivative
couplings and the theory is invariant under constant shifts of the 
axion field,~$a(x)\rightarrow a(x)+\mbox{constant}$. 
The presence of the 
term~$f_{\pi}^{3}B_{0}{\rm tr\,}\big(M^{\dagger}\boldsymbol{\Sigma}+\boldsymbol{\Sigma}^{\dagger}
M\big)$, however, induces a potential that can be read off eq.~\eqref{eq:V(theta)_gen}
with~$\theta$ replaced by~$a/f_{a}$. Expanding around the minimum at~$a=0$, we find
\begin{align}
V(a)=
{m_{\pi}^{2}f_{\pi}^{2}\over 2f_{a}^{2}}{m_{u}m_{d}\over (m_{u}+m_{d})^{2}}a^{2}+\ldots,
\end{align}
where we have dropped constant terms and the ellipsis indicates higher-order axion self-interactions.
This gives the axion mass
\begin{align}
m_{a}={m_{\pi}f_{\pi}\over f_{a}}{\sqrt{m_{u}m_{d}}\over m_{u}+m_{d}}
=5.7\left({10^{9}\mbox{ GeV}\over f_{a}}\right)\mbox{ meV}.
\label{eq:QCD_axion_mass}
\end{align}
The field redefinition~\eqref{eq:axion_field_red} also induces axion interactions with 
mesons, baryons, leptons, and photons. 
For example,
\begin{align}
S_{\rm axion}\supset -\int d^{4}x\,\left({i\over 2}g_{ap\gamma}a\overline{p}\sigma^{\mu\nu}\gamma_{5}p
F_{\mu\nu}+{i\over 2}g_{an\gamma} a\overline{n}\sigma^{\mu\nu}\gamma_{5}n F_{\mu\nu}
+{g_{a\gamma\gamma}\over 4}a F_{\mu\nu}\widetilde{F}^{\mu\nu}\right),
\end{align} 
where~$g_{an\gamma}=-g_{ap\gamma}\sim f_{a}^{-2}$ and~$g_{a\gamma\gamma}\sim f_{a}^{-1}$.
The last non-minimal electromagnetic coupling of the axion comes from 
the anomaly-induced term in the chiral Lagrangian 
pointed out in the footnote on page~\pageref{page:anomaly_pion_footnote}.
In a strong magnetic field, this term allows the conversion of a photon 
into an axion and vice versa, one of the main astrophysical signatures of the axion and also the 
target process of the light-shining-through-walls experiments~\cite{Redondo_Ringwald}.

\begin{figure}[t]
\centerline{\includegraphics[scale=0.28]{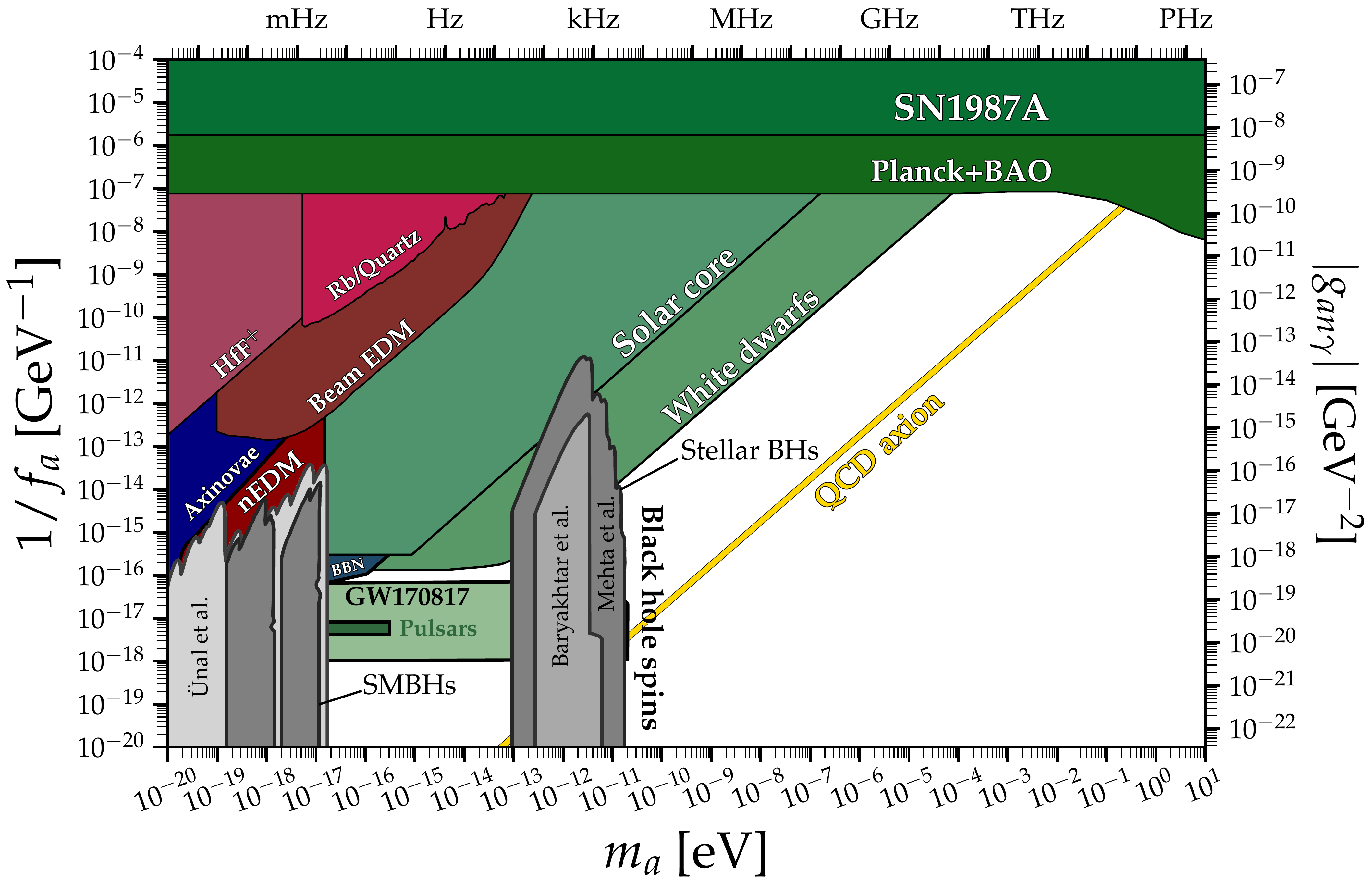}}
\caption[]{Exclusion plot from~\cite{Ohare} for the axion parameters~$f_{a}$ 
(resp.~$g_{an\gamma}$) and~$m_{a}$.
The yellow line represents the relation~\eqref{eq:QCD_axion_mass}.}
\label{fig:axion_plot}
\end{figure}

Among other candidates for dark matter (sterile neutrinos, supersymmetric particles,...)
axions are currently one of the most popular candidates to account for the
missing matter in the universe~\cite{Sigl,Marsh}. 
Cosmological and astrophysical phenomena provide a wide class of observational windows
for these kind of particles, ranging from CMB physics to stellar astrophysics and black holes
(see fig.~\ref{fig:axion_plot}). Observations so far have been used to constraint
the parameter space for axion-like particles, leaving a wide 
allowed region including most of the values of the QCD axion. A comprehensive
overview of current axion experiments and the bounds on different parameters 
can be found in the review~\cite{Hook}, as well as in~\cite{PDG} (see
also~\cite{Ohare} for a collection of exclusion plots for various parameters).

\section{The electroweak theory}
\label{sec:EW}

It is time we look into the electroweak sector of the SM. As already mentioned several times
in these lectures, our current understanding of the electromagnetic and weak forces is based
on a gauge theory with group~$\mbox{SU(2)}\times\mbox{U(1)}_{Y}$. This theory has subtle differences
with respect to
the color~$\mbox{SU(3)}$ QCD gauge group used to describe strong interactions. 
The basic one is that it is a chiral theory in which left- and right-handed fermions transform
in different representations of the gauge group. Closely related to this is
that the~$\mbox{SU(2)}\times\mbox{U(1)}_{Y}$
gauge invariance is spontaneously broken at low energies by an implementation of the BEH mechanism
explained in section~\ref{sec:tale_of_symmetries}.
This feature, that for decades was the shakiest part of the electroweak theory, was finally
confirmed in July 2012 when
the detection of the Higgs boson was announced at CERN, thus fitting the final piece into
the jigsaw puzzle.

Whereas only hadrons (i.e., quarks) partake of the strong interaction, the weak force affects 
both quarks and leptons. Its chiral character is reflected in that the weak interaction violate
parity, a fact discovered in the late 1950s in the study of 
$\beta$-decay and other processes mediated by the weak force~\cite{Lee_Yang,Wu,Garwin,Friedman}. Unlike gluons, coupling to quarks
through a vector current~$J_{\rm QCD}^{\mu}=\overline{q}\gamma^{\mu}q$, 
the carriers of the weak force interact with matter via
the $\mbox{V}-\mbox{A}$ current~$J_{\rm weak}^{\mu}=\overline{\psi}\gamma^{\mu}
(\mathbbm{1}-\gamma_{5})\psi$, with~$\psi$ either a lepton or a 
quark field~\cite{Sudarshan_Marshak,Feynman_Gell-MannV-A}.

\subsection{Implementing~$\mbox{SU(2)}\times \mbox{U(1)}_{Y}$}

To be more precise, $\beta$-decay transmutes left-handed electrons into left-handed electron neutrinos
(an vice versa), while $u$-quarks (resp. $d$-quarks) transform into $d$~quarks (resp. $u$-quarks).
This suggests grouping left-handed electrons/neutrinos and quarks into doublets
\begin{align}
\mathbf{L}=\left(
\begin{array}{c}
\nu_{e} \\
e^{-}
\end{array}
\right)_{L}, \hspace*{1cm}
\mathbf{Q}=\left(
\begin{array}{c}
u \\
d
\end{array}
\right)_{L},
\label{eq:lepton_quarks_doublet_def}
\end{align}
and assume they transform in the fundamental representation~$\mathbf{2}$ of the $\mbox{SU(2)}$~algebra.
At the same time, since right-handed electrons and quarks 
do not undergo $\beta$-decay, their components are taken to be $\mbox{SU(2)}$ singlets
\begin{align}
\ell_{R}\equiv e^{-}_{R}, \hspace*{1.5cm} U_{R}\equiv u_{R}, \hspace*{1.5cm}
D_{R}\equiv d_{R}.
\end{align} 
Moreover, since there is no experimental evidence of the existence of right-handed neutrinos, we do not
include them in the description (at least for now; we will return to this issue later).

The whole picture is complicated because the weak force mixes with the electromagnetic 
interaction. In fact, the~$\mbox{U(1)}_{Y}$ of the electroweak gauge group is not 
the~$\mbox{U(1)}$ of Maxwell's theory. The generator~$Y_{\mathbf{R}}$ 
of the former, called the {\em weak hypercharge},
satisfies the Gell-Mann--Nishijima relation
\begin{align}
Q=Y_{\mathbf{R}}+t^{3}_{\mathbf{R}},
\label{eq:Gell-Mann-Nishijima}
\end{align}
where~$Q$ is the charge of the field in units of~$e$ and~$t^{3}_{\mathbf{R}}$ 
is the Cartan generator of~$\mbox{SU(2)}$ in the 
representation~$\mathbf{R}$. As an example, for~$\mathbf{L}$ in eq.~\eqref{eq:lepton_quarks_doublet_def} we 
have~$t^{3}_{\mathbf{2}}\equiv {1\over 2}\sigma^{2}
=\mbox{diag}({1\over 2},-{1\over 2})$ and~$Q=\mbox{diag}(0,-1)$, so we
have~$Y(\mathbf{L})=-{1\over 2}\mathbbm{1}$. Repeating this for all leptons and quark fields, we find
\begin{align}
Y(\mathbf{L})=-{1\over 2}\mathbbm{1}, \hspace*{0.5cm} 
Y(\ell)=-1, \hspace*{0.5cm}
Y(\mathbf{Q})=-{1\over 6}\mathbbm{1}, \hspace*{0.5cm}
Y(U_{R})={2\over 3}, \hspace*{0.5cm}
Y(D_{R})=-{1\over 3},
\end{align}
where for the~$\mbox{SU(2)}$ singlets we have~$t^{3}_{\mathbf{1}}=0$. Notice that
for~$\mbox{U(1)}_{Y}$
we have~$Y_{\mathbf{R}}
=Y\mathbbm{1}$, so 
the representation of~$\mbox{U(1)}_{Y}$
is fully determined by the {\em hypercharge}~$Y$.

\begin{table}[t]
\begin{eqnarray*}
\begin{array}{c|c|c|c"c|c}
\multicolumn{6}{c}{\mbox{\bf Leptons}} \\[0.3cm]
\multicolumn{6}{c}{ \hspace*{2.5cm} i=1 \hspace*{1.4cm} i=2 
\hspace*{1.5cm} i=3 
\hspace*{1.5cm} t_{\mathbf{R}}^{3} \hspace*{1.3cm} Y_{\mathbf{R}} } \\[0.1cm]
\thickhline
\hspace*{2cm} & \hspace*{2cm} & \hspace*{2cm} & \hspace*{2cm} & \hspace*{1.3cm} & \hspace*{1.3cm} \\
\mathbf{L}^{i} & \left(\begin{array}{c} \nu_{e} \\ e^{-} \end{array}\right)_{L}
& \left(\begin{array}{c} \nu_{\mu} \\ \mu^{-} \end{array}\right)_{L} & 
\left(\begin{array}{c} \nu_{\tau} \\ \tau^{-} \end{array}\right)_{L} &
{1\over 2}\sigma^{3} & -{1\over 2}\mathbbm{1}  \\
  & & & &  & \\ 
  & & & & & \\
\ell^{i}_{R} & e_{R}^{-} & \mu^{-}_{R} & \tau^{-}_{R} & 0  & -1 \\
 & & & & &  \\
 \thickhline
\end{array}
\\[0.4cm]
\begin{array}{c|c|c|c"c|c}
\multicolumn{6}{c}{\mbox{\bf Quarks}} \\[0.3cm]
\multicolumn{6}{c}{ \hspace*{2.5cm} i=1 \hspace*{1.4cm} i=2 
\hspace*{1.5cm} i=3 
\hspace*{1.5cm} t_{\mathbf{R}}^{3} \hspace*{1.3cm} Y_{\mathbf{R}} } \\[0.1cm]
\thickhline 
\hspace*{2cm} & \hspace*{2cm} & \hspace*{2cm} & \hspace*{2cm} & \hspace*{1.3cm} & \hspace*{1.3cm} \\
\mathbf{Q}^{i} & \left(\begin{array}{c} u \\ d \end{array}\right)_{L} 
& \left(\begin{array}{c} c \\ s \end{array}\right)_{L} 
& \left(\begin{array}{c} t \\ b \end{array}\right)_{L} &
{1\over 2}\sigma^{3} & {1\over 6}\mathbbm{1}   \\
  & & & & &  \\ 
  & & & & &  \\
U^{i}_{R} & u_{R} & c_{R}  & t_{R} & 0  & {2\over 3} \\
 & & & & & \\
  & & & & &\\
D^{i}_{R} & d_{R} & s_{R}  & b_{R} & 0  & -{1\over 3} \\
 & & & & & \\
 \thickhline
\end{array}
\end{eqnarray*}
\caption{Transformation properties of leptons and quarks in the electroweak sector of the SM.
In addition to the indicated representations of~$\mbox{SU(2)}\times\mbox{U(1)}_{Y}$, quarks
transform in the fundamental~$\mathbf{3}$ irrep of~$\mbox{SU(3)}$, whereas leptons
are singled under this group.}
\label{table:leptons+quarks}
\end{table}

We might be tempted to believe that with this we have determined how {\em all} matter fields in the SM
transform under the gauge group~$\mbox{SU(2)}\times \mbox{U(1)}_{Y}$. 
However, for reasons that we so far ignore, nature has decided to have three copies of the structure
just described. In addition to the electron, its neutrino, and the $u$- and $d$-quarks there are 
two more replicas or {\em families}. 
The second family includes the muon~($\mu^{-}$) and its
neutrino~($\nu_{\mu}$), together with the charm ($c$) and strange ($s$) quarks. 
The third family, on the other hand, contains the~$\tau^{-}$ lepton, its neutrino ($\nu_{\tau})$, and
the top ($t$) and bottom ($t$) quarks. Apart from an increasing hierarchy of masses, 
each extra family exactly replicates the transformation properties of the fields in the
first one. To include this feature in our description, we add an index~$i=1,2,3$ to the 
doublet~$\{\mathbf{L}^{i},\mathbf{Q}^{i}\}$ and
singlet~$\{\ell_{R}^{i},U_{R}^{i},D_{R}^{i}\}$ fields introduced above, sumarizing in 
table~\ref{table:leptons+quarks} the three-family
structure with the corresponding representations of~$\mbox{SU(2)}\times \mbox{U(1)}_{Y}$.
We should not forget that, besides the electroweak quantum numbers, leptons are singlets with 
respect to color~$\mbox{SU(3)}$, whereas quarks are triplets transforming in the fundamental 
representation of this group.

Once the matter content of the SM is determined, as well as how the fields transform under 
the electroweak gauge group, we
fix our attention on the gauge bosons. In the case of~$\mbox{SU(2)}$, it is convenient to use  
the~$\{t_{\mathbf{R}}^{\pm},t_{\mathbf{R}}^{3}\}$ basis, so the corresponding gauge 
field is written as\footnote{In terms of the generators~$t^{\pm}_{\mathbf{R}}\equiv
t^{1}_{\mathbf{R}}\pm it^{2}_{\mathbf{R}}$, the~$\mbox{SU(2)}$ algebra reads~$[t^{3}_{\mathbf{R}},
t^{\pm}_{\mathbf{R}}]=\pm t^{\pm}_{\mathbf{R}}$, $[t^{+}_{\mathbf{R}},t^{-}_{\mathbf{R}}]=
2t^{3}_{\mathbf{R}}$. This is just the algebra of ladder operators familiar from the 
theory of angular momentum in quantum mechanics.}
\begin{align}
\mathbf{W}_{\mu}=W^{+}_{\mu}t^{-}_{\mathbf{R}}+W^{-}_{\mu}t^{+}_{\mathbf{R}}+W^{3}_{\mu}t^{3}_{\mathbf{R}},
\end{align}
whereas for the Abelian gauge field associated with~$\mbox{U(1)}_{Y}$, we have
\begin{align}
\mathbf{B}_{\mu}=B_{\mu}Y\mathbbm{1}.
\end{align}
The covariant derivative needed to construct the matter action is then given by
\begin{align}
D_{\mu}&=\partial_{\mu}-ig\mathbf{W}_{\mu}-ig'\mathbf{B}_{\mu} \nonumber \\[0.2cm]
&=\partial_{\mu}-igW^{+}_{\mu}t^{-}_{\mathbf{R}}-igW^{-}_{\mu}t^{+}_{\mathbf{R}}
-igW^{3}_{\mu}t^{3}_{\mathbf{R}}-ig'B_{\mu}Y\mathbbm{1},
\label{eq:covariant_der_SM}
\end{align}
where~$g$ and~$g'$ are the coupling constant associated with the two factors of the electroweak 
gauge group.

We should not forget however that the electric charge~$Q$, the hypercharge~$Y\mathbbm{1}$,
and the~$\mbox{SU(3)}$ Cartan generator~$t^{3}_{\mathbf{R}}$ are not independent, but
connected by the Gell-Mann--Nishijima relation~\eqref{eq:Gell-Mann-Nishijima}.
It is therefore useful to
consider combinations
\begin{align}
A_{\mu}&=B_{\mu}\cos\theta_{w}+W^{3}_{\mu}\sin\theta_{w}, \nonumber \\[0.2cm]
Z_{\mu}&=-B_{\mu}\sin\theta_{w}+W^{3}_{\mu}\cos\theta_{w},
\label{eq:AZvsBW3}
\end{align}
where~$A_{\mu}$ is to be identified with the electromagnetic field, whose gauge group will
be denoted by~$\mbox{U(1)}_{\rm em}$ to distinguish it from the one associated with the gauge
field~$\mathbf{B}_{\mu}$. The
parameter~$\theta_{w}$ is called the {\em weak mixing angle} and sometimes also
the Weinberg angle, although it was first introduced by Glashow in~\cite{Glashow_SM}.
Expressing the covariant derivative~\eqref{eq:covariant_der_SM} in terms of 
the~$\{W_{\mu}^{\pm},A_{\mu},Z_{\mu}\}$ gauge fields, 
we find
\begin{align}
D_{\mu}&=\partial_{\mu}-igW^{+}_{\mu}t^{-}_{\mathbf{R}}-igW^{-}_{\mu}t^{+}_{\mathbf{R}}
-iA_{\mu}\big(g\sin\theta_{w}t^{3}_{\mathbf{R}}+g'\cos\theta_{w} Y\mathbbm{1}\big) \nonumber \\[0.2cm]
&-iZ_{\mu}\big(g\sin\theta_{w}t^{3}_{\mathbf{R}}-g'\cos\theta_{w}Y\mathbbm{1}\big).
\end{align}  
Now, if~$A_{\mu}$ is to be identified with the electromagnetic field, it has to couple
to the electric charge matrix~$eQ$. Consistency with the Gell-Mann--Nishijima 
relation~\eqref{eq:Gell-Mann-Nishijima} implies then
\begin{align}
g\sin\theta_{w}=g'\cos\theta_{w}=e \hspace*{1cm} \Longrightarrow \hspace*{1cm}
\tan\theta_{w}={g\over g'}.
\end{align}
This relation shows that the weak mixing angle not only measures the mixing among 
the Abelian gauge fields associated with the~$\mbox{U(1)}_{Y}$ and the Cartan generator of~$\mbox{SU(2)}$,
but also of the relative strength of the interactions
associated with the two factors of the electroweak gauge group.
Implementing all the previous relations, the covariant derivative reads
\begin{align}
D_{\mu}&=\partial_{\mu}-{ie\over \sin\theta_{w}}W^{+}_{\mu}t^{-}_{\mathbf{R}}
-{ie\over \sin\theta_{w}}W^{-}_{\mu}t^{+}_{\mathbf{R}}
-ieA_{\mu}Q-{2ie\over \sin(2\theta_{w})}Z_{\mu}\big(t^{3}_{\mathbf{R}}
-Q\sin^{2}\theta_{w}\big),
\label{eq:gauge_cov_der_SM}
\end{align}
where we have eliminated~$Y$,~$g$, and~$g'$ in favor of~$Q$,~$e$, and~$\theta_{w}$.
With this, the SM matter action reads
\begin{align}
S_{\rm matter}=\sum_{k=1}^{3}\int d^{4}x\,\Big(
i\overline{\mathbf{L}}^{k}{D\!\!\!\!/\,}\mathbf{L}^{k}+i\overline{\ell}^{k}_{R}{D\!\!\!\!/\,}
\ell^{k}_{R}+i\overline{\mathbf{Q}}^{k}{D\!\!\!\!/\,}\mathbf{Q}^{k}
+i\overline{U}_{R}^{k}{D\!\!\!\!/\,}U^{k}_{R}+i\overline{D}^{k}_{R}{D\!\!\!\!/\,}
D^{k}_{R}\Big).
\label{eq:S_matter_SM}
\end{align}

Next we now look at the gauge action
\begin{align}
S_{\rm gauge}=-{1\over 2}\int d^{4}x\left[{\rm tr\,}\big(\mathbf{W}_{\mu\nu}\mathbf{W}^{\mu\nu}\big)
+{\rm tr\,}\big(\mathbf{B}_{\mu\nu}\mathbf{B}^{\mu\nu}\big)\right],
\end{align}
where~$\mathbf{W}_{\mu\nu}$ and~$\mathbf{B}_{\mu\nu}$ are the field strength of~$\mathbf{W}_{\mu}$ 
and~$\mathbf{B}_{\mu}$ respectively. Recasting it in terms of the electromagnetic and~$Z_{\mu}$ gauge fields
defined in eq.~\eqref{eq:AZvsBW3}, we have
\begin{align}
S_{\rm gauge}&=-\int d^{4}x\left\{{1\over 4}W_{\mu\nu}^{+}W^{-\mu\nu}+{1\over 4}Z_{\mu\nu}Z^{\mu\nu}
+{1\over 4}F_{\mu\nu}F^{\mu\nu}-{ie\over 2}\cot\theta_{w} W_{\mu}^{+}W_{\nu}^{-}Z^{\mu\nu}
\right.\nonumber \\[0.2cm]
&\left.-{ie\over 2}W_{\mu}^{+}W_{\nu}^{-}F^{\mu\nu}
+{e^{2}\over 2\sin\theta_{w}}\Big[(W_{\mu}^{+}W^{+\mu})(W^{-}_{\mu}W^{-\mu})-(W^{+}_{\mu}W^{-\mu})^{2}\Big]
\right\},
\label{eq:S_gauge_SM}
\end{align}
where~$Z_{\mu\nu}=\partial_{\mu}Z_{\nu}-\partial_{\nu}Z_{\mu}$,~$F_{\mu\nu}=\partial_{\mu}A_{\nu}-
\partial_{\nu}A_{\mu}$, and we have defined
\begin{align}
W^{\pm}_{\mu\nu}&= \partial_{\mu}W^{\pm}_{\nu}-\partial_{\nu}W^{\pm}_{\mu}
\mp e\big(W^{\pm}_{\mu}A_{\nu}-W^{\pm}_{\nu}A_{\mu}\big)\mp ie\cot\theta_{w}
\big(W^{\pm}Z_{\nu}-W^{\pm}_{\nu}Z_{\mu}\big).
\label{eq:W_field_strength}
\end{align}

The~SM gauge couplings can be now read off 
eqs.~\eqref{eq:gauge_cov_der_SM},~\eqref{eq:S_matter_SM},~\eqref{eq:S_gauge_SM},
and~\eqref{eq:W_field_strength}.
The first thing to notice from the last two equations is that the $W^{\pm}_{\mu}$ gauge 
fields have electric charge~$\pm e$ and also couple to
the $Z_{\mu}$ gauge field, which has itself zero electric charge. A look a the matter action also shows
that the two components of the~$\mbox{SU(2)}$
doublets are transmuted into one another by the emission/absorption of a~$W$ boson. As to the $Z^{0}$, 
it can be emitted/absorbed by quarks and leptons with couplings that depend 
on their~$\mbox{SU(2)}\times \mbox{U(1)}_{Y}$ 
quantum numbers~(see chapter 5 of~\cite{AG_VM} or any other SM textbook for the details). As 
a practical example, the neutron 
$\beta$-decay~$n\rightarrow p^{+}e^{-}\overline{\nu}_{e}$
proceeds by the emission of a~$W^{-}$ by one of the neutron's~$d$ quarks, turning itself
into a $u$~quark (and the neutron into a proton). 
The~$W^{-}$ then decays into an electron and an electronic antineutrino. 
\begin{eqnarray}
\nonumber \\[-0.3cm]
n[udd]\longrightarrow p^{+}[uud]+e^{-}+\overline{\nu}_{e}\hspace*{1cm} \Longrightarrow \hspace*{1.2cm}
\parbox{40mm}{
\begin{fmfgraph*}(100,90)
\fmfleft{i1}
\fmfright{o1,o2,o3}
\fmf{fermion}{i1,v1}
\fmf{fermion}{v1,o1}
\fmf{photon,label=$W^{-}$,tension=0.5}{v1,v2}
\fmf{fermion,tension=0.5}{o2,v2}
\fmf{fermion,tension=0.5}{v2,o3}
\fmflabel{$d$}{i1}
\fmflabel{$u$}{o1}
\fmflabel{$\overline{\nu}_{e}$}{o2}
\fmflabel{$e^{-}$}{o3}
\end{fmfgraph*}
}\label{eq:neutron_beta_dec}
\\[-0.2cm]
\nonumber
\end{eqnarray}
As a second example, we also have lepton-neutrino scattering mediated by the 
interchange of a~$Z^{0}$
\begin{eqnarray}
\nonumber \\[-0.2cm]
\ell^{-}+\nu_{\ell}\longrightarrow \ell^{-}+\nu_{\ell} \hspace*{1cm} \Longrightarrow \hspace*{1cm}
\parbox{30mm}{
\begin{fmfgraph*}(100,90)
\fmfleft{i1,i2}
\fmfright{o1,o2}
\fmf{fermion}{i1,v1,o1}
\fmf{photon,label=$Z^{0}$,tension=0.5}{v1,v2}
\fmf{fermion}{i2,v2,o2}
\fmflabel{$\ell^{-}$}{i1}
\fmflabel{$\ell^{-}$}{o1}
\fmflabel{$\nu_{\ell}$}{i2}
\fmflabel{$\nu_{\ell}$}{o2}
\end{fmfgraph*}
}
\\[-0.1cm]
\nonumber
\end{eqnarray}
where~$\ell$ stands for~$e$,~$\mu$ or~$\tau$.  The existence of weak processes 
without transfer of electric charge is
a distinctive prediction of the Glashow-Weinberg-Salam model. The discovery of these so-called
neutral weak currents in the Gargamelle bubble chamber at CERN in 1973~\cite{Gargamelle} 
was solid experimental evidence 
in favor of the electroweak theory (see also~\cite{WNC_review} for a historical account).
Let us also mention that~$S_{\rm matter}+S_{\rm gauge}$
includes QED, and therefore describes all electromagnetic-mediated processes among leptons and quarks.

\begin{mdframed}[backgroundcolor=lightgray,hidealllines=true]
\vspace*{0.2cm}
\centerline{\greybox{\bf Hypercharges and anomaly cancellation}}
\vspace*{0.2cm}
\label{pag:box_anomaliesSM}

Our discussion in section~\ref{sec:anomalies} has very much stressed the
need to eliminate anomalies affecting gauge invariance.   
Gauge anomalies come from the
same triangle diagrams we encountered in our discussion of the chiral anomaly, namely
those shown in eq.~\eqref{eq:triangle_diagrams}. The only difference is that instead of
having an axial-vector current on the left and two vector currents on the right, now we have
three gauge currents, one at each vertex. 

Fortunately, to decide whether the SM is anomaly free we do not need to compute the 
diagrams themselves. It is enough to look at the group theory factor and check that 
the result is zero once 
we sum over all chiral fermions running in the loop. To compute this
factor we consider the gauge generator at each vertex~$(T^{a}_{\mathbf{R}})_{ij}$, 
where the indices~$i$,~$j$ are associated with the gauge index of the incoming/outgoing fermion
entering/leaving the vertex, while~$a$ is the index of the gauge field attached to it. 
Thus, for a given fermion species in the loop, the
group theory factor multiplying the sum of the two triangles
in~\eqref{eq:triangle_diagrams} is given by
\begin{align}
(T^{a}_{\mathbf{R}})_{ij}(T^{b}_{\mathbf{R}})_{jk}(T^{c}_{\mathbf{R}})_{ki}
+(T^{a}_{\mathbf{R}})_{ij}(T^{c}_{\mathbf{R}})_{jk}(T^{b}_{\mathbf{R}})_{ki}
={\rm tr\,}\big(T^{a}_{\mathbf{R}}\{T^{b}_{\mathbf{R}},T^{c}_{\mathbf{R}}\}\big).
\end{align}
Notice how the second term on the left-hand side is obtained from the first one by interchanging 
the two right vertices, as it happens in the second triangle diagram. Next, we have to sum
over all fermion species, taking into account that left- and right-handed fermions contribute with
opposite signs. Thus, the condition for anomaly cancellation is 
\begin{align}
\sum_{L}{\rm tr\,}\big(T^{a}_{\mathbf{R}}\{T^{b}_{\mathbf{R}},T^{c}_{\mathbf{R}}\}\big)_{L}
-\sum_{R}{\rm tr\,}\big(T^{a}_{\mathbf{R}}\{T^{b}_{\mathbf{R}},T^{c}_{\mathbf{R}}\}\big)_{R}=0,
\label{eq:anomaly_cancellation_cond}
\end{align}
where the sums are respectively over all left- and right-handed fermions in their corresponding
representations. In checking anomaly cancellation it is important to keep in
mind that if the gauge group has several semisimple factors, like the
case of the SM, the generator~$T^{a}_{\mathbf{R}}$ is the tensor product of the generators of
each factors.

There is a simple way to summarize the group-theoretical information contained in table~\ref{table:leptons+quarks} by just
indicating the representations of the different fermion species with respect to~$\mbox{SU(3)}\times\mbox{SU(2)}\times\mbox{U(1)}_{Y}$,
including also now the gauge group factor associated with the strong force. Using the 
notation~$(\mathbf{N}_{c},\mathbf{N})_{Y}$,
with~$\mathbf{N}_{c}$, $\mathbf{N}$, and~$Y$ the representations of~$\mbox{SU(3)}$,~$\mbox{SU(2)}$, 
and~$\mbox{U(1)}_{Y}$, we write for a single family \label{page:(p,q)_Y_notation}
\begin{align}
\mathbf{L}^{i}&: (\mathbf{1},\mathbf{2})_{-{1\over 2}}^{L}, \hspace*{1cm}
\ell_{R}^{i}: (\mathbf{1},\mathbf{1})_{-1}^{R}, \nonumber \\[0.2cm]
\mathbf{Q}^{i}&: (\mathbf{3},\mathbf{2})_{1\over 6}^{L}, \hspace*{1.2cm}
U^{i}_{R}:(\mathbf{3},\mathbf{1})_{2\over 3}^{R}, \hspace*{1cm}
D^{i}_{R}:(\mathbf{3},\mathbf{1})_{-{1\over 3}}^{R},
\label{eq:fermion_shorthand_trans}
\end{align}
and we also introduced a superscript to remind ourselves whether they are left- or
right-handed fermions (a useful information to decide what sign they come with in the 
anomaly cancellation condition).
In this notation, the generators of the
representation~$(\mathbf{N}_{c},\mathbf{N})_{Y}$ are given by
\begin{align}
T^{(I,a)}_{(\mathbf{N}_{c},\mathbf{N})_{Y}}=t^{I}_{\mathbf{N}_{c}}\otimes\mathbf{1}\otimes\mathbf{1}
+\mathbf{1}\otimes t^{a}_{\mathbf{N}}\otimes 1+\mathbf{1}\otimes\mathbf{1}\otimes Y,
\end{align}
where~$I=1,\ldots,8$ and~$a=1,2,3$ respectively label the generators of~$\mbox{SU(3)}$ 
and~$\mbox{SU(2)}$. At a practical level, in order to check anomaly cancellation in the~SM
we attach a group factor to each vertex of the triangle and compute the left-hand side
of~\eqref{eq:anomaly_cancellation_cond} to check whether it vanishes. Since we have three
different factors and three vertices, there are ten inequivalent possibilities
\begin{align*}
\includegraphics[width=12.5cm]{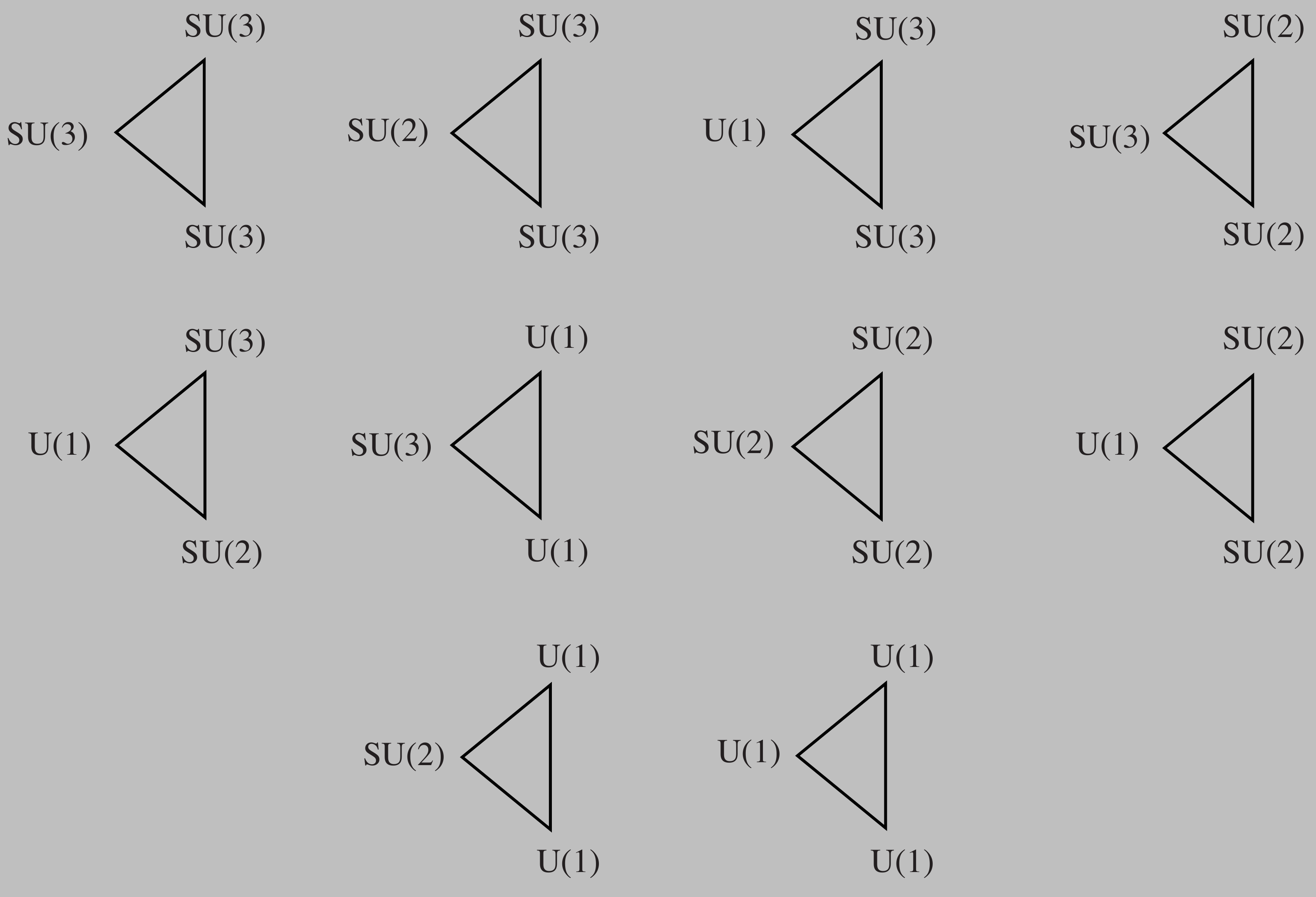}
\end{align*}
Some of the possibilities are rather trivial. For example, the triangle with three~$\mbox{SU(3)}$ factors
gives zero since the strong interaction does not distinguishes left- from right-handed quarks and
the two terms on the left-hand side 
of~\eqref{eq:anomaly_cancellation_cond} are equal. The same happens whenever we have
a single~$\mbox{SU(3)}$ or~$\mbox{SU(2)}$ factor, since the generators of these groups are traceless.
At the end of the day, there are just four nontrivial cases. Using an 
obvious notation, they are:~$\mbox{SU(2)}^{3}$, 
 $\mbox{SU(2)}^{2}\mbox{U(1)}$, $\mbox{SU(3)}^{2}\mbox{U(1)}$,
and~$\mbox{U(1)}^{3}$. In the first case, since only left-handed fermions couple to~$\mbox{SU(2)}$,
anomaly cancellation follows directly from the properties of the Pauli matrices
\begin{align}
{\rm tr\,}\big(\sigma^{i}\{\sigma^{j},\sigma^{k}\}\big)=2\delta_{jk}{\rm tr\,}\sigma_{i}=0.
\end{align}
For~$\mbox{SU(2)}^{2}\mbox{U(1)}$, again the~$\mbox{SU(2)}$ factors only allow left-handed
fermions in the loop, and the anomaly cancellation condition reads
\begin{align}
\sum_{L}Y_{L}=0,
\label{eq:ACCSM1}
\end{align}
while in the~$\mbox{SU(3)}^{2}\mbox{U(1)}$ triangle the color factor rules out leptons,
so we have
\begin{align}
\sum_{{\rm quarks},L}Y_{L}-\sum_{{\rm quarks},R}Y_{R}=0.
\label{eq:ACCSM2}
\end{align} 
Finally, we are left with the triangle with one~$\mbox{U(1)}$ at each vertex, leading to the
condition
\begin{align}
\sum_{L}Y_{L}^{3}-\sum_{R}Y_{R}^{3}=0,
\label{eq:ACCSM3}
\end{align}
where the sum in this case extends to all fermion species.

But this is not all. Since the SM model couples to gravity, it turns out that
we might have gauge anomalies triggered by triangle diagrams with one gauge boson and
two gravitons. The condition to avoid this is 
\begin{align}
\sum_{L}{\rm tr\,}(T^{a}_{\mathbf{R}})_{L}-\sum_{R}{\rm tr\,}(T^{a}_{\mathbf{R}})_{R}=0.
\end{align}  
In this case there are just three possibilities, corresponding to having a~$\mbox{SU(3)}$,
~$\mbox{SU(2)}$ or~$\mbox{U(1)}$ factor in the non-graviton vertex. For the first two cases,
the condition for anomaly cancellation is automatically satisfied, 
again because the generators of~$\mbox{SU(3)}$ 
and~$\mbox{SU(2)}$ are traceless. The third possibility, on the other hand, gives a nontrivial
condition
\begin{align}
\sum_{L}Y_{L}-\sum_{R}Y_{R}=0,
\label{eq:ACCSM4}
\end{align}
where the sum runs over both leptons and quarks.

We have found the four 
conditions~\eqref{eq:ACCSM1},~\eqref{eq:ACCSM2},~\eqref{eq:ACCSM3}, and~\eqref{eq:ACCSM4} 
to ensure the cancelation of anomalies, all of them involving the hypercharges
of the chiral fermion fields in the SM. Now, instead of checking whether the hypercharges
in eq.~\eqref{eq:fermion_shorthand_trans} satisfy this condition, we are going to see
to what extend anomaly cancellation determines the fermion hypercharges. 
Let us therefore write the representations of leptons and quarks in each family 
as~$(\mathbf{1},\mathbf{2})_{Y_{1}}^{L}$, $(\mathbf{1},\mathbf{1})_{Y_{2}}^{R}$, 
$(\mathbf{3},\mathbf{2})_{Y_{3}}^{L}$, $U^{i}_{R}:(\mathbf{3},\mathbf{1})_{Y_{4}}^{R}$, 
and~$D^{i}_{R}:(\mathbf{3},\mathbf{1})_{Y_{5}}^{R}$, reading now 
the anomaly cancellation conditions as equations to determine~$Y_{1},\ldots,Y_{5}$.
These are
\begin{align}
2Y_{1}+6Y_{3}&=0, \nonumber \\[0.2cm]
6Y_{3}-3Y_{4}-3Y_{5}&=0, \nonumber\\[0.2cm]
2Y_{1}^{3}+6Y_{3}^{3}-Y_{2}^{3}-3Y_{4}^{3}-3Y_{5}^{3}&=0, \\[0.2cm]
2Y_{1}+6Y_{3}-Y_{2}-3Y_{4}-3Y_{5}&=0.
\nonumber
\end{align}
Now, since these are homogeneous equations there exists the freedom to fix the overall 
normalization of the five hypercharges or, equivalently, to choose the value of one of them.
Taking for example~$Y_{2}=-1$, we are left
with four equations for the four remaining unknowns. They have a single solution given by
\begin{align}
Y_{1}=-{1\over 2},\hspace*{0.75cm} Y_{2}=-1, \hspace*{0.75cm}
Y_{3}={1\over 6}, \hspace*{0.75cm} Y_{4}=-{1\over 3}, \hspace*{0.75cm}
Y_{5}={2\over 3},
\end{align}
up to the interchange of~$Y_{4}$ and~$Y_{5}$ (notice that the associated
fields~$U^{i}_{R}$ and~$D^{i}_{R}$ transform in the same representation with respect to the other
two gauge group factors). This solution precisely reproduces the hypercharges shown
in eq.~\eqref{eq:fermion_shorthand_trans}. 

With this calculation we have learned two things.
One is that all gauge anomalies (and also the so-called mixed gauge-gravitational anomalies) cancel 
in the SM, and that they do within each family. And second, that anomaly cancellation condition 
is a very powerful way of constraining viable models in particle physics: in the~SM it 
fixes, up to a global normalization, the~$\mbox{U(1)}_{Y}$ charges of all chiral fermions in the theory.

\end{mdframed}

\subsection{But, where are the masses?} 

Adding together eqs.~\eqref{eq:S_matter_SM} and~\eqref{eq:S_gauge_SM},
we still do not get the full action of the electroweak sector of the SM model. The reason is that
all fermion species in the SM have nonvanishing masses and, therefore, we need to add the corresponding 
mass terms to the matter action. This is, however, a very risky business in a chiral theory like the
electroweak model. As we learned in Box~7 (see page~\pageref{box:Dirac_spinors}), fermion mass terms
mix left- and right-handed components. In our case, since they transform in different representations
of the~$\mbox{SU(2)}\times\mbox{U(1)}_{Y}$ gauge group, adding such terms spoils
gauge invariance and with that all hell breaks loose.
 
Fermion masses are not the only problem. Weak interactions are short ranged, something that can only 
be explained if the intermediate bosons~$W^{\pm}$ and~$Z^{0}$ have masses of 
the order of
tens of GeV. Mass terms of the form~$m_{W}^{2}W_{\mu}^{\mp}W^{\pm\mu}$ and~$m_{Z}^{2}Z_{\mu}Z^{\mu}$ 
also violate gauge invariance, so it seems that we are facing double trouble.

The theory resulting from adding all needed mass terms to~$S_{\rm matter}
+S_{\rm gauge}$ is the original model proposed in 1961 by Glashow~\cite{Glashow_SM}, where
gauge invariance in {\em explicitly broken}. The inclusion of masses in the SM in a manner compatible 
with gauge invariance was achieved by Weinberg and Salam~\cite{Weinberg_SM,Salam_SM} and requires the 
implementation of the BEH mechanism~\cite{Brout_Englert,Higgs1,Higgs2}
studied in section~\ref{sec:tale_of_symmetries} in its Abelian version. 
In the case at hand, 
we need to introduce is a~$\mbox{SU(2)}$ complex scalar doublet
\begin{align}
\mathbf{H}=\left(
\begin{array}{cc}
H^{+} \\
H^{0}
\end{array}
\right),
\label{eq:Higgs_doublet_general}
\end{align}
with~$Y(\mathbf{H})={1\over 2}\mathbbm{1}$, 
so using the Gell-Mann--Nishijima relation~\eqref{eq:Gell-Mann-Nishijima} 
we find that~$H^{+}$ has charge~$e$ and~$H^{0}$ is neutral. We consider then the action
\begin{align}
S_{\rm Higgs}=\int d^{4}x\,\left[(D_{\mu}\mathbf{H})^{\dagger}D^{\mu}\mathbf{H}
-{\lambda\over 4}\left(\mathbf{H}^{\dagger}\mathbf{H}-{v^{2}\over 2}\right)^{2}\right],
\label{eq:higgs_actionSM}
\end{align}
where the covariant derivative is defined in~\eqref{eq:gauge_cov_der_SM}.
Although the action is fully~$\mbox{SU(2)}\times\mbox{U(1)}_{Y}$ invariant, the potential has the
Mexican hat shape shown in fig.~\ref{fig:higgs_potential} and the field~$\mathbf{H}$ gets a 
nonzero vev, that by a suitable gauge transformation can always be brought to the form
\begin{align}
\langle\mathbf{H}\rangle={1\over \sqrt{2}}\left(
\begin{array}{cc}
0 \\
v
\end{array}
\right).
\label{eq:higgs_vev}
\end{align}
This vev obviously 
breaks~$\mbox{SU(2)}$ and, having nonzero hypercharge, also~$\mbox{U(1)}_{Y}$. However,
since~$\langle H^{+}\rangle=0$ it nevertheless preserve the gauge invariance of electromagnetism. 
We have then the SSB pattern
\begin{align}
\mbox{SU(2)}\times\mbox{U(1)}_{Y}\longrightarrow \mbox{U(1)}_{\rm em}.
\end{align}

The masses of the gauge bosons are obtained by substituting the vev~\eqref{eq:higgs_vev} into the 
action~\eqref{eq:higgs_actionSM} and collecting the terms quadratic in the gauge fields. With this, we
see that the~$W$ and~$Z$ bosons acquire a nonzero masses given respectively by
\begin{align}
m_{W}={ev\over 2\sin\theta_{w}}, \hspace*{1cm} m_{Z}={ev\over \sin(2\theta_{w})},
\label{eq:massWZ}
\end{align}
and satisfying the custodial relation~$m_{W}=m_{Z}\cos\theta_{w}$. 

Interestingly, the scale~$v$ is related to the Fermi constant~$G_{F}$, a quantity that can be measured
at low energies. Considering the neutron $\beta$-decay process in eq.~\eqref{eq:neutron_beta_dec} 
at energies below the mass of the~$W$ boson and comparing with the result obtained from the
Fermi interaction
\begin{align}
S_{\rm Fermi}={G_{F}\over \sqrt{2}}\int d^{4}x\,\overline{\nu}_{e}\gamma_{\mu}(1-\gamma_{5})e
\,\overline{d}\gamma^{\mu}(1-\gamma_{5})u,
\end{align}
we get the relation
\begin{align}
G_{F}={\sqrt{2}\over 8}{e^{2}\over m_{W}^{2}\sin^{2}\theta_{w}}={1\over \sqrt{2}v^{2}},
\end{align}
where the expression of~$m_{W}$ given in eq.~\eqref{eq:massWZ} has been used.
Substituting now the experimental value of the Fermi 
constant~$G_{F}=1.166 \times 10^{-5}\mbox{ GeV}^{2}$~\cite{PDG}, we find
\begin{align}
v\approx 246\mbox{ GeV}.
\label{eq:value_v_Higgs}
\end{align}

In order to give mass to the fermions, we need to follow the strategy explained in
page~\pageref{page:Yukawa_coupling} and 
write the appropriate Yukawa couplings, which in this case read
\begin{align}
S_{\rm Yukawa}&=-\sum_{i,j=1}^{3}\int d^{4}x\Big(
C_{ij}^{(\ell)}\overline{\mathbf{L}}^{i}\mathbf{H}\ell_{R}^{j}+C_{ji}^{(\ell)*}\overline{\ell}^{i}_{R}
\mathbf{H}^{\dagger}\mathbf{L}^{j}
+C_{ij}^{(q)}\overline{\mathbf{Q}}^{i}\mathbf{H}D_{R}^{j}+C^{(q)*}_{ji}\overline{D}_{R}^{i}\mathbf{H}^{\dagger}\mathbf{Q}^{j}
\nonumber \\[0.2cm]
&+\widetilde{C}^{(q)}_{ij}\overline{\mathbf{Q}}\widetilde{\mathbf{H}}U_{R}^{j}+\widetilde{C}_{ji}^{(q)*}\overline{U}^{i}_{R}
\widetilde{\mathbf{H}}^{\dagger}\mathbf{Q}^{j}\Big).
\label{eq:Yukawa_actionSM}
\end{align}
The two terms in the
second line involve the conjugate field
\begin{align}
\widetilde{\mathbf{H}}\equiv i\sigma^{2}\left(
\begin{array}{cc}
H^{+*} \\
H^{0*}
\end{array}
\right)=\left(
\begin{array}{cc}
H^{0*} \\
-H^{+*}
\end{array}
\right),
\end{align}
which has~$Y(\widetilde{\mathbf{H}})=-{1\over 2}\mathbbm{1}$ and can be seen to transform also as a~$SU(2)$ doublet.
Given the transformation properties of all fields involved, 
it is very easy to check that the action~\eqref{eq:Yukawa_actionSM} is~$\mbox{SU(2)}\times\mbox{U(1)}_{Y}$ gauge invariant. 
Notice that here 
we are assuming that neutrino masses are not due to the BEH mechanism. This is the reason why lepton doublets only couple 
to the Higgs doublet~$\mathbf{H}$, whose upper component has zero vev. In the case of quarks, however, we need to generate masses for
both the upper and lower components of~$\mathbf{Q}$. This is why they couple to the conjugate field~$\widetilde{\mathbf{H}}$,
whose upper component acquires a nonzero vev
\begin{align}
\langle \widetilde{\mathbf{H}}\rangle 
={1\over \sqrt{2}}\left(
\begin{array}{cc}
v \\
0
\end{array}
\right).
\label{eq:HtildevevSM}
\end{align} 
To find the expression of 
the fermions masses generated by the BEH mechanism, we substitute in the Yukawa action the field~$\mathbf{H}$ and
its conjugate~$\widetilde{\mathbf{H}}$ by their vevs~\eqref{eq:higgs_vev} and~\eqref{eq:HtildevevSM}. The resulting 
mass terms have the form
\begin{align}
S_{\rm mass}&=-\int d^{4}x\,\left[\big(\overline{e}_{L},\overline{\mu}_{L},\overline{\tau}_{L}\big)M^{(\ell)}\left(
\begin{array}{c}
e_{R} \\
\mu_{R} \\
\tau_{R}
\end{array}
\right)
+\big(\overline{d}_{L},\overline{s}_{L},\overline{b}_{L}\big)M^{(q)}\left(
\begin{array}{c}
d_{R} \\
s_{R} \\
b_{R}
\end{array}
\right)
\right. \nonumber \\[0.2cm]
&+\left.\big(\overline{u}_{L},\overline{c}_{L},\overline{t}_{L}\big)\widetilde{M}^{(q)}\left(
\begin{array}{c}
u_{R} \\
c_{R} \\
t_{R}
\end{array}
\right)+\mbox{H.c.}
\right],
\end{align}
where the mass matrices are given in term of the couplings in eq.~\eqref{eq:Yukawa_actionSM} by
\begin{align}
M_{ij}^{(\ell)}={v\over \sqrt{2}}C^{(\ell)}_{ij}, 
\hspace*{1cm}
M_{ij}^{(q)}={v\over \sqrt{2}}C^{(q)}_{ij}, 
\hspace*{1cm} 
\widetilde{M}_{ij}^{(q)}={v\over \sqrt{2}}\widetilde{C}_{ij}^{(q)}.
\end{align}
These complex matrices are however not necessarily diagonal, although they can be diagonalized through bi-unitary 
transformations
\begin{align}
U_{L}^{(\ell)\dagger}M^{(\ell)}U_{R}^{(\ell)}&=\mbox{diag}(m_{e},m_{\mu},m_{\tau}), \nonumber \\[0.2cm]
V_{L}^{(q)\dagger}M^{(q)}V_{R}^{(q)}&=\mbox{diag}(m_{d},m_{s},m_{b}), 
\label{eq:V_L,R,tilde}\\[0.2cm]
\widetilde{V}_{L}^{(q)\dagger}\widetilde{M}^{(q)}\widetilde{V}_{R}^{(q)}&=\mbox{diag}(m_{u},m_{c},m_{t}),
\nonumber
\end{align}
where the eigenvalues are the leptons and quarks masses. Notice that fermion masses 
are determined by both the Higgs vev scale~$v$ and the dimensionless Yukawa 
couplings~$C^{(\ell)}_{ij}$,~$C^{(q)}_{ij}$, and~$\widetilde{C}_{ij}^{(q)}$, which are
experimentally determined.

Let us focus for the time being on the quark sector (leptons will be dealt with below in
section~\ref{sec:neutrino_masses}). Since~$V^{(q)}_{L,R}$, $\widetilde{V}^{(q)}_{L,R}$ 
are constant unitary matrices
we could use them to redefine the quark and lepton triplets in the total action 
\begin{align}
\left(
\begin{array}{c}
u'_{L,R} \\
c'_{L,R} \\
t'_{L,R} 
\end{array}
\right)=\widetilde{V}^{(q)\dagger}_{L,R}
\left(
\begin{array}{c}
u_{L,R} \\
c_{L,R} \\
t_{L,R} 
\end{array}
\right), \hspace*{1cm}
\left(
\begin{array}{c}
d'_{L,R} \\
s'_{L,R} \\
b'_{L,R} 
\end{array}
\right)=V^{(q)\dagger}_{L,R}
\left(
\begin{array}{c}
d_{L,R} \\
s_{L,R} \\
b_{L,R} 
\end{array}
\right),
\end{align}
in such a way that the new fields are mass eigenstates, i.e., their free kinetic 
terms in the action have the standard diagonal form. A problem however arises when implementing this
field redefinition in the interaction terms between the quarks 
and the~$W^{\pm}$ gauge bosons, mixing the lower with upper components of the~$SU(2)$ doublets. 
The issue is that, unlike in the kinetic terms,  
the matrices implementing the field redefinition 
do not cancel
\begin{align}
S\supset \int d^{4}x\,\big(\overline{u}_{L},\overline{c}_{L},\overline{t}_{L}\big)\gamma^{\mu}
\left(
\begin{array}{c}
d_{L} \\
s_{L} \\
b_{L}
\end{array}
\right)W^{+}_{\mu}=
\int d^{4}x\,\big(\overline{u}'_{L},\overline{c}'_{L},\overline{t}'_{L}\big)
\widetilde{V}_{L}^{(q)\dagger}V_{L}^{(q)}\gamma^{\mu}
\left(
\begin{array}{c}
d'_{L} \\
s'_{L} \\
b'_{L}
\end{array}
\right)W^{+}_{\mu},
\label{eq:coupling_W+quarks}
\end{align}
where, to simplify the expression 
the overall coupling is omitted and the corresponding coupling of the quarks to the~$W^{-}$ boson
is obtained by taking the Hermitian conjugate of this term. 
The combination
\begin{align}
\widetilde{V}_{L}^{(q)\dagger}V_{R}^{(q)}\equiv V_{\rm CKM}
\label{eq:CKM_matrix}
\end{align} 
defines
the {\em Cabibbo-Kobayashi-Maskawa (CKM) matrix}~\cite{Kobayashi_Maskawa} and determines the mixing 
among the quarks families. It is an experimental fact that this matrix is nondiagonal,
so the emission/absorption of 
a~$W^{\pm}$ boson does not merely transform the upper into the lower fields (or vice versa) {\em within} a single~$\mbox{SU(2)}$ quark doublet, 
but can also ``jump'' into another family. This gives rise to processes  
known as flavor changing charged currents.
For example, there is a nonzero probability that a~$u$ quark turns into a~$s$ quark by the 
emission of a~$W^{+}$, or vice versa with a~$W^{-}$, accounting for
decays like~$\Lambda^{0}\rightarrow p^{+}e^{-}\overline{\nu}_{e}$. 
What happens inside the~$\Lambda^{0}$ baryon~($uds$)
is that the strange quark emits
a~$W^{-}$ and transforms into a~$u$-quark, thus converting the~$\Lambda^{0}$ into a proton ($uud$). 
The~$W^{-}$ then decays into an electron and 
its antineutrino. 

It is an interesting feature of the electroweak sector of the SM that there are no flavor changing
{\em neutral} currents at tree level. In the case of electromagnetic-mediated processes, this follows
from the fact that the field redefinitions induced by the 
matrices~$V^{(q)}_{L,R}$ and~$\widetilde{V}^{(q)}_{L,R}$ mix fields with the same electric charge, 
so they commute with the charge matrix~$Q$ and cancel from the quark electromagnetic couplings.
In the case of the weak neutral currents (mediated by the~$Z^{0}$) the same happens, though maybe 
it is less obvious.
Indeed, looking at the form of the covariant derivative~\eqref{eq:gauge_cov_der_SM} we find 
the following couplings between the quarks and the~$Z^{0}$ 
\begin{align}
S&\supset \int d^{4}x\left[\left({1\over 2}-{2\over 3}\sin^{2}\theta_{w}\right)
(\overline{u}_{L},
\overline{c}_{L},\overline{t}_{L})\gamma^{\mu}\left(
\begin{array}{c}
u_{L} \\
c_{L} \\
t_{L}
\end{array}
\right)
-\left({1\over 2}-{1\over 3}\sin^{2}\theta_{w}\right)
(\overline{d}_{L},
\overline{s}_{L},\overline{b}_{L})\gamma^{\mu}\left(
\begin{array}{c}
d_{L} \\
s_{L} \\
b_{L}
\end{array}
\right)
\right.
\nonumber \\[0.2cm]
&+\left.{2\over 3}\sin^{2}\theta_{w}(\overline{u}_{R},
\overline{c}_{R},\overline{t}_{R})\gamma^{\mu}\left(
\begin{array}{c}
u_{R} \\
c_{R} \\
t_{R}
\end{array}
\right)
-{1\over 3}\sin^{2}\theta_{w}
(\overline{d}_{R},
\overline{s}_{R},\overline{b}_{R})\gamma^{\mu}\left(
\begin{array}{c}
d_{R} \\
s_{R} \\
b_{R}
\end{array}
\right)\right],
\end{align}
where again we have dropped an overall constant
which is irrelevant
for the argument. What matters for
our discussion is that after the field redefinition we get the combinations~$V^{(q)\dagger}_{L,R}
V^{(q)}_{L,R}=\mathbbm{1}=\widetilde{V}^{(q)\dagger}_{L,R}\widetilde{V}^{(q)}_{L,R}$
and no mixing matrix is left behind. This shows that there are 
no flavor changing neutral currents at tree level\footnote{Once quantum effects
are included, flavor changing neutral are suppressed due to the flavor mixing brought
about by the Cabibbo-Kobayashi-Maskawa matrix, via the so-called GIM (Glashow-Iliopoulos-Maiani) 
mechanism~\cite{GIM}.}.
 
\begin{mdframed}[backgroundcolor=lightgray,hidealllines=true]
\vspace*{0.2cm}
\centerline{\greybox{\bf SSB or QCD?}}
\vspace*{0.2cm}

We have seen how the BEH mechanism provides the rationale to understand how the particles
in the~SM acquire their masses, 
a scenario ultimately confirmed by the experimental detection of the Higgs boson. But, does the BEH
mechanism really explains the mass of everything we see around us, from the paper in our hands to
the sun over our heads? The answer is no. As we will see, 
the fraction of the mass of macroscopic objects that 
we can assign to the Higgs boson acquiring a vev is really tiny. 

We know that the masses of protons
and neutrons are very similar to one another, and much larger than the mass of the electron
\begin{align}
m_{p}\simeq m_{n}\simeq 1836\,m_{e}.
\label{eq:mpmne}
\end{align}
In turn, the mass of a~$(A,Z)$ nucleus is
\begin{align}
M(A,Z)=Zm_{p}+(A-Z)m_{n}+\Delta M(A,Z),
\end{align}
with~$\Delta M(A,Z)$ the binding energy, which varies from a bit over~$1\%$ for deuterium
to around~$10\%$ for~$_{28}^{62}\mbox{Ni}$. Taking eq.~\eqref{eq:mpmne} into account and
to a fairly good approximation, the mass of an atom can be written in terms of its mass number alone
\begin{align}
m(A,Z)\simeq Am_{p}.
\end{align}
The point of this argument is to show that in order to explain the 
mass around us we essentially need to explain the mass of the proton. But here we run into trouble
if we want to trace back~$m_{p}$ to the BEH mechanism. The values of the masses of
the~$u$ and $d$~quarks accounted for by the BEH mechanism (the so-called current algebra masses)
are
\begin{align}
m_{u}\simeq 2.2\mbox{ MeV} , \hspace*{1cm} m_{d}=4.7\mbox{ MeV}.
\end{align}
Comparing with~$m_{p}[uud]\simeq 938.3\mbox{ MeV}$ and~$m_{d}[udd]= 939.6\mbox{ MeV}$, we see that
quark masses only explain about~$1\%$ of the nucleon mass. Thus, close to~$99\%$ of the 
mass in atomic form in the universe is not due to the BEH mechanism.

Where does this mass/energy come from? Actually, from QCD effects. Protons and neutrons 
are not only made out of their three valence quarks, but they are filled with a plethora of 
virtual quarks and gluons fluctuating in and out of existence whose energy make up the missing~$99\%$.
These effects can be computed numerically using lattice field theory~\cite{Fodor_Hoelbling,Hatsuda}. 
Here, however, we just want to offer some general arguments pointing to the origin of the 
difficulties in describing protons and neutrons in terms of their constituent quarks. 

Let us begin
with a very simple argument. We know that because of the strong dynamics of QCD at low energies
quarks get confined into hadrons in a region whose linear size is of the order~$\Lambda_{\rm QCD}^{-1}$.  
Applying Heisenberg's uncertainty principle, we can estimate the size of their momentum fluctuations
to be about
\begin{align}
\Delta p\sim \Lambda_{\rm QCD}.
\end{align}
If fluctuations are isotropic the statistical average
of the quark momentum vanishes, $\langle \mathbf{p}\rangle=0$. Since~$(\Delta p)^{2}\equiv
\langle \mathbf{p}^{2}
\rangle-\langle \mathbf{p}\rangle^{2}$, we determine the averaged quark momentum squared to be
\begin{align}
\langle\mathbf{p}^{2}\rangle\sim \Lambda_{\rm QCD}^{2}.
\end{align}
Now,~$\Lambda_{\rm QCD}$ is of the order of a few hundred MeV, so the masses of the~$u$ and $d$~quarks 
satisfy~$m_{u},m_{d}\ll \Lambda_{\rm QCD}$. This means that the linear momenta of the
valence quarks inside protons and neutrons is much larger than their masses, so  
they are relativistic particles. Moreover, since their typical energy is 
of order~$\Lambda_{\rm QCD}$, they are in the low energy regime of QCD where the dynamics is
strongly coupled.

What we said about the~$u$ and $d$~quarks does not apply however to the top~($m_{t}\simeq 
173.7\mbox{ GeV}$), bottom ($m_{b}\simeq\mbox{4.6 GeV}$), and charm ($m_{c}\simeq\mbox{1.3 GeV}$)  
quarks, which under the same conditions
would behave as nonrelativistic particles. Besides, since their energies are dominated by their masses which
are well above~$\Lambda_{\rm QCD}$, their QCD interactions are weakly coupled. 
This is why 
heavy quark bounds states (quarkonium) can be analytically studied using perturbation theory, 
unlike the bound states of
light quarks~($u$, $d$, and~$s$) that have to be treated numerically.  
The difficulties in describing quarks inside protons and neutrons boils down to 
them being untrarelativistic particles.

The moral of the story is that the popular line that the BEH mechanism ``explains''
mass is simply not correct. Most of our own mass and the mass of every objects we see around us 
(and this includes
the Earth, the Sun, the Moon, and the stars in the sky) has nothing to do with the Higgs field
and is the result of the 
quantum behavior of the strong interaction. Even in a universe where the up and down quarks
were massless, the proton and the neutron would still have nonzero masses
and moreover very similar to the ones in our world. 

\end{mdframed}

\subsection{The Higgs boson}

In order to analyze mass generation in the electroweak sector of the SM, it was enough to replace the
scalar doublet~$\mathbf{H}$ by its vev. 
However, as we learned in section~\ref{sec:BEH_mechanism} for the Abelian case, 
the system has excitations around the
minimum of the potential corresponding to a propagating scalar degree of freedom. 
To analyze the dynamics of this field, the {\em Higgs boson}, we write the Higgs
doublet~$\mathbf{H}$ as
\begin{align}
\mathbf{H}(x)={1\over \sqrt{2}}e^{ia^{I}(x)t^{I}_{\mathbf{2}}}\left(
\begin{array}{cc}
0 \\
v+h(x)
\end{array}
\right),
\label{eq:Higgs_field_expansion}
\end{align} 
where~$a^{I}(x)$ and~$h(x)$ are the four real degrees of freedom encoding the two
complex components in~\eqref{eq:Higgs_doublet_general}. In fact, as in the Abelian case
of section~\ref{sec:BEH_mechanism}, we can use
the gauge invariance of~$S_{\rm Higgs}+S_{\rm Yukawa}$ to eliminate the 
global~$\mbox{SU(2)}$ global factor, after which we are left with a single real degree of freedom
representing the Higgs boson~\cite{Higgs2}. 
Substituting into~\eqref{eq:higgs_actionSM} and expanding, we get
\begin{align}
S_{\rm Higgs}&=\int d^{4}x\,\left[{1\over 2}\partial_{\mu}h\partial^{\mu}h-{\lambda v^{2}\over 4}h^{2}
-{\lambda v\over 4}h^{3}-{\lambda\over 16}h^{4}
+{2m_{W}^{2}\over v} W^{-}_{\mu}W^{+\mu}h \right. \label{eq:Higgs_action_h} \\[0.2cm]
&+{m_{W}^{2}\over v^{2}}W^{-}_{\mu}
W^{+\mu}h^{2}+{m_{Z}^{2}\over v} Z_{\mu}Z^{\mu}h
+{m_{Z}^{2}\over 2v^{2}} Z_{\mu}Z^{\mu}h^{2}
+m_{W}^{2}W_{\mu}^{+}W^{-\mu}+{m_{Z}^{2}\over 2}Z_{\mu}Z^{\mu}\bigg]
\nonumber,
\end{align}
where in the last two terms we recognize the 
masses for the $W^{\pm}$ and~$Z^{0}$ gauge bosons. The first thing
to be noticed is that the mass of the Higgs boson
is determined by the vev~$v$ and the strength~$\lambda$ of the Higgs 
quartic self-couplings
\begin{align}
m_{H}=v\sqrt{\lambda\over 2}=(125.25 \pm 0.17)\mbox{ GeV},
\label{eq:higgs_mass}
\end{align}
where the current average experimental value is quoted~\cite{PDG}.
The action~\eqref{eq:Higgs_action_h} also contains the coupling between the Higgs boson and 
the~$W^{\pm}$ and~$Z^{0}$ intermediate bosons, giving rise to the interaction vertices
\begin{eqnarray} 
\nonumber\\
\hspace*{1cm}\parbox{30mm}{
\begin{fmfgraph*}(60,40)
\fmfleft{i1,o1}
\fmfright{i2}
\fmflabel{$W^{\pm},Z^{0}$}{i1}
\fmflabel{$h$}{i2}
\fmflabel{$W^{\pm},Z^{0}$}{o1}
\fmf{photon}{i1,v1,o1}
\fmf{dashes}{i2,v1}
\end{fmfgraph*}
} \hspace*{-0.2cm}\sim \hspace*{0.2cm} {m^{2}_{W,Z}\over v} \hspace*{3cm}  
\parbox{30mm}{
\begin{fmfgraph*}(60,40)
\fmfleft{i1,o1}
\fmfright{i2,o2}
\fmflabel{$W^{\pm},Z^{0}$}{i1}
\fmflabel{$h$}{i2}
\fmflabel{$h$}{o2}
\fmflabel{$W^{\pm},Z^{0}$}{o1}
\fmf{photon}{i1,v1,o1}
\fmf{dashes}{i2,v1,o2}
\end{fmfgraph*}
} \hspace*{-0.2cm}\sim \hspace*{0.2cm} {m^{2}_{W,Z}\over v^{2}}.
\\[0.25cm]
\nonumber
\end{eqnarray} 
In both cases, the strength of the coupling is proportional
to the mass squared of the corresponding intermediate bosons.

As to the coupling of the Higgs boson to fermions, this is obtained by 
replacing~\eqref{eq:Higgs_field_expansion} into the Yukawa action~\eqref{eq:Yukawa_actionSM}
\begin{align}
S_{\rm Yukawa}&=-
\int d^{4}x\,\left[\big(\overline{e}_{L},\overline{\mu}_{L},\overline{\tau}_{L}\big)
\left({1\over v}M^{(\ell)}\right)\left(
\begin{array}{c}
e_{R} \\
\mu_{R} \\
\tau_{R}
\end{array}
\right)h
\right. \\[0.2cm]
&+\big(\overline{d}_{L},\overline{s}_{L},\overline{b}_{L}\big)\left({1\over v}M^{(q)}\right)\left(
\begin{array}{c}
d_{R} \\
s_{R} \\
b_{R}
\end{array}
\right)h
+\left.\big(\overline{u}_{L},\overline{c}_{L},\overline{t}_{L}\big)
\left({1\over v}\widetilde{M}^{(q)}\right)\left(
\begin{array}{c}
u_{R} \\
c_{R} \\
t_{R}
\end{array}
\right)h+\mbox{H.c.}
\right],
\nonumber
\end{align}
This, upon switching to mass eigenstates, takes the general form
\begin{align}
S_{\rm Yukawa}&=-\sum_{f}{m_{f}\over v}\int d^{4}x\,\overline{f}fh,
\end{align}
where~$f=(e',\mu',\tau',u',d',c',s',t',b')$ 
runs over are all the fermion mass eigenstates, apart from the three neutrinos that
we will treat separately. The corresponding interaction vertices are
\begin{eqnarray} 
\nonumber\\
\hspace*{1cm}\parbox{30mm}{
\begin{fmfgraph*}(60,40)
\fmfleft{i1,o1}
\fmfright{i2}
\fmflabel{$f$}{i1}
\fmflabel{$h$}{i2}
\fmflabel{$f$}{o1}
\fmf{fermion}{i1,v1,o1}
\fmf{dashes}{i2,v1}
\end{fmfgraph*}
} \hspace*{-0.2cm}\sim \hspace*{0.2cm} {m_{f}\over v}.
\label{eq:Higgs_coup_fer}\\[0.25cm]
\nonumber
\end{eqnarray} 
That the coupling of the Higgs boson to the fermions is proportional to their masses has 
important experimental consequences. Given the value of the Higgs vev energy scale
found in~\eqref{eq:value_v_Higgs}, only the heaviest fermions have sizeable Higgs couplings, 
in particular the top quark with mass~$m_{t}=173.3\mbox{ GeV}$.
This fact is at the heart of the experimental strategy culminated with the detection 
of the Higgs boson at CERN. In a hadron collider such as LHC, there are plenty of
gluons produced during the collision that can fuse through a top quark loop to produce a Higgs
boson
\begin{align}
\nonumber \\[-0.3cm]
\parbox{40mm}{
\begin{fmfgraph*}(100,65)
\fmfright{i1,i2,i3}
\fmfleft{o1,o2}
\fmf{phantom}{i1,v1,v2,o1}
\fmf{phantom}{i3,v4,v3,o2}
\fmf{phantom,tension=0}{v1,v5,v2}
\fmf{phantom,tension=0}{v2,v3}
\fmf{gluon,tension=0.6}{v3,o2}
\fmf{gluon,tension=0.6}{v2,o1}
\fmffreeze
\fmf{dashes,tension=0.3}{v5,i2}
\fmf{fermion,label=$t$,tension=0.1}{v5,v3,v2,v5}
\fmflabel{$g$}{o2}
\fmflabel{$g$}{o1}
\fmflabel{$h$}{i2}
\end{fmfgraph*}
} \\[-0.4cm]
\nonumber
\end{align}
The Higgs boson produced in the gluon fusion process can decay in various distinctive ways. One
of them is
by a second top loop with emission of two photons 
\begin{align}
\nonumber \\[-0.3cm]
\parbox{40mm}{
\begin{fmfgraph*}(100,65)
\fmfleft{i1,i2,i3}
\fmfright{o1,o2}
\fmf{phantom}{i1,v1,v2,o1}
\fmf{phantom}{i3,v4,v3,o2}
\fmf{phantom,tension=0}{v1,v5,v2}
\fmf{phantom,tension=0}{v2,v3}
\fmf{photon,tension=0.6}{v3,o2}
\fmf{photon,tension=0.6}{v2,o1}
\fmffreeze
\fmf{dashes,tension=0.3}{v5,i2}
\fmf{fermion,label=$t$,tension=0.1}{v5,v2,v3,v5}
\fmflabel{$\gamma$}{o2}
\fmflabel{$\gamma$}{o1}
\fmflabel{$h$}{i2}
\end{fmfgraph*}
}
\end{align}
Alternatively, the Higgs boson may produce  
a pair of~$Z^{0}$ bosons that in turn decay into two lepton-antilepton pairs
\begin{align}
\nonumber \\[-0.3cm]
\parbox{40mm}{
\begin{fmfgraph*}(75,65)
\fmfleft{i1}
\fmfright{o1,o2,o3,o4}
\fmflabel{$h$}{i1}
\fmflabel{$\ell$}{o1}
\fmflabel{$\overline{\ell}$}{o2}
\fmflabel{$\ell$}{o3}
\fmflabel{$\overline{\ell}$}{o4}
\fmf{dashes,tension=1}{i1,v1}
\fmf{photon,label=$Z^{0}$,tension=0.4}{v2,v1}
\fmf{photon,label=$Z^{0}$,tension=0.4}{v1,v3}
\fmf{fermion,tension=0.5}{o2,v2,o1}
\fmf{fermion,tension=0.5}{o4,v3,o3}
\end{fmfgraph*}
} 
\\[-0.4cm]
\nonumber
\end{align}
These were precisely the decay channels that led to the discovery of the Higgs boson
by the ATLAS and CMS collaborations at the LHC~\cite{Higgs_ATLAS,Higgs_CMS}.

\subsection{Neutrino masses}
\label{sec:neutrino_masses}

We have been postponing the issue of neutrinos masses. 
It is however an experimental fact that neutrinos have nonzero masses and this is something we 
have to incorporate in the SM action.
One way to do it is to extend the SM to include right-handed {\em sterile}
neutrinos~$\nu_{R}^{i}$ transforming as~$(\mathbf{1},\mathbf{1})_{0}$ 
under~$\mbox{SU(3)}\times\mbox{SU(2)}\times\mbox{U(1)}_{Y}$
(see the notation introduced page~\pageref{page:(p,q)_Y_notation}),
adding then the following terms
to the Yukawa action
\begin{align}
\Delta S_{\rm Yukawa}=-\sum_{i=1}^{3}\int d^{4}x\,\Big(\widetilde{C}^{(\nu)}\overline{\mathbf{L}}^{i}
\widetilde{\mathbf{H}}\nu_{R}^{i}+\widetilde{C}^{(\nu)*}_{ji}\overline{\nu}_{R}^{i}
\widetilde{\mathbf{H}}\mathbf{L}^{j}\Big).
\end{align}
Once the Higgs field gets a vev, this term generate a mass term of the form
\begin{align}
\Delta S_{\rm Yukawa}=-\int d^{4}x\,\left[
(\overline{\nu}_{e L},\overline{\nu}_{\mu L},\overline{\nu}_{\tau L})
\widetilde{M}^{(\nu)}\left(
\begin{array}{c}
\nu_{1R} \\
\nu_{2R} \\
\nu_{3R}
\end{array}
\right)+\mbox{H.c.}
\right],
\end{align}
with
\begin{align}
M^{(\nu)}_{ij}={v\over \sqrt{2}} \widetilde{C}^{(\nu)}_{ij}.
\label{eq:Dirac_neut_mass_matrix}
\end{align} 
Being singlets under all SM gauge groups, the sterile neutrinos only interact
gravitationally with other particles.

\begin{mdframed}[backgroundcolor=lightgray,hidealllines=true]
\vspace*{0.2cm}
\centerline{\greybox{\bf Dirac vs. Majorana fermions}}
\vspace*{0.2cm}

In previous sections, we have shown how antiparticles in QFT are somehow related to complex fields,
for example in the complex scalar field discussed in Box~6 
(see page~\pageref{page:box6_complex_fields}). In this case, particles are interchanged with 
antiparticles by replacing the field~$\varphi(x)$ with its complex conjugate~$\varphi(x)^{*}$.
To make things more elegant, we may call this operation {\em charge conjugation} and
the result the {\em charge conjugated field}
\begin{align}
\mbox{C}:\varphi(x) \longrightarrow \eta_{C}\varphi(x)^{*}\equiv \varphi^{c}(x),
\end{align}
where~$\eta_{C}$ is some phase that we are always free to add while keeping the
action~\eqref{eq:action_free_scalar_field} invariant. At the quantum level,~C
does indeed interchange particles and antiparticles
\begin{align}
\mbox{C}|\mathbf{p};0\rangle =\eta_{C}^{*}|0;\mathbf{p}\rangle, \hspace*{1cm}
\mbox{C}|0;\mathbf{p}\rangle =\eta_{C}|\mathbf{p};0\rangle.
\end{align}
From this perspective, a {\em real} scalar field is one identical to its charge 
conjugate,~$\varphi(x)=\varphi^{c}(x)$. After quantization, 
its elementary excitations are their own antiparticles.

Let us try to make something similar with the Dirac field. In the scalar field case,
replacing~$\varphi(x)$ by $\varphi(x)^{*}$ does not change the field's Lorentz transformation properties, 
after all complex conjugate or not 
both fields are {\em scalars}. Not so for a Dirac fermion. 
The spinor~$\psi(x)$ and its complex conjugate~$\psi(x)^{*}$ do not transform the same
way under the Lorentz group and neither satisfy the same Dirac equation. This means that 
we cannot define a ``real'' Dirac
spinor just
requiring~$\psi(x)=\psi(x)^{*}$. We have to work a little bit more and consider 
\begin{align}
\mbox{C}:\psi(x)\longrightarrow \eta_{C}(-i\gamma^{2})\psi(x)^{*}\equiv \psi^{c}(x),
\end{align}
where~$\eta_{C}$ again is a complex phase.
This charge conjugate spinor transforms in the same way as the original field and also satisfies
the same free Dirac equation. Moreover, its action on the multi-particle states
generated by the creation operators~$\widehat{b}(\mathbf{k},s)^{\dagger}$
and~$\widehat{d}(\mathbf{k},s)^{\dagger}$ in eq.~\eqref{eq:Dirac_op_bdscaop}
is given by
\begin{align}
\mbox{C}|\mathbf{k},s;0\rangle=\eta_{C}^{*}|0;\mathbf{k},s\rangle, \hspace*{1cm}
\mbox{C}|0;\mathbf{k},s\rangle=\eta_{C}|\mathbf{k},s;0\rangle,
\label{eq:Cfermion_oneparticle}
\end{align} 
and interchanges particles and antiparticles.

The spinor analog of the real scalar field is a 
{\em Majorana spinor}, which equals its charge conjugate
\begin{align}
\psi(x)=\psi^{c}(x).
\label{eq:Majorana_cond_general}
\end{align}
Upon quantization, this identifies particles and antiparticles,
as follows from eq.~\eqref{eq:Cfermion_oneparticle}.
It is interesting to implement the Majorana condition expressing the Dirac fermion in terms of
its chiral components and using
the representation~\eqref{eq:Dirac_matr_repschiral} of the Dirac matrices
\begin{align}
\left(
\begin{array}{c}
\chi_{+} \\
\chi_{-}
\end{array}
\right)=\eta_{C}
\left(
\begin{array}{c}
i\sigma^{2}\chi_{-}^{*} \\
-i\sigma^{2}\chi_{+}
\end{array}
\right) \hspace*{1cm} \Longrightarrow \hspace*{1cm}
\psi={1\over \sqrt{2}}
\left(
\begin{array}{c}
\chi_{+} \\
-i\eta_{C}\sigma^{2}\chi_{+}^{*}
\end{array}
\right).
\end{align} 
In the second identity we wrote a solution to~\eqref{eq:Majorana_cond_general}, 
and a similar expression can be written
in terms of the negative chirality component~$\chi_{-}$. Here we see how the Majorana condition
halves the four complex components of a Dirac field down to two.
In fact, the Majorana 
spinor can be written as the sum of a Weyl fermion and its charge conjugate as
\begin{align}
\psi={1\over \sqrt{2}}
\left(
\begin{array}{c}
\chi_{+} \\
0
\end{array}
\right)+
{1\over \sqrt{2}}
\left(
\begin{array}{c}
0 \\
-i\eta_{C}\sigma^{2}\chi_{+}
\end{array}
\right)\equiv {1\over \sqrt{2}}\big(\psi_{+}+\psi_{+}^{c}\big).
\label{eq:Weyl+CWeyl}
\end{align}
Using this expression, we write the Dirac action for a Majorana fermion
\begin{align}
S&=\int d^{4}x\,\left[i\overline{\psi}_{+}{\partial\!\!\!/\,}
\psi_{+}
-{m\over 2}\big(\overline{\psi^{c}}_{\!\!\!+}\psi_{+}+\overline{\psi}_{+}\psi^{c}_{+}\big)
\right],
\label{eq:action_Majorana_final}
\end{align}
Unlike Weyl fermions, Majorana spinors admit a mass term
without doubling the number of degrees of freedom.

An important point concerning Majorana fermions is that they cannot be coupled to 
the electromagnetic field. This is to be expected, since the Majorana condition identifies particles
with antiparticles that, as we saw in Box~7, have opposite electric charge. In more precise
terms what happens is that the associated Noether current vanishes
\begin{align}
j^{\mu}=\overline{\psi}\gamma^{\mu}\psi={1\over 2}\Big(\chi_{+}^{\dagger}\sigma_{+}^{\mu}\chi_{+}
+\chi_{+}^{T}\sigma^{\mu T}_{+}\chi_{+}^{*}\Big)=0.
\end{align}
This can be also seen as a consequence of the incompatibility of the Majorana 
condition~\eqref{eq:Majorana_cond_general} with a global~$\mbox{U(1)}$ phase
rotation of the spinor~$\psi\rightarrow e^{i\vartheta}\psi$. In particular, the Majorana mass
term in~\eqref{eq:action_Majorana_final} does not conserve the~U(1) charge
\begin{align}
\overline{\psi^{c}}_{\!\!\!+}\psi_{+}+\overline{\psi}_{+}\psi^{c}_{+}
\longrightarrow e^{2i\theta}
\overline{\psi^{c}}_{\!\!\!+}\psi_{+}+e^{-2i\theta}\overline{\psi}_{+}\psi^{c}_{+},
\label{eq:U(1)_viol_Majorana_term}
\end{align}
a very important feature for
the accidental symmetries of the SM such as lepton number.

\end{mdframed}

The addition of sterile neutrinos to generate neutrino masses is only partly satisfactory. One 
obvious problem is 
its lack of economy, since it requires the addition of extra species to the SM
that nevertheless do not partake in its interactions. But the solution is also unnatural. 
Due to the smallness of the neutrino masses, 
the new Yukawa couplings have
to be many orders of magnitude smaller than the ones for charged leptons.

Generating a Dirac mass term is not the only possibility of accounting for neutrino masses. 
Having zero electric charge, they are the only fermions in the SM that
can be of Majorana type. If this were the case, 
their mass terms in the action would be build from the left components
alone, as we saw in Box~15
\begin{align}
\Delta S=-\sum_{i,j=1}^{3}
\int d^{4}x\left({1\over 2}M_{ij}\overline{\nu^{ic}}_{\!\!L}\nu_{L}^{j}+\mbox{H.c.}\right),
\end{align} 
where because of Fermi statistics~$\overline{\nu^{ic}}_{\!\!L}\nu_{L}^{j}
=\overline{\nu^{jc}}_{\!\!L}\nu_{L}^{i}$ and the mass matrix~$M^{(\nu)}_{ij}$ can be taken to be symmetric.
The problem now lies in how to generate a Majorana mass from a 
coupling of the neutrinos to the Higgs field, since 
both~$\mathbf{L}^{i}$ and
its charge conjugate are both~$\mbox{SU(2)}$ doublets and there is
no way to construct a gauge invariant {\em dimension four} 
operator involving~$\mathbf{L}^{i}$,~$\mathbf{L}^{ic}$,
and~$\mathbf{H}$ (or~$\widetilde{\mathbf{H}}$). A group-theoretical way to
see this is by noticing that the product representation~$\mathbf{2}\otimes
\mathbf{2}\otimes\mathbf{2}=\mathbf{4}\oplus\mathbf{2}\oplus\mathbf{2}$ does not
contain any $\mbox{SU(2)}$~singlet. This changes if we admit a dimension-five 
operator with two Higgs doublets,
a left-handed fermion and its charge conjugate. Now it is possible to 
construct a gauge invariant term since~$\mathbf{2}\otimes\mathbf{2}\otimes
\mathbf{2}\otimes\mathbf{2}=\mathbf{5}\oplus\mathbf{3}\oplus\mathbf{3}\oplus\mathbf{3}
\oplus\mathbf{1}\oplus\mathbf{1}$. For example
\begin{align}
\Delta S
&=-{1\over M}\sum_{i,j=1}^{3}\int d^{4}x\left[C_{ij}^{(\nu)}\left(\overline{\mathbf{L}^{ic}}
\,\widetilde{\mathbf{H}}^{*}\right)\left(\widetilde{\mathbf{H}}^{\dagger}\mathbf{L}^{j}\right)+\mbox{H.c.}\right],
\label{eq:Higgs_dim5_neutrinomass}
\end{align}
is invariant under~$\mbox{SU(2)}\times\mbox{U(1)}_{Y}$. This operator in the action has to be understood,
in the spirit of EFT, as the result of some new physics appearing
at the energy scale~$M\gg v$, with~$v$ the
Higgs vev.

When the Higgs field acquires its vev, the coupling~\eqref{eq:Higgs_dim5_neutrinomass}
generates a Majorana mass term for the neutrinos
\begin{align}
\Delta S
=-{1\over 2}\sum_{i,j=1}^{3}\int d^{4}x\Big(M_{ij}^{(\nu)}
\overline{\nu^{ic}}_{\!\!L}\nu_{L}^{j}+\mbox{H.c.}\Big),
\label{eq:Majorana_mass_neutrinos_afterSSB}
\end{align}
where the neutrino mass matrix is given by
\begin{align}
M_{ij}^{(\nu)}={v^{2}\over M}C_{ij}^{(\nu)}.
\label{eq:mass_matrix_Majorana_neut}
\end{align}
The entries of this matrix are suppressed by the factor~$v/M\ll 1$, naturally producing neutrinos with masses
well below the ones of the charged leptons. Thus, Majorana neutrinos not only are the most economical
solution, making unnecessary adding new fermion species,
but also avoids the unnaturalness of the neutrino Yukawa couplings. 
Incidentally, the Majorana mass term~\eqref{eq:Majorana_mass_neutrinos_afterSSB}
violates lepton number, since~$\nu_{L}^{j}$ and~$\overline{\nu^{ic}}_{\!\!L}$ transform with 
the same phase~[cf.~\eqref{eq:U(1)_viol_Majorana_term}].

Neutrinos are regarded as one of the most promising windows to physics beyond the~SM,
being the main reason why neutrino physics has remained for decades 
one of the most exciting fields in (astro)particle
physics and cosmology~\cite{Giunti_Kim,Lesgourges,Bilenky}.
As to the question of whether the neutrino is a Dirac or a Majorana particle, however,
the jury is still out.
Some processes can only take place if the neutrino is its own
antiparticle, most notably neutrinoless double $\beta$~decay~\cite{Bilenky_Giunti,Jones_ndbd}. 
A nucleus with mass and atomic 
numbers~$(A,Z)$ can undergo double $\beta$-decay and transmute into the 
nucleus~$(A,Z+2)$ with emission of
two electrons and two antineutrinos:
\begin{align}
\begin{array}{rlllll}
(A,Z) & \longrightarrow & (A,Z+1) & \hspace*{-0.2cm} 
+\,\,\,e^{-} & \hspace*{-1.2cm}\!+\,\,\,\overline{\nu}_{e} & \\
   &   & \hspace*{1cm}\mbox{$|\hspace{-0.09cm}$\raisebox{-0.5 em}{$\longrightarrow$}} & 
   \raisebox{-0.5 em}{$(A,Z+2)$} &\raisebox{-0.5 em}{$\!\!+\,\,\,e^{-}$} & 
   \raisebox{-0.5 em}{$\!\!\!+\,\,\,\overline{\nu}_{e}$}
\end{array}.
\end{align}
If the neutrino is a Majorana particle there is an alternative. The neutrino
produced in the first decay may interact with a neutron in the nucleus, turning it into a proton
with the emission of an electron
\begin{align}
\overline{\nu}_{e}(\equiv \nu_{e})+n\longrightarrow p^{+}+e^{-},
\end{align}
so no neutrino is emitted in the process~$(A,Z)\rightarrow (A,Z+2)+2e^{-}$.
This is describe by the diagram
\begin{align}
\nonumber \\[-0.4cm]
\parbox{35mm}{
\begin{fmfgraph*}(110,90)
\fmfleft{i1,i2}
\fmfright{o1,o2}
\fmflabel{$(A,Z)$}{i1}
\fmflabel{$e^{-}$}{o1}
\fmflabel{$(A,Z+2)$}{i2}
\fmflabel{$e^{-}$}{o2}
\fmf{heavy,tension=0.5}{i1,v1}
\fmf{photon,tension=.15,label=$W^{-}$,l.side=right}{v1,v4}
\fmf{fermion,tension=0.5}{v4,o1}
\fmf{fermion,tension=0.5}{v3,o2}
\fmf{photon,tension=.15,label=$W^{-}$,l.side=left}{v2,v3}
\fmf{heavy,tension=0.5}{v2,i2}
\fmfblob{.10w}{v1}
\fmfblob{.10w}{v2}
\fmf{heavy,tension=.2,label=$(A,,Z+1)$,l.side=left}{v1,v2}
\fmf{alt_majorana,tension=.2,label=$\nu_{e}$,l.side=left}{v3,v4}
\fmffreeze
\end{fmfgraph*}
}
\\[-0.4cm]
\nonumber
\end{align}
where the double-arrowed line represents the Majorana neutrino. The detection of neutrinoless 
double $\beta$-decay would decide the question of the Dirac or Majorana character of the neutrino.
A lot of experimental effort is being dedicated to this problem, 
so far without definite results (see~\cite{GGY}
for an updated overview of past, present, and future experiments).

\begin{mdframed}[backgroundcolor=lightgray,hidealllines=true]
\vspace*{0.2cm}
\centerline{\greybox{\bf CP violation and the CKM and PMNS matrices}}
\vspace*{0.2cm}

When studying the strong CP problem in section~\ref{sec:CP_violation}, we hinted at the fact the
CP violation is associated with the existence of complex couplings in the action.  
This is shown easily, taking into account that the
CP~transformation acting on an operator~$\mathscr{O}$ transforms it into its Hermitian 
conjugate,~$\mbox{CP}\mathscr{O}(\mbox{CP})^{-1}=\mathscr{O}^{\dagger}$. Hence, a term
in the Hamiltonian of the form~$g\mathscr{O}+g^{*}\mathscr{O}^{\dagger}$, although being
Hermitian, leads to CP violation
unless the coupling is real~$g=g^{*}$. This is why when exploiting the axial anomaly to move the $\theta$~dependence 
in the QCD action from the $\theta$-term into a complex phase in the fermion mass matrix we said that we were {\em shifting} the source of
CP-violation to a complex coupling. 

Besides the $\theta$-term in the QCD action, 
it is a fact that CP symmetry is broken in the electroweak sector of the SM, 
for example in neutral kaon decays. Its origin is found in the unitary CKM matrix
\begin{align}
V_{\rm CKM}=\left(
\begin{array}{ccc}
V_{ud} & V_{us} & V_{ub} \\
V_{cd} & V_{cs} & V_{cb} \\
V_{td} & V_{ts} & V_{tb}
\end{array}
\right),
\label{eq:CKM_matrix_general}
\end{align}
introduced in~\eqref{eq:CKM_matrix} since,
as we will see now, it contains a complex phase that cannot be removed by redefinition of the quark fields. 
Let us be general and analyze the case of a SM with~$n$ 
families. An~$n\times n$ unitary matrix depends on~$n^{2}$ real parameters (the~$2n^{2}$ real parameter
of a general complex matrix reduced by the~$n^{2}$ conditions imposed by unitarity). 
In addition to this, we can play with the phases of the~$2n$ quarks, keeping in mind
the invariance of the action
under a common phase redefinition of all quark fields leading
to (perturbative) baryon number conservation. This means that~$2n-1$ of the~$n^{2}$ real 
parameters can be absorbed in the phases of the quark fields, and we are left with~$n^{2}-2n+1=(n-1)^{2}$
independent ones. The question is how many of them correspond to complex phases. To decide this, 
let us recall that were the~CKM matrix real it would be an~$\mbox{SO($N$)}$ matrix depending 
on~${1\over 2}n(n-1)$ real angles. Subtracting this number from the total number of independent real parameters computed above, 
we get the final number of complex phases in the CKM matrix to be
\begin{align}
n^{2}-2n+1-{1\over 2}n(n-1)={1\over 2}(n-1)(n-2).
\label{eq:numbercomplexCKM}
\end{align} 

For three families~$(n=3)$ the matrix depends on a single complex phase~$e^{i\delta}$ and three real 
angles~$\theta_{12}$, $\theta_{13}$, and~$\theta_{23}$. In terms of them,
the CKM~matrix is usually parametrized as
\begin{align}
V_{\rm CKM}=\left(
\begin{array}{ccc}
c_{12}c_{13} & s_{12}c_{13} & s_{13}e^{-i\delta} \\
-s_{12}c_{23}-c_{12}s_{23}s_{13}e^{i\delta} & c_{12}c_{23}-s_{12}s_{23}s_{13}e^{i\delta} & s_{23}c_{13} \\
s_{12}s_{23}-c_{12}c_{23}s_{13}e^{i\delta} & -c_{12}s_{23}-s_{12}c_{23}s_{13}e^{i\delta} & c_{23}c_{13}
\end{array}
\right),
\label{eq:CKM_parametrization1}
\end{align}
where~$s_{ij}\equiv \sin\theta_{ij}$ and~$c_{ij}\equiv \cos\theta_{ij}$.
The modulus of the entries can be measured through the observation of various
weak interaction mediated decays and scattering processes (see for example~\cite{Sozzi}),
with the result~\cite{PDG}
\begin{align}
|V_{\rm CKM}|=\left(
\begin{array}{ccc}
0.97435\pm 0.00016 &  0.22500 \pm 0.00067 & 0.00369 \pm 0.00011 \\
0.22486\pm 0.00067 & 0.97349 \pm 0.00016 & 0.04182^{+0.00085}_{-0.00074} \\
0.00857^{+0.00020}_{-0.00018} & 0.04110^{+0.00083}_{-0.00072} & 0.999118^{+0.000031}_{-0.000036}
\end{array}
\right).
\end{align}
while the value of the CP-violating phase is~$\delta=1.144\pm 0.027$.
The experimental measurement of~$|V_{\rm CKM}|$ exhibits a clear hierarchy among its entries, derived 
from~$s_{13}\ll s_{23}\ll s_{12}\ll 1$. This 
is manifest in the so-called Wolfenstein parametrization~\cite{Wolfenstein}
\begin{align}
V_{\rm CKM}=\left(
\begin{array}{ccc}
1-{1\over 2}\lambda^{2} & \lambda & A\lambda^{3}(\rho-i\eta) \\
-\lambda & 1-{1\over 2}\lambda^{2} & A\lambda^{2} \\
A\lambda^{3}(1-\rho-i\eta) & -A\lambda^{2} & 1 
\end{array}
\right)+\mathcal{O}(\lambda^{4}),
\label{eq:CKM_Wolfenstein}
\end{align}
where~$\lambda\equiv s_{12}$. The diagonal elements are all of order one, 
whereas the size of the other entries decreases as we move away from it.

A look at~\eqref{eq:numbercomplexCKM} shows that with just two families 
the corresponding flavor mixing
matrix would contain no complex phases and depend on a single real parameter, 
the Cabibbo angle~$\theta_{C}\equiv \theta_{12}$~\cite{Cabibbo}. Thus, CP violation in the 
electroweak sector, like the one showing up in for example kaon decay,
requires the existence of at least three~SM families. 

CP-violation in the SM is of major importance, since it
is a basic ingredient to explain why there is such a tiny amount of antimatter in our 
universe. However, the amount of CP~violation produced by the single complex phase
of the CKM matrix is far too small to account for the 
observed matter-antimatter asymmetry~\cite{Cline_baryogenesis}. Finding additional sources
in or beyond the SM is one of the big open problems in contemporary high energy physics.

Maybe the lepton sector is a good place to look for more CP~violation. 
As with quarks, lepton masses appear when switching from interaction 
to mass eigenstates by diagonalizing the lepton mass matrix. 
Redefining the massive lepton fields
\begin{align}
\left(
\begin{array}{c}
e'_{L,R} \\
\mu'_{L,R} \\
\tau'_{L,R}
\end{array}
\right)=U^{(\ell)}_{L,R}
\left(
\begin{array}{c}
e_{L,R} \\
\mu_{L,R} \\
\tau_{L,R}
\end{array}
\right)
\end{align}
with~$U^{(\ell)}_{L,R}$ defined in eq.~\eqref{eq:V_L,R,tilde}, 
the interaction terms with
the~$W^{\pm}$ bosons take the form
\begin{align}
S\supset 
\int d^{4}x\,\left[\big(\overline{e}'_{L},\overline{\mu}'_{L},\overline{\tau}'_{L}\big)
U_{L}^{(\ell)\dagger}\gamma^{\mu}
\left(
\begin{array}{c}
\nu_{eL} \\
\nu_{\mu L} \\
\nu_{\tau L}
\end{array}
\right)W^{+}_{\mu}+\mbox{H.c.}\right].
\label{eq:Winter_lepton_sec_nnm}
\end{align}
Here, the Hermitian conjugate term contains the interaction with the~$W^{-}$ and we have
dropped the global normalization. 
In the original version of the SM there are no right-handed neutrinos and therefore we can 
reabsorb the matrix~$U_{L}^{(\ell)\dagger}$ in a redefinition of the left-handed
neutrino fields,
without it appearing elsewhere in the SM action. As a result, if the
the neutrino were massless there would be no flavor mixing in the lepton sector.

Things are drastically different once we add the neutrino mass terms. Let us consider first
the case of Dirac masses. As with quarks and charged leptons, 
the mass matrix in eq.~\eqref{eq:Dirac_neut_mass_matrix}
can be diagonalized by a bi-unitary transformation
\begin{align}
U^{(\nu)\dagger}_{L}M^{(\nu)}U^{(\nu)}_{R}=\mbox{diag}(m_{1},m_{2},m_{3}),
\end{align}
and the interaction term~\eqref{eq:Winter_lepton_sec_nnm} is recast in terms of
neutrino mass eigenstates as
\begin{align}
S\supset 
\int d^{4}x\,\left[\big(\overline{e}'_{L},\overline{\mu}'_{L},\overline{\tau}'_{L}\big)
U_{L}^{(\ell)\dagger}U_{L}^{(\nu)}\gamma^{\mu}
\left(
\begin{array}{c}
\nu_{1L} \\
\nu_{2L} \\
\nu_{3L}
\end{array}
\right)W^{+}_{\mu}+\mbox{H.c.}\right],
\label{eq:mixing_leptonic_SMaction}
\end{align}
where
\begin{align}
U\equiv U_{L}^{(\ell)\dagger}U_{L}^{(\nu)}=\left(
\begin{array}{ccc}
U_{e1} & U_{e2} & U_{e3} \\
U_{\mu 1}& U_{\mu 2} & U_{\mu 3} \\
U_{\tau 1} & U_{\tau 2} & U_{\tau 3}
\end{array}
\right),
\end{align}
is the Pontecorvo-Maki-Nakagawa-Sakata (PMNS) unitary matrix~\cite{Pontecorvo,MNS}.
Similarly to what the CKM matrix does for quarks, 
the PMNS matrix introduces flavor mixing in the leptonic sector.
Moreover, following the same reasoning as with the CKM matrix, 
we see that for three families the PMNS matrix also depends on three real angles and a single 
complex phase, representing an additional source of CP~violation. It also admits a parametrization
similar to the one shown in eq.~\eqref{eq:CKM_parametrization1} for the CKM matrix where
the phase is denoted by~$\delta_{\rm CP}$.

For Majorana neutrinos, however, the mass matrix~\eqref{eq:mass_matrix_Majorana_neut} is symmetric
and can be diagonalized by a {\em unitary} transformation
\begin{align}
U^{(\nu)T}_{L}MU^{(\nu)}_{L}=\mbox{diag}(m_{1},m_{2},m_{3}),
\end{align}
so switching to neutrino mass eigenstates we find again an interaction term of
the form~\eqref{eq:mixing_leptonic_SMaction}. The big
difference with respect to the Dirac case is that
since the Majorana mass term~\eqref{eq:Majorana_mass_neutrinos_afterSSB} is not invariant
under phase rotations of the neutrino fields, we cannot get rid of two of three phases 
in the PMNS matrix. As a consequence, besides the three 
angles~$\theta_{12}$,~$\theta_{13}$,~$\theta_{23}$ and the phase~$e^{i\delta_{\rm CP}}$
of the Dirac case, the matrix depends now on two additional complex 
phases~$e^{i\lambda_{1}}$ and~$e^{i\lambda_{2}}$, known as Majorana phases.
The three angles and~$\delta_{\rm CP}$ can be measured from the neutrino oscillations, whereas
the measurement of the two Majorana phases would be possible through the observation of
neutrinoless double $\beta$~decay~\cite{GGY}.
Fits of neutrino data (including the Super-Kamiokande
atmospheric neutrino data) give the following $3\sigma$~ranges for
the absolute values of the entries of the PMNS matrix~\cite{EGGMSZ}
\begin{align}
|U|=
\left(
\begin{array}{ccc}
0.801 \rightarrow 0.845 & 0.513 \rightarrow 0.579 & 0.143 \rightarrow 0.155 \\
0.234 \rightarrow 0.500 & 0.471 \rightarrow 0.689 & 0.637 \rightarrow 0.776 \\
0.271 \rightarrow 0.525 & 0.477 \rightarrow 0.694 & 0.613 \rightarrow 0.756
\end{array}
\right).
\label{eq:PMNS_values}
\end{align}
It is interesting to compare the textures of the matrices~\eqref{eq:CKM_Wolfenstein} 
and~\eqref{eq:PMNS_values}. As already mentioned,
for quarks the matrix is of order~1 at the diagonal,~$\lambda$ 
for the second diagonal, and~$\lambda^{2}$ in the upper right and lower left corners. 
There seems to be a hierarchical pattern (this is a bit of wishful thinking clearly). In the case of 
neutrinos, however, it seems that there is democracy in all its entries, and a crude approximation 
to~\eqref{eq:PMNS_values} would be to set all its entries to 1. This is a matrix with a single nonzero 
eigenvalue and two degenerate zeros, reminiscent of the normal or inverted hierarchies in the fit of the 
neutrino masses. Both textures are so different, that it is difficult to imagine that they have a common 
origin. A major mystery, whose clarification is beyond the SM.

\end{mdframed}

\section{Scale invariance and renormalization}
\label{sec:renormalization}

Renormalization appeared in physics as a way to make sense of
the divergent results in QFT. In quantum mechanics, infinities
are usually handled by invoking a normal ordering prescription, and even
in QFT, they are absent when computing semiclassical contributions to processes
in perturbation theory\footnote{Here
we are going to be concerned with UV divergences associated with the high energy 
regime of the theory. IR divergences, which appear in the limit of low momenta, cancel once
the physical question is properly posed and
all contributions to the given process are taken into account.}.
The trouble comes when calculating quantum corrections, associated in the perturbative
expansion to
Feynman diagrams with closed loops. These contain integrals 
over all independent momenta running in the loops that are frequently divergent.

We will not enter into the many details and subtleties involved in the study 
of divergences in QFT and the philosophy and practicalities of renormalization. They are explained 
in all major textbooks on the subject and a concise and not too technical overview can be found
in chapter~8 of~\cite{AG_VM}. The first step is to make the divergent integrals finite in order
to handle them mathematically. This is done by introducing a proper regulator, that can 
either be a scale where loop momenta are cut off or a more abstract procedure to
render the integrals finite, such as playing with the dimension of spacetime or introducing
PV fermions. In any case, regularization implies the introduction of 
an energy scale~$\Lambda$, called the cutoff for short. The basic point is that this cutoff 
is an artefact of the calculation and cannot appear in any {\em physical} quantity that we compute.

Roughly speaking, renormalization consists on getting rid of the cutoff. The key point to do  
this is the realization that the masses, couplings, and the fields themselves 
appearing in the classical action are 
not physical quantities. Therefore, there is nothing wrong with them depending on~$\Lambda$.
What must be cutoff independent are the physical quantities that we compute and can (and will) 
be compared with experiments. This quantities are {\em operationally defined}, in the sense that
their definition within the theory's framework is given in terms of the process to be used to
measure them. An example is the self-interacting scalar theory
\begin{align}
S=\int d^{4}x\,\left({1\over 2}\partial_{\mu}\varphi\partial^{\mu}\varphi-{m^{2}\over 2}\varphi^{2}
-{\lambda\over 4!}\varphi^{4}\right),
\label{eq:lambdaphi4}
\end{align}
where we would like to define the physical coupling~$\lambda_{\rm phys}$. We could identify it
as the value of the scattering amplitude for four scalar particles when all~$\mathbf{p}_{i}^{2}$
are equal
\begin{align}
\nonumber \\[-0.5cm]
\lambda_{\rm phys}\equiv \hspace*{0.2cm}
\hspace*{0.4cm}
\left.
\vphantom{
\begin{array}{c}
\\[1.5cm]
\end{array}
}
\parbox{30mm}{
\begin{fmfgraph*}(55,30)
\fmfleft{i1,i2}
\fmfright{o1,o2}
\fmflabel{$p_{2}$}{i1}
\fmflabel{$p_{3}$}{o1}
\fmflabel{$p_{1}$}{i2}
\fmflabel{$p_{4}$}{o2}
\fmfblob{.25w}{v1}
\fmf{plain}{i1,v1,i2}
\fmf{plain}{o1,v1,o2}
\end{fmfgraph*}
}
\hspace*{-0.5cm}\right|_{\mathbf{p}_{i}^{2}=\mu^{2}},
\label{eq:physical_coupling_def}
\end{align}
where the blob stands for all diagrams contributing at a given order in perturbation theory
and~$\mu$ is the energy scale of the process.
The dependence of the action parameters on~$\Lambda$ is then chosen so this renormalization condition 
remains cutoff independent. Once this is done not just for the coupling constant but also
for {\em all} physical quantities (e.g., masses), the theory is renormalized and everything can be computed
in terms of experimentally defined physical couplings and masses.

In the case of the scalar theory defined by the action~\eqref{eq:lambdaphi4}, as well as in other physically
relevant theories like QED, QCD or the SM as a whole, it is possible to get rid of the cutoff dependence
in any physical process by ``hiding'' it in 
a {\em finite} number of parameters. Those theories for which this can 
be accomplished are called renormalizable. Nonrenormalizable theories, on the other hand, require
the introduction of an infinite number of parameters 
to absorb the cutoff dependence, that in turn means that we need to specify 
an infinite number of operationally-defined physical quantities. In this picture, nonrenormalizability
seems quite a disaster, since it seems that to compute physical observables we need 
to specify an infinite number of physical renormalization conditions. This is the reason why, 
historically, nonrenormalizable theories were considered to be no good for physics. 

Regularization and renormalization may have important consequences for classical symmetries, 
and we have seen
examples of this in section~\ref{sec:anomalies}. One of the immediate consequences of regularization is the
necessity of introducing a cutoff in the theory and therefore an energy scale. This has the 
result that after renormalization, the physical couplings acquire a dependence on the energy scale 
where they are measured. This scale dependence is codified in 
the $\beta$~function, containing information on how the coupling constant~$g$ depends
on the scale where it is measured
\begin{align}
\beta(g)\equiv \mu{dg\over d\mu}.
\end{align}
This function can be computed order by order in perturbation theory.
In QCD~$\beta(g)<0$, which means that the coupling constant decreases as the energy grows, 
a property known as asymptotic freedom.
Besides, the theory dynamically generates an energy scale~$\Lambda_{\rm QCD}$ 
below which it becomes strongly coupled, with quarks and gluons confined into mesons and baryons.
Asymptotic freedom is the reason behind QCD's success as a description of strong interactions.
It allows to understand, for example, why in deep inelastic scattering experiments electrons seem to interact 
with quasifree partons inside the proton. 

To summarize, we can say that
generically classical scale invariance is anomalous,
in the sense that it disappears as the result of renormalization\footnote{This happens, 
for example, in QCD with massless quarks. There are however a few examples of theories
for which this does not happen, most notably~$\mathcal{N}=4$ supersymmetric Yang Mills theory
in four dimensions. Due to its large symmetry, 
classical conformal invariance is preserved by quantization.}.
The $\beta$-function is just one example of a set of functions describing how couplings and masses
change with the energy scale. Together, they build the coefficients of a set of first-order 
differential equations satisfied by the theory's correlation functions and other quantities and known as
the {\em renormalization group equations}. 

The cartoon description of renormalization presented above 
might lead to thinking that it is just a smart trick, somehow
justifying Feynman's dictum that renormalization is sweeping the
infinities under the rug~\cite{Feynman_rug}. We have come however a long way from there.
The current understanding of renormalization, dating back to the groundbreaking work
of Kenneth Wilson~\cite{Wilson,Wilson2,RGbook1}, goes much deeper and beyond the mere mathematics of shifting 
cutoff dependence from one place to another. It is also closely related to the idea of EFTs, 
so now we can revisit our discussion on pages~\pageref{page:EFTs}-\pageref{page:EFTsf}
in more precise terms.

Everything boils down to making a physical interpretation of the cutoff. Instead of seeing it as
an artificial scale introduced to render integrals finite, we can regard it as the upper energy scale
at which our theory is defined. At energies above~$\Lambda$ new physics may pop up, but we do not 
really care too much, since all we need to know are the values of the
masses~$m_{i}(\Lambda)$ and dimensionless 
couplings~$g_{i}(\Lambda)$.  

Now we ask ourselves how the theory looks at some 
lower energy scale~$\mu<\Lambda$. To answer, we need to ``integrate out'' 
all physical processes
taking place in the range~$\mu\leq E\leq \Lambda$,
which result in a new field theory
now defined at scale~$\mu$ and expressed in terms of some ``renormalized'' fields. Generically, 
the masses and
couplings of this theory will differ from the original ones, so we 
have~$m_{i}(\mu)\neq m_{i}(\Lambda)$ and~$g_{i}(\mu)
\neq g_{i}(\Lambda)$. But, in addition
to this, the new theory might also contain additional couplings not present
at the scale~$\Lambda$, in principle an infinite number of them.
Using the language of path integrals, we 
symbolically summarize all this by writing
\begin{align}
\int\limits_{\mu\leq E\leq \Lambda} \mathscr{D}\Phi_{0}\, e^{iS_{0}[\Phi_{0}]}
=e^{iS[\Phi]},
\end{align} 
where~$\Phi_{0}$ collectively denotes the fields of the original theory 
and~$\Phi$ their renormalized counterparts, while~$S[\Phi]$ is the action of the new theory
defined at the energy scale~$\mu$. On general grounds, it can be written as
\begin{align}
S[\Phi]=S_{0}[\Phi]+\sum_{n}{g'_{n}(\mu)\over 
\Lambda^{{\rm dim\,}\mathcal{O}_{n}-4}}
\int d^{4}x\,\mathcal{O}_{n}[\Phi].
\end{align} 
In this expression~$S_{0}[\Phi]$ is
the action of the original theory with all fields, masses, and couplings
replaced by the corresponding renormalized quantities, and~$\mathcal{O}_{i}[\Phi]$ are
new operators with dimensiones greater than or equal to four
induced by the physics integrated out between the scales~$\Lambda$ 
and~$\mu$. Their couplings~$g'_{n}(\mu)$ are dimensionless and we see that higher-dimensional 
operators are suppressed by inverse powers of the high energy scale~$\Lambda$.

In this Wilsonian picture of renormalization the dependence of the coupling constants with the scale
has a clear physical meaning: as we go to lower energies, their changing values incorporate
the physics that we are integrating out at intermediate scales. But not only this, also the
difference between renormalizable and nonrenormalizable theories gets blurred. All theories
are defined at a given energy scale~$\Lambda$. In order to describe the physics above this scale,
the theory would have to be ``completed'' with additional degrees of freedom and/or interactions.
What is special in renormalizable theories is that they are their own UV completion, in the sense that
they can be extended to arbitrarily high energies without running into trouble, although technically
this only makes sense for asymptotically free theries.

Nonrenormalizable theories need to be completed in the~UV to make sense of
them above~$\Lambda$. Let us look at the example of Fermi's theory of weak interaction. 
It has a natural cutoff given by~$\Lambda=m_{W}$, and if we try to go beyond this energy we run into
trouble. For example, the theory violates unitarity at high energies. The theory however can
be completed in the~UV by the electroweak model studied in section~\ref{sec:EW}, which being renormalizable
can in principle be extended to higher energies without inconsistencies. 

Another case of nonrenormalizable theories encountered in 
section~\ref{sec:tale_of_symmetries} is the chiral Lagrangian 
(see page~\pageref{page:chiral_lagrangian}). Again, the theory is endowed with a physical cutoff, in 
this case~$\Lambda_{\rm QCD}$, above which the description in terms of pions is no longer valid.
In fact, we can see the chiral Lagrangian as resulting from Wilsonian renormalization applied
to QCD: by integrating out the physics of strongly coupled quarks and gluons we get a low energy
action for the new fields (the pions) and their interactions. Since the resulting theory does not
make sense above~$\Lambda_{\rm QCD}$ there is no problem with the divergences appearing in loops. After all,
the before the momenta running in them can reach infinity the pion as such ceases to exist.

The final instance of a nonrenormalizable theory we discuss is gravity, which, as explained in
section~\ref{sec:preliminaries}, has to be completed above the Planck scale~\eqref{eq:Planck_scale}.
But here we have to remember that 
everything couples to gravity, including the SM. Thus, we are led to conclude that despite
being renormalizable, the SM itself has to be regarded as an effective description to be
supplemented at the Planck scale, if not earlier. In fact, phenomena like the nonzero neutrino masses 
strongly indicate
new physics lurking somewhere between the electroweak scale and the Planck scale. 

The bottomline of our discussion is that nonrenormalizability is just
a sign that we are dealing with an EFTs and that the ubiquitous presence of gravity 
in nature forces us to regard {\em all} QFTs as EFTs (have a look again at fig.~\ref{fig:EFT_scheme}
in page~\pageref{fig:EFT_scheme}). Nonrenormalizable theories are not anymore those sinister 
objects they were when renormalization was seen as nothing but infinites removal. They are perfectly 
reasonable theories, provided we are aware what they are and what they are good for (and they are
indeed {\em very}
good for quite many things!).

\begin{mdframed}[backgroundcolor=lightgray,hidealllines=true]
\vspace*{0.2cm}
\centerline{\greybox{\bf The Planck chimney}}
\vspace*{0.2cm}

Let us go back to the Higgs action~\eqref{eq:higgs_actionSM} and particularly to the potential
\begin{align}
V(\mathbf{H},\mathbf{H}^{\dagger})=
{\lambda\over 4}\left(\mathbf{H}^{\dagger}\mathbf{H}-{v^{2}\over 2}\right)^{2}.
\end{align}
We have seen that after symmetry breaking the 
parameter~$\lambda$ directly relates to the Higgs mass~\eqref{eq:higgs_mass} and determines  
its self couplings in the action~\eqref{eq:Higgs_action_h}. Since after quantization
masses and couplings get a dependence on the energy scale, we would like
to know how~$\lambda(\mu)$ or the Higgs mass~$m_{H}(\mu)$ depend on the scale~$\mu$. 
At this point we should recall that the strength
of the coupling of the Higgs to fermions is proportional to the latter's masses [see eq.~\eqref{eq:Higgs_coup_fer}], so its
interactions with the matter fields are dominated by the top quark. Thus 
the renormalization group equations determining the evolution of~$\lambda(\mu)$ and~$\mu_{H}(\mu)$
with the energy scale should also involve the top quark mass~$m_{t}(\mu)$.

An important question is whether the evolution of these parameters with the scale 
changes in a significative way 
the shape of the Mexican hat potential changes and, most importantly, whether  
this jeopardizes the existence of a stable Higgs vacuum
(see~\cite{sher_PR} and references therein). It might be that the sombrero's 
brim get flattened at higher energies, or even inverted like in the case shown here
\begin{align*}
\includegraphics[width=8.5cm]{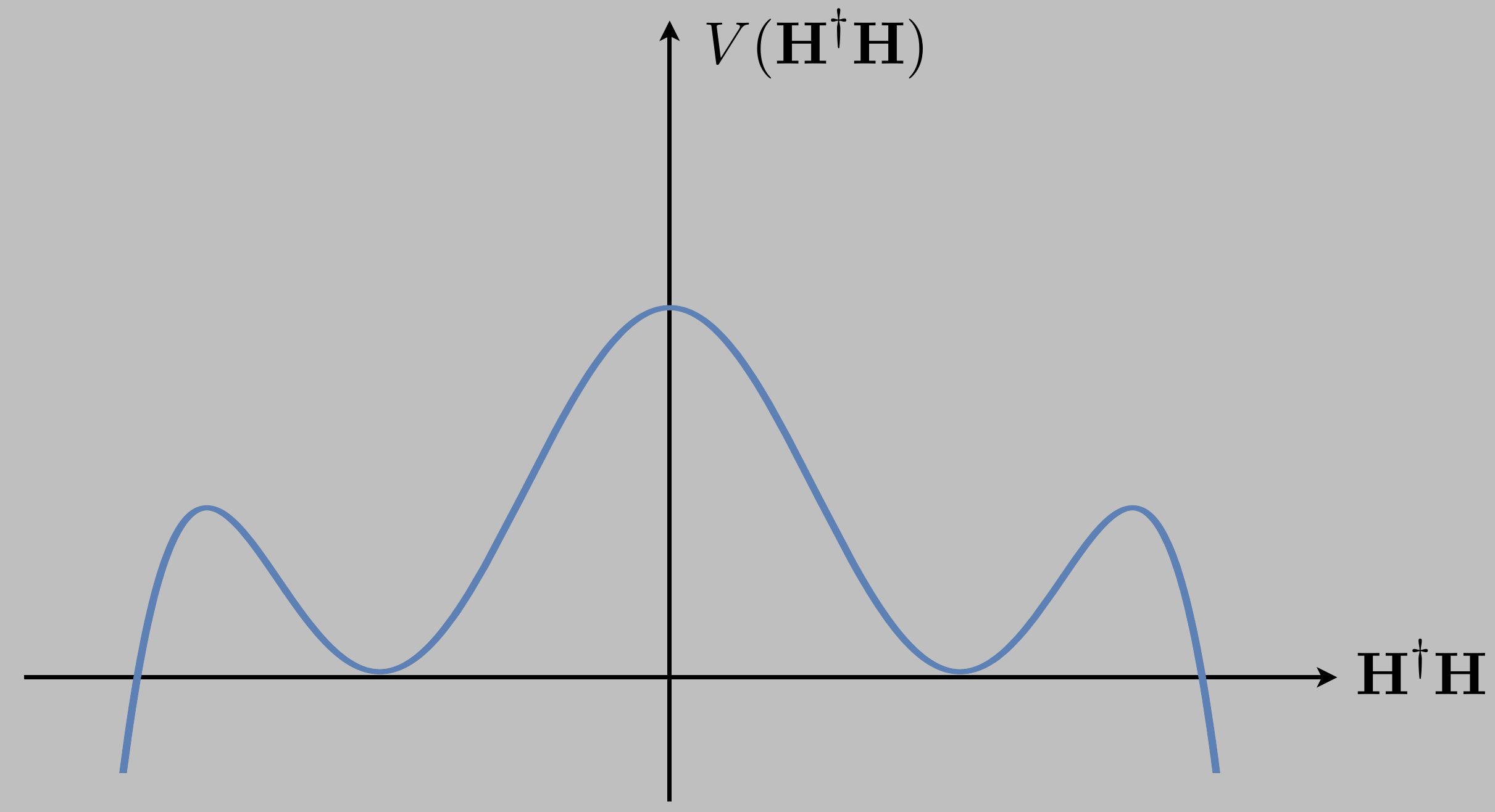}
\end{align*}
If his happens, the Higgs vacuum becomes
metastable or outright unstable.

Since the renormalization group equations
are first order, we need to specify some ``initial conditions''. In this case they are
the values of the Higgs and top masses measured at LHC. Assuming the SM correctly describes
the physics all the way to~$\Lambda_{\rm Pl}$, the bounds to be satisfied by the 
masses in order to 
preserve the stability of the Higgs vacuum are~\cite{Higgsinst1,Higgsinst2,Higgsinst3}
\begin{align}
m_{H}&> (129.1\pm 1.5)\mbox{ GeV},\nonumber \\[0.2cm]
m_{t}&< (171.53\pm 0.42)\mbox{ GeV}.
\label{eq:stability_bounds}
\end{align}
Comparing with the experimental values~$m_{H}=(125.25\pm 0.17)\mbox{ GeV}$ 
and~$m_{t}=(172.69\pm 0.30)\mbox{ GeV}$~\cite{PDG}, 
we see that the SM lies slightly outside the stability zone. 
In fact, the SM seems to be metastable, with the Higgs boson trapped in a false vacuum.
The energy scale where the instability appears turns out to be of the order of
the geometric mean of the $W$~mass and the Planck 
scale~$\Lambda_{\rm inst}\sim \sqrt{m_{W}\Lambda_{\rm Pl}}$. This we can say to be quite
a discovery made at the LHC!

The instability of the Higgs vacuum 
is indeed no good news. Of course, living in a metastable universe is no major problem 
if its tunneling probability is so low that its decay time turns out to be 
much larger than the age of the universe, around~$13.6\mbox{ Gyr}$. But we have to remember that
the bounds~\eqref{eq:stability_bounds} are obtained with the proviso that there are no new degrees of
freedom between the electroweak and the Planck scales. This is yet another reason to expect
some physics beyond the SM making the universe stable.

The apparent metastability of the Higgs vacuum
highlights a very important feature of the renormalization group. We can run it from 
high to low energies with total confidence. Knowing the degrees of freedom and interactions
at a certain scale~$\Lambda$,
everything is determined at energies~$\mu<\Lambda$. The worst thing that may happen is that
the degrees of freedom get ``rearranged'', as it happens in QCD where 
mesons and baryons replace quarks a gluons at low energies. But if the aim is 
getting information about what is going on at~$\mu>\Lambda$, additional assumptions are required: 
either that no new degrees of freedom emerge above~$\Lambda$,
or that there is some
UV~completion whose details are necessarily an educated guess. 
After all, this is why particle physics is hard. Whatever happens above the energies
we explore is blurred in the parameters of the theory we test. The best we can do is play the
model building game to reproduce this blurriness, and hopefully predict distinct 
signals that could be detected in some future facility.

\end{mdframed}

\section{Closing remarks}

The SM is a vast and complex subject, providing the best description of particle physics and its applications 
at energies below a few TeV. It explains a large amount of phenomena in microphysics and in cosmology. 
However, its precise formulation delineates some of its limitations. For instance;
\begin{itemize}

\item[-] The values for the masses and mixing angles of quarks and leptons (including neutrino masses).

\item[-] The~SM does not provide adequate candidates to explain dark matter.

\item[-] The only real progress in the study of dark energy has been to change its name 
from the previous one: the cosmological constant.

\item[-] We know that~CP needs to be violated in the universe in order to generate a matter-antimatter 
asymmetry. Thus, three families are the minimum needed to generate a~CP violating angle, apart from the~QCD 
vacuum angle. Unfortunately, CP violation from the CKM~matrix is not enough to 
generate the observed asymmetry. The equivalent angle in the neutrino sector has not yet been measured. It 
would be ironical if the ultimate origin of ``humans'' was related to properties of the ghostly neutrinos. 
Theories beyond the standard model provide many scenarios with larger amounts of CP~violation.

\item[-] The currently preferred paradigm in cosmology is inflation. We still do not have a convincing 
candidate for what the inflaton is, or how the big bang was triggered, if that question makes any 
sense at all. There are still many open questions in cosmology, including what is the correct paradigm.

\end{itemize}
This is just  
a sample of the most pressing issues for which the SM cannot provide a satisfactory answer.
For decades now the scientific community has been trying to address these problems through
extensions of the SM, from minimal ones inspired by supersymmetry to radical proposals rethinking
the very structure of the elementary constituents, like string theory.

So far the experiments are refusing to give 
any positive indication as to where the answers to the
open questions might lie. Despite transient anomalies or data bumps, 
the more we probe the Higgs particle the more it looks like its ``vanilla version''.
It is truly fascinating that in order to give
masses to the SM particles nature has chosen the simplest solution we came up with, the Higgs field.
The SM's definite triumph, the discovery of the Higgs particle in 2012, was also  
a disappointment, because it apparently closed the door to more exciting possibilities with a clear bearing
on new physics. 

One of the reasons for the impasse might be that we are at the end of a cycle and the current
conceptual framework based on symmetry and locality has been exhausted, or maybe the idea of
naturalness, a basic guiding principle in our understanding of particle physics, 
is after all a red herring. We still need to bring gravity into the SM and this opens
a plethora of problems and questions, some of them touching notions like landscapes or multiverses
loaded with philosophical or just
metascientific ideas.  

Cosmology and astroparticle physics might offer some hope. In recent years, we have witnessed
important discoveries, from the first direct detection of gravitational waves in 2015~\cite{LIGO}
to the ``photo''
of the black hole at the center of the M87 galaxy~\cite{EHT} in 2019. 
The rapidly developing field of gravitational wave astronomy opens up new windows to  
phenomena up to now out of observational reach, and it may allow unprecedented 
glimpses into the physics
of compact astrophysical objects or the very early universe.

We should not give up hope. Maybe we are on the verge of a golden era of discoveries that will leave us 
gasping with awe and laughing with joy in amazement of a new visions of the universe. 
One never knows, and dreaming
is for free.

\section*{Acknowledgments}

These lecture notes contain an extended version of 
courses taught by the authors at the 2022 European School for High Energy Physics
(L.A.-G.), the TAE 2017 and 2019 schools, and graduate courses at Madrid Aut\'onoma University (M.A.V.-M.).
L.A.-G. would like to thank Markus Elsing, Martijn Mulders, Gilad Perez, and Kate Ross for 
their invitation to present 
the lectures at the 2022 ESHEP Jerusalem school, 
and fun moments together. We would also like to thank Het Joshi, student 
assistant at the Simons Center for Geometry and Physics, 
for her excellent work editing the first draft of these lecture notes. M.A.V.-M. 
acknowledges financial support from the
Spanish Science Ministry through research grant PID2021-123703NB-C22
(MCIN/AEI/FEDER, EU), as well as from Basque Government grant
IT1628-22.


\begin{flushleft}
\interlinepenalty=10000

\end{flushleft}


\cleardoublepage

\end{fmffile}


\begin{thebibliography}{99}

\bibitem{B_D}
J.~D.~Bjorken and S.~D.~Drell, {\it Relativistic Quantum Fields}, McGraw-Hill 1965.

\bibitem{I_Z}
C.~Itzykson and J.~B.~Zuber, {\it Quantum Field Theory}, McGraw-Hill 1980.

\bibitem{Ramond}
P.~Ramond, {\it Field Theory: A Modern Primer}, Addison-Wesley 1990.

\bibitem{Peskin}
M.~E.~Peskin and D.~V.~Schroeder, {\it An Introduction to Quantum Field Theory}, Addison-Wesley 1995.

\bibitem{Weinberg}
S.~Weinberg, 
\href{https://www.cambridge.org/core/books/quantum-theory-of-fields/22986119910BF6A2EFE42684801A3BDF}{\it The Quantum Theory of Fields, vols. 1-3}, Cambridge 1995, 1996, and 2000.

\bibitem{DeWitt}
B.~DeWitt, {\it The Global Approach to Quantum Field Theory}, Oxford 2003.

\bibitem{Maggiore}
M.~Maggiore, {\it A Modern Introduction to Quantum Field Theory}, Oxford 2005.

\bibitem{Nair}
V.~P.~Nair, 
\href{https://link.springer.com/book/10.1007/b106781}{\it Quantum Field Theory: A Modern Perspective}, Springer 2005.

\bibitem{B_M}
C.~Burgess and G.~Moore, 
\href{https://www.cambridge.org/core/books/standard-model/64B3382C8BEB95293DF1F09F83411822}{\it The Standard Model: A Primer}, Cambridge 2006.

\bibitem{Paschos}
E.~A.~Paschos, \href{https://www.cambridge.org/core/books/electroweak-theory/BC9D2F11A73D907D29211735FC70553F}{\it Electroweak Theory}, Cambridge 2007.

\bibitem{C_G}
W.~N.~Cottingham and D.~A.~Greenwood, 
\href{https://www.cambridge.org/core/books/an-introduction-to-the-standard-model-of-particle-physics/82428B02F3EE6E45113421D2386B29A1}{\it An Introduction to the Standard Model of Particle
Physics (2nd edition)}, Cambridge 2007.

\bibitem{Banks}
T.~Banks, 
\href{https://www.cambridge.org/core/books/modern-quantum-field-theory/8774E010700BD97772AC7D5C75438197}{\it Modern Quantum Field Theory: A Concise Introduction}, Cambridge 2008.

\bibitem{Zee}
A.~Zee, {\it Quantum Field Theory in a Nutshell (2nd edition)}, Princeton 2010. 

\bibitem{AG_VM}
L.~\'Alvarez-Gaum\'e and M.~\'A.~V\'azquez-Mozo, 
\href{https://link.springer.com/book/10.1007/978-3-642-23728-7}{\it An Invitation to Quantum Field Theory}, Springer 2012.

\bibitem{Schwartz}
M.~D.~Schwartz, 
\href{https://www.cambridge.org/highereducation/books/quantum-field-theory-and-the-standard-model/A4CD66B998F2C696DCC75B984A7D5799#overview}{\it Quantum Field Theory and the Standard Model}, Cambridge 2013.

\bibitem{Kane}
G.~Kane, 
\href{https://www.cambridge.org/highereducation/books/modern-elementary-particle-physics/C62040562C8DF2879A07DD6B649B95F0#overview}{\it Modern Elementary Particle Physics: Explaining and Extending the Standard Model
(2nd edition)}, Cambridge 2017.

\bibitem{Goldberg}
D.~Goldberg, {\it The Standard Model in a Nutshell}, Princeton 2017.

\bibitem{Raby}
S.~Raby, 
\href{https://www.cambridge.org/highereducation/books/introduction-to-the-standard-model-and-beyond/D89CE7DEBA789B749F52CA72C4ACE0B7#overview}{\it Introduction to the Standard Model and Beyond: Quantum Field Theory, Symmetries
and Phenomenology}, Cambridge 2021.

\bibitem{Higgs_ATLAS}
G.~Aad \textit{et al.} [ATLAS],
{\it Observation of a new particle in the search for the Standard Model Higgs boson with the ATLAS detector at the LHC},
\href{https://www.sciencedirect.com/science/article/pii/S037026931200857X?via%3Dihub}{Phys. Lett. B \textbf{716} (2012) 1}
[\href{https://arxiv.org/abs/1207.7214}{\tt arXiv:1207.7214 [hep-ex]}].

\bibitem{Higgs_CMS}
S.~Chatrchyan \textit{et al.} [CMS],
{\it Observation of a New Boson at a Mass of 125 GeV with the CMS Experiment at the LHC},
\href{https://www.sciencedirect.com/science/article/pii/S0370269312008581?via%3Dihub}{Phys. Lett. B \textbf{716} (2012) 30}
[\href{https://arxiv.org/abs/1207.7235}{\tt arXiv:1207.7235 [hep-ex]}].

\bibitem{AGD}
A.~A. Abrikosov, L.~P. Gorkov and I.~E. Dzyaloshinski, \emph{{Methods of
  Quantum Field Theory in Statistical Physics}}. Dover, 1963.

\bibitem{Fetter_Walecka}
A.~L. Fetter and J.~D. Walecka, \emph{{Quantum Theory of Many-Particle
  Systems}}. Dover, 2003.

\bibitem{Bruus_Frensberg}
H.~Bruss and K.~Flensberg, \emph{{Many-Body Quantum Theory in Condensed Matter
  Physics}}. Oxford, 2004.

\bibitem{Burgess_LH}
C.~P.~Burgess,
\href{https://academic.oup.com/book/44246/chapter-abstract/372563709?redirectedFrom=fulltext}{\it Introduction to Effective 
Field Theories and Inflation},
in:~``Effective Field Theory in Particle Physics and Cosmology'', eds. 
S.~Davidson, P.~Gambino, M.~Laine, M.~Neubert and C.~Salomon, 
Oxford 2020.~[\href{https://arxiv.org/abs/1711.10592}{arXiv:1711.10592 [hep-th]}]

\bibitem{Baldauf_LH}
T.~Baldauf,
\href{https://academic.oup.com/book/44246/chapter-abstract/372565490?redirectedFrom=fulltext&login=false}{\it Effective Field Theory of Large-Scale Structure},
in:~``Effective Field Theory in Particle Physics and Cosmology'', eds. 
S.~Davidson, P.~Gambino, M.~Laine, M.~Neubert and C.~Salomon, 
Oxford 2020.

\bibitem{snowmass_EFTC}
G.~Cabass, M.~M.~Ivanov, M.~Lewandowski, M.~Mirbabayi and M.~Simonovi\'c,
{\it Snowmass white paper: Effective field theories in cosmology,}
\href{https://www.sciencedirect.com/science/article/abs/pii/S2212686423000274?via%3Dihub}{Phys. 
Dark Univ. \textbf{40} (2023) 101193}
[\href{https://arxiv.org/abs/2203.08232}{arXiv:2203.08232 [astro-ph.CO]}].

\bibitem{Pich_EFT}
A.~Pich,
{\it Effective Field Theory}, in: ``Probing the Standard Model of Particle Interactions'', 
eds. R.~Gupta, A.~Morel, E.~de Rafael and F.~David, North Holland 1999
[\href{https://arxiv.org/abs/hep-ph/9806303}{arXiv:hep-ph/9806303 [hep-ph]}]

\bibitem{Kaplan_EFT}
D.~B.~Kaplan,
{\it Five lectures on effective field theory},
[\href{https://arxiv.org/abs/nucl-th/0510023}{arXiv:nucl-th/0510023 [nucl-th]}]

\bibitem{FLG}
R.~Feynman, {\it Feynman Lectures on Gravitation}, Addison-Wesley 1995.

\bibitem{EA}
E.~\'Alvarez, {\it Quantum gravity: an introduction to some recent results}, 
\href{https://journals.aps.org/rmp/abstract/10.1103/RevModPhys.61.561}{Rev. Mod. Phys. 
{\bf 61} (1989) 561}. 

\bibitem{Hamber}
H.~W.~Hamber, 
\href{https://link.springer.com/book/10.1007/978-3-540-85293-3}{\it Quantum Gravitation: The Feynman Path Integral Approach}, Springer 2008.

\bibitem{I_U}
L.~E.~Ib\'a\~nez and \'A.~M.~Uranga, 
\href{https://www.cambridge.org/core/books/string-theory-and-particle-physics/7D005A97DA657F6675C2A62E449FC62E}{\it String Theory and Particle Physics}, Cambridge 2012.

\bibitem{Kiritsis}
E.~Kiritsis, {\it String Theory in a Nutshell (2nd edition)}, Princeton 2021.

\bibitem{Brout_Englert}
F.~Englert and R.~Brout,
{\it Broken Symmetry and the Mass of Gauge Vector Mesons},
\href{https://journals.aps.org/prl/abstract/10.1103/PhysRevLett.13.321}{Phys. Rev. Lett. \textbf{13} (1964) 321}.

\bibitem{Higgs1}
P.~W.~Higgs,
{\it Broken symmetries, massless particles and gauge fields},
\href{https://www.sciencedirect.com/science/article/abs/pii/0031916364911369?via%3Dihub}{Phys. Lett. \textbf{12} (1964) 132}.

\bibitem{Higgs2}
P.~W.~Higgs,
{\it Broken Symmetries and the Masses of Gauge Bosons},
\href{https://journals.aps.org/prl/abstract/10.1103/PhysRevLett.13.508}{Phys. Rev. Lett. \textbf{13} (1964) 508}.

\bibitem{Glashow_SM}
S.~L.~Glashow,
{\it Partial Symmetries of Weak Interactions},
\href{https://www.sciencedirect.com/science/article/abs/pii/0029558261904692?via%3Dihub}{Nucl. Phys. \textbf{22} (1961) 579}.

\bibitem{Weinberg_SM}
S.~Weinberg,
{\it A Model of Leptons},
\href{https://journals.aps.org/prl/abstract/10.1103/PhysRevLett.19.1264}{Phys. Rev. Lett. \textbf{19} (1967) 1264}.

\bibitem{Salam_SM}
A.~Salam,
{\it Weak and Electromagnetic Interactions}, in:
``Proceedings of the 8th Nobel Symposium'', 1968.

\bibitem{tHooft}
G.~'t Hooft,
{\it Renormalizable Lagrangians for Massive Yang-Mills Fields},
\href{https://www.sciencedirect.com/science/article/abs/pii/0550321371901398?via%3Dihub}{Nucl. Phys. B \textbf{35} (1971) 167}.

\bibitem{tHooft_Veltman}
G.~'t Hooft and M.~J.~G.~Veltman,
{\it Regularization and Renormalization of Gauge Fields},
\href{https://www.sciencedirect.com/science/article/abs/pii/0550321372902799?via%3Dihub}{Nucl. Phys. B \textbf{44} (1972) 189}.

\bibitem{klein}
F.~Klein, {\it Vergleichende Betrachtungen \"uber neuere geometrische Forschungen},
\href{https://doi.org/10.1007/BF01446615}{Math. Ann.~{\bf 43} (1893) 63}. (For an English translation, 
see~\href{https://arxiv.org/abs/0807.3161}{\tt arXiv:0807.3161})

\bibitem{kline_book}
M.~Kline, {\it Mathematical Thought from Ancient to Modern Times}, Oxford 1990.

\bibitem{curie1894}
P.~Curie, {\it Sur la sym\'etrie dans les ph\'enomenes physiques, 
sym\'etrie d'un champ \'electrique et d'un champ magn\'etique}, 
\href{https://jphystap.journaldephysique.org/articles/jphystap/abs/1894/01/jphystap_1894__3__393_0/jphystap_1894__3__393_0.html}{J. de Phys. {\bf 3} (1894) 26}.

\bibitem{Noether}
E.~Noether, {\it Invariante Variationsprobleme}, 
\href{https://gdz.sub.uni-goettingen.de/id/PPN252457811_1918?tify=%7B%22view%22:%22info%22,%22pages%22:%5B241%5D%7D}{Nachr. v. d. Kgl. Ges. d. Wiss. zu
G\"ottingen, Math.-phys. Kl. (1918) 235.}
(For an English translation, 
see~\href{https://arxiv.org/abs/physics/0503066}{\tt arXiv:physics/0503066}) 

\bibitem{pais}
A.~Pais, {\it ``Subtle is the Lord...'': The Science and Life of Albert Einstein}, Oxford 1982.

\bibitem{Holstein}
B.~R.~Holstein, {\it Klein's paradox},~\href{https://aapt.scitation.org/doi/10.1119/1.18891}{Am. J. Phys. 66 (1999) 507}.

\bibitem{Stueckelberg_antiparticles}
E.~C.~G.~Stueckelberg, {\it La Mecanique du point materiel en theorie de relativite et en theorie des quanta,}
\href{https://www.e-periodica.ch/cntmng?pid=hpa-001:1942:15::790}{Helv. Phys. Acta \textbf{15} (1942) 23}.

\bibitem{Feynman_antiparticles}
R.~P.~Feynman,
{\it A Relativistic Cutoff for Classical Electrodynamics,}
\href{https://journals.aps.org/pr/abstract/10.1103/PhysRev.74.939}{Phys. Rev. \textbf{74} (1948) 939}.

\bibitem{Feynman_antiparticles2}
R.~P.~Feynman, {\it The reason for antiparticles}, in: R.~P.~Feynman and S.~Weinberg, {\it Elementary Particles and the Laws of Physics. The 
1986 Dirac Memorial Lectures}, Cambridge 1987.

\bibitem{classicalED}
J.~D.~Jackson, {\it Classical Electrodynamics (3rd edition)}, Wiley 1999.

\bibitem{AB}
Y.~Aharonov and D.~Bohm,
{\it Significance of electromagnetic potentials in the quantum theory,}
\href{https://journals.aps.org/pr/abstract/10.1103/PhysRev.115.485}{Phys. Rev. \textbf{115} (1959) 485}.

\bibitem{dirac_monopole}
P.~A.~M.~Dirac,
{\it Quantised singularities in the electromagnetic field,}
\href{https://royalsocietypublishing.org/doi/10.1098/rspa.1931.0130}{Proc. Roy. Soc. Lond. A 
\textbf{133} (1931) 60}.

\bibitem{cabrera_monopole}
B.~Cabrera,
{\it First Results from a Superconductive Detector for Moving Magnetic Monopoles,}
\href{https://journals.aps.org/prl/abstract/10.1103/PhysRevLett.48.1378}{Phys. Rev. Lett. \textbf{48} (1982) 1378}.

\bibitem{price_et_al_monopole}
P.~B.~Price, E.~K.~Shirk, W.~Z.~Osborne and L.~S.~Pinsky,
{\it Evidence for Detection of a Moving Magnetic Monopole,}
\href{https://journals.aps.org/prl/abstract/10.1103/PhysRevLett.35.487}{Phys. Rev. Lett. \textbf{35} (1975) 487}.

\bibitem{wu_yang}
T.~T.~Wu and C.~N.~Yang,
{\it Dirac's Monopole Without Strings: Classical Lagrangian Theory,}
\href{https://journals.aps.org/prd/abstract/10.1103/PhysRevD.14.437}{Phys. Rev. D \textbf{14} (1976), 437}.

\bibitem{Azcarraga_Izquierdo}
J.~A.~Azc\'arraga and J.~M.~Izquierdo, 
\href{https://www.cambridge.org/core/books/lie-groups-lie-algebras-cohomology-and-some-applications-in-physics/B570D04EC2EAA2A2C21BA23F245D2457}{\it Lie groups, Lie algebras, cohomology and some
applications in physics}, Cambridge 1995.

\bibitem{Nakahara}
M.~Nakahara, {\it Geometry, Topology and Physics (2nd edition)}, CRC Press 2017.

\bibitem{Nash}
C.~Nash and S.~Sen, {\it Topology and Geometry for Physicists}, Dover 2011.

\bibitem{Frankel}
T.~Frankel, {\it The Geometry of Physics (3rd edition)}, Cambridge 2011.

\bibitem{WeinbergCC}
S.~Weinberg,
{\it The Cosmological Constant Problem,}
\href{https://journals.aps.org/rmp/abstract/10.1103/RevModPhys.61.1}{Rev. Mod. Phys. \textbf{61} (1989) 1}.

\bibitem{PadmanabhanCC}
T.~Padmanabhan, {\it Cosmological Constant: The Weight of the Vacuum,}
\href{https://www.sciencedirect.com/science/article/pii/S0370157303001200?via%3Dihub}{Phys. Rept. \textbf{380} (2003) 235.}
[\href{https://arxiv.org/abs/hep-th/0212290}{arXiv:hep-th/0212290 [hep-th]}].

\bibitem{BoussoCC}
R.~Bousso,
{\it TASI Lectures on the Cosmological Constant,}
\href{https://link.springer.com/article/10.1007/s10714-007-0557-5}{Gen. Rel. Grav. \textbf{40} (2008) 607}
[\href{https://arxiv.org/abs/0708.4231}{arXiv:0708.4231 [hep-th]}].

\bibitem{Haag_onQFT}
R.~Haag, {\it On Quantum Field Theories}, 
\href{https://cds.cern.ch/record/212242}{Danske  Vid. Selsk. Mat.-Fys. Medd. {\bf 29} (1955) 669}.

\bibitem{Streater_Wightman}
R.~F.~Streater and A.~S.~Wightman, {\it PCT, Spin and Statistics, and All That}, Princeton 1989.

\bibitem{Haag_book}
R.~Haag, {\it Local Quantum Physics (2nd edition)}, Springer 1996.

\bibitem{Strocchi}
F.~Strocchi, {\it An Introduction to Non-Perturabative Foundations of Quantum Field Theory}, Oxford 2013.

\bibitem{Georgi}
H.~Georgi, {\it Lie Algebras in Particle Physics: From Isospin to Unified Field Theories (2nd edition)}, 
Perseus Books 1999.

\bibitem{Ramond_groups}
P.~Ramond, {\it Group Theory: A Physicist's Survey}, Cambridge 2010.

\bibitem{Gell-Mann_SU3}
M.~Gell-Mann,
{\it The Eightfold Way: A Theory of strong interaction symmetry,} Caltech report
\href{https://www.osti.gov/biblio/4008239/}{CTSL-20/TID-12608} (1961).

\bibitem{Neeman_SU3}
Y.~Ne'eman,
{\it Derivation of strong interactions from a gauge invariance,}
\href{https://www.sciencedirect.com/science/article/abs/pii/0029558261901341?via%3Dihub}{Nucl. Phys. \textbf{26} (1961) 222}.

\bibitem{Gell-Mann_quarks}
M.~Gell-Mann,
{\it A Schematic Model of Baryons and Mesons,}
\href{https://linkinghub.elsevier.com/retrieve/pii/S0031916364920013}{Phys. Lett. \textbf{8} (1964) 214}.

\bibitem{Zweig_quarks}
G.~Zweig,
{\it An SU(3) model for strong interaction symmetry and its breaking (versions 1 \& 2),}
CERN reports \href{https://cds.cern.ch/record/352337?ln=en}{CERN-TH-401} 
and \href{https://cds.cern.ch/record/570209?ln=en}{CERN-TH-412} (1964).

\bibitem{Wigner_th}
E.~Wigner, \href{https://link.springer.com/book/10.1007/978-3-663-02555-9}{\it Gruppentheorie und ihre Anwendung auf die Quantenmechanik der Atomspektren}, 
Vieweg+Teubner Verlag 1931. 
English translation: {\it Group Theory and its Applications to the Quantum Mechanics
of Atomic Spectra}, Academic Press 1959.

\bibitem{Belinfante}
F.~J.~Belinfante, {\it On the Current and the Density of the Electric Charge, the Energy,
the Linear Momentum and the Angular Momentum of Arbitrary Fields}, 
\href{https://www.sciencedirect.com/science/article/abs/pii/S003189144090091X?via%3Dihub}{Physica~{\bf 7} (1940) 29}.

\bibitem{Rosenfeld}
L.~Rosenfeld, {\it Sur le tenseur d'impulsion-energie}, Mem. Acad. Belgique cl. sc. {\bf 18} (1940) 1.
(for an English translation, see \href{http://neo-classical-physics.info/uploads/3/4/3/6/34363841/rosenfeld_-_on_the_energy-momentum_tensor.pdf}{here})

\bibitem{Merzbacher}
E.~Merzbacher, {\it Quantum Mechanics (3rd edition)}, 1998. 

\bibitem{Goldstone}
J.~Goldstone, {\it Field Theories with `Superconductor' Solutions,}
\href{https://link.springer.com/article/10.1007/BF02812722}{Nuovo Cim. \textbf{19} (1961) 154}.

\bibitem{GSW}
J.~Goldstone, A.~Salam and S.~Weinberg,
{\it Broken Symmetries},
\href{https://journals.aps.org/pr/abstract/10.1103/PhysRev.127.965}{Phys. Rev. \textbf{127} (1962) 965}.

\bibitem{Nambu_NG}
Y.~Nambu,
\href{https://inspirehep.net/files/7da93219e7f727069950f4e103a8affc}{\it Dynamical 
theory of elementary particles suggested by superconductivity,}
in: ``Proceedings of the 10th International Conference on High-Energy Physics'', 1960.

\bibitem{NJL}
Y.~Nambu and G.~Jona-Lasinio,
{\it Dynamical Model of Elementary Particles Based on an Analogy with Superconductivity. 1.,}
\href{https://journals.aps.org/pr/abstract/10.1103/PhysRev.122.345}{Phys. Rev. \textbf{122} (1961) 345}.

\bibitem{Goldenfeld}
N.~Goldenfeld, {\it Lectures on Phase Transitions and the Renormalization Group}, Addison-Wesley 1992.

\bibitem{Annett}
J.~F.~Annett, {\it Superconductivity, Superfluids and Condensates}, Oxford 2004.

\bibitem{Schakel}
A.~M.~J.~Schakel, {\it Boulevar of Broken Symmetries: Effective Field Theories of Condensed
Matter}, World Scientific 2008.

\bibitem{Pich}
A.~Pich,
{\it Chiral perturbation theory},
\href{https://iopscience.iop.org/article/10.1088/0034-4885/58/6/001}{Rept. Prog. Phys. \textbf{58} (1995) 563}.

\bibitem{Scherer_Schindler}
S.~Scherer and M.~R.~Schinler, 
\href{https://link.springer.com/book/10.1007/978-3-642-19254-8}{\it A Primer for Chiral Perturbation Theory}, Springer 2012.

\bibitem{Anderson}
P.~W.~Anderson,
{\it Plasmons, Gauge Invariance, and Mass,}
\href{https://journals.aps.org/pr/abstract/10.1103/PhysRev.130.439}{Phys. Rev. \textbf{130} (1963) 439}.

\bibitem{AG_E}
L.~\'Alvarez-Gaum\'e and J.~Ellis, {\it Eyes on a prize particle}, 
\href{https://www.nature.com/articles/nphys1874}{Nature Phys. {\bf 7} (2011) 2}.

\bibitem{Yukawa}
H.~Yukawa,
{\it On the Interaction of Elementary Particles I,}
\href{https://academic.oup.com/ptps/article/doi/10.1143/PTPS.1.1/1878532?login=false}{Proc. Phys. Math. Soc. Jap. \textbf{17} (1935) 48}.

\bibitem{Littlewood_Varma}
P.~B.~Littlewood and C.~M.~Varma, {\it Amplitude collective modes in superconductors and 
their coupling to charge-density waves}, 
\href{https://journals.aps.org/prb/abstract/10.1103/PhysRevB.26.4883}{Phys. 
Rev. B~{\bf 26} (1982) 4883}.

\bibitem{Shimano_Tsuji}
R.~Shimano and N.~Tsuji, {\it Higgs Mode in Superconductors},
\href{https://www.annualreviews.org/doi/10.1146/annurev-conmatphys-031119-050813}{Ann. Rev. Condensed Matter Phys. \textbf{11} (2020) 103}
[\href{https://arxiv.org/abs/1906.09401}{arXiv:1906.09401 [cond-mat.supr-con]}].

\bibitem{YM}
C.~N.~Yang and R.~L.~Mills,
{\it Conservation of Isotopic Spin and Isotopic Gauge Invariance,}
\href{https://journals.aps.org/pr/abstract/10.1103/PhysRev.96.191}{Phys. Rev. \textbf{96} (1954) 191}.

\bibitem{AG_anomalies}
L.~\'Alvarez-Gaum\'e, \emph{{An Introduction to Anomalies}},  in
  \emph{{``Fundamental Problems of Gauge Field Theory''}}, {Plenum Press},
  1985.
  
\bibitem{Bertlmann}
R.~A.~Bertlmann, \emph{{Anomalies in Quantum Field Theory}}, Oxford 1996.

\bibitem{Fujikawa_Suzuki}
K.~Fujikawa and H.~Suzuki, \emph{{Path Integrals and Quantum Anomalies}},
  Oxford 2004.

\bibitem{AG_VM_GS}
L.~Alvarez-Gaum\'e and M.~\'A.~V\'azquez-Mozo,
{\it Anomalies and the Green-Schwarz Mechanism,} in: ``Handbook of Quantum Gravity'', 
eds. C.~Bambi, L.~Modesto and I.~L.~Shapiro, Springer 2023 (to appear)
[\href{https://arxiv.org/abs/2211.06467}{arXiv:2211.06467 [hep-th]}].

\bibitem{Faddeev}
L.~D.~Faddeev,
{\it Operator Anomaly for the Gauss Law,}
\href{https://www.sciencedirect.com/science/article/abs/pii/0370269384909523?via%3Dihub}{Phys. Lett. B \textbf{145} (1984) 81}.

\bibitem{Nelson_AG}
P.~Nelson and L.~\'Alvarez-Gaum\'e,
{\it Hamiltonian Interpretation of Anomalies,}
\href{https://link.springer.com/article/10.1007/BF01466595}{Commun. Math. Phys. \textbf{99} (1985) 103}.

\bibitem{KSS}
M.~Kobayashi, K.~Seo and A.~Sugamoto,
{\it Commutator Anomaly for the Gauss Law Constraint Operator,}
\href{https://www.sciencedirect.com/science/article/abs/pii/0550321386903809?via%3Dihub}{Nucl. Phys. B \textbf{273} (1986) 607}.

\bibitem{Jackiw_ABJ}
R.~Jackiw, {\it Field Theoretic Investigations in Current Algebra}, in~``Current Algebras and Anomalies'', 
eds. S.~B.~Treiman, R.~Jackiw, B.~Zumino and E.~Witten, Princeton 1985.

\bibitem{Steinberger}
J.~Steinberger, {\it On the use of subtraction fields and the lifetimes of some types of meson decay,}
\href{https://journals.aps.org/pr/abstract/10.1103/PhysRev.76.1180}{Phys. Rev. \textbf{76} (1949) 1180}.

\bibitem{Schwinger}
J.~Schwinger,
{\it On gauge invariance and vacuum polarization,}
\href{https://journals.aps.org/pr/abstract/10.1103/PhysRev.82.664}{Phys. Rev. \textbf{82} (1951) 664}.

\bibitem{Fukuda_Miyamoto}
H.~Fukuda and Y.~Miyamoto, {\it On the $\gamma$-Decay of Neutral Mesons}, 
\href{https://academic.oup.com/ptp/article/4/3/347/1888473}{Prog. Theor. Phys.~{\bf 4}
(1949) 347}.

\bibitem{OOS}
S.~Ozaki, S.~Oneda and S.~Sasaki, {\it On the Decay of Heavy Mesons I}, 
\href{https://academic.oup.com/ptp/article/4/4/524/1844123}{Prog. Theor. Phys.~{\bf 4}
(1949) 524}.

\bibitem{Adler}
S.~L.~Adler,
{\it Axial vector vertex in spinor electrodynamics,}
\href{https://journals.aps.org/pr/abstract/10.1103/PhysRev.177.2426}{Phys. Rev. \textbf{177} (1969) 2426}.

\bibitem{Bell_Jackiw}
J.~S.~Bell and R.~Jackiw,
{\it A PCAC puzzle: $\pi^0 \to \gamma \gamma$ in the $\sigma$ model,}
\href{https://link.springer.com/article/10.1007/BF02823296}{Nuovo Cim. A \textbf{60} (1969) 47}.

\bibitem{pion_width}
I.~Larin \textit{et al.} [PrimEx-II Collaboration],
{\it Precision measurement of the neutral pion lifetime,}
\href{https://www.science.org/doi/10.1126/science.aay6641}{Science \textbf{368} (2020) 506}.

\bibitem{Sutherland}
D.~G.~Sutherland,
{\it Current algebra and some nonstrong mesonic decays,}
\href{https://www.sciencedirect.com/science/article/abs/pii/0550321367901800?via%3Dihub}{Nucl. Phys. B \textbf{2} (1967) 433}.

\bibitem{Veltman}
M.~J.~G.~Veltman, {\it Theoretical aspects of high-energy neutrino interactions}, 
\href{https://royalsocietypublishing.org/doi/10.1098/rspa.1967.0193}{Proc. R. Soc. A {\bf 301} (1967) 107}.

\bibitem{Fujikawa}
K.~Fujikawa,
{\it Path Integral Measure for Gauge Invariant Fermion Theories,}
\href{https://journals.aps.org/prl/abstract/10.1103/PhysRevLett.42.1195}{Phys. Rev. Lett. \textbf{42} (1979) 1195}.

\bibitem{Fujikawa2}
K.~Fujikawa, {\it Path Integral for Gauge Theories with Fermions,}
\href{https://journals.aps.org/prd/abstract/10.1103/PhysRevD.21.2848}{Phys. Rev. D \textbf{21} (1980) 2848}.

\bibitem{Achucarro_Townsend}
A.~Ach\'ucarro and P.~K.~Townsend,
{\it A Chern-Simons Action for Three-Dimensional anti-De Sitter Supergravity Theories,}
\href{https://www.sciencedirect.com/science/article/pii/0370269386901401?via%3Dihub}{Phys. Lett. B \textbf{180} (1986) 89}.

\bibitem{Witten_CS}
E.~Witten,
{\it (2+1)-Dimensional Gravity as an Exactly Soluble System,}
\href{https://www.sciencedirect.com/science/article/pii/0550321388901435?via%3Dihub}{Nucl. Phys. B \textbf{311} (1988) 46}.

\bibitem{Ezawa}
Z.~F.~Ezawa, {\it Quantum Hall Effects (3rd edition)}, World Scientific 2013.

\bibitem{Villadoro}
G.~Villadoro, \href{https://www.youtube.com/playlist?list=PLDxsZU4NC6Z4kL18PhWTeHicRP13OfHYI}{{\it Axions}},
Lectures at the Galileo Galilei Institute, Arcetri 2015.

\bibitem{Hook}
A.~Hook, {\it TASI Lectures on the Strong CP Problem and Axions}, 
\href{https://pos.sissa.it/333/004/}{PoS {\bf TASI2018} (2019) 004} 
[\href{https://arxiv.org/abs/1812.02669}{arXiv:1812.02669 [hep-ph]}].

\bibitem{PDG}
R.~L.~Workman \textit{et al.} [Particle Data Group],
\href{https://pdg.lbl.gov/}{\it Review of Particle Physics,}
PTEP \textbf{2022} (2022) 083C01.

\bibitem{Snowmass_EDM}
R.~Alarcon, J.~Alexander, V.~Anastassopoulos, T.~Aoki, R.~Baartman, S.~Bae\ss{}ler, L.~Bartoszek, D.~H.~Beck, F.~Bedeschi and R.~Berger, \textit{et al.}
{\it Electric dipole moments and the search for new physics,}
[\href{https://arxiv.org/abs/2203.08103}{arXiv:2203.08103 [hep-ph]}]

\bibitem{BKM}
V.~Bernard, N.~Kaiser and U.~G.~Mei\ss ner,
{\it Chiral dynamics in nucleons and nuclei,}
\href{https://www.worldscientific.com/doi/abs/10.1142/S0218301395000092}{Int. J. Mod. Phys. E \textbf{4} (1995) 193}.

\bibitem{Georgi_WI}
H.~Georgi, {\it Weak Interactions and Modern Particle Physics (Revised and Updated Edition)}, Dover 2009.

\bibitem{CWZ}
S.~Coleman, J.~Wess and B.~Zumino,
{\it Structure of phenomenological Lagrangians. 1.},
\href{https://journals.aps.org/pr/abstract/10.1103/PhysRev.177.2239}{Phys. Rev. \textbf{177} (1969) 2239}.

\bibitem{CCWZ}
C.~G.~Callan, Jr., S.~Coleman, J.~Wess and B.~Zumino,
{\it Structure of phenomenological Lagrangians. 2.},
\href{https://journals.aps.org/pr/abstract/10.1103/PhysRev.177.2247}{Phys. Rev. \textbf{177} (1969) 2247}.

\bibitem{GSS}
J.~Gasser, M.~E.~Sainio and A.~Svarc,
{\it Nucleons with Chiral Loops,}
\href{https://www.sciencedirect.com/science/article/abs/pii/0550321388901083?via%3Dihub}{Nucl. Phys. B \textbf{307} (1988) 779}.

\bibitem{Goldberger_Treiman}
M.~L.~Goldberger and S.~B.~Treiman,
{\it Decay of the pi meson,}
\href{https://journals.aps.org/pr/abstract/10.1103/PhysRev.110.1178}{Phys. Rev. \textbf{110} (1958) 1178}.

\bibitem{Crewther_et_al}
R.~J.~Crewther, P.~Di Vecchia, G.~Veneziano and E.~Witten,
{\it Chiral Estimate of the Electric Dipole Moment of the Neutron in Quantum Chromodynamics,}
\href{https://www.sciencedirect.com/science/article/abs/pii/037026937990128X?via%3Dihub}{Phys. Lett. B \textbf{88} (1979) 123} [Erratum: \href{https://www.sciencedirect.com/science/article/pii/0370269380910254?via%3Dihub}{Phys. Lett. B \textbf{91} (1980) 487}].

\bibitem{Vafa_Witten}
C.~Vafa and E.~Witten,
{\it Parity Conservation in QCD},
\href{https://journals.aps.org/prl/abstract/10.1103/PhysRevLett.53.535}{Phys. Rev. Lett. \textbf{53} (1984) 535}.

\bibitem{Peccei_Quinn1}
R.~D.~Peccei and H.~R.~Quinn,
{\it CP Conservation in the Presence of Instantons},
\href{https://journals.aps.org/prl/abstract/10.1103/PhysRevLett.38.1440}{Phys. Rev. Lett. \textbf{38} (1977) 1440}.

\bibitem{Peccei_Quinn2}
R.~D.~Peccei and H.~R.~Quinn,
{\it Constraints Imposed by CP Conservation in the Presence of Instantons,}
\href{https://journals.aps.org/prd/abstract/10.1103/PhysRevD.16.1791}{Phys. Rev. D \textbf{16} (1977) 1791}.

\bibitem{Weinberg_axion}
S.~Weinberg,
{\it A New Light Boson?},
\href{https://journals.aps.org/prl/abstract/10.1103/PhysRevLett.40.223}{Phys. Rev. Lett. \textbf{40} (1978) 223}.

\bibitem{Wilczek_axion}
F.~Wilczek,
{\it Problem of Strong  $P$  and  $T$  Invariance in the Presence of Instantons},
\href{https://journals.aps.org/prl/abstract/10.1103/PhysRevLett.40.279}{Phys. Rev. Lett. \textbf{40} (1978) 279}.

\bibitem{Redondo_Ringwald}
J.~Redondo and A.~Ringwald,
{\it Light shining through walls},
\href{https://www.tandfonline.com/doi/abs/10.1080/00107514.2011.563516}{Contemp. Phys. \textbf{52} (2011) 211}
[\href{https://arxiv.org/abs/1011.3741}{arXiv:1011.3741 [hep-ph]}]

\bibitem{Sigl}
G.~Sigl, \href{https://link.springer.com/book/10.2991/978-94-6239-243-4}{\it Astroparticle Physics: Theory and Phenomenology}, Springer 2017.

\bibitem{Marsh}
D.~J.~E.~Marsh,
{\it Axion Cosmology},
\href{https://www.sciencedirect.com/science/article/abs/pii/S0370157316301557?via%3Dihub}{Phys. Rept. \textbf{643} (2016) 1}
[\href{https://arxiv.org/abs/1510.07633}{arXiv:1510.07633 [astro-ph.CO]}].

\bibitem{Ohare}
C.~O'Hare, \href{https://zenodo.org/record/3932430}{\it cajohare/Axionlimits: AxionLimits},
Zenodo 2020.

\bibitem{Lee_Yang}
T.~D.~Lee and C.~N.~Yang,
{\it Question of Parity Conservation in Weak Interactions},
\href{https://journals.aps.org/pr/abstract/10.1103/PhysRev.104.254}{Phys. Rev. \textbf{104} (1956) 254}.

\bibitem{Wu}
C.~S.~Wu, E.~Ambler, R.~W.~Hayward, D.~D.~Hoppes and R.~P.~Hudson,
{\it Experimental Test of Parity Conservation in $\beta$ Decay,}
\href{https://journals.aps.org/pr/abstract/10.1103/PhysRev.105.1413}{Phys. Rev. \textbf{105} (1957) 1413}.

\bibitem{Garwin}
R.~L.~Garwin, L.~M.~Lederman and M.~Weinrich,
{\it Observations of the Failure of Conservation of Parity and Charge Conjugation in Meson Decays: The Magnetic Moment of the Free Muon},
\href{https://journals.aps.org/pr/abstract/10.1103/PhysRev.105.1415}{Phys. Rev. \textbf{105} (1957) 1415}.

\bibitem{Friedman}
J.~I.~Friedman and V.~L.~Telegdi,
{\it Nuclear Emulsion Evidence for Parity Nonconservation in the Decay Chain $\pi^+ \to \mu^+ \to e^+$,}
\href{https://journals.aps.org/pr/abstract/10.1103/PhysRev.106.1290}{Phys. Rev. \textbf{106} (1957) 1290}.

\bibitem{Sudarshan_Marshak}
E.~C.~G.~Sudarshan and R.~E.~Marshak,
{\it Chirality Invariance and the Universal Fermi Interaction},
\href{https://journals.aps.org/pr/abstract/10.1103/PhysRev.109.1860.2}{Phys. Rev. \textbf{109} (1958) 1860}.
(Originally published in ``Proceedings of the Padua-Venice
Conference on Mesons and Recently Discovered Particles'', 1957.)

\bibitem{Feynman_Gell-MannV-A}
R.~P.~Feynman and M.~Gell-Mann,
{\it Theory of the Fermi Interaction},
\href{https://journals.aps.org/pr/abstract/10.1103/PhysRev.109.193}{Phys. Rev. \textbf{109} (1958) 193}.

\bibitem{Gargamelle}
F.~J.~Hasert \textit{et al.} [Gargamelle Neutrino],
{\it Observation of Neutrino Like Interactions Without Muon Or Electron in the Gargamelle Neutrino Experiment},
\href{https://www.sciencedirect.com/science/article/pii/0370269373904991?via%3Dihub}{Phys. Lett. B \textbf{46} (1973) 138}.

\bibitem{WNC_review}
D.~Haidt, {\it The Discovery of Weak Neutral Currents}, in ``\href{https://www.worldscientific.com/worldscibooks/10.1142/9441#t=aboutBook}{60 Years 
of CERN Experiments and Discoveries}'', eds. H.~Schopper and L.~Di~Lella, 
World Scientific 2015.

\bibitem{Kobayashi_Maskawa}
M.~Kobayashi and T.~Maskawa,
{\it CP Violation in the Renormalizable Theory of Weak Interaction},
\href{https://academic.oup.com/ptp/article/49/2/652/1858101?login=false}{Prog. Theor. Phys. \textbf{49} (1973) 652}.

\bibitem{GIM}
S.~L.~Glashow, J.~Iliopoulos and L.~Maiani,
{\it Weak Interactions with Lepton-Hadron Symmetry},
\href{https://journals.aps.org/prd/abstract/10.1103/PhysRevD.2.1285}{Phys. Rev. D \textbf{2} (1970) 1285}.

\bibitem{Fodor_Hoelbling}
Z.~Fodor and C.~Hoelbling,
{\it Light Hadron Masses from Lattice QCD},
\href{https://doi.org/10.1103/RevModPhys.84.449}{Rev. Mod. Phys. \textbf{84} (2012) 449}
[\href{https://arxiv.org/abs/1203.4789}{\tt arXiv:1203.4789 [hep-lat]}].

\bibitem{Hatsuda}
T.~Hatsuda, \href{https://link.springer.com/chapter/10.1007/978-3-319-53336-0_3}{\it Lattice 
Quantum Chromodynamics}, 
in: ``An Advanced Course in Computational Nuclear
Physics'', eds. M.~Hjorth-Jensen, M.~P.~Lombardo and U.~van~Kolck, Springer~2017.

\bibitem{Lesgourges}
J.~Lesgourges, G.~Mangano, G.~Miele and~S.~Pastor, 
\href{https://www.cambridge.org/core/books/neutrino-cosmology/44AF52C5F02A1943850F3B239B2F9588}{\it Neutrino Cosmology}, Cambridge 2013.

\bibitem{Giunti_Kim}
C.~Giunti and C.~W.~Kim, {\it Fundamental of Neutrino Physics and Astrophysics}, 
Oxford 2007.

\bibitem{Bilenky}
S.~Bilenky, 
\href{https://link.springer.com/book/10.1007/978-3-319-74802-3}{\it Introduction to the Physics of Massive and Mixed Neutrinos (2nd edition)}, 
Springer 2018.

\bibitem{Bilenky_Giunti}
S.~M.~Bilenky and C.~Giunti,
{\it Neutrinoless Double-Beta Decay: a Probe of Physics Beyond the Standard Model},
\href{https://doi.org/10.1142/S0217751X1530001X}{Int. J. Mod. Phys. A \textbf{30} (2015) 1530001}
[\href{https://arxiv.org/abs/1411.4791}{arXiv:1411.4791 [hep-ph]}].

\bibitem{Jones_ndbd}
B.~J.~P.~Jones,
{\it The Physics of Neutrinoless Double Beta Decay: A Beginners Guide},
\href{https://pos.sissa.it/388/007}{PoS \textbf{TASI2020} (2021) 007}
[\href{https://arxiv.org/abs/2108.09364}{arXiv:2108.09364 [nucl-ex]}].

\bibitem{GGY}
M.~C.~Gonz\'alez-Garc\'{\i}a and M.~Yokoyama, {\it Neutrino Masses, Mixing, and Oscillations}, 
in~\cite{PDG}.

\bibitem{Sozzi}
M.~S.~Sozzi, {\it Discrete Symmetries and CP Violation: From Experiment to Theory}, Oxford 2008.

\bibitem{Wolfenstein}
L.~Wolfenstein,
{\it Parametrization of the Kobayashi-Maskawa Matrix},
\href{https://doi.org/10.1103/PhysRevLett.51.1945}{Phys. Rev. Lett. \textbf{51} (1983) 1945}.

\bibitem{Cabibbo}
N.~Cabibbo,
{\it Unitary Symmetry and Leptonic Decays},
\href{https://journals.aps.org/prl/abstract/10.1103/PhysRevLett.10.531}{Phys. Rev. Lett. \textbf{10} (1963) 531}.

\bibitem{Cline_baryogenesis}
J.~M.~Cline,
{\it TASI Lectures on Early Universe Cosmology: Inflation, Baryogenesis and Dark Matter},
\href{https://pos.sissa.it/333/001/}{PoS \textbf{TASI2018} (2019) 001}
[\href{https://arxiv.org/abs/1807.08749}{arXiv:1807.08749 [hep-ph]}].

\bibitem{Pontecorvo}
B.~Pontecorvo,
{\it Neutrino Experiments and the Problem of Conservation of Leptonic Charge},
\href{http://www.jetp.ras.ru/cgi-bin/e/index/e/26/5/p984?a=list}{Zh. Eksp. Teor. Fiz. \textbf{53} (1967) 1717}.

\bibitem{MNS}
Z.~Maki, M.~Nakagawa and S.~Sakata,
{\it Remarks on the unified model of elementary particles},
\href{https://doi.org/10.1143/PTP.28.870}{Prog. Theor. Phys. \textbf{28} (1962) 870}.

\bibitem{EGGMSZ}
I.~Esteban, M.~C.~Gonz\'alez-Garc\'{\i}a, M.~Maltoni, T.~Schwetz and A.~Zhou,
{\it The fate of hints: updated global analysis of three-flavor neutrino oscillations},
\href{https://link.springer.com/article/10.1007/JHEP09(2020)178}{JHEP \textbf{09} (2020) 178}
[\href{https://arxiv.org/abs/2007.14792}{\tt arXiv:2007.14792 [hep-ph]}].

\bibitem{Feynman_rug}
\href{http://caltechcampuspubs.library.caltech.edu/662/1/1965_10_22_67_5_.05.pdf}{\it Dr. Richard Feynman Nobel Laureate!}, California Tech (October 22, 1965).

\bibitem{Wilson}
K.~G.~Wilson,
{\it The Renormalization Group: Critical Phenomena and the Kondo Problem},
\href{https://journals.aps.org/rmp/abstract/10.1103/RevModPhys.47.773}{Rev. Mod. Phys. \textbf{47} (1975) 773}.

\bibitem{Wilson2}
K.~G.~Wilson,
{\it The renormalization group and critical phenomena},
\href{https://journals.aps.org/rmp/abstract/10.1103/RevModPhys.55.583}{Rev. Mod. Phys. \textbf{55} (1983) 583}.

\bibitem{RGbook1}
J.~Zinn-Justin, {\it Phase Transitions and Renormalization Group}, Oxford 2013.

\bibitem{sher_PR}
M.~Sher,
{\it Electroweak Higgs Potentials and Vacuum Stability},
\href{https://www.sciencedirect.com/science/article/abs/pii/0370157389900616?via%3Dihub}{Phys. Rept. \textbf{179} (1989) 273}.

\bibitem{Higgsinst1}
G.~Degrassi, S.~Di Vita, J.~Elias-Mir\'o, J.~R.~Espinosa, G.~F.~Giudice, G.~Isidori and A.~Strumia,
{\it Higgs mass and vacuum stability in the Standard Model at NNLO},
\href{https://link.springer.com/article/10.1007/JHEP08(2012)098}{JHEP \textbf{08} (2012) 098}
[\href{https://arxiv.org/abs/1205.6497}{arXiv:1205.6497 [hep-ph]}].

\bibitem{Higgsinst2}
S.~Alekhin, A.~Djouadi and S.~Moch,
{\it The top quark and Higgs boson masses and the stability of the electroweak vacuum},
\href{https://www.sciencedirect.com/science/article/pii/S0370269312008611?via%3Dihub}{Phys. Lett. B \textbf{716} (2012) 214}
[\href{https://arxiv.org/abs/1207.0980}{arXiv:1207.0980 [hep-ph]}].

\bibitem{Higgsinst3}
D.~Buttazzo, G.~Degrassi, P.~P.~Giardino, G.~F.~Giudice, F.~Sala, A.~Salvio and A.~Strumia,
{\it Investigating the near-criticality of the Higgs boson},
\href{https://link.springer.com/article/10.1007/JHEP12(2013)089}{JHEP \textbf{12} (2013) 089}
[\href{https://arxiv.org/abs/1307.3536}{arXiv:1307.3536 [hep-ph]}].

\bibitem{LIGO}
B.~P.~Abbott \textit{et al.} [LIGO Scientific and Virgo],
{\it Observation of Gravitational Waves from a Binary Black Hole Merger},
\href{https://journals.aps.org/prl/abstract/10.1103/PhysRevLett.116.061102}{Phys. Rev. Lett. \textbf{116} (2016) 061102}
[\href{https://arxiv.org/abs/1602.03837}{arXiv:1602.03837 [gr-qc]}].

\bibitem{EHT}
K.~Akiyama \textit{et al.} [Event Horizon Telescope],
{\it First M87 Event Horizon Telescope Results. I. The Shadow of the Supermassive Black Hole},
\href{https://iopscience.iop.org/article/10.3847/2041-8213/ab0ec7}{Astrophys. J. Lett. \textbf{875} (2019) L1}
[\href{https://arxiv.org/abs/1906.11238}{arXiv:1906.11238 [astro-ph.GA]}].

\end{thebibliography}
\end{document}